\documentclass[oneside,12pt]{Classes//jiit-phd-thesis}
\usepackage{titlesec}
\usepackage[centertags]{amsmath}
\usepackage[chapter]{algorithm}
\usepackage{algorithmic}
\usepackage{titlesec}
\usepackage{longtable}
\usepackage{ulem}
\usepackage{multicol}
\usepackage{multirow}
\usepackage{amsfonts}
\usepackage{amssymb}
\usepackage{amsthm}
\usepackage{tabu}
\usepackage{mathtools}
\usepackage{graphicx}
\usepackage{tabularx}
\usepackage{multirow}
\usepackage[center]{caption}
\usepackage{rotating}
\usepackage{fancyhdr}
\usepackage{amsmath}
\usepackage{epstopdf}
\usepackage{pdflscape}
\usepackage[T1]{fontenc}
\usepackage[latin9]{inputenc}
\usepackage{geometry}
\usepackage{amssymb} 
\usepackage{physics}
\usepackage{color,graphicx}
\usepackage{pdfpages}
\usepackage{subfig}
\usepackage{float}
\usepackage{wrapfig}
\usepackage{array}
\usepackage{ulem}
\usepackage{tocloft}
\usepackage{hyphenat}
\usepackage{ragged2e}
\usepackage{blindtext}
\usepackage[nopar]{lipsum}
\usepackage{blindtext}
\usepackage{hyperref}
\usepackage{cleveref}
\usepackage{nicefrac}
\usepackage{dsfont}
\usepackage{mathrsfs}
\usepackage{booktabs}
\usepackage{longtable}
\usepackage{tabularx}
\usepackage{xspace}
\usepackage{csquotes}
\usepackage{caption}
\usepackage{afterpage}

\newcommand\blankpage{%
	\null
	\thispagestyle{empty}%
	\addtocounter{page}{-1}%
	\newpage}

\titleformat{\section}
  {\normalfont\fontsize{14}{16}\bfseries}
  {\thesection}{1em}
  {\MakeUppercase}

\titleformat{\subsection}
  {\normalfont\fontsize{12}{16}\bfseries}
  {\thesubsection}{1em}
  {\MakeUppercase}
  
  \titleformat{\subsubsection}
  {\normalfont\fontsize{12}{16}\bfseries}
  {\thesubsubsection}{1em}
  {\MakeUppercase}
  
\DeclareCaptionFont{tenpt}{\fontsize{10pt}{12pt}\selectfont}
\makeatletter
\renewcommand*\l@figure{\@dottedtocline{1}{1.5em}{2.3em}}
\addtocontents{lof}{\def\string\@dotsep{1000}}
\addtocontents{lot}{\def\string\@dotsep{1000}}
\let\l@table\l@figure
\makeatother

\Title{Design and Analysis of a Set of Discrete Variable Protocols for Secure Quantum Communication}

\Author{Arindam Dutta}
\Roll{ENROLLMENT NO. 20410009}

\SupervisorA{Prof. (Dr.) Anirban Pathak}

\Department{Department of Physics and Materials Science and Engineering }

\Institute{JAYPEE INSTITUTE OF INFORMATION TECHNOLOGY}
\InstituteTag{(Declared Deemed to be University U/S 3 of UGC Act)}
\InstituteAdd{A-10, SECTOR-62, NOIDA, INDIA}

\InstituteLogo{\includegraphics[scale=0.10]{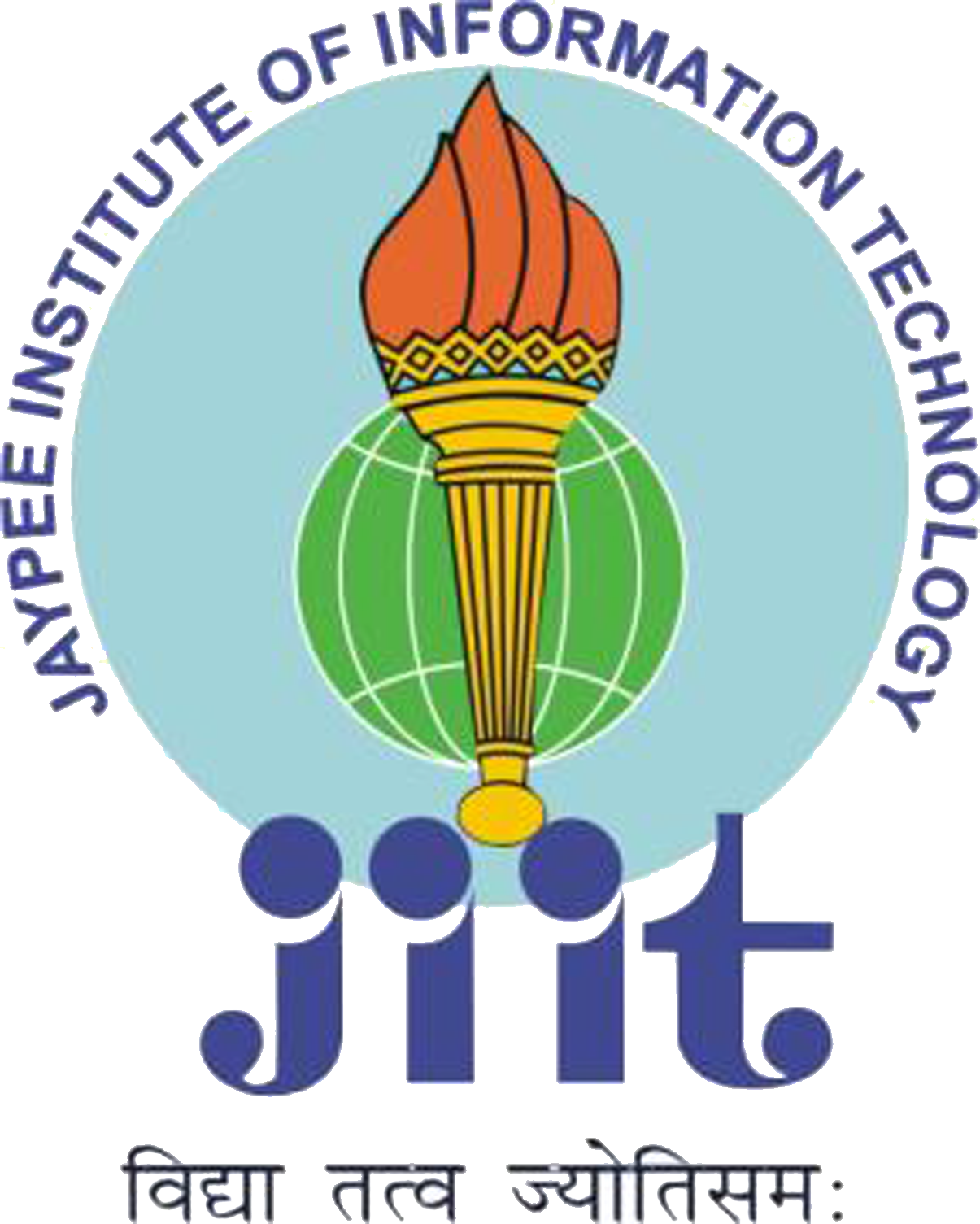}}

\Degree{DOCTOR OF PHILOSOPHY}

\StartDegreeDate{(August), (2020)}

\EndDegreeDate{March, 2025}

\begin{document}


\addtocontents{toc}{\textbf{}\hfill\textbf{\fontsize{12}{10}\selectfont Page Number}\par}

\maketitle
\addcontentsline{toc}{chapter}{\fontsize{12.5}{10}\selectfont{INNER FIRST PAGE}}
\newpage

\textbf{}\\
\textbf{}\\
\textbf{}\\
\textbf{}\\
\textbf{}\\
\textbf{}\\
\textbf{}\\
\textbf{}\\
\textbf{}\\
\textbf{}\\
\textbf{}\\
\textbf{}\\
\textbf{}\\
\textbf{}\\
\textbf{}\\
\textbf{}\\
\textbf{}\\
\textbf{}\\
\textbf{}\\
\textbf{}\\
\textbf{}\\
\textbf{}\\
\textbf{}\\
\textbf{}\\
\textbf{}\\
\textbf{}\\
 
\begin{Large} 
	\centering
	\singlespacing
	\fontsize{10}{12}\selectfont{
		{\copyright \ Copyright  JAYPEE INSTITUTE OF INFORMATION TECHNOLOGY, NOIDA}\\
		{(Declared Deemed to be University U/S 3 of UGC Act)}\\
		March, 2025\\
		ALL RIGHTS RESERVED\\
	}
\end{Large}

\newpage
\thispagestyle{empty} 
\begin{center}
    \vspace*{4cm} 
    {\large \textit{With deepest love and gratitude,}} \\[1cm]
    {\large \textit{This thesis is dedicated to the cherished memories of my grandparents,}} \\[0.3cm]
    {\large \textbf{Bimal Krishna Dutta} and \textbf{Gouri Bala Dutta},} \\[0.5cm]
    
    {\large \textit{and my mother,}} \\[0.3cm]
    {\large \textbf{Sulata Dutta}.}
    
    \vspace*{\fill}
\end{center}


\afterpage{\blankpage}
\newpage
\cleardoublepage




\addcontentsline{toc}{chapter}{\fontsize{12}{10}\selectfont{TABLE OF CONTENTS}}
 \renewcommand{\contentsname}{\hfill Table of Contents\hfill}
\renewcommand{\cfttoctitlefont}{\fontsize{16}{10}\bfseries\MakeUppercase}
 \tableofcontents

\newpage

\addcontentsline{toc}{chapter}{\fontsize{12}{10}\selectfont{DECLARATION BY SCHOLAR}}

\begin{declaration}

\doublespacing
I hereby declare that the work reported in the Ph.D. thesis entitled {\fontsize{14}{10}{\bfseries{{``\TITLE''}}}} submitted at  \ \fontsize{14}{10}{\bfseries{Jaypee Institute of Information Technology, Noida, India,}} is an authentic record of my work carried out under the supervision of \fontsize{14}{10}{\bfseries{{\SUPERVISORA}}}. I have not submitted this work elsewhere for any other degree or diploma. I am fully responsible for the contents of my Ph.D. thesis. 
\\
\\
\\
\fontsize{14}{16}{{{(Signature of the Scholar)}}}\\
\fontsize{14}{16}{{{(\AUTHOR)}}}\\
\fontsize{14}{10}{{{Department of Physics and Materials Science and Engineering}}}\\
\fontsize{14}{16}{{{Jaypee Institute of Information Technology, Noida, India}}}\\
\fontsize{14}{16}{{{Date: \hspace{1cm}June, 2025}}}
\end{declaration} 


\afterpage{\blankpage}
\newpage
\cleardoublepage

\newpage


\addcontentsline{toc}{chapter}{\fontsize{12}{10}\selectfont{SUPERVISOR'S CERTIFICATE}}


\begin{Supdeclaration}
\doublespacing
This is to certify that the work reported in the Ph.D. thesis entitled {\fontsize{14}{10}{\bfseries{{``\TITLE''}}}}, submitted by \fontsize{14}{10}{\bfseries{{\AUTHOR}}} at \fontsize{14}{10}{\bfseries{{Jaypee Institute of Information Technology, Noida, India}}}, is a bonafide record of his original work carried out under my supervision. This work has not been submitted elsewhere for any other degree or diploma.
\newline
\\
\\
\\
\\
\\
\fontsize{14}{16}{{{(Signature of Supervisor)}}}\\
\fontsize{14}{16}{{{(\SUPERVISORA)}}}\\
\fontsize{14}{16}{{{Professor and HOD}}}\\
\fontsize{14}{16}{{{Department of Physics and Materials Science and Engineering}}}\\
\fontsize{14}{16}{{{Jaypee Institute of Information Technology, Noida, India}}}\\
\fontsize{14}{16}{{{Date: \hspace{1cm}June, 2025}}}
\end{Supdeclaration} 

\newpage


\afterpage{\blankpage}
\newpage
\cleardoublepage

\addcontentsline{toc}{chapter}{\fontsize{12}{10}\selectfont{ACKNOWLEDGEMENT}}


\begin{acknowledgements}

Throughout my thesis journey, I have received invaluable support and
guidance from many individuals, and I would like to take this opportunity
to express my heartfelt gratitude to them.

First and foremost, I extend my deepest appreciation to my supervisor,
Prof. (Dr.) Anirban Pathak, for giving me the opportunity to conduct
research under his guidance. His mentorship has been instrumental
not only in shaping my research but also in teaching me how science
should be learned and effectively communicated. His unwavering focus
and dedication have been a source of inspiration, motivating me to
approach my work with diligence and enthusiasm. One of his memorable
Bengali phrases, ``Sob korte hobe'' (which translates to ``Everything has to be done together''), has encouraged me to multitask and persevere
through challenges. I sincerely hope that his guidance will continue
to inspire me beyond these years.

In addition, I extend my heartfelt thanks to Prof. (Dr.) Subhashish
Banerjee, Dr. V. Narayanan, Dr. R. Srikanth, Dr. Anindita Banerjee,
and Dr. Sandeep Mishra for their insightful discussions and valuable
suggestions that helped shape my work. I am also deeply grateful to
my senior group members, Dr. Kishore Thapliyal and Dr. Priya Malpani,
for their continuous support and motivation throughout this journey.
My sincere appreciation goes to my colleague and friend, Mr. Satish
Kumar. Furthermore, I would like to thank Dr. Sukhamoy Bhattacharyya
(Sukhamoy Sir), Surajit Sir, Sumit Da, and Didhiti for their encouragement
and for imparting valuable life lessons during this period. I extend
my sincere gratitude to my DPMAC members, Prof. (Dr.) Bhagwati Prasad
Chamola and Dr. Anuraj Panwar, for their continuous guidance, constructive
feedback, and keen observation of my work, which have been instrumental
in my academic growth. I am also grateful to all the faculty members
and departmental staff for fostering a supportive and enriching research
environment. A special note of appreciation goes to the housekeeping
staff, whose dedication ensures the smooth functioning of the department.
I firmly believe that knowledge should be freely accessible to all,
and scientific journals should be open to readers worldwide. In this
regard, I sincerely acknowledge Sci-Hub and arXiv for its open-access initiative,
without which the completion of this thesis would not have been possible.

I express my heartfelt gratitude to the memory of my beloved mother
(Maa), Mrs. Sulata Dutta, whose unwavering desire for my progress
in education and science continues to inspire me. Her absence and
cherished memories shape my growth in various aspects of life. I also
extend my sincere thanks to my father (Baba), Mr. Asit Baran Dutta,
for his constant support and motivation at every stage of my life.
His encouragement has been instrumental in shaping my journey in research
and beyond. Words are insufficient to express my gratitude to my best
friend, Monika. Her incredible and enthusiastic support, from the
very beginning of my research to the completion of this thesis, has
been invaluable. I sincerely hope that not only her support but also
her presence remains a lifelong companion to both me and my work.
I would also like to extend my heartfelt appreciation to my dear friend
and cousin, Mr. Avishek Audhya (Damu Da). Our conversations, often
unpredictable in direction, always bring a sense of healing and inner
peace, for which I am truly grateful.

Lastly, I find it difficult to determine whether it is appropriate
to express gratitude to Nature---the all-encompassing force of the
Universe---but I feel compelled to acknowledge its ever-changing
attributes. Nature continuously transforms forms and energies, shaping
existence through cycles of destruction and creation. If destruction
can be perceived as ``death'' and creation as ``birth'', then
I, too, am a manifestation of Nature's constant transformation. In
essence, Nature evolves within my consciousness just as I change within
it. I would find true peace and fulfillment in wholeheartedly dedicating myself to the wisdom and harmony of Nature.

\begin{flushright}
    (Arindam Dutta)
\end{flushright}

\end{acknowledgements}

\newpage 

\addcontentsline{toc}{chapter}{\fontsize{12}{10}\selectfont ABSTRACT}


\begin{abstract}
The advent of quantum key distribution (QKD) has revolutionized secure
communication by providing unconditional security, unlike classical
cryptographic methods. However, its effectiveness relies on robust
identity authentication, as vulnerabilities in the authentication process can
cause a compromise with the security of the entire communication system. Over the past three decades,
numerous quantum identity authentication (QIA) protocols have been
proposed. This thesis first presents a chronological review of these protocols,
categorizing them based on quantum resources and computational tasks {involved}
while analyzing their strengths and limitations. Subsequently, by recognizing inherent
symmetries present in the existing protocols, we design novel QIA schemes based
on secure computational and communication tasks. Specifically, this work introduces
a set of new QIA protocols that utilize controlled secure direct quantum communication. The proposed scheme facilitates
mutual authentication between two users, Alice and Bob, with assistance
from a third party, Charlie, using Bell states. A comprehensive security
analysis demonstrates its robustness against impersonation, intercept-resend,
and fraudulent authentication attacks. The comparative evaluation highlights
its advantages over existing schemes. Additionally, this thesis presents
two novel QKD protocols that eliminate the need for entanglement or
ideal single-photon sources, making them feasible with commercially
available photon sources. These protocols are rigorously proven to
be secure against various attacks, including intercept-resend and
certain collective attacks. Key rate bounds are established, demonstrating
that specific classical pre-processing enhances the tolerable error
threshold. A trade-off between quantum resource utilization and information
leakage to an eavesdropper (Eve) is identified, with greater efficiency
achieved through increased use of quantum resources. Notably, these protocols
outperform the SARG04 protocol in efficiency, albeit at the cost of
additional quantum resources. Furthermore, their critical distances
under photon number splitting (PNS) attacks surpass those for BB84
and SARG04 protocols under similar conditions. Moreover, this thesis introduces
a controlled quantum key agreement protocol and another key agreement
scheme with a strong focus on security. Rigorous security analysis
establishes their resilience against impersonation and collective
attacks while ensuring fairness and correctness. Comparative analysis
reveals that, unlike existing schemes, the proposed approach does
not require quantum memory. Additionally, the controlled quantum key
agreement protocol leverages Bell and single-photon states, which
are easier to generate and maintain than the GHZ states used in Tang
et al.'s protocol, offering a more practical and efficient alternative.
Furthermore, Nash equilibrium is employed to establish a game-theoretic
security bound on the quantum bit error rate (QBER) for the DL04 protocol, which is a protocol for quantum secure direct communication that has been experimentally realized recently. The sender, receiver, and eavesdropper are modeled as quantum
players, with Eve capable of executing quantum attacks (e.g., W{\'o}jcik\textquoteright s
original and symmetrized attacks, and Pavi{\v{c}}i{\'c}\textquoteright s
attack) and classical intercept-resend attacks. A game-theoretic security
analysis reveals the absence of a Pareto optimal Nash equilibrium,
leading to the identification of mixed-strategy Nash equilibrium points
that define upper and lower bounds for QBER. Additionally, the DL04
protocol\textquoteright s vulnerability to Pavi{\v{c}}i{\'c}\textquoteright s
attack in message mode is established, and it is observed that quantum
attacks are more effective than classical ones, as they result in
lower QBER values and a reduced probability of Eve\textquoteright s
detection. Overall, this thesis contributes to the advancement of
quantum cryptographic protocols by addressing some of the existing issues with identity authentication, key
distribution, key agreement, and security analysis through novel schemes and rigorous
theoretical frameworks.

\end{abstract}

\newpage


\addcontentsline{toc}{chapter}{\fontsize{12}{10}\selectfont LIST OF ACRONYMS \& ABBREVIATIONS}


\begin{nomenclature}

\fontsize{12}{10}
\doublespacing 
 \begin{tabbing}

 AAAAAAA \= AAAAAAAAAAAAA \kill

{	AD}	\>	\quad	\quad\quad	Amplitude Damping 	\\
{	BQC}	\>	\quad	\quad\quad	Blind Quantum Computing	\\
{	CA}	\>	\quad	\quad\quad	Certification Authority	\\
{	CDSQC}	\>	\quad	\quad\quad	Controlled Deterministic Secure Quantum Communication	\\
{	CNOT}	\>	\quad	\quad\quad	Controlled-NOT	\\
{	CPBS}	\>	\quad	\quad\quad	Controlled Polarizing Beam Splitter 	\\
{	CQKA}	\>	\quad	\quad\quad	Controlled Quantum Key Agreement 	\\
{	CQSDC}	\>	\quad	\quad\quad	Controlled Quantum Secure Direct Communication	\\	
{	CV}	\>	\quad	\quad\quad	Continuous Variable	\\
{	DBS}	\>	\quad	\quad\quad	Dichroic Beam Splitter 	\\
{	DFS}	\>	\quad	\quad\quad	Decoherence-Free Subspace 	\\
{	DH}    	\>	\quad	\quad\quad	Diffie-Hellman\\
{	DSQC}	\>	\quad	\quad\quad	Deterministic Secure Quantum Communication	\\
{	DV}	\>	\quad	\quad\quad	Discrete Variable	\\
{	EPR}       \>	\quad	\quad\quad	Einstein-Podolsky-Rosen	\\
{	GHZ}	\>	\quad	\quad\quad	Greenberger-Horne-Zeilinger	\\
{	GV}	\>	\quad	\quad\quad	Goldenberg Vaidman	\\
{	HWP}	\>	\quad	\quad\quad	Half-Wave Plate 	\\
{	IoT}	\>	\quad	\quad\quad	Internet of Things	\\
{	IR}	\>	\quad	\quad\quad	Intercept Resend 	\\
{	IRUD}	\>	\quad	\quad\quad	Intercept-Resend Unambiguous Discrimination  	\\
{	KA}	\>	\quad	\quad\quad	Key Agreement 	\\
{	MAC}	\>	\quad	\quad\quad	Message Authentication Code	\\
{	MDI}	\>	\quad	\quad\quad	Measurement Device Independent  	\\
{	MQKA}	\>	\quad	\quad\quad	Multiparty Quantum Key Agreement 	\\
{	OT}	\>	\quad	\quad\quad	Oblivious Transfer	\\
{	PBS}	\>	\quad	\quad\quad	Polarizing Beam Splitter 	\\
{	PD}	\>	\quad	\quad\quad	Phase Damping 	\\
{	PNS}	\>	\quad	\quad\quad	Photon Number Splitting 	\\
{	POVM}	\>	\quad	\quad\quad	Positive Operator-Valued Measure 	\\
{	PUF}	\>	\quad	\quad\quad	Physical Unclonable Function	\\
{	QBER}   	\>	\quad	\quad\quad	Quantum Bit Error Rate\\
{	QD}	\>	\quad	\quad\quad	Quantum Dialogue	\\
{	QECC}	\>	\quad	\quad\quad	Quantum Error-Correcting Code	\\
{	QFT}	\>	\quad	\quad\quad	Quantum Fourier Transform 	\\
{	QIA}   	\>	\quad	\quad\quad	Quantum Identity Authentication\\
{	QKA}	\>	\quad	\quad\quad	Quantum Key Agreement	\\
{	QKD}	\>	\quad	\quad\quad	Quantum Key Distributione	\\
{	QND}	\>	\quad	\quad\quad	Quantum Non-Demolition 	\\
{	QSDC}   	\>	\quad	\quad\quad	Quantum Secure Direct Communication\\
{	QSS}   	\>	\quad	\quad\quad	Quantum Secret Sharing\\
{	RSA}	        \>	\quad	\quad\quad	Rivest-Shamir-Adleman \\		
{	SFG}	\>	\quad	\quad\quad	Sum-Frequency Generation 	\\
{	SPDC}	\>	\quad	\quad\quad	Spontaneous Parametric Down-Conversion 	\\
{	TFQKD}       \>	\quad	\quad\quad	Twin-Field Quantum Key Distribution	\\
{	WCP}	\>	\quad	\quad\quad	Weak Coherent Pulse	\\

\end{tabbing}

\end{nomenclature} 
\newpage


\addcontentsline{toc}{chapter}{\fontsize{12}{10}\selectfont LIST OF FIGURES}
\section*{\hfill LIST OF FIGURES\hfill}
\begin{tabular}{p{1.5cm}p{12cm}p{1.5cm}}
	\textbf{Figure Number} &\centering \textbf{Caption} & \textbf{Page Number}\\
	1.1 & \centering Eve may attempt to mimic Alice (or Bob) while interacting with Bob (or Alice)
by substituting the direct communication channel between Alice and Bob with
two separate channels: one connecting Alice to Eve and the other linking Eve
to Bob. & 17\\
	1.2 & \centering Categorization of QIA schemes. & 41\\
	2.1 & \centering The graph presents the correlation between the probability of detecting
Eve\textquoteright s presence, $P(n)$, and the number of classical
bits used, $n$. & 59\\
	2.2 & \centering Flowchart depicting the operational workflow of the proposed QIA
protocol. & 70\\
	2.3 & \centering The probability $P\left(n\right)$ of detecting Eve's presence is
correlated with the number of classical authentication keys $n$ that
are pre-shared. & 72\\
	2.4 & \centering The IR attack strategy utilized by Eve is structured such that qubits
labeled 1, 2, 3, and 4 correspond to sequences $S_{A1}$, $S_{A2}$,
$S_{B3}$ and $S_{B4}$, respectively. & 74\\
	2.5 & \centering The collective error probability associated with the noise parameter
can be classified into: (a) Collective dephasing error probability,
characterized by the noise parameter $\phi$, and (b) Collective rotation
error probability, defined by the noise parameter $\theta$. & 78\\
	2.6 & \centering A flowchart visually represents the operational framework of the proposed QIA
protocol. & 89\\
	2.7 & \centering The probability of detecting Eve\textquoteright s interference denoted
as $P(n)$, is correlated with the number of classical bits, $n$,
utilized as the pre-shared authentication key. & 91\\
	2.8 & \centering Eve employs an intercept-and-resend strategy. In this context, the
qubits labeled as 1, 2, 3, 4, and 5 correspond to the sequences $S_{1}$,
$S_{2}$, $S_{3}$, $S_{4}$, and $S_{5}$, respectively. & 93\\

\end{tabular}
\newpage

\begin{tabular}{p{1.5cm}p{12cm}p{1.5cm}}

	3.1 & \centering The secret key rate is plotted as a function of the quantum bit error
rate ($\mathcal{E}$): $(a)$ a plot assessing the maximum tolerable
error threshold (security limit) for sequence $S_{B1}$, where the
solid (black) line and dashed (blue) line represent scenarios with
and without the incorporation of the new variable $\mathcal{Y}$,
respectively; (b) a plot evaluating the maximum tolerable error threshold
(security limit) for sequence $S_{B2}$ without introducing the new
variable $\mathcal{X}$; and $(c)$ a plot assessing the maximum tolerable
error threshold (security limit) for sequence $S_{B2}$ with the inclusion
of the new variable $\mathcal{X}$. & 117\\
	3.2 & \centering The variation of the secret key rate with respect to bit-flip probability
($q$) and QBER ($\mathcal{E}$) is illustrated: $(a)$ the lower
bound on the secret key rate as a function of bit-flip probability
and QBER, $(b)$ a contour plot depicting the lower bound error limit,
correlating QBER with bit-flip probability, $(c)$ the upper bound
on the secret key rate as a function of bit-flip probability and QBER,
and $(d)$ a contour plot illustrating the upper bound error limit
with respect to QBER and bit-flip probability. & 124\\
	3.3 & \centering The variation in Eve's information as a function of distance is analyzed
to determine the critical distance ($l_{c}$): $(a)$ Eve's information
plotted against distance to estimate the critical threshold where
the attacker extracts maximum key information via a PNS attack on
Protocol 3.1, and $(b)$ Eve's information plotted against distance
to estimate the critical threshold where the attacker gains maximum
key information using an IRUD attack on Protocol 3.2. & 128\\
	4.1 & \centering The flowchart illustrates the operational framework of the proposed
CQKA protocol. & 139\\ 
	4.2 & \centering The correlation between Eve\textquoteright s detection probability
$({\rm d})$ and the parameter $\alpha$, which characterizes non-orthogonality. & 146\\
	4.3 & \centering The dependence of Eve\textquoteright s accessible information $({\rm I})$
on the angle ${\rm \alpha}$, representing non-orthogonality. & 148\\

\end{tabular}
\newpage

\begin{tabular}{p{1.5cm}p{12cm}p{1.5cm}}
	4.4 & \centering The interrelation between ${\rm Pr,d}$ and ${\rm n}$ is depicted
as follows. (a) The dependence of ${\rm Pr}$ on ${\rm n}$ is examined
for varying values of ${\rm d}$. In particular, the ``circle''
line corresponds to ${\rm d=0\%}$, the ``continuous'' line represents
${\rm d=12.5\%}$, the ``plus sign'' line denotes ${\rm d=25\%}$,
and the ``asterisk'' line signifies ${\rm d=50\%}$. (b) A histogram
illustrates Eve's probability of success for ${\rm n=6}$ across different
values of ${\rm d}$. & 149\\
	4.5 & \centering The plot illustrates the relationship between tolerable QBER ($\epsilon$)
and the angle characterizing non-orthogonality $({\rm \alpha})$. & 153\\
	4.6 & \centering A sketch to visualize the QB scheme of Yan Yu et alThe variation of average fidelity under a noisy quantum channel is
analyzed. (a) Depicts the dependence of average fidelity on the channel
parameter $\eta_{j}$ for the CQKA scheme within the Markovian framework
for AD and PD channels, as described by Equations (\ref{eq:Amplitude_Damping})
and (\ref{eq:Phase_Damping}). (b) and (c) illustrate the variation
in average fidelity for NMDPH and NMDPO channels within the non-Markovian
regime, plotted against the time-like parameter $p$ for various values
of $\alpha$, respectively (refer to Equations (\ref{eq:DePhasing})
and (\ref{eq:DePolarizing})). & 155\\
	4.7 & \centering The optical design of the CQKA scheme integrates entangled photons,
a CNOT operation, and a full Bell state measurement. The entangled
photon pairs (Bell states) are generated using two thick $\beta$-barium
borate (BBO) nonlinear crystals arranged in a cross-configuration
\cite{PPH+22}. Bell state measurement is performed via nonlinear
interactions, specifically sum-frequency generation (SFG) of type-I
and type-II. A dichroic beam splitter (DBS) is incorporated, along
with two 45$^{\circ}$ projectors (PP1 and PP2) \cite{SKP21}. The system further
utilizes polarizing beam splitter(s) (${\rm PBS}_{1}$ and ${\rm PBS}_{2}$),
each configured at a 45$^{\circ}$ orientation with half-wave plate (HWP) integrated
at the input and output \cite{ZZC+05}. & 160\\
	5.1 & \centering The circuit representation of DL04 quantum game. & 174\\
\end{tabular}

\begin{tabular}{p{1.5cm}p{12cm}p{1.5cm}}
	5.2 & \centering The figure illustrates the Nash equilibrium points where the optimal
response functions of the three participants intersect in our mixed-strategy
game framework. The response functions are represented as follows:
the low-density layer for Alice, the medium-density layer for Bob,
and the high-density layer for Eve. The subfigures depict different
game scenarios: (a) $E_{1}$-$E_{2}$, (b) $E_{1}$-$E_{3}$, (c)
$E_{2}$-$E_{3}$, and (d) $E_{1}$-$E_{4}$. & 206 \\
	
\end{tabular}

\newpage



\addcontentsline{toc}{chapter}{\fontsize{12}{10}\selectfont LIST OF TABLES}
\section*{\hfill LIST OF TABLES\hfill}
\begin{tabular}{p{1.5cm}p{12cm}p{1.5cm}}
	\textbf{Table Number} &\centering \textbf{Caption} & \textbf{Page Number}\\
	1.1 & \centering The classification of existing schemes for QIA is presented. When
the use of a pre-shared key is referred to as ``By QKD'', as indicated
by the authors of the respective studies, it can be interpreted that
a classical identity sequence was initially exchanged as a prerequisite
for performing QKD for the first time. The following abbreviations
are employed in this table: C=classical, B=Bell state, FC=five-particle
cluster state, CS=classical identity sequence, N=no, HF=single one-way
hash function, QPKC=quantum public key cryptography, Q=quantum, T=trusted,
SP=single photon, ST=semi-trusted, Y=yes, UT=un-trusted. & 31\\
	2.1 & \centering The preparation method for the decoy sequence (Protocol 2.1). & 54\\
	2.2 & \centering The preparation method of the authentication sequence (Protocol 2.1). & 55\\
	2.3 & \centering Bob employs a predetermined measurement basis to extract results
from the authentication sequence (Protocol 2.1). & 55\\
	2.4 & \centering The preparation method of the authentication sequence (Protocol 2.2). & 56\\
	2.5 & \centering The measurement basis employed by Bob to extract results for the
authentication sequence (Protocol 2.2). & 56\\
	2.6 & \centering This outlines the possible measurement results for the legitimate
participants within the proposed QIA scheme. & 66\\
	2.7 & \centering A comprehensive comparison of Protocol 2.3 is conducted with various
existing QIA protocols. The column labeled ``Minimum secret key required''
presents multiple values corresponding to different protocols discussed
within the same reference. The following abbreviations are employed
in this table: TC=two-way channel, IR=intercept-resend, IF=impersonated
fraudulent, EM=entangled measure. & 82\\
	2.8 & \centering This outlines the potential measurement results observed by all participants
within Protocol 2.4 for QIA. & 84\\
	2.9 & \centering Possible results of measurements. & 91\\
\end{tabular}

\newpage

\begin{tabular}{p{1.5cm}p{12cm}p{1.5cm}}
	2.10 & \centering Detailed comparison of Protocol 2.4 with previous protocols. & 96\\
	3.1 & \centering This table outlines the encoding and decoding principles for both
Protocol 3.1 and Protocol 3.2, while also presenting the measurement
results obtained after the classical sifting subprotocol. & 102\\
	3.2 & \centering Table mapping between measurement outcomes and Alice's determined
results in Protocol 3.1. & 105\\
	3.3 & \centering Table mapping between measurement results and Alice's determined
outcomes in Protocol 3.2. & 107\\
	4.1 & \centering The table below illustrates the relationship between the announcements
made by Charlie, Alice, and Bob, and the final key. & 137\\
	4.2 & \centering The table below illustrates the relationship between the announcements
made by Alice, Bob, and the final key. & 140\\
	4.3 & \centering Explicit comparison with earlier protocols. Y - required, N - not
required, QC - quantum channel, QR - quantum resources, TR - third
party, QM - quantum memory, NoP - number of parties, QE - quantum
efficiency. & 159\\
	5.1 & \centering The payoff matrix corresponding to the $\mathcal{M}$ game is presented,
where the three values within parentheses signify the payoffs of Alice,
Bob, and Eve, respectively. Here, the probabilities assigned to an
eavesdropper Eve's choice of attack strategies $E_{i}$ and $E_{j}$
are denoted by $r$ and $1-r$, respectively. & 176\\
	5.2 & \centering The table presents Nash equilibrium points for different strategic
interactions and provides qualitative payoff assessments for all involved
entities. & 209\\
	
\end{tabular}

\newpage


\setcounter{secnumdepth}{3}
\setcounter{tocdepth}{3}
\frontmatter 
\pagenumbering{roman}


\mainmatter 



\chapter{INTRODUCTION}
\label{Ch1:Chapter1_Introduction}
\graphicspath{{Chapter1/Chapter1Figs/}{Chapter1/Chapter1Figs/}}


\section{Introductory overview}\label{secChapter1_Sec1}

Quantum physics, quantum mechanics, and quantum technologies constitute
some of the most transformative advancements in modern science. Collectively,
they have redefined our understanding of the natural world and driven
significant technological innovations. This section provides a concise
historical overview of these fields, highlighting key discoveries,
theoretical advancements, and the development of technological applications.
Before delving into these topics to establish the background of this
thesis, it is important to note that quantum technologies have been
in existence for several decades. However, the quantum principles that explain early technologies such as lasers and transistors
are relatively intuitive. In contrast, recent advancements leverage
more counterintuitive---often described as ``weird''---quantum
phenomena. For instance, the measurement-induced wavefunction collapse
has been utilized in the development of quantum random number generators.
Technologies based on such phenomena are often classified under the
so-called second quantum revolution. This new generation of quantum
technologies can be broadly categorized into three domains: quantum
sensing, quantum computing, and quantum communication. The present
thesis focuses on secure quantum communication, an area of particular
interest for multiple reasons. First, quantum cryptography offers
unconditional security, a feature unattainable by classical cryptographic
methods. Second, the field has witnessed significant advancements
in recent years. Notably, China and Austria have demonstrated satellite-assisted
secure quantum communication over a distance of 7,600 km \cite{LCH+18}.
Additionally, Liu et al. have implemented twin-field quantum key distribution
(TFQKD) over a 1,000 km fiber link \cite{LZJ+23}, and several commercial
quantum cryptography solutions have emerged \cite{ZTA+22,SR23,LSH+22}.
These developments will be discussed in detail in the subsequent sections
of this chapter and throughout this thesis. However, this section
provides a brief overview of key aspects and advancements in quantum
communication to establish the focus and context for the following
discussions.

\subsection{The birth of quantum physics}

By the late 19th century, physics was arguably composed of three core
areas: classical mechanics, electromagnetism, and thermodynamics.
Classical mechanics provided the tools to predict the motion of material
bodies, while Maxwell's theory of electromagnetism served as the foundation
for studying radiation. Matter and radiation were understood in terms
of particles and waves, respectively. Interactions between matter
and radiation were effectively explained using the Lorentz force or
thermodynamic principles. It appeared that all known physical phenomena
could be accounted for within this framework of matter and radiation. However, the beginning of the 20th century brought significant challenges to traditional physics in two key areas: the ``relativistic'' and
``microscopic'' domains. In the relativistic domain, Einstein's
1905 work on the special theory of relativity demonstrated that Newtonian
mechanics breaks down at velocities approaching the speed of light.
In the microscopic domain, advances in experimental techniques that
allowed investigation of atomic and subatomic structures revealed
that classical physics was inadequate for explaining newly observed
phenomena. This inadequacy made it clear that classical physics was
not valid at the microscopic scale, necessitating the development
of new concepts to describe atomic and molecular structures and their
interactions with light. The inability of classical physics to explain
phenomena such as blackbody radiation, the photoelectric effect, atomic
stability, and atomic spectra highlighted its limitations and paved
the way for the emergence of new theories beyond its scope.

The foundations of quantum physics were laid with Max Planck's pioneering
concept of energy quantization, introduced in 1901 to resolve the
``ultraviolet catastrophe'' in black-body radiation \cite{planck1901law}.
Planck suggested that energy transferred from oscillators in a cavity
wall to the radiation field occurs in discrete packets or quanta
($\Delta E=h\nu$, where $h$ is Plank's constant). Albert Einstein
extended this idea in 1905 to explain the photoelectric effect \cite{einstein1905erzeugung},
proposing that energy is not only transferred in quantized packets but
also traverses in this discrete form (for a detailed overview, see
\cite{PG18}). Niels Bohr's 1913 model of the atom introduced quantized
orbits for electrons, explaining atomic spectra. While the model had
limitations, it incorporated Planck's quantum ideas into atomic structure
\cite{bohr1913constitution}. In 1924, Louis de Broglie extended
``wave-particle duality'' to matter, proposing that particles such
as electrons have an associated wavelength ($\lambda=\frac{h}{p}$,
where $p$ is momentum). This hypothesis was later confirmed experimentally
by electron diffraction \cite{de1924recherches}. Quantum mechanics
took mathematical shape with Erwin Schr{\"o}dinger's wave equation
(1926) and Werner Heisenberg's matrix mechanics (1925). Schr{\"o}dinger's
approach modeled particles as wavefunctions, while Heisenberg's formalism
relied on observable quantities \cite{schrodinger1926quantisierung,heisenberg1925quantum}.
In 1927, Heisenberg formulated the uncertainty principle, stating
that certain pairs of observables, like position and momentum, cannot
be simultaneously measured with arbitrary precision. This principle
underscored the probabilistic nature of quantum mechanics \cite{heisenberg1927physical}.
This foundation paved the way for major developments, such as the
Bohr model, Einstein's work on stimulated emission, de Broglie's matter
waves, Heisenberg's uncertainty principle, Dirac's theory of radiation,
matrix mechanics, and Schr{\"o}dinger equation; each transforming
our understanding of the microscopic world and establishing quantum
mechanics as an advanced model of natural phenomena. 

Quantum mechanics introduced numerous counterintuitive concepts that
were initially met with skepticism, even by some of its founding figures.
For instance, while Einstein's work on the photoelectric effect \cite{einstein1905erzeugung}
was instrumental in the development of quantum mechanics, he remained
critical of the Copenhagen interpretation. Notably, his objections,
as articulated in the Einstein-Podolsky-Rosen (EPR) paper \cite{EPR1935},
contributed significantly to the evolution of the field and the advancement
of future technologies central to this thesis. It is worth emphasizing
that initially, several foundational principles of quantum mechanics---such
as nonlocality and various no-go theorems (including Heisenberg's
uncertainty principle, the no-cloning theorem, and the no-broadcasting
theorem)---were perceived as intrinsic limitations of the theory.
However, over the past four decades, it has become evident that these
so-called limitations can be harnessed to develop quantum technologies
that either surpasses classical efficiency and performance constraints
or enable entirely novel paradigms with no classical analog. Given
that schemes for unconditionally secure quantum communication are
central to this thesis, the following section provides a brief introduction
to quantum technologies, highlighting the key motivations behind the
research presented herein.

\subsection{Brief introduction quantum technologies and motivation of this thesis}

From foundational experiments to revolutionary technologies, quantum
physics and mechanics have profoundly shaped our understanding of
the universe and practical capabilities. As the second quantum revolution
accelerates, quantum technologies hold promise for transformative
applications across computing, communication, and sensing, ensuring
their place at the forefront of scientific and technological innovation.
With recent advancements in quantum computation, communication, and
foundational studies, the inherent limitations of quantum mechanics---long
regarded as constraints---are increasingly recognized as valuable
resources for diverse applications. These limitations, first connected
through seminal no-go theorems such as Heisenberg's uncertainty principle,
and later expanded to include concepts like the no-cloning, no-broadcasting,
and no-deletion theorems are central to the unique capabilities of
quantum systems. For example, Heisenberg's uncertainty principle and
the no-cloning theorem plays a key role in quantum key distribution
(QKD). The uncertainty principle asserts that measurements of non-commuting
operators (e.g., position and momentum) cannot achieve arbitrary precision
simultaneously, while the no-cloning theorem prohibits the creation
of identical copies of an arbitrary unknown quantum state. Together,
these principles underpin the security of QKD protocols based on conjugate
coding, such as the BB84 and B92 protocols \cite{BB84,B92}. These
protocols leverage quantum mechanics to ensure that any eavesdropping
attempt introduces detectable disturbances. In addition, to secure
communication, other important quantum properties like entanglement
and nonlocality enable groundbreaking tasks such as quantum teleportation
\cite{BBC+93}, dense coding \cite{BW92}, and measurement device-independent
QKD \cite{YLW+22,XLW+22}. These capabilities transcend classical
limitations, opening new frontiers in information transfer and processing.
Moreover, the principles underlying quantum mechanics are crucial
to quantum computing. Quantum entanglement and superposition empower
algorithms, such as Shor's algorithm and Grover's search, to outperform
their classical counterparts in specific computational tasks. These
advancements are not confined to computing and communication; they
also extend to quantum sensing, where quantum-enhanced precision measurements
redefine the limits of sensitivity and accuracy. In summary, the unique,
nonclassical properties of quantum systems, along with the constraints
defined by quantum no-go theorems, have become indispensable tools
across three major domains: (i) quantum communication, (ii) quantum
computing, and (iii) quantum sensing. These advancements underscore
the duality of quantum mechanics, where what were once considered
limitations now serve as the foundation for transformative technologies.

Cryptography has been a critical tool for protecting information since
the beginning of human civilization. Historically, cryptographic methods
were developed to disguise secret information, yet cryptanalysts frequently
devised more advanced techniques to reveal these secrets. A major
advancement occurred in the 1970s with the development of public key
cryptography, exemplified by the Rivest-Shamir-Adleman (RSA) scheme
\cite{RSA78} and the Diffie-Hellman (DH) scheme \cite{DH76}. The
security of these, along with other classical key distribution schemes,
rests on the computational difficulty of the problems they employ.
For example, the RSA scheme's security relies on the complexity of
factoring a large bi-prime number, while DH security is grounded in
the discrete logarithm problem \cite{P13}. In a groundbreaking development
in 1994, Peter W. Shor demonstrated that quantum computers could efficiently
solve both problems in polynomial time \cite{S94}. This discovery revealed that many classical cryptographic methods would be vulnerable if scalable quantum computing became feasible, presenting a profound challenge to conventional cryptography. QKD offers a promising solution to this issue, as it secures key distribution using the principles
of quantum mechanics rather than computational complexity. The first QKD protocol (BB84) was proposed a decade before Shor's
findings highlighted the vulnerability of classical cryptography \cite{BB84}.
This protocol leverages fundamental quantum properties, such as the
no-cloning theorem \cite{WZ82}, the measurement postulate, and Heisenberg's
uncertainty principle, to ensure security, allowing it to be implemented
with polarization-encoded single photons or other forms of photonic
qubits. In an ideal setting, any eavesdropping attempt on a QKD channel
leaves detectable traces. Beyond QKD, the principles of quantum mechanics
enable other secure communication methods, including quantum key agreement
(QKA), quantum digital signatures, quantum secure direct communication
(QSDC), quantum secret sharing (QSS), and so on. Moreover, before
initiating any secure communication, it is essential to authenticate
the identities of legitimate participants, a process called identity
authentication. When quantum techniques provide unconditional security
in this verification, the process is termed quantum identity authentication
(QIA). This expanding suite of quantum-based protocols thus addresses
a variety of cryptographic needs, leveraging the unique properties
of quantum physics to offer security resilient to quantum computational
threats.

The security of communication critically depends on the security of
the key used. Quantum technology enables the distribution of keys
with unconditional security through QKD. However, initiating a QKD
protocol requires verifying the authenticity of the communicating
parties. In many cases, classical methods are used for authentication,
which is vulnerable to attacks where an eavesdropper (Eve) could
impersonate a legitimate party, creating a significant loophole in
achieving unconditional security. To address this issue, we analyze
existing QIA schemes and propose new QIA protocols to enhance security.
These QIA schemes utilize various quantum resources, making them feasible
for implementation with current technology. Once legitimate user identities
are verified, quantum communication protocols like QKD can be performed
to distribute symmetric keys. Our analysis of previous QKD schemes
reveals that the process involves splitting information into quantum
and classical components. We hypothesize that increasing the quantum
component while reducing the classical component could enhance both
efficiency and security bounds. To test this, we propose new QKD protocols
and analyze their security under different scenarios, determining
their tolerable security bounds. In QKD schemes, one party may introduce
bias in the distributed symmetric key. Furthermore, extending QKD
schemes to multiparty scenarios presents significant challenges. To
address this, we propose a new QKA scheme that operates without requiring
quantum memory. We also conceptualize quantum communication protocols
as games, where legitimate parties aim to conceal secure information
from Eve and increase her detection probability, while Eve seeks to
maximize her information gain and minimize her detection probability.
Using game-theoretic principles, we develop a technique to determine
the tolerable quantum bit error rate (QBER) based on Nash equilibrium.
This approach, referred to as ``gaming the quantum'', provides a novel
perspective on quantum protocol analysis.

This thesis focuses on advancing quantum communication and cryptography,
addressing key challenges, and proposing innovative solutions. The
security of the proposed protocols is rigorously analyzed and that
forms the central contribution. To ensure the thesis is comprehensive
and self-contained, foundational concepts of quantum mechanics and
quantum communication are discussed in Section \ref{sec:Chapter1_Sec2},
with a focus on those concepts that are directly relevant to the presented
research. A significant portion of this chapter focuses on QIA schemes.
An initial discussion on QIA, along with the motivation for exploring
QIA schemes are presented in detail in Section \ref{sec:Chapter1_Sec3}.
Subsequently, in Section \ref{sec:Chapter1_Sec4}, a chronological
analysis of previously proposed QIA schemes is conducted, highlighting
their evolution in terms of security considerations, quantum resources
utilization, and application perspectives. Following this historical
overview, the structural symmetry among these schemes is analyzed,
enabling their classification based on the quantum resources utilized
and their corresponding communication and computational tasks. This
classification is presented in Section \ref{sec:Chapeter1_Sec5},
which also serves as the foundation for introducing novel QIA schemes
in Chapter \ref{Ch2:Chapter2_Authentication}. An introductory description and motivation for the research
presented in Chapters 3, 4, and 5---focusing on QKA, QKD, and quantum
game theory for securing quantum communication protocols---are provided
in Sections \ref{sec:Chapter1_Sec6}, \ref{sec:Chapter1_Sec7}, and
\ref{sec:Chapter1_Sec8}, respectively. Finally, the chapter concludes
with a summary and an overview of the thesis in Section \ref{sec:Chapter1_Sec9}.

\section{Basic idea of quantum mechanics and quantum communication}\label{sec:Chapter1_Sec2}

\subsection{Postulates of quantum mechanics}

In classical mechanics, the state of a particle at any given time
$t$ is characterized by two primary dynamical variables: its position
$r(t)$ and momentum $p(t)$. These variables provide the foundation
for deriving any other physical quantity relevant to the system. Furthermore,
if these variables are known at a specific time $t$, their values
for that specific time can be determined using equations such as Hamilton's
equations:

\[
\frac{dx}{dt}=\frac{\partial H}{\partial p},\,\,\,\,\frac{dp}{dt}=-\frac{\partial H}{\partial x}.
\]
In quantum mechanics, analogous principles are established through
postulates that describe the following:

\textit{State of a Quantum System}: The state of a quantum system at
time $t$ is represented by a state vector $|\psi(t)\rangle$ in a
Hilbert space $\mathcal{H}$. This state vector contains all the information
required to describe the system. Additionally, any linear superposition
of state vectors also represents a valid state vector.

\textit{Quantum Operators}: For each measurable physical quantity ($A$)
(referred to as an observable or dynamical variable), there exists
a corresponding linear \textit{Hermitian operator} $\hat{A}$ . The
eigenvectors of $\hat{A}$ form a complete basis in the Hilbert space.

\textit{Eigenvalue of Operator}: The measurement of an observable $A$
is mathematically represented by the action of its corresponding operator
$\hat{A}$ on the state vector $|\psi(t)\rangle$. The possible outcomes
of the measurement are limited to the eigenvalues $a_{n}$ of $\hat{A}$,
which are real. Following a measurement that yields the result $a_{n}$,
the state of the system collapses to the corresponding eigenvector
$|\psi_{n}\rangle$.

\[
\begin{array}{lcl}
\hat{A}|\psi(t)\rangle & = & a_{n}|\psi_{n}\rangle,\end{array}
\]
here, $a_{n}=\langle\psi_{n}|\psi(t)\rangle$, which represents the
component of $|\psi(t)\rangle$ when projected onto the eigenvector
$|\psi_{n}\rangle$.

\textit{Collapse on measurement}: When measuring an observable $A$ for
a system in the state $|\psi\rangle$, the probability of obtaining
one of the nondegenerate eigenvalues $a_{n}$ of the associated operator
$\hat{A}$ is expressed as:

\[
P_{n}(a_{n})=\frac{\left|\left\langle \psi_{n}|\psi\right\rangle \right|^{2}}{\left\langle \psi|\psi\right\rangle }=\frac{\left|a_{n}\right|^{2}}{\left\langle \psi|\psi\right\rangle },
\]
where $|\psi_{n}\rangle$ denotes the eigenstate of $\hat{A}$ corresponding
to the eigenvalue $a_{n}$. For cases where $a_{n}$ has $m$-fold
degeneracy, the probability $P_{n}$ is generalized as:

\[
P_{n}(a_{n})=\frac{\sum_{j=1}^{m}\left|\left\langle \psi_{n}^{j}|\psi\right\rangle \right|^{2}}{\left\langle \psi|\psi\right\rangle }=\frac{\sum_{j=1}^{m}\left|a_{n}^{(j)}\right|^{2}}{\left\langle \psi|\psi\right\rangle }.
\]
The measurement process alters the system's state from $|\psi\rangle$
to $|\psi_{n}\rangle$. If the system is initially in the eigenstate
$|\psi_{n}\rangle$ of $\hat{A}$, a measurement of $A$ will always
yield the eigenvalue $a_{n}$ with certainty, satisfying: $a_{n}:\hat{A}|\psi_{n}\rangle=a_{n}|\psi_{n}\rangle$.

\subsection{Preliminaries of quantum communication}\label{subsec:Preliminaries-of-quantum-communication}

\subsubsection{Qubit and measurement basis}

The ``qubit'' serves as the quantum analogue of the ``classical
bit'', fundamental to quantum computation and information. Unlike a
classical bit, which exists exclusively in states $0$ or $1$, a
qubit can simultaneously occupy a superposition of these states. These
basis states are denoted in Dirac notation as $|0\rangle=\left(\begin{array}{c}
1\\
0
\end{array}\right)$ and $|1\rangle=\left(\begin{array}{c}
0\\
1
\end{array}\right)$, referred to as ``ket-zero'' and ``ket-one''. The conjugate transpose
of a state vector $|\psi\rangle$ is written as $\langle\psi|$, termed
``bra-$\psi$''. State vectors are represented as column matrices
($|\psi\rangle$), while their conjugates are row matrices ($\langle\psi|$).
A qubit state is typically represented as $|\psi\rangle=\alpha|0\rangle+\beta|1\rangle$,
where $\alpha$ and $\beta$ are complex probability amplitudes satisfying
$\left|\alpha\right|^{2}+\left|\beta\right|^{2}=1$. The terms $\left|\alpha\right|^{2}$
and $\left|\beta\right|^{2}$ correspond to the probabilities of finding
the qubit in states $|0\rangle$ and $|1\rangle$, respectively, when
measured in the computational basis. Measurement causes the quantum
state to collapse into one of the basis states, depending on the measurement
outcome.

\textit{Basis Sets}: A set of vectors $\left\{ |v_{1}\rangle,|v_{2}\rangle,|v_{3}\rangle,\ldots,|v_{n}\rangle\right\} $
is a basis if the vectors are linearly independent ($\sum_{i=1}^{n}a_{i}|v_{i}\rangle=0$
implies $a_{i}=0$ for all $i$), orthogonal $\langle v_{i}|v_{j}\rangle=\delta_{ij}$),
and satisfy the completeness relation $\sum_{i=1}^{n}|v_{i}\rangle\langle v_{i}|=1$.
This thesis work primarily employs the computational, diagonal, and
Bell basis sets, are defined as follows:

\textit{Computational basis $\left\{ |0\rangle,|1\rangle\right\} $}:\textit{
}In quantum communication, $|0\rangle$ and $|0\rangle$ are interpreted
as horizontal and vertical polarization states of photons, respectively.
For multi-qubit systems, the computational basis extends to combinations
like $\left\{ |00\rangle,|01\rangle,|10\rangle,|11\rangle\right\} $,
constructed as tensor products of individual qubit states.

\textit{Diagonal basis $\left\{ |+\rangle,|-\rangle\right\} $}: Here,
$|+\rangle=\frac{|0\rangle+|1\rangle}{\sqrt{2}}$ and $|-\rangle=\frac{|0\rangle-|1\rangle}{\sqrt{2}}$
correspond to photons polarized at $45^{\circ}$ and $135^{\circ}$
relative to the horizontal direction. This basis is used to exploit
superposition principles and complements the computational basis,
ensuring the security of quantum protocols through measurements in
both bases.

\textit{Bell basis $\left\{ |\phi^{+}\rangle,|\phi^{-}\rangle,|\psi^{+}\rangle,|\psi^{-}\rangle\right\} $}:
The Bell basis comprises maximally entangled two-qubit states, central
to quantum communication protocols involving entanglement. The Bell
states are:

\[
\begin{array}{ccccccc}
|\phi^{+}\rangle & = & \frac{1}{\sqrt{2}}\left(|00\rangle+|11\rangle\right), &  & |\phi^{-}\rangle & = & \frac{1}{\sqrt{2}}\left(|00\rangle-|11\rangle\right),\\
|\psi^{+}\rangle & = & \frac{1}{\sqrt{2}}\left(|01\rangle+|10\rangle\right), &  & |\psi^{-}\rangle & = & \frac{1}{\sqrt{2}}\left(|01\rangle-|10\rangle\right).
\end{array}
\]
These states are crucial for quantum teleportation and superdense
coding. For instance, in quantum teleportation, an entangled Bell
pair shared between two parties allows the transmission of an unknown
quantum state using classical communication and quantum operations.

\subsubsection{Pure state}

If the quantum system's state can be expressed as a linear combination
of the basis states $|n\rangle$:

\[
\begin{array}{lcl}
|\psi\rangle & = & \sum_{n}c_{n}|n\rangle,\end{array}
\]
the state is described as ``pure''. The density operator $\rho$
for a pure state is defined as:

\[
\begin{array}{lcl}
\rho & = & |\psi\rangle\langle\psi|.\end{array}
\]
By substituting the relevant expressions, we arrive at:

\[
\rho=\sum_{n}\sum_{m}c_{n}c_{m}^{*}|n\rangle\langle m|=\sum_{n,m}\rho_{nm}|n\rangle\langle m|.
\]
The matrix elements of the density operator for a pure quantum state
are defined as $\rho_{nm}=\langle n|\rho|m\rangle$. For a measurement
performed on the state $|\psi\rangle$, the probability of observing
the system in the state $|n\rangle$ is given by $\left|c_{n}\right|^{2}$.
This interpretation connects the diagonal elements of the density
matrix to physical probabilities, where $\rho_{nn}=\left|c_{n}\right|^{2}$.
Since the diagonal elements are nonnegative, the density operator
is inherently positive. The density operator for a pure state possesses
specific mathematical properties.

\[
Tr(\rho)=\sum_{n}\rho_{nn}=\sum_{n}\left|c_{n}\right|^{2}=1.
\]
Since $\rho^{2}=|\psi\rangle\langle\psi|\psi\rangle\langle\psi|=|\psi\rangle\langle\psi|=\rho$
so $Tr(\rho^{2})=1$ as for a normalized state $|\psi\rangle\langle\psi|=1$.

\subsubsection{Mixed state}

A quantum system, however, may not always exist in a pure state. Instead,
it might be represented as a mixture of states $|\psi_{i}\rangle$,
which are not necessarily orthogonal. Each state $|\psi_{i}\rangle$
has its unique expansion in terms of the eigenvector basis $|n\rangle$,
with a probability $p_{i}\geq0$ associated with finding the system
in the pure state $|\psi_{i}\rangle$. For such mixed states, the
density operator is expressed as 

\[
\rho=\sum_{i}p_{i}|\psi_{i}\rangle\langle\psi_{i}|.
\]
This operator satisfies the conditions of positivity and a unit trace.
A quantum state is classified as pure if it satisfies the condition
$Tr(\rho^{2})=1$, whereas for a mixed state, $Tr(\rho^{2})<1$. Both
the state vector representation and the density operator representation
are equivalent, with the choice of representation dictated by convenience.

\subsubsection{Quantum gate}

Except for the NOT and Identity gates, all other standard ``classical
gates''---such as AND, OR, NOR, and NAND---are irreversible ($\longrightarrow$).
This means that the input state cannot be uniquely reconstructed from
the output state. The operation of these irreversible gates results
in the erasure of one bit of information since they consistently map
a 2-bit input to a 1-bit output. Landauer's groundbreaking work \cite{Landauer1961irreversibility}
highlighted that erasing a bit of information leads to a minimum energy
dissipation of $kT\log2$. However, this energy loss can be avoided
by utilizing reversible logic gates. Although it is technically feasible
to construct classical reversible gates, they are less significant
because they do not perform tasks that cannot already be achieved
by irreversible circuits. In contrast, quantum gates are inherently
reversible ($\longleftrightarrow$), enabling the unique reconstruction
of input states from output states (a \textit{bijective} mapping). It
is essential to note that while all quantum gates are reversible,
not all reversible gates qualify as quantum gates. Typically, the
term ``quantum gates'' refers specifically to quantum logic gates,
while ``reversible gates'' pertains to classical reversible gates.
This convention will be adhered to here as well. Fundamentally, a
``quantum gate'' or ``quantum logic gate'' represents a unitary operation
that uniquely maps one quantum state to another. The mapping is inherently
bijective. Mathematically, an $N$-qubit quantum gate is represented
as a $2^{N}\times2^{N}$ matrix. In the following sections, important
single-qubit and two-qubit quantum gates relevant to this thesis will
be discussed. Additionally, certain quantum gates not directly used
in this thesis will be briefly summarized.

A single-qubit state is represented as a vector $\left(\begin{array}{c}
\alpha\\
\beta
\end{array}\right)$ , while a ``single-qubit quantum gate'' is described by a $2\times2$
unitary matrix. Such a gate, denoted as a unitary operator $U$, transforms
an input single-qubit state $|\Psi\rangle_{{\rm inp}}$ into an output
single-qubit state $|\Psi\rangle_{{\rm out}}=U|\Psi\rangle_{{\rm inp}}$.
The unitarity of $U$ ensures the existence of its inverse $U^{-1}$,
which enables the inverse operation $U^{-1}|\Psi\rangle_{{\rm out}}=|\Psi\rangle_{{\rm imp}}$.

\textit{Pauli-X gate or NOT gate}: The Pauli-X gate, represented as
the unitary operator $X$, has the matrix form:

\[
\begin{array}{lcl}
X & = & \left(\begin{array}{cc}
0 & 1\\
1 & 0
\end{array}\right)\end{array}.
\]
Alternatively, it can be expressed as $X=|0\rangle\langle1|+|1\rangle\langle0|$.
The action of the $X$ gate on a single-qubit state is described as
follows:

\[
\begin{array}{lcl}
X\left(\alpha|0\rangle+\beta|1\rangle\right) & = & \alpha|1\rangle+\beta|0\rangle\end{array}.
\]
\textit{Pauli-Z gate}: The Pauli-Z gate, represented as the unitary
operator $Z$, has the matrix form:

\[
\begin{array}{lcl}
Z & = & \left(\begin{array}{cc}
1 & 0\\
0 & -1
\end{array}\right)\end{array}.
\]
It can also be expressed as $Z=|0\rangle\langle0|-|1\rangle\langle1|$.
The action of the $Z$ gate on a single-qubit state is given by:

\[
\begin{array}{lcl}
Z\left(\alpha|0\rangle+\beta|1\rangle\right) & = & \alpha|0\rangle-\beta|1\rangle\end{array}.
\]
\textit{Pauli-Y gate}: The matrix form of the unitary operator for the
$Y$-gate is:

\[
\begin{array}{lcl}
iY & = & \left(\begin{array}{cc}
0 & 1\\
-1 & 0
\end{array}\right)\end{array}.
\]
It can also be expressed as $Y=|0\rangle\langle1|-|1\rangle\langle0|$.
The action of the $Y$-gate on a single-qubit system can be described
as follows. The $Y$-gate is a combination of the $X$-gate and the
$Z$-gate, functioning as both a bit-flip and phase-flip operator:

\[
\begin{array}{lcl}
Y\left(\alpha|0\rangle+\beta|1\rangle\right) & = & -\alpha|1\rangle+\beta|0\rangle\end{array}.
\]
\textit{Hadamard gate}: The matrix representation of the unitary operator
for the $H$-gate is:

\[
\begin{array}{lcl}
H & = & \frac{1}{\sqrt{2}}\left(\begin{array}{cc}
1 & 1\\
1 & -1
\end{array}\right)\end{array}.
\]
It can also be represented as $H=|+\rangle\langle0|-|-\rangle\langle1|$.
The Hadamard transformation is defined as:

\[
\begin{array}{lcl}
H|0\rangle & = & \frac{1}{\sqrt{2}}\left(|0\rangle+|1\rangle\right),\\
\\H|1\rangle & = & \frac{1}{\sqrt{2}}\left(|0\rangle-|1\rangle\right).
\end{array}
\]
\textit{Phase gate}: The matrix representation of the phase gate is
given by:

\[
\begin{array}{lcl}
P(\theta) & = & \left(\begin{array}{cc}
1 & 0\\
0 & \exp(i\theta)
\end{array}\right)\end{array}.
\]
Since $\theta$ can take infinitely many values, specific cases are
often considered. For $\theta=\frac{\pi}{2}$, the phase gate corresponds
to the $S$-gate:

\[
S=P\left(\frac{\pi}{2}\right)=\left(\begin{array}{cc}
1 & 0\\
0 & i
\end{array}\right).
\]
The action of the $S$-gate is:

\[
S|0\rangle=|0\rangle,\,S|1\rangle=i|1\rangle.
\]
For $\theta=\frac{\pi}{4}$, the phase gate becomes the $T$-gate,
represented as:

\[
T=P\left(\frac{\pi}{4}\right)=\left(\begin{array}{cc}
1 & 0\\
0 & \frac{1}{\sqrt{2}}\left(1+i\right)
\end{array}\right).
\]
It is important to note that the $P$-gate is generally not self-inverse.

A two-qubit quantum state is represented as a column vector: $\left(\begin{array}{c}
\alpha\\
\beta\\
\gamma\\
\delta
\end{array}\right)$ , while a two-qubit quantum gate is expressed as a $4\times4$ matrix,
since $2^{2}\times2^{2}=4\times4$. When a ``two-qubit gate'' acts on
a two-qubit state, it transforms the input state into another two-qubit
state, with the mapping uniquely defined by the specific gate's properties.

\textit{Controlled-NOT gate}: The Controlled-NOT (CNOT) gate is a two-qubit
gate. Its matrix representation is:

\[
\begin{array}{lcl}
{\rm CNOT} & = & \left(\begin{array}{cccc}
1 & 0 & 0 & 0\\
0 & 1 & 0 & 0\\
0 & 0 & 0 & 1\\
0 & 0 & 1 & 0
\end{array}\right),\end{array}
\]
In the bra-ket notation, the CNOT gate can be written as:

\[
\begin{array}{lcl}
{\rm CNOT} & = & |00\rangle\langle00|+|01\rangle\langle01|+|11\rangle\langle10|+|10\rangle\langle11|.\end{array}
\]
In this gate, the first qubit serves as the control, and the second
as the target. If the control qubit is $|0\rangle$, the target qubit
remains unchanged. Conversely, if the control qubit is $|1\rangle$,
the target qubit is flipped. The CNOT gate performs the following
mappings:

\[
\begin{array}{lcl}
|00\rangle & \longrightarrow & |00\rangle\\
|01\rangle & \longrightarrow & |01\rangle\\
|10\rangle & \longrightarrow & |11\rangle\\
|11\rangle & \longrightarrow & |10\rangle
\end{array}.
\]
\textit{SWAP gate}: The SWAP gate is another two-qubit gate. Its matrix
representation is:

\[
\begin{array}{lcl}
{\rm SWAP} & = & \left(\begin{array}{cccc}
1 & 0 & 0 & 0\\
0 & 0 & 1 & 0\\
0 & 1 & 0 & 0\\
0 & 0 & 0 & 1
\end{array}\right).\end{array}
\]
In bra-ket notation, the SWAP gate is expressed as:

\[
\begin{array}{lcl}
{\rm SWAP} & = & |00\rangle\langle00|+|10\rangle\langle01|+|01\rangle\langle10|+|11\rangle\langle11|.\end{array}
\]
The SWAP gate exchanges the states of the two qubits, performing the
following mappings:

\[
\begin{array}{lcl}
|00\rangle & \longrightarrow & |00\rangle\\
|01\rangle & \longrightarrow & |10\rangle\\
|10\rangle & \longrightarrow & |01\rangle\\
|11\rangle & \longrightarrow & |10\rangle
\end{array}.
\]
The fundamental concepts introduced thus far are crucial for understanding
quantum technologies such as quantum sensing \cite{TOL+23}, quantum
computing \cite{K23}, and quantum communication. However, this thesis
specifically focuses on secure quantum communication---an area where
quantum cryptographic schemes provide unconditional security, a feature
unattainable by classical counterparts. Notably, current implementations
of quantum cryptography, particularly QKD, are not entirely quantum
mechanical. A critical pre-QKD phase requires the sender and receiver
to authenticate each other, a process that is typically conducted
classically. In the next section, we introduce the core topic of this
thesis by addressing a key question: Why is quantum identity authentication
necessary? The subsequent sections of this chapter systematically
explore foundational concepts related to this work, covering QIA,
QKD, QKA, and game theory in the context of quantum communication
in Sections \ref{sec:Chapter1_Sec3}--\ref{sec:Chapeter1_Sec5},
\ref{sec:Chapter1_Sec6}, \ref{sec:Chapter1_Sec7}, and \ref{sec:Chapter1_Sec8},
respectively.

\section{Why is quantum identity authentication necessary?}\label{sec:Chapter1_Sec3}

Identity authentication, often referred to as authentication, is a
structured process aimed at verifying the identities of authorized
users or devices/components. This procedure is critical in mitigating
various types of attacks targeting secure computation and communication
frameworks. The significance of identity authentication protocols
has grown substantially in recent years, driven by the increased adoption
of e-commerce, online banking, the Internet of Things (IoT), online
voting, and similar applications. These use cases fundamentally rely
on robust identity verification mechanisms. For instance, IoT primarily
focuses on authenticating devices or components, while online voting
emphasizes verifying the identities of users (voters). From a cryptographic
standpoint, the distinction between a user and a device/component
is minimal, and thus, they will be treated equivalently in subsequent
discussions.

In a two-party framework, authentication can be conceptualized as
a process enabling a legitimate sender, Alice, to convey a message
$X$ (or a key in the context of key distribution schemes) to a receiver,
Bob, ensuring that Bob can verify the message's integrity and confirm
it was not altered during transmission through the channel \cite{DL99}.
Essentially, this process serves to authenticate the origin of message
$X$ and guarantee its integrity upon reception by Bob. This concept
is closely associated with digital signatures, which additionally
permit a third party, Charlie, to later verify that the message $X$
received by Bob from Alice has not been tampered with \cite{DL99}.
Although authentication and digital signatures are related, they are
distinct tasks. This discussion is confined to authentication schemes,
specifically those utilizing quantum resources, referred to as QIA.
To understand the significance of QIA schemes, it is helpful to briefly
examine the evolution of quantum cryptography, which has historically
redefined the concept of security.

In 1984, Bennett and Brassard introduced the first QKD protocol \cite{BB84},
now referred to as the BB84 protocol \footnote{The origins of quantum cryptography are significantly attributed to
Wiesner's pioneering work \cite{W83}. For an insightful account
of the historical developments in this field, see \cite{B05}.}. This groundbreaking proposal asserted the potential for unconditional
security---a feature unattainable through classical key distribution
methods, which rely on the computational complexity of specific mathematical
problems for security. This assertion captured significant attention
within the cryptographic community, leading to the development of
several notable QKD protocols, such as the E91 protocol by Ekert (1991)
\cite{E91} and the B92 protocol by Bennett (1992) \cite{B92}.
The importance of authenticating the identities of legitimate participants
was acknowledged even in the foundational BB84 paper. Bennett and
Brassard highlighted, ``The need for the public (non-quantum) channel
in this scheme to be immune to active eavesdropping can be relaxed
if Alice and Bob have agreed beforehand on a small secret key, which
they use to create Wegman-Carter authentication tags \cite{WC81}
for their messages over the public channel''. Prior to the development
of quantum identity authentication protocols in 1995, classical methods
like the Wegman-Carter authentication scheme were inherently incorporated
into early QKD protocols \cite{TKK+24}, though this was not always
explicitly stated. Across these protocols, as well as in subsequent
authentication schemes, the ``pre-shared small key'' mentioned above
remains a fundamental and indispensable component \footnote{All QKD protocols can be interpreted as key amplification mechanisms.
Each QKD session begins with a small pre-shared key for authentication,
generates an extended key, and subsequently uses a portion of the
extended key as the pre-shared key for the next session.}. This concept will be elaborated further in this chapter. The inclusion
of the Wegman-Carter scheme, or similar classical methods, in early
QKD protocols arose from the practical consideration that no communication
channel could be assumed entirely trustworthy. An adversary (Eve)
could potentially intercept and replace the communication channel
between the sender (Alice) and receiver (Bob) with two separate channels---one
connecting Alice to Eve and the other connecting Eve to Bob. In such
a case, Eve could generate separate keys with both Alice and Bob,
effectively isolating them. To prevent such scenarios, employing identity
authentication protocols became a critical necessity.

\begin{figure}
\centering{}\includegraphics[scale=0.3]{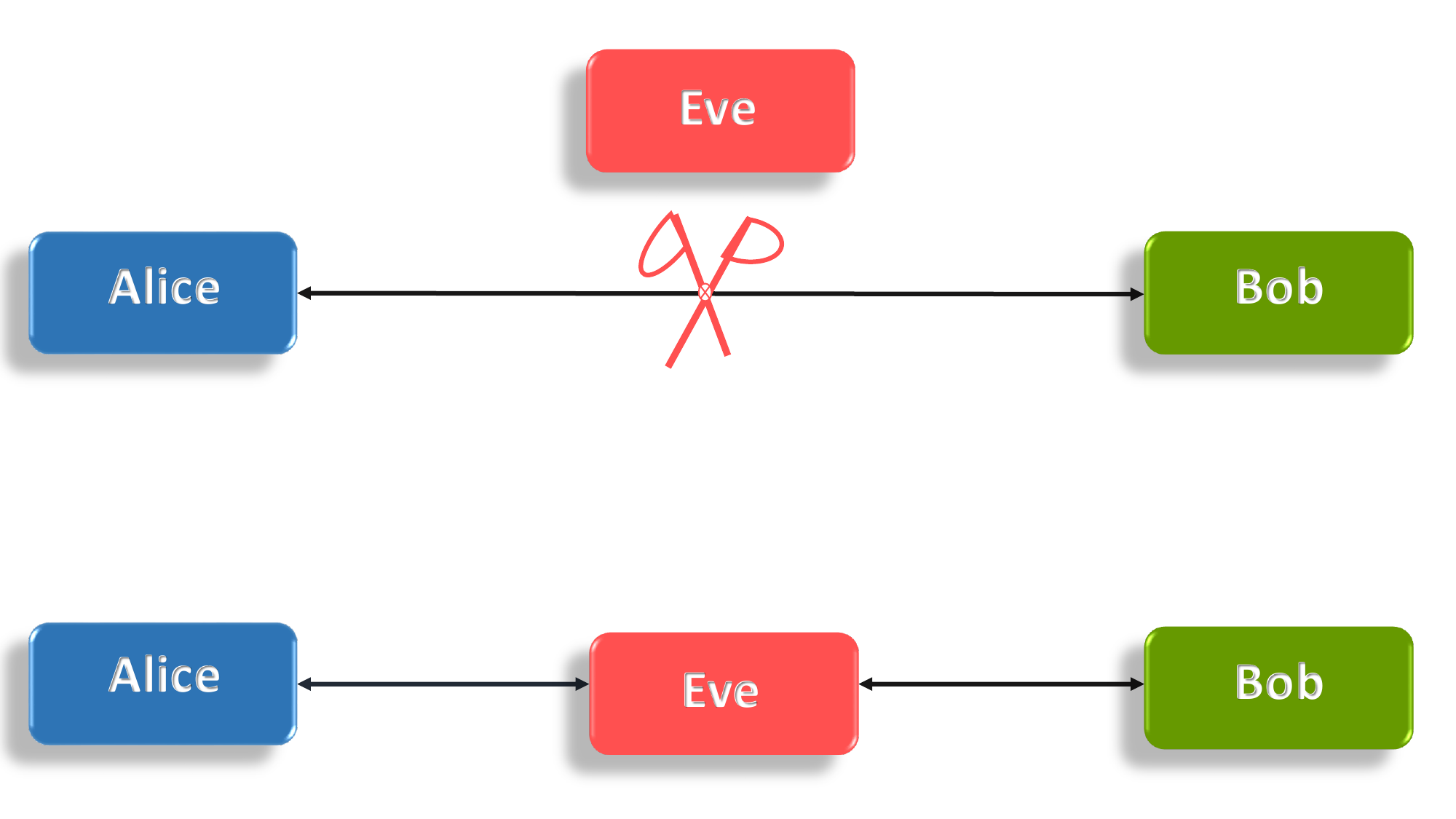}\caption{Eve may attempt to mimic Alice (or Bob) while interacting with Bob
(or Alice) by substituting the direct communication channel between
Alice and Bob with two separate channels: one connecting Alice to
Eve and the other linking Eve to Bob.}\label{fig:1=000020Chapter1_Fig1}
\end{figure}

Since the security of classical authentication methods relies on the
computational complexity of certain tasks, they cannot guarantee unconditional
security. This limitation introduces potential vulnerabilities in
so-called unconditionally secure QKD schemes. Achieving true unconditional
security in QKD necessitates unconditionally secure authentication
methods, which in turn require the use of quantum resources. Authentication
schemes leveraging quantum principles are commonly referred to as
QIA schemes. The first such scheme was introduced by Cr{\'e}peau
and Salvail in 1995 \cite{CS_1995}. Subsequent advancements followed.
For instance, in 1998, Zeng et al. proposed a QKD protocol that simultaneously
enabled key distribution and identity verification \cite{ZW_1998}.
Shortly after, in 1999, Du{\v{s}}ek et al. presented two hybrid QIA
protocols \cite{DHHM_1999}. These protocols combined classical authentication
methods with QKD, often emphasizing the one-time pad. Over time, numerous
QIA schemes emerged, each with distinct advantages and limitations
\cite{LB_2004,WZT_2006,ZLG_2000,ZZZX_2006,ZCSL_2020,KHHYHM_2018,CXZY_2014,T.Mihara_2002}.
Given the critical role of authentication in secure quantum communication
and computation protocols, this chapter aims to systematically review
the existing QIA schemes.

Before delving into a formal chronological review of schemes for QIA,
it is essential to briefly introduce the concept of quantum communication
complexity in the context of quantum authentication protocols. This
is because quantum communication complexity serves as a measure for
assessing the amount of communication necessary between two parties,
Alice and Bob, to complete a distributed communication task \cite{B_03}.
Consider a scenario where Alice and Bob aim to jointly compute a function
$f(a,b)$ using an error-free quantum channel, where Alice possesses
input $a$ and Bob holds input $b$. To accomplish this task, they
exchange information according to a specific protocol derived from
distributed algorithms. The communication required to execute the
task using such a protocol defines the communication complexity for
the given scenario. There are three primary models of quantum communication
complexity relevant to this discussion. The first is Yao's model,
wherein the parties communicate through a quantum channel that facilitates
qubit exchange \cite{Y_93}. Both Alice and Bob perform unitary operations
and interact by placing qubits into a shared channel that each can
access via further unitary operations. The protocol concludes when
they successfully compute the precise value of $f(a,b)$. The quantum
communication complexity, $Q(f)$, in this model, is determined by
the number of communications required to complete the task. The second
model, introduced by Cleve and Buhrman, involves parties sharing entangled
pairs and exchanging classical information over a classical channel
\cite{CB_97}. In this framework, the communication required for
pre-sharing entangled states is excluded from the computation of quantum
communication complexity. Instead, only the number of classical bits
exchanged contributes to the complexity. This model is particularly
suitable for analyzing protocols utilizing teleportation and superdense
coding, with the quantum communication complexity denoted as $C^{*}(f)$.
The third model, known as the hybrid model, incorporates features
of the previous two approaches. It employs a quantum channel, a classical
channel, and pre-shared entangled pairs for communication. In this
model, the entangled pairs used are not counted again during the evaluation
of communication complexity; only the classical bits exchanged are
considered. The quantum communication complexity in this case is represented
as $Q^{*}(f)$. The relevance of these models to QIA becomes evident
in subsequent sections. The suitability of a particular quantum communication
complexity model depends on how a specific QIA scheme is implemented.
Recent studies, such as those by Guedes and de Assis (in 2011) and
Ghilen et al. (in 2014) \cite{GA_11,GBB_14}, have examined quantum
communication complexity for various authentication protocols, providing
systematic methodologies for analyzing the complexity of such protocols. 

\section{A short chronological history of the protocols for quantum identity
authentication }\label{sec:Chapter1_Sec4}

Classical authentication schemes, such as the Wegman-Carter scheme,
were highlighted in Section \ref{sec:Chapter1_Sec3} as having been
employed in the early protocols of QKD. These message authentication
schemes can be effectively conceptualized as operating in two distinct
phases. In the initial phase, specific functions are utilized to generate
identity authentication, while in the subsequent phase, these functions
are used to verify the authenticity of a message. Classical methods
like message encryption, message authentication codes (MACs), and
hash functions are widely employed for such purposes. Although the
primary focus here is not on classical authentication schemes, it
is worthwhile to briefly touch upon these concepts before delving
into the chronological development of QIA. This brief discussion is
pertinent, as hybrid schemes and purely classical methods remain integral
to secure quantum communication and computation. Additionally, many
classical authentication techniques can be adapted to design schemes
for QIA. To illustrate how classical methods can be used for QIA scheme
design, consider a pre-shared key $x_{1},x_{2},\cdots,x_{n}$ between
Alice and Bob, where $x_{i}\in\{0,1\}$. Alice can divide this key
into two parts, using one half (e.g., all odd-indexed bits) as the
message and the other half (e.g., all even-indexed bits) as the key.
She can then construct a new sequence of length $\left\lceil \frac{n}{2}\right\rceil $
where the $i$-th bit of the sequence is $x_{2i-1}\oplus x_{2i}$.
This sequence can function as a tag or cryptographic element. Alice
transmits it using a conventional method, and Bob, employing the same
key subset $\{x_{2i}\}$, can decode it. The ability to convey a message
inherently includes the ability to transmit a Tag. As a result, every
protocol for QSDC \cite{YSP14,NP+23}, quantum dialogue (QD) \cite{SKB+13},
and deterministic secure quantum communication (DSQC) \cite{SP14}
can accomplish this function. In other words, the classical method
described can be seamlessly adapted to create straightforward schemes
for QIA using various QSDC, DSQC, and QD protocols. Furthermore, in
the MAC approach, both participants share a common secret key, $K$.
When Alice intends to send a message to Bob, she computes the MAC
as a function of the message and the shared key, then transmits the
message along with the MAC to Bob. Bob subsequently calculates the
MAC for the received message using the same procedure and compares
it with the received MAC. A match verifies the message's authenticity
and integrity. A hash function, as a variant of MAC, processes input
messages of variable lengths to produce a fixed-size output known
as a hash code. With its one-way property, the hash function is particularly
effective for generating a message fingerprint. The no-cloning theorem,
which asserts that it is impossible to duplicate an unknown quantum
state with perfect fidelity, alongside the ``collapse on measurement''
attribute of quantum states, plays a pivotal role in demonstrating
the advantages of QIA schemes over their classical counterparts. To
illustrate this, imagine a scenario where Eve breaks into Alice's
house and secretly replicates her authentication key without leaving
any evidence. In the classical domain, such an act would enable Eve
to impersonate Alice and communicate with Bob undetected. In contrast,
within the quantum realm, such a situation is prevented if the authentication
key is encoded and stored in a quantum state, as the no-cloning theorem
prohibits perfect copying. Furthermore, the ``collapse-on-measurement''
principle ensures that any eavesdropping attempt would leave detectable
traces, thereby enhancing the security of quantum cryptographic protocols.
This serves as the foundation for developing QIA protocols.

In the prior section, it was noted that the first protocol for QIA
was introduced by Cr{\'e}peau et al. \cite{CS_1995}. This protocol
was grounded in the concept of oblivious transfer (OT), a fundamental
cryptographic primitive. However, a pivotal result by Lo and Chau\cite{LC_1997}
demonstrated the impossibility of achieving unconditionally secure
two-party OT. The groundbreaking contribution by Cr{\'e}peau et al.
was succeeded by early explorations into QIA, including those by Zeng
and Wang (1998), Du{\v{s}}ek et al. (1999), and Barnum (1999) \cite{ZW_1998,DHHM_1999,B99},
as briefly discussed earlier. Despite these initial efforts, progress
in quantum or hybrid identity authentication during the 20th century
was limited. The landscape has evolved significantly in the 21st century.
This section aims to trace the chronological development of QIA protocols.
Given the extensive literature, this effort does not strive to cover
all contributions but instead highlights representative works that
illustrate the progression of quantum and hybrid identity authentication
\footnote{For readers seeking a comprehensive list of publications on quantum
authentication from 2009 to 2019, Ref. \cite{MD21} provide an extensive
review. Notably, the classification of QIA schemes presented in \cite{MD21}
aligns with the framework established in this chapter, even though
their study was encountered after completing this review.}. It also emphasizes that different types of QKD protocols---such
as continuous variable QKD (CV-QKD), discrete variable QKD (DV-QKD),
counterfactual QKD, and semi-QKD---necessitate tailored QIA schemes
to enable the use of shared resources and devices for both communication
and authentication tasks. While early QIA protocols were predominantly
tied to conventional ``conjugate-coding-based'' DV cryptographic
schemes, their applicability has since expanded significantly.
\begin{description}
\item [{2000:}] The year was notably productive, with the introduction
of numerous new protocols for QIA, numbering at least ten. The majority
of these protocols, with the exception of three proposed by Ljunggren
et al. \cite{LBK_2000} were based on Bell states.
In early 2000, Zeng and Guo introduced a Bell state-based approach
for QIA \cite{ZG_2000}. Shortly thereafter, Zhang
et al. presented another Bell state-based protocol \cite{ZLG_2000},
utilizing a pre-shared Bell state as either a pre-shared secret or
an authentication key. Between these developments, Jensen and Schack \cite{JS_2000} proposed a protocol founded on
bipartite entangled states \footnote{Although it is not explicitly stated that the particles are maximally
entangled, we can, without any loss of generality, interpret this
as a Bell-state-based approach for QIA.}, asserting its security even under conditions where Eve has full
control over both classical and quantum channels. Their work built
upon Barnum's study, providing a generalization while also raising
several open research questions \cite{B99}.\\
In 2000, Ljunggren et al. introduced a set of five protocols for authenticated
QKD between two parties, Alice and Bob \cite{LBK_2000}.
Their approach incorporated a third party, Trent, serving as an arbitrator.
Of these protocols, three utilized single-photon techniques \cite{MD21},
while the remaining two were based on Bell states. Notably, one of
the entangled state-based protocols exhibited similarities to the
B92 protocol. The symmetry identified between QKD and QIA schemes
in this protocol suggests the potential for transforming other QKD
schemes into frameworks for QIA or authenticated QKD. Furthermore,
Trent's role parallels that of a controller in a cryptographic switch,
as discussed in Ref. \cite{SOS+14,TP15}. Leveraging
the findings of these studies, the three-party protocols proposed
by Ljunggren et al. can be generalized to derive novel protocols for
QIA.
\item [{2001:}] Curty et al. introduced an optimal protocol for quantum
authentication of classical messages, referred to as the CS01 protocol
\cite{CS_01}. This protocol was optimal in that it enabled the authenticated
transmission of a two-bit classical message using a single qubit as
the authentication key, effectively serving as an optimal pre-shared
secret. Notably, they categorized potential attacks on authentication
schemes into two types: (i) no-message attacks and (ii) message attacks.
In the CS01 protocol and several other message authentication schemes,
perfect deterministic decoding of the classical message is achieved.
This means that the protocol fails only if Bob erroneously accepts
an unauthenticated message as legitimate. Curty et al. observed that
this failure can occur through two distinct mechanisms, corresponding
to the two categories of attacks. In a no-message attack, an adversary
(Eve) prepares a quantum state before Alice transmits any message,
intending to trick Bob's decoding algorithm with a certain success
probability $p_{f}$. This attack is termed ``no-message'' because
it operates in the absence of any transmitted message from Alice.
Conversely, in a message attack, Eve intercepts the message sent by
Alice to extract information. Using this intercepted data, she constructs
a counterfeit message and attempts to bypass Bob's decoding algorithm.
This classification of attack strategies significantly facilitated
the security analysis of later protocols. However, the security analysis
in \cite{CS_01} and many subsequent works relied on the assumption
of an idealized noiseless quantum channel, which does not accurately
reflect real-world conditions.
\item [{2002:}] In the CS01 protocol, a classical message was authenticated
using an optimal amount of quantum resources. This naturally raised
intriguing questions, such as: How can a quantum message be authenticated
using a quantum key? Is it possible to authenticate a qubit---a minimal-sized
quantum message---using another qubit as a minimal-sized pre-shared
quantum secret? These questions were explored by Curty et al. \cite{CSPF_02}),
who presented a no-go theorem stating that a single qubit cannot be
authenticated using only another qubit. T. Mihara proposed protocols
for QIA and message authentication \cite{T.Mihara_2002} that utilized
pre-shared entanglement. Their QIA scheme involved a trusted third
party, while their message authentication protocol combined quantum
resources with a hash function, making it a hybrid approach. Notably,
the methodology adopted in their message authentication scheme can
be generalized, allowing all existing QSDC protocols to design analogous
frameworks. However, it was later demonstrated that this scheme while
employing quantum correlations, does not leverage them effectively,
and the same task can be achieved using classically correlated states
\cite{V03}. Additionally, the quantum component of Mihara's QIA
protocol \cite{T.Mihara_2002} was found to be susceptible to vulnerabilities.
Specifically, during the entanglement distribution phase (see Steps
2--3 in the section ``Identifying Alice to Bob'' in \cite{T.Mihara_2002}),
neither the BB84 subroutine nor the Goldenberg Vaidman (GV) subroutine
\cite{sharma2016verification} was utilized. Instead, all subsequent
measurements were performed in the computational basis, making the
protocol vulnerable to a ``man-in-the-middle attack''. This issue
could be mitigated by incorporating decoy qubits and implementing
one of the aforementioned subroutines, as is commonly done in QSDC
protocols \cite{Z12,Z13,ZPM13}. For example, one might refer to
the ping-pong protocol described in Ref. \cite{P13}.
\item [{2004:}] Li et al. introduced a QIA protocol utilizing Bell states
\cite{LB_2004}. In their approach, a pre-shared Bell state serves
as an identification token between Alice and Bob. During the authentication
process, Alice generates an auxiliary Bell state to interact with
the identification token and transmits the resulting auxiliary information
to Bob. Bob then performs specific operations and concludes the process
with a Bell state measurement to authenticate. Notably, this Bell
state-based protocol does not leverage the \textit{non-local} properties
of Bell states, as the violation of Bell inequalities is not employed
as a quantum resource in the QIA design.
\item [{2005:}] Zhou et al. introduced a QIA framework based on quantum
teleportation and entanglement swapping \cite{ZZZZ_2005}. Their
approach aimed to address the limitations of straightforward point-to-point
QIA systems, leveraging entanglement swapping to increase the range
over which authentication could occur. This method, widely recognized
and often employed in QKD, helps overcome distance restrictions imposed
by noise and the absence of quantum repeaters. However, it also introduces
a new security challenge, as the intermediary device facilitating
entanglement swapping must be trusted.\\
One way to assess the efficiency of an authentication protocol is
by evaluating the amount of pre-shared information it requires during
the authentication process---schemes that use less information are
considered more efficient. Using this metric, Peev et al. proposed
an efficient QIA scheme \cite{ZZZZ_2005}.
\item [{2006:}] Lee et al. introduced a pair of quantum direct communication
protocols incorporating user authentication \cite{LLY_2006}. In
their approach, Alice transmits a confidential message to Bob without
relying on a ``pre-shared secret''. The protocols are structured into
two distinct components: one designated for authentication and the
other for direct message transmission. A trusted third party, Trent,
possessing greater authority than the communicating parties, is responsible
for authenticating participants by distributing tripartite ``Greenberger-Horne-Zeilinger''
(GHZ) states. \\
Wang et al. introduced an authentication protocol that enables a trusted
third party to authenticate multiple users simultaneously by utilizing
entanglement swapping and GHZ states \cite{ZZZX_2006}. In a separate
approach, Zhang et al. presented a one-way QIA scheme \cite{ZZZX_2006}
based on the ping-pong protocol of QSDC as proposed by Bostr{\"o}m
and Felbinger \cite{ping-pong_BF_02}. Their scheme qualifies as
one-way QIA because it designates Alice as a trusted certification
authority (CA), responsible for verifying Bob's identity when he attempts
communication with her (Alice). Many existing authentication frameworks
share this one-way nature, although it is not always explicitly stated,
since one-way authentication inherently allows for two-way verification
through repeated application of the process. The authors demonstrated
that their protocol effectively resists specific attacks, including
impersonation attacks---where an adversary, Eve, without knowledge
of the authentication key, attempts to impersonate Bob by forging
qubits---and substitution attacks, in which Eve endeavors to obtain
the authentication key by leveraging a newly established quantum channel
and intercepting qubits transmitted from Bob to Alice. It is important
to highlight that while attacks such as message and no-message attacks
are not consistently categorized under these labels, the majority
of attacks examined can be classified into one of these two categories.
\item [{2007:}] Zhang et al. identified vulnerabilities in the protocol
introduced by Lee et al. in 2006 \cite{ZLWS_2007}, highlighting
that Trent could exploit it through two distinct types of attacks.
To address these security flaws, Zhang et al. proposed an enhanced
version of Lee et al.'s protocol, effectively mitigating the identified
weaknesses. In the domain of cryptography, such analyses---wherein
a scheme's vulnerabilities are uncovered, followed by the development
of a more robust version---are a common practice. This scenario serves
as a representative instance within the context of QIA.
\item [{2009:}] Guang et al. introduced a multiparty QIA protocol utilizing
GHZ states \cite{GY_2009}, which closely resembled the approach
previously proposed by Wang et al. \cite{WZT_2006}. However, their
scheme demonstrated greater efficiency as it required fewer quantum
resources and operations compared to the protocol by Wang et al.\\
In another development, Rass et al. proposed a novel approach by shifting
the authentication phase to the conclusion of the QKD process \cite{RSG09},
diverging from the conventional practice of performing it at the outset.
Their scheme integrated concepts from both QKD and quantum coin-flipping,
and they successfully derived tight bounds on the amount of pre-shared
secret information necessary for QIA.
\item [{2010:}] Dan et al. proposed a straightforward QSDC protocol incorporating
authentication, utilizing Bell states with qubits encoded through
polarization \cite{DXXN_2010}. Although authentication schemes do
not always explicitly specify this detail, it is inherently presumed
that the qubits involved are photonic. This assumption arises because
other forms of qubits, such as ``transmon qubits'' used in superconducting
quantum computing and other qubit implementations, are not readily
transferable over long distances. Furthermore, Dan et al. demonstrated
the security of their protocol against common attacks, including intercept-and-resend
attacks, Trojan horse attacks, and entanglement-based attacks.
\item [{2011:}] In a subsequent development, Huang et al. introduced a
CV-QIA protocol based on Gaussian-modulated squeezed states \cite{HZLZ_2011}.
This protocol, along with similar approaches, enables CV-QKD schemes
to authenticate using CV states. This advancement is crucial, as employing
DV quantum authentication within CV quantum protocols can lead to
decreased efficiency and increased operational complexity. The reason
for this is that handling both CV and DV states requires additional
devices and resources for performing respective operations and measurements.
\item [{2012:}] Gong et al. explored the application of quantum one-way
functions to formulate a QIA protocol that incorporated a trusted
third party, which they referred to as a trusted server in their study
\cite{GTZ12}.\\
Additionally, a hybrid QIA scheme within a quantum cryptographic network
was patented by MagiQ Technologies under US Patent number US 8,340,298
B2 \cite{GB12}.\\
Skoric introduced an intriguing concept that integrates quantum challenges
with classical physical unclonable functions (PUFs). The core idea
involved utilizing a quantum state to challenge a classical PUF. Due
to the no-cloning theorem and the principle of ``collapse on measurement'',
any interception attempt would leave detectable traces. Consequently,
the verifier, who issues the challenge through a quantum state, could
reliably confirm the prover's identity if the expected quantum state
is received in response. It is important to highlight that the concept
of PUFs was initially proposed by Pappu et al. at the beginning of
the 21st century \cite{PRT+02}. A PUF is characterized as a physical
entity embedded in a specific structure that is straightforward to
evaluate but challenging to characterize comprehensively. It is also
considered an entity that is inherently resistant to duplication due
to its complex, uncontrollable physical parameters (degrees of freedom)
\cite{GHMSP_2014}. Notably, PUFs can be implemented using various
devices such as FPGAs, RFIDs, and other hardware components, enabling
their application in classical authentication protocols \cite{ZPD+14}.
However, the focus here is on PUF-based schemes specifically designed
for QIA. The growing interest in such schemes will be examined in
the subsequent discussion.
\item [{2013:}] Yang et al. introduced two authenticated direct secure
quantum communication protocols utilizing Bell states \cite{YTXZ_2013}.
These protocols incorporated a third party, Trent, and employed a
hybrid approach by integrating Hash functions. The scheme enabled
mutual authentication between Alice and Bob through Trent, leveraging
Bell states combined with their secret identity sequences and one-way
Hash functions $h_{A}$ and $h_{B}$.
\item [{2014:}] In the context of controlled quantum teleportation, an
intermediary, Charlie, controls the process, allowing Alice to transmit
a quantum state to Bob \cite{P13,TVP15}. To resolve the challenge
of identity verification in controlled teleportation, Tan et al. proposed
identity authentication protocols utilizing entanglement swapping
\cite{TJ_2014}. However, the efficiency of this approach is limited,
as it necessitates two (three) entanglement swapping operations for
Alice to verify the identity of Charlie (Bob). \\
Goorden et al. experimentally demonstrated a quantum authentication
method for a physical unclonable key \cite{GHMSP_2014}. In this
approach, the key, which is inherently classical in nature, was verified
by exposing it to a light pulse with fewer photons than the available
spatial degrees of freedom. Authentication was achieved by analyzing
the spatial characteristics of the reflected light. \\
Shi et al. introduced a ``quantum deniable'' authentication protocol leveraging
the characteristics of unitary transformations and quantum one-way
functions, utilizing the GHZ state \cite{SZY_2014}. Their protocol
ensures both authentication completeness and deniability. Authentication
completeness refers to the principle that, provided both the sender
and receiver adhere to the protocol, the intended recipient can always
verify the origin of the message. In contrast to traditional authentication
schemes, deniable authentication guarantees that only the intended
recipient can verify the message's source, while preventing the recipient
from proving its origin to a third party. \\
Yuan et al. proposed a practical QIA scheme \cite{YLP+14} based
on a single-particle ping-pong protocol, similar to the LM05 protocol
of QSDC \cite{LM_05}. This approach requires minimal quantum resources
and utilizes simple devices. \\
Fountain codes, also known as ``rateless erasure codes'', possess
a unique feature: starting with $n$ source symbols, they can produce
$m>n$ encoded symbols, where $m$ can be arbitrarily large. In a
distributed setting, distributed fountain codes exploit this property.
Using such codes, \cite{LXO+14} devised a QIA scheme within the
framework of quantum secret sharing.
\item [{2015:}] Shi et al. enhanced their earlier research \cite{SZY_2014}
by utilizing a single-photon source in place of a GHZ state \cite{SZZY_2015}.
This revised protocol demonstrated greater efficiency and required
fewer quantum resources. This improvement is intuitive since generating
and maintaining single-photon states is significantly simpler compared
to photonic GHZ states.
\item [{2016:}] Ma et al. introduced a CV-QIA protocol based on teleportation,
employing a two-mode squeezed vacuum state and a coherent state. Notably,
CV-QKD protocols outperform their discrete-variable counterparts over
short distances \cite{POS+15,XCQ+15,POS++15}. Thus, for urban networks,
integrating CV-QKD with CV-QIA is anticipated to provide a more efficient
solution. \\
Rass et al. proposed a quantum information authentication scheme linked
to the BB84 protocol \cite{RSS+16}. The innovative aspect of this
scheme lies in the use of a secondary public channel, distinct from
the channel employed for the primary BB84 protocol. \\
The security of PUF-based authentication schemes, as explored by Goorden
et al. and {\v{S}}kori{\'c} \cite{GHMSP_2014,S12}, was challenged
in \cite{YGL+16}. It was demonstrated that earlier analyses, such
as those by Ref. \cite{SMP13} and related works, were incomplete,
as they only addressed security against challenge-estimation attacks.
However, it was shown that cloning attacks could surpass the effectiveness
of challenge-estimation strategies.\\
Composable security for QIA schemes was introduced by Hayden et al.
\cite{HLM16}, marking a significant advancement in enabling secure
quantum communication across networks. Notably, composable security
ensures that a real-world implementation of a protocol remains $\epsilon$-indistinguishable
from an ideal protocol designed for the same purpose. This framework
allows for the derivation of information-theoretic security for composite
schemes, such that $\epsilon\leq\epsilon_{p}+\epsilon_{a}$, where
$\epsilon_{p}$-secure cryptographic protocols are combined with $\epsilon_{a}$-secure
applications, as discussed by Renner in \cite{R08}.
\item [{2017:}] Hong et al. proposed a QIA protocol utilizing single-photon
states \cite{HCJ+17}. This approach was both resource-efficient
and practical, as it required fewer resources and facilitated the
authentication of two bits of a key sequence using a single qubit.
\\
Abulkasim et al. introduced an authenticated QSS protocol that leveraged
Bell states as its foundational mechanism \cite{AHKB_2017}. To enhance
security, the protocol incorporated a widely employed technique in
secure quantum communication, where the pre-shared key was encrypted
prior to usage---a method frequently adopted in quantum communication
frameworks. \\
Nikolopoulos and Diamanti proposed a CV authentication scheme utilizing
coherent light states, wavefront shaping, and homodyne detection methods
\cite{ND_2017}.\\
Portmann presented a quantum authentication scheme featuring key recycling,
demonstrating its resilience against noisy channels and shared secret
key vulnerabilities \cite{Portmann_2017}. Additionally, the study
established that the number of recycled key bits is optimal for certain
authentication protocols.
\item [{2018:}] Kang et al. introduced a mutual authentication protocol
enabling Alice and Bob to verify each other's identities despite the
involvement of an untrusted third party, Trent \cite{KHHYHM_2018}.
This protocol relies on entanglement correlation verification, utilizing
GHZ-like states to confirm that the state distributed by Trent is
genuinely entangled. To prevent Trent from deducing the key, Alice
and Bob incorporate random numbers in their process. This approach
can be extended to other QIA protocols involving a trusted third party,
potentially leading to the development of new QIA protocols.\\
Additionally, a US patent (Patent Number US 9,887,976 B2) was granted
for a multi-factor authentication method employing quantum communication.
This patent describes a two-stage process, encompassing enrollment
and identification, implemented through a computer system acting as
a trusted authority \cite{HPT+18}.
\item [{2019:}] Semi-quantum protocols represent a category of cryptographic
protocols designed to extend quantum-level security to certain classical
users. The first semi-quantum protocol for QKD was introduced by Boyer
et al. in 2007 \cite{boyer2007quantum}. Since then, semi-quantum
schemes have been developed for a variety of purposes \cite{STP17,thapliyal2018orthogonal,mishra2021quantum,asagodu2021quantum}.
However, until 2019, no semi-quantum authentication protocol had been
proposed, despite its fundamental importance for semi-quantum QKD
and other semi-quantum cryptographic applications. This gap was addressed
by Wen et al. in 2019 when they introduced a semi-quantum authentication
protocol \cite{WZGZ_2019} that does not require all users to possess
quantum capabilities. Specifically, this protocol enables a quantum-capable
Alice (or Bob) to authenticate a classical Bob (or Alice). \\
 In the same year, another significant development in quantum cryptography
emerged when Liu et al. proposed a QIA protocol \cite{LGXHZX_2019},
which can be integrated into counterfactual QKD schemes.\\
 Zheng et al. introduced a controlled QSDC (CQSDC) protocol with authentication
in 2019, utilizing a five-qubit cluster state \cite{ZL_2019}. While
the protocol was demonstrated to be secure even in noisy environments,
the preparation and maintenance of the five-qubit cluster state pose
significant challenges. Although it is conceptually straightforward
to extend Bell state or GHZ state-based protocols to those employing
$n$-partite entangled states for $n>3$, such extensions often lack
practical advantages. Notably, this was not the first instance where
a five-qubit cluster state was utilized for quantum information applications.
For instance, a QIA scheme employing a five-qubit cluster state and
one-time pad was proposed in 2014 \cite{CXZY_2014} and later subjected
to cryptanalysis in 2016 \cite{GH_2016}. Moreover, an entanglement
swapping-based protocol using a six-qubit cluster state was proposed
in 2012, and in 2015, a scheme for authenticated CQSDC was designed
with a six-qubit entangled state \cite{YSL+15}. However, none of
these approaches relying on multi-partite entangled states are currently
practical for real-world applications.
\item [{2020:}] Many of the QIA protocols referenced above have not accounted
for the impact of noise, an inevitable factor in practical scenarios.
To address this limitation, Qu et al. introduced a QIA protocol utilizing
a ``three-photon error-avoidance code'' \cite{QLW_2020}. This approach
effectively mitigates the influence of noise on the information transmission
through a quantum channels.\\
Zhang et al. proposed a QIA protocol based on Bell states and entanglement
swapping \cite{ZCSL_2020}, which offers security against the strategies
of a semi-honest third party. It is important to note that a semi-honest
party adheres to the protocol's rules while attempting to exploit
the system within those bounds. As a result, semi-honest attacks are
generally weaker than those of a dishonest third party. Therefore,
when designing quantum cryptographic protocols, it is preferable to
account for the potential threats posed by dishonest adversaries.
\item [{2021:}] Existing protocols are either subjected to cryptanalysis
or modified to propose enhanced versions of QIA schemes tailored to
specific applications. For instance, \cite{XL21} and \cite{DPM21}
examine QIA schemes in the contexts of quantum private query and QSDC,
respectively. Additionally, a previously proposed hybrid authentication
scheme by Zawadzki \cite{Z19}, which improved upon a single-photon-based
protocol introduced by Ho Hong et al. \cite{HCJ+17}, underwent systematic
cryptanalysis by Gonz{\'a}lez-Guill{\'e}n et al. \cite{GCM+21}.
In their analysis \cite{GCM+21}, they designed two attacks, demonstrating
vulnerabilities not only in the protocols from Ho Hong et al. and
Zawadzki \cite{HCJ+17,Z19} but also in several other existing QIA
protocols. Furthermore, significant interest has been observed in
PUF-based quantum authentication schemes \cite{JSR21,DKS+21}, and
closely related applications explored by \cite{PAA+21}.
\item [{2022-2024}] Li et al. introduced a multiparty simultaneous identity
authentication protocol utilizing the GHZ state \cite{LZZ+22}. This
protocol assumes that the third-party Trent is both honest and a classical
participant, which presents a limitation for communication security.
In the same year, Das et al. \cite{DP+22} proposed a measurement-device-independent
(MDI) QSDC protocol incorporating user authentication \cite{DP+22},
wherein authentication is conducted prior to a secure message exchange
between authenticated parties. Similarly, He et al. \cite{HPD22}
developed a QKA protocol in which quantum authentication is performed
first. Leveraging the structural symmetry of these two protocols,
Li et al. introduced an MDI-QKA scheme with integrated identity authentication
\cite{LCW+23,KL23}.\\
In 2023, Chen et al. proposed a QIA scheme employing quantum rotation
and public-key cryptography \cite{CWJ+23}. In subsequent work,
the same researchers designed a QIA protocol using quantum homomorphic
encryption \cite{CWJ++23}. Additionally, Zhang et al. introduced
a three-party QSDC scheme with user authentication, utilizing a single-photon
source and implementing hyper-encoding techniques across different
degrees of freedom \cite{ZDZ+24}.\\
Most recently, Mawlia et al. proposed a QSS scheme incorporating user
identity authentication via mutually unbiased bases, where each participant
performs quantum Fourier transform (QFT) and unitary transformations.
This methodology ensures that during the secret recovery phase, the
confidential information remains undisclosed and is never directly
transmitted \cite{MSB+24}.
\end{description}
Before concluding this section, it is important to highlight that
authentication protocols come in several variations. One notable variant
is deniable authentication, which prevents a receiver from proving
the message's origin to a third party. For instance, Shi et al. introduced
a quantum protocol for deniable authentication \cite{SZY_2014}.
Additionally, most QIA protocols focus on two-party interactions,
typically involving a sender (Alice) and a receiver (Bob). However,
some schemes incorporate a third party, often called an authenticator
(Trent). For example, Lee et al. proposed two three-party QIA schemes
using GHZ states \cite{LLY_2006}. Later, Zhang et al. identified
that Trent could eavesdrop on both protocols but also provided a solution
to mitigate these vulnerabilities \cite{ZLWS_2007}. In \cite{LBK_2000},
a scenario involving three parties was analyzed, where an arbitrator
named Trent handled the authentication process. Additionally, it is
worth noting that the QIA protocols outlined here in chronological
order differ significantly in various technical aspects, such as the
inclusion or exclusion of a third party and the use of entangled versus
separable states. These differences provide a basis for categorizing
the protocols according to such criteria. In fact, the next section
focuses on classifying the existing schemes. However, prior to this
classification, we presented a chronological summary of the technical
features of several representative QIA protocols in Table \ref{tab:Chapter1_Tab1}.
This table not only highlights the uniqueness and progression of these
protocols but also suggests the potential for classification-based on the characteristics listed in the third column and those to its
right. The following section will expand on this idea, systematically
categorizing the QIA protocols.

\begin{table}
\caption{The classification of existing schemes for QIA is presented. When
the use of a pre-shared key is referred to as ``By QKD'', as indicated
by the authors of the respective studies, it can be interpreted that
a classical identity sequence was initially exchanged as a prerequisite
for performing QKD for the first time. The following abbreviations
are employed in this table: C=classical, B=Bell state, FC=five-particle
cluster state, CS=classical identity sequence, N=no, HF=single one-way
hash function, QPKC=quantum public key cryptography, Q=quantum, T=trusted,
SP=single photon, ST=semi-trusted, Y=yes, UT=un-trusted.}\label{tab:Chapter1_Tab1}

\centering{}%
\begin{tabular}{|>{\centering}p{1.25cm}|>{\centering}p{0.8cm}|>{\centering}p{1.75cm}|>{\centering}p{1.75cm}|>{\centering}p{1.5cm}|>{\centering}p{1.25cm}|>{\centering}p{1.4cm}|>{\centering}p{2.5cm}|}
\hline 
Protocol proposed in & Year & Quantum resources used & Pre-shared key used & Was there a third party? & Channel used & Is quantum memory used? & Computation or communication task used\tabularnewline
\hline 
\cite{DHHM_1999} & 1999 & SP & CS & N & C, Q & N & QKD\tabularnewline
\hline 
\cite{ZG_2000} & 2000 & B, SP & CS & N & C, Q & N & QSDC/DSQC\tabularnewline
\hline 
\cite{ZLG_2000} & 2000 & B, SP & B & N & Q & Y & QSDC/DSQC\tabularnewline
\hline 
\cite{LBK_2000} & 2000 & B, SP & By QKD & T & C, Q & Y & QKD\tabularnewline
\hline 
\cite{CS_01}\cite{CSPF_02} & 2001 & B, SP & B & N & Q & Y & QSDC/DSQC\tabularnewline
\hline 
\cite{T.Mihara_2002} & 2002 & B & B & T & C, Q & Y & QSS\tabularnewline
\hline 
\cite{LB_2004} & 2004 & B & B & N & Q & Y & QSDC/DSQC\tabularnewline
\hline 
\cite{ZZZZ_2005} & 2005 & B & B & T & C, Q & Y & Teleportation\tabularnewline
\hline 
\cite{ZZZX_2006} & 2006 & B, SP & By QKD & N & Q & Y & QSDC/DSQC\tabularnewline
\hline 
\cite{WZT_2006} & 2006 & GHZ & CS, HF & T & C, Q & Y & QSDC/DSQC\tabularnewline
\hline 
\cite{LLY_2006} & 2006 & GHZ & CS, HF & UT & C, Q & Y & QSDC/DSQC\tabularnewline
\hline 
\cite{GY_2009} & 2009 & GHZ & CS, HF & T & C, Q & N & QSS\tabularnewline
\hline 
\cite{DXXN_2010} & 2010 & B, SP & CS & N & C, Q & Y & QSDC/DSQC\tabularnewline
\hline 
\cite{YTXZ_2013} & 2013 & B & CS, HF & T & C, Q & Y & QSDC/DSQC\tabularnewline
\hline 
\cite{CXZY_2014} & 2014 & FC & CS & UT & Q & N & QSDC/DSQC\tabularnewline
\hline 
\cite{SZY_2014} & 2014 & GHZ & By QKD & T & C, Q & N & QSS \& MAC\tabularnewline
\hline 
\cite{YLP+14} & 2014 & SP & CS & N & Q & N & QSDC/DSQC\tabularnewline
\hline 
\cite{SZZY_2015} & 2015 & SP & By QKD & T & Q & N & QSS \& MAC\tabularnewline
\hline 
\cite{HCJ+17} & 2017 & SP & CS & N & C, Q & N & QSDC/DSQC\tabularnewline
\hline 
\cite{KHHYHM_2018} & 2018 & GHZ-like & CS & UT & C, Q & Y & QSDC/DSQC\tabularnewline
\hline 
\cite{LGXHZX_2019} & 2019 & SP & CS & N & Q & N & QKD\tabularnewline
\hline 
\cite{WZGZ_2019} & 2019 & GHZ-like, W & CS, HF & N & C, Q & Y & Teleportation\tabularnewline
\hline 
\cite{ZL_2019} & 2019 & FC & By QKD & T & C, Q & N & QSS\tabularnewline
\hline 
\cite{ZCSL_2020} & 2020 & B & CS & ST & C, Q & N & QSDC/DSQC\tabularnewline
\hline 
\cite{QLW_2020} & 2020 & GHZ-like & CS & N & C, Q & N & QECC\tabularnewline
\hline 
\cite{ZWZ_2020} & 2020 & SP & CS & N & C, Q & N & QSDC/DSQC\tabularnewline
\hline 
\cite{HPD22} & 2022 & B & CS & N & C, Q & Y & QSDC/DSQC\tabularnewline
\hline 
\cite{CWJ+23} & 2023 & SP & CS & N & Q & N & QPKC\tabularnewline
\hline 
\cite{ZDZ+24} & 2024 & SP & By QKD & N & C, Q & Y & QSDC\tabularnewline
\hline 
\end{tabular}
\end{table}

\section{Classification of the protocols for quantum identity authentication }\label{sec:Chapeter1_Sec5}

The classification of any concept depends on the selected criteria.
In this context, the existing schemes for QIA are categorized based
on two specific criteria: (i) the quantum resources utilized in the
design of the schemes (as detailed in the third column of Table \ref{tab:Chapter1_Tab1}
and (ii) the computational or communication tasks inherently employed
in their design (as described in the final column of Table \ref{tab:Chapter1_Tab1}).
It is important to note that this classification is neither exhaustive
nor exclusive. Alternative criteria, such as a technical feature highlighted
in another column of Table \ref{tab:Chapter1_Tab1}, could be chosen
for classification. The primary motivation for using these two criteria
is to uncover the underlying symmetry among existing QIA protocols
and to establish a foundation for developing new protocols.

\subsection{Classification based on the quantum resources used}\label{sec:Chapter1_Sec3.1}

When classified by quantum resources, the existing QIA schemes can
be divided into two main categories: (a) those utilizing entangled
states (e.g., \cite{LB_2004,WZT_2006,ZLG_2000,ZZZX_2006,ZCSL_2020,KHHYHM_2018,CXZY_2014,T.Mihara_2002})
and (b) those not relying on entangled states (e.g., \cite{Z19,HCJ+17,SZZY_2015,YLP+14,DHHM_1999,ZWZ_2020}).

\subsubsection{Entangled state based schemes of QIA}

Bell states represent the simplest form of entangled states and are
straightforward to generate and preserve. Consequently, numerous QIA
schemes have been developed based on Bell states. Protocols utilizing
solely Bell states are outlined in several studies, including \cite{ZLG_2000,ZG_2000,LBK_2000,CS_01,CSPF_02,T.Mihara_2002,LB_2004,ZZZZ_2005,ZZZX_2006,DXXN_2010,YTXZ_2013,AHKB_2017,ZCSL_2020}.
These protocols vary significantly, with some employing Bell states
for entanglement swapping (e.g., \cite{TJ_2014,WZT_2006}) and others
leveraging Bell states to design authentication schemes analogous
to the ping-pong protocol for QSDC \cite{ping-pong_BF_02}. Despite
their versatility, the full potential of Bell states, particularly
their non-local properties, remains underexplored. For instance, the
inherent symmetry discussed in this review could inspire the development
of device-independent or semi-device-independent QIA schemes. Additionally,
QIA protocols exploiting violations of Bell inequalities could be
proposed, though their implementation would likely be more challenging
than existing schemes while offering distinct advantages. Some QIA
schemes also rely on multipartite entangled states. For example, three-qubit
GHZ states are employed for authentication in studies such as \cite{WZT_2006,LLY_2006,GY_2009,TJ_2014,SZY_2014}.
Similarly, GHZ-like states have been used for QIA in \cite{KHHYHM_2018,WZGZ_2019},
while five-qubit cluster states are utilized in \cite{CXZY_2014}.
However, the creation and stabilization of such multipartite entangled
states remain technically demanding. Furthermore, many entanglement-based
QIA schemes necessitate quantum memory, a technology that is currently
unavailable.

\subsubsection{Schemes of QIA which don't use entanglement}

Several QIA schemes utilize separable states, specifically single-photon
states. Examples include the protocols proposed in \cite{Z19,HCJ+17,SZZY_2015,YLP+14,DHHM_1999,ZWZ_2020},
among others. These approaches hold an advantage over entangled-state-based
schemes due to the relative simplicity of preparing single-qubit states.
Furthermore, most entangled-state-based protocols require quantum
memory for storing home qubits, a technology that remains unavailable,
whereas single-qubit-based schemes generally do not depend on quantum
memory. This characteristic makes single-qubit-based protocols more
practical for implementation. Such schemes often employ BB84-style
encoding, where travel qubits are prepared in the $Z=\{\vert0\rangle,\vert1\rangle\}$
basis or the $X=\{|+\rangle,|-\rangle\}$ basis. The selection of
the basis is determined by a pre-shared authentication key and a predefined
rule linking the key's bit values to the corresponding basis for travel
qubit preparation. A notable example of this type is Zawadzki's QIA
protocol \cite{Z19}. However, Gonz{\'a}lez-Guill{\'e}n et al. \cite{GCM+21}
recently demonstrated that Zawadzki's scheme is fundamentally insecure.
They showed that the logarithmic security claimed for the protocol
is compromised under a key space reduction attack, as explicitly described
in their work. Many such schemes incorporate hash functions to secure
the authentication process. While relying on hash functions moves
away from strictly quantum-based security, current cryptographic confidence
in specific hash functions remains high. In Chapter \ref{Ch2:Chapter2_Authentication}, we propose
two novel QIA protocols based on single qubits. These protocols address
vulnerabilities, including key space reduction attacks and other known
strategies, thereby enhancing their robustness.

\subsection{Classification of the protocols based on the computational or communication
tasks used to design the schemes }\label{sec:Chapter1_Sec3.2}

Here, we categorize existing protocols for QIA based on the computational
or communication tasks required for their implementation. To begin,
we will outline QIA schemes that rely on specific quantum communication
tasks.

\subsubsection{Protocols based on the schemes for QKD}

Numerous authentication protocols are fundamentally rooted in QKD
schemes. In principle, any QKD scheme can be adapted into a QIA scheme,
provided a pre-shared secret key is available. However, slight modifications
to the original QKD protocol may be necessary to achieve this adaptation.
For instance, Sobota et al. \cite{SKB11} demonstrated how the BB84
QKD protocol could be modified to establish a QIA scheme. More recently,
Liu et al. \cite{LGXHZX_2019} proposed a QIA scheme by altering
a counterfactual QKD protocol. In earlier research by Du{\v{s}}ek
et al., a hybrid classical-quantum authentication scheme was introduced,
where the quantum component was derived from QKD. Specifically, this
approach leveraged the fact that QKD schemes are inherently designed
for key amplification. In this method, Alice and Bob begin with a
``pre-shared sequence'' for identity authentication, which is used only
once to authenticate their identities and generate a secure key through
QKD. In subsequent rounds of identity authentication, a portion of
the key generated by QKD is employed. While this method primarily
uses QKD to refresh the authentication sequence, it does not fully
qualify as a QIA scheme. Nonetheless, this early work was significantly
influenced by QKD. Today, the integration of QKD in authentication
processes is widely observed.

\subsubsection{Protocols based on the schemes for QSDC and DSQC }\label{subsec:Chapter1_Sec3.2.2}

Secure quantum direct communication protocols are categorized into
two types: QSDC and DSQC. Numerous QSDC and DSQC protocols have been
proposed (see \cite{STP20,STP17,YSP14,BP12,SP14}, among others),
all enabling the direct transmission of information from a sender
(Alice) to a receiver (Bob) with unconditional security. When Alice
and Bob share a ``pre-shared secret'', Alice can transmit part or all
of this secret using a QSDC or DSQC protocol. Bob, in turn, can decode
the message and compare it with his own secret to authenticate Alice's
identity. Consequently, any QSDC or DSQC protocol can, in principle,
be adapted for QIA. An example of an early QSDC protocol is the ping-pong
protocol, which Zhang et al. (in 2006) extended to develop a QIA scheme
\cite{ZZZX_2006}. It is noteworthy that even ``secure direct
quantum communication'' protocols unsuitable for long-distance communication
due to noise can still serve as the foundation for QIA. In the case
of CV-secure direct quantum communication, this adaptability was highlighted
in \cite{PBM+08}. CV-QIA schemes, whether derived from CV-QSDC
or CV-DSQC protocols, leverage existing infrastructure for their implementation,
making them particularly advantageous as indicated in \cite{PBM+08}.

LM05, introduced by Lucamarini and Mancini in 2005 \cite{LM_05},
is a single-photon-based protocol for QSDC, conceptually similar to
the ping-pong protocol. Building upon LM05, Yuan et al. (2014) proposed
another QIA scheme \cite{YLP+14}, describing it as a non-entangled
ping-pong protocol. However, it is essentially a modified version
of the LM05 protocol. Notably, neither Zhang et al., Yuan et al.,
nor subsequent authors have acknowledged the fundamental symmetry
that enables the transformation of existing QSDC and DSQC protocols
into QIA schemes.

The transformation of QSDC or DSQC protocols into schemes tailored
for QIA often necessitates certain modifications that move the resulting
protocol beyond the conventional definitions of QSDC or DSQC. Consequently,
what we achieve are QSDC- or DSQC-inspired protocols for QIA. For
instance, the structure of Zhang et al.'s scheme \cite{ZLG_2000}
resembles the original ping-pong protocol but differs in that the
travel qubit sent by Bob to Alice is not subsequently returned by
Alice. Specifically, in Zhang et al.'s scheme, a Bell state $\psi_{AB}^{+}$($=\frac{1}{\sqrt{2}}(\vert0\rangle_{A}\vert1\rangle_{B}+\vert1\rangle_{A}\vert0\rangle_{B})$)
is pre-shared between Alice and Bob, where the subscripts $A$ and
$B$ denote the qubits held by Alice and Bob, respectively. In the
original ping-pong protocol, Bob generates the Bell state and transmits
one of its qubits (the travel qubit) to Alice, which may be considered
an analogous step. In Zhang et al.'s scheme, however, Bob makes
any single-qubit pure state $|\psi_{1}\rangle=\alpha|0\rangle+\beta|1\rangle$
and sends it to Alice. Alice then performs a CNOT operation, using
her portion of the Bell state as the control qubit and $|\psi_{1}\rangle$
as the target qubit. Alice subsequently returns the qubit labeled
as qubit $1$ (establishing the analogy with the ping-pong protocol).
Upon receiving qubit $1$, Bob applies a CNOT operation, with his
share of the Bell state (qubit $B$) serving as the control qubit
and qubit 1 as the target qubit. Bob then measures qubit $1$ in a
basis where $|\psi_{1}\rangle=\alpha|0\rangle+\beta|1\rangle$ is
a basis element. If the measurement yields $|\psi_{1}\rangle$, the
authentication is deemed successful. This type of QSDC-inspired protocol
garners additional attention for two specific reasons. Firstly, this
work introduces novel QSDC/DSQC-based protocols. Secondly, several
QSDC/DSQC-based protocols for QIA have recently been developed, many
of which can be classified within the ping-pong protocol framework.
For instance, Li et al. proposed a QIA protocol in 2004 \cite{LB_2004}
that shares similarities with Zhang et al.'s scheme, but with a key
difference: instead of Bob preparing and sharing a single-qubit state
$|\psi_{1}\rangle$, Alice prepares an auxiliary Bell state and executes
a CNOT operation. In this operation, the first qubit of the auxiliary
Bell state acts as the control qubit, while Alice's qubit (qubit $A$)
from the pre-shared Bell state serves as the target qubit. Alice then
transmits the auxiliary Bell pair to Bob, who subsequently performs
a CNOT operation, using the second qubit of the auxiliary pair as
the control qubit and his qubit (qubit $B$) as the target. Authentication
is deemed successful if Bob's Bell measurement on the auxiliary Bell
state yields the same state initially prepared by Alice. Both protocols
necessitate quantum memory; however, commercial quantum memory devices
are not yet available, and the current laboratory-grade quantum memory
technologies require significant advancement before they can be deployed
in practical applications.

Moreover, Zhang et al. \cite{ZZZX_2006} introduced a one-way QIA
protocol, which bears similarities to the protocol proposed by Zhang
et al. \cite{ZLG_2000}. This was followed by Li et al.'s protocol
\cite{LC_2007}, which can be regarded as a modified version of their
earlier work \cite{LB_2004}. In this updated version, the auxiliary
state is independently prepared by Alice and Bob in single-qubit states
instead of Bell states. From the overview of these protocols, it is
evident that they all belong to the ``ping-pong'' family of QIA
protocols. This family also includes Yuan et al.'s single-qubit-based
protocol \cite{YLP+14}.

\subsubsection{Protocols based on teleportation}

Teleportation is recognized as a method for enabling secure communication,
provided an ideal (noise-free) quantum channel is available, as noted
by Lo and Chau \cite{LC99}. This particular challenge, along with
related concerns, will be explored in further detail in Section \ref{subsec:Protocols-based-on-QECC}.
Despite these challenges, including issues associated with entanglement
concentration \cite{LMR+23,BSB+16} and purification, numerous researchers
have proposed QIA schemes leveraging teleportation. Examples include
the works of Tan and Jiang (2014), Ma et al. (2016), Zhou et al. (2005)
\cite{TJ_2014,MHBZ_2016,ZZZZ_2005}, among others cited therein.
Notably, a teleportation-based approach has also been suggested for
CV-QIA by Ma et al. \cite{MHBZ_2016}. However, the effectiveness
of these schemes is constrained by the susceptibility of shared entanglement
to noise, which necessitates entanglement purification or concentration
to restore the desired entangled state. This process requires interaction
between Alice and Bob prior to authentication. It should be emphasized
that this issue also affects a wide range of QIA protocols that rely
on shared entanglement but do not explicitly involve teleportation,
such as those discussed in \cite{ZLG_2000} and similar approaches.
The need for such interaction is a notable limitation of QIA protocols
that rely directly on teleportation or shared entanglement. Nevertheless,
innovative techniques, like the approach presented by Barnum et al.
\cite{BCC02}, offer potential solutions to address this limitation.
A brief discussion of these techniques can be found in Section \ref{subsec:Protocols-based-on-QECC}.

\subsubsection{Protocols based on quantum secret sharing}

Quantum secret sharing (QSS) protocols \cite{BDP+23} have been adapted
to develop schemes for QIA, allowing for the authentication of legitimate
users either individually or collectively by a group of users or a
trusted third party \cite{YWZ_08}. Furthermore, various QSS schemes
incorporating identity authentication to enhance security have been
introduced \cite{AHBR_16,AHKB_2017,shi2019useful,YC_21}. These schemes
employ a range of quantum resources and operations, such as Bell states,
phase-shift operations, and GHZ states. Notably, Abulkasim et al.
explored the simultaneous integration of QSS and QD concepts to develop
protocols for mutual QIA \cite{AHBR_16}, as also discussed in Banerjee
et al. (2017) and Shukla et al. (2013) \cite{BST+17,SKB+13}. In
the multiparty version of the protocol by Abulkasim et al., Alice
was designated as the boss, with Bob and Charlie acting as her agents.
However, Gao et al. identified a security vulnerability in this scheme
\cite{GWWY_18}, demonstrating that collusion between agents (Bob
and Charlie) compromised its security. In response to this issue,
Abulkasim et al. introduced an improved QSS-based protocol in 2018,
designed to withstand collusion attacks by agents \cite{AHE_18}.

The protocols reviewed and categorized in this section are primarily
derived from those developed for various quantum communication applications,
such as quantum teleportation, QKD, QSDC, and DSQC. Subsequently,
it will be shown that similar derivations can also be made from protocols
intended for quantum computing tasks.

\subsection*{Protocols based on quantum computation tasks}

As discussed earlier, existing protocols for various quantum computing
tasks can be adapted to develop schemes for QIA, and this approach
has been employed since the inception of QIA. Notably, the foundational
effort by Cr{\'e}peau et al. \cite{CS_1995} in designing QIA utilized
OT. We begin by acknowledging this contribution and proceed to examine
schemes that have been developed---or could potentially be developed---by
leveraging protocols designed for other quantum computing tasks, such
as ``blind quantum computing'' and ``quantum private comparison''.

\subsubsection{Protocols Utilizing oblivious transfer}

As outlined in Section \ref{sec:Chapter1_Sec3}, the first QIA protocol,
proposed by Cr{\'e}peau and Salvail (in 1995) \cite{CS_1995}, was
based on OT. This protocol aimed to verify the identities of legitimate
participants by comparing their mutual knowledge of pre-shared common
information. This pioneering work introduced a QIA scheme whose security
relied on the fundamental principles of quantum mechanics, a foundation
that has since been expanded upon by many subsequent efforts. Indeed,
the security of all QIA protocols reviewed here stems from quantum
mechanical properties. However, OT has been infrequently employed
as a resource in designing QIA schemes.

\subsubsection{Protocols based on blind quantum computing}

Blind quantum computing (BQC), introduced by Childs in 2005 \cite{childs_2005},
allows a client with limited quantum resources to delegate computational
tasks to a quantum server or user with greater quantum capabilities.
Crucially, this delegation ensures the privacy of the client's input,
output, and computational process, such that the more powerful
quantum server remains unaware of these details. Various BQC schemes
incorporating identity authentication have been proposed, including
works by Li et al. (2018), Quan et al. (2021), and Shan et al. (2021)
\cite{LLCZL_18,QLLSP_21,SCY_21}. However, the potential of BQC in
designing QIA schemes remains underutilized. A closely related concept,
blind quantum signatures, was explored in Li et al. (in 20127) \cite{LSG_17})
using BQC principles. By mapping this approach to QIA, there is significant
potential to develop QIA schemes by leveraging both existing and novel
BQC frameworks. This represents a promising area for further investigation.

\subsubsection{Protocols based on quantum error detection or correction code }\label{subsec:Protocols-based-on-QECC}

Authentication can be achieved by employing a shared entangled state
and a private quantum key, which Alice teleports to Bob using the
shared entanglement to verify her identity. However, the entangled
state may degrade due to noise, necessitating a purity-testing protocol
to confirm whether the shared entangled state remains in the intended
configuration. Unlike entanglement concentration and purification
schemes, such protocols focus on verification rather than correction.
Barnum et al. in 2002 observed that purity testing protocols are typically
interactive \cite{BCC02}, a feature undesirable in QIA and quantum
message authentication when one-way authentication is preferred. To
address this, Barnum et al. demonstrated that purity-testing could
be implemented using a family of quantum error-correcting codes (QECCs)
characterized by their ability to detect any Pauli error in most codes
within the family. They further established that secure non-interactive
quantum authentication schemes could be derived from purity-testing
protocols constructed from QECCs. This marked a significant milestone
in using QECCs for quantum authentication. Building on this foundation,
Qu et al. later introduced a QIA protocol leveraging a three-photon
``error-avoidance code'' \cite{QLW_2020,QLW19}.

\subsubsection{Protocols based on quantum private comparison}

In 2017, Hong et al. introduced a highly efficient single-qubit protocol
for QIA, incorporating decoy qubits to ensure security against eavesdropping
\cite{HCJ+17}. The efficiency of the protocol stemmed from its ability
to verify two bits of authentication information using just a single
qubit. Subsequently, Zawadzki conducted a cryptanalysis of this protocol
and proposed enhancements \cite{Z19}. Zawadzki's primary critique
was that each execution of Hong et al.'s protocol resulted in information
leakage, allowing an eavesdropper to accumulate data over multiple
protocol runs. To address this issue, Zawadzki presented an improved
protocol that incorporated a hash function for added security. However,
recent research by Gonz{\'a}lez-Guill{\'e}n et al. (in 2021) \cite{GCM+21}
revealed that Zawadzki's protocol is vulnerable to key space reduction
attacks. While a detailed examination of this vulnerability is beyond
the scope of this discussion, it is worth noting that in Zawadzki's
protocol, as well as many other QIA protocols, the core objective
is to perform quantum private comparison. This involves computing
a function $f(A,B)$, where $f(A,B)=0$ if $A\neq B$ and $f(A,B)=1$
if $A=B$, using quantum resources without revealing the actual values
of $A$ and $B$ to the party performing the computation. Naturally,
some logical information can be inferred from the computed value of
$f(A,B)$. This issue is a well-established challenge in secure multiparty
computation, as discussed in \cite{SaTP20,STP20,CGS02}. The task
is also strongly connected to the socialist millionaire problem, as
noted by Shukla et al. \cite{SKB+13}. To better understand this,
we examine Zawadzki's protocol in greater detail. In Zawadzki's protocol,
as well as in many other similar protocols, Alice and Bob share a
secret authentication key $k$. Alice begins by generating a random
number $r_{A}$ and computes a session secret $s_{A}=H(k,r_{A})$
using the shared secret key $k$ and a hash function $H$. Alice then
sends $r_{A}$ to Bob, who may receive a different value $r_{B}$.
Bob uses $r_{B}$ to compute his own session secret $s_{B}=H(k,r_{B})$.
The task subsequently reduces to leveraging quantum resources to determine
whether $s_{A}=s_{B}$, effectively performing a quantum private comparison
by evaluating $f(s_{A},s_{B}).$. If $s_{A}$ and $s_{B}$ are equal,
successful authentication is achieved. Recognizing the connection
between quantum private comparison and QIA allows us to infer that
existing protocols for quantum private comparison and the socialist
millionaire problem can be transformed into QIA protocols. Protocols
proposed by some of the recent authors and detailed in Shukla et al.
(2013), Shukla et al. (2017), Banerjee et al. (2017), Saxena et al.
(2020), and Thapliyal et al. (2018) can be straightforwardly adapted
into schemes for QIA \cite{SKB+13,STP17,BST+17,SaTP20,thapliyal2018orthogonal}.
Notably, several of these adapted QIA schemes remain functional within
the semi-quantum framework \cite{STP17,thapliyal2018orthogonal},
wherein not all participants require access to quantum resources.
Additionally, it is worth highlighting that the ability to perform
QD, enabling simultaneous two-way communication between Alice and
Bob, or quantum conferencing \cite{BTS+18}, inherently enables QSDC
or DSQC, which can be leveraged for QIA (as elaborated in Section
\ref{subsec:Chapter1_Sec3.2.2}). Consequently, without converting
QD or quantum conference protocols into quantum private comparison
schemes, one can adopt a more direct approach by designing novel QSDC/DSQC-based
schemes for QIA.

\begin{figure}[h]
\begin{centering}
\includegraphics[scale=0.45]{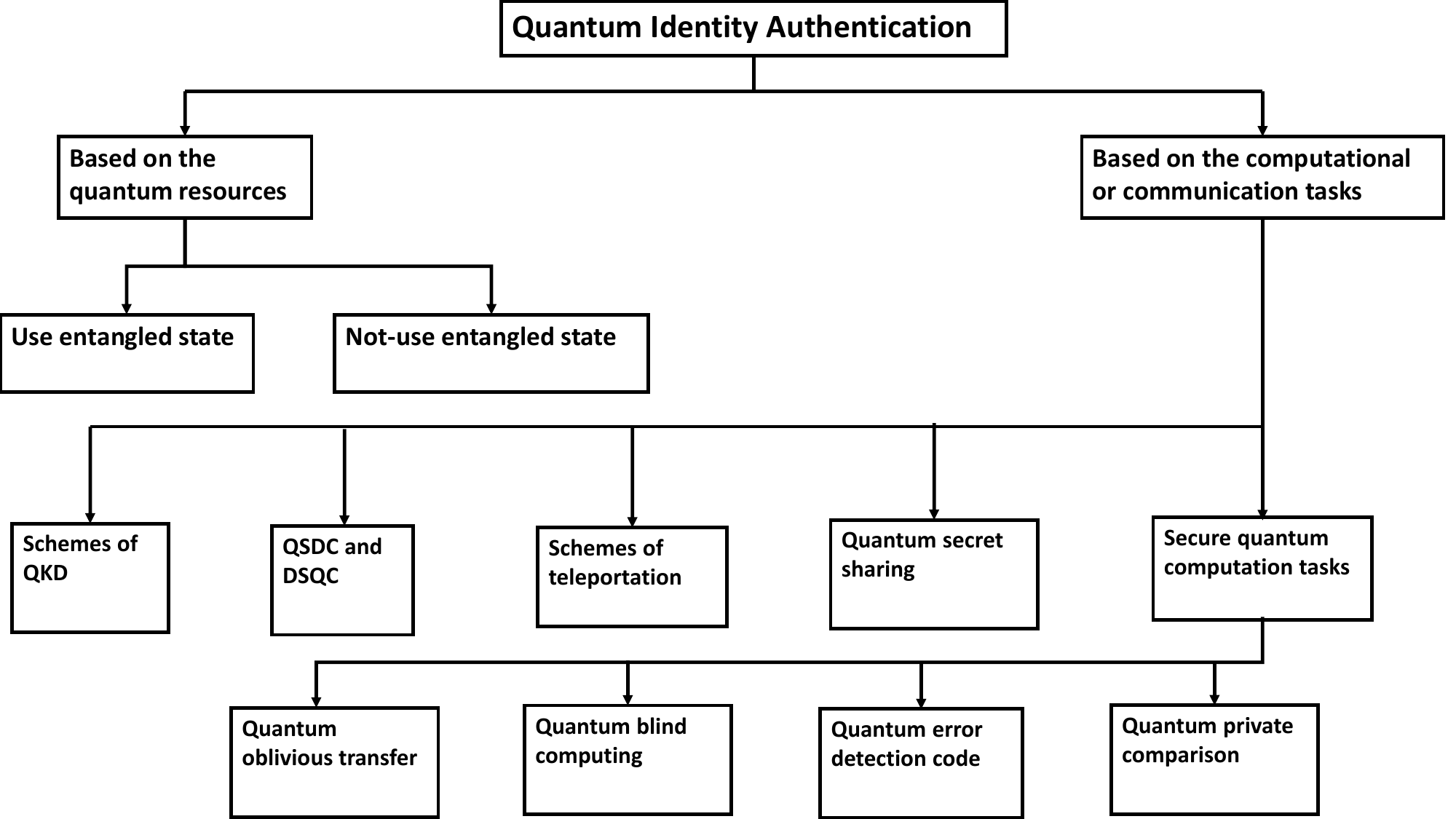}
\par\end{centering}
\caption{Categorization of QIA schemes.}\label{fig:Chapter1_Fig2}
\end{figure}

The classification of schemes for QIA discussed in this section is
concisely depicted in Fig. (\ref{fig:Chapter1_Fig2}). This figure
demonstrates the diverse methodologies and resources available for
implementing QIA. The selection of a specific scheme is influenced
by the task's requirements and the resources at hand. For example,
existing schemes can be categorized based on factors such as the necessity
of classical communication channels, quantum memory, the involvement
of a third party, the type of pre-shared key (classical or quantum),
and whether one-way or two-way authentication is desired.

It is important to highlight that Table \ref{tab:Chapter1_Tab1} serves
as a complement to Figure \ref{fig:Chapter1_Fig2}, detailing the
placement of representative schemes within the classification framework
depicted in the figure. Each row of Table \ref{tab:Chapter1_Tab1}
provides comprehensive information about a specific scheme, thereby
encapsulating the discussion of this section and substantiating the
proposed classification.


\section{Quantum key distribution following quantum identity authentication}\label{sec:Chapter1_Sec6}

Cryptography has been a critical and invaluable tool for humanity
since the dawn of civilization. Historically, cryptographic techniques
have been employed to conceal confidential information, though cryptanalysts
have frequently developed more advanced methods to decrypt such messages.
A significant transformation in cryptography occurred in the 1970s
with the advent of public-key cryptography, exemplified by schemes
such as RSA \cite{RSA78} and the DH protocol \cite{DH76}. The
security of these and similar classical key distribution methods is
fundamentally based on the computational difficulty of specific mathematical
problems integral to their design. For instance, the RSA scheme derives
its security from the complexity of factoring large semi-prime numbers,
while the DH protocol relies on the computational challenge of solving
the discrete logarithm problem \cite{P13}. In a groundbreaking study
published in 1994, Peter W. Shor \cite{S94} demonstrated that both
the factorization of large semi-prime numbers and the discrete logarithm
problem could be efficiently addressed using quantum computers, specifically
in polynomial time. This revelation highlights the potential vulnerabilities
in many classical key distribution methods should scalable quantum
computers become a reality. Consequently, cryptography is significantly
challenged by the advent of quantum computers, or more precisely,
by quantum algorithms that outperform classical algorithms in solving
various computational problems. Notably, a resolution to this challenge
has long existed in the form of QKD. Unlike classical methods, QKD
leverages quantum resources for key distribution, ensuring security
that relies on the fundamental principles of quantum mechanics rather
than the computational intractability of certain problems. The earliest
QKD protocol was introduced a decade prior to Shor's groundbreaking
work, which challenged the security of classical cryptography. Specifically,
Bennett and Brassard \cite{BB84} developed the first QKD scheme,
leveraging fundamental physical principles such as the no-cloning
theorem \cite{WZ82}, the ``collapse on measurement'', and Heisenberg's
uncertainty principle. These principles underpin the security of the
protocol, which is based on single qubits and can be implemented using
polarization-encoded single photons or other forms of photonic qubits.
Notably, under ideal conditions, any eavesdropping attempt in a QKD
protocol leaves a detectable trace. However, in practical scenarios,
device imperfections may enable undetected eavesdropping.

Following the introduction of the BB84 protocol, numerous QKD protocols
were proposed, including those by \cite{B92,E91,GV95,STP20}, along
with protocols addressing related cryptographic applications (e.g.,
\cite{YSP14,SKB+13,LXS_18,BST+17,STP17,thapliyal2018orthogonal,TP15,STP20,DP+23}).
For comprehensive reviews, refer to Refs. \cite{SPR17,GRT+02}.
These protocols present various benefits and limitations. While most
are theoretically unconditionally secure, practical implementations
often encounter imperfections in devices, which introduce vulnerabilities
exploitable via quantum hacking techniques \footnote{Quantum identity authentication, as discussed by Curty and Santos,
Dutta and Pathak \cite{CS_01,DP22,DP23}, is a fundamental prerequisite
for ensuring secure communication prior to the implementation of a
QKD protocol.}. For instance, the BB84 protocol, along with similar protocols like
B92 \cite{B92}, theoretically necessitates the use of a single-photon
source, as Alice needs to transmit single-photon states to Bob. Significant
experimental progress has been made in developing reliable single-photon
sources (refer to \cite{LP21,TS21}), but commercially available
systems typically rely on weak coherent pulses (WCPs), produced by
attenuating laser outputs \cite{BBL+22}, as approximate single-photon
sources. The quantum state of such WCPs is represented as follows:

\begin{equation}
|\alpha\rangle=|\sqrt{\mu}\exp(i\theta)\rangle=\sum_{n=0}^{\infty}\left(\frac{e^{-\mu}\mu^{n}}{n!}\right)^{\frac{1}{2}}\exp(in\theta)|n\rangle,\label{eq:coherent}
\end{equation}
Alice prepares a quantum state characterized by a superposition of
Fock states, where $|n\rangle$ represents an $n$-photon Fock state,
and the mean photon number is $\mu=|\alpha|^{2}\ll1$. The photon
number follows a Poissonian distribution, given by $p(n,\mu)=\frac{e^{-\mu}\mu^{n}}{n!}$.
Consequently, the probability of generating a single-photon state
is $p(1,\mu)$, while the probability of emitting multi-photon pulses
is $1-p(0,\mu)-p(1,\mu)$. The presence of multi-photon states introduces
a vulnerability, as an adversary, Eve, could exploit this to execute
a photon-number-splitting (PNS) attack, as discussed by Huttner et
al. \cite{HIG+95}. Furthermore, in long-distance quantum communication,
channel loss presents a critical security risk. An eavesdropper with
advanced technological capabilities may replace the lossy channel
with an ideal, lossless one, facilitating an undetectable interception
attack, as demonstrated by Brassard et al. \cite{BLM+2000}. To mitigate
this vulnerability, Scarani et al. \cite{SAR+04} introduced the
SARG04 QKD protocol, which enhances security against PNS attacks.
In Chapter \ref{Ch3:Chapter3_QKD}, we propose two novel QKD protocols designed to be resilient
against PNS attacks, akin to SARG04, while also offering improved
resistance to a broader class of adversarial strategies. These protocols
provide distinct advantages over SARG04 and similar QKD schemes in
terms of security and implementation.

In every QKD protocol, an inherent process of information partitioning
occurs. For instance, in the BB84 \cite{BB84} and B92 \cite{B92}
protocols, the transmitted information is divided into a classical
component (detailing the basis used for qubit preparation) and a quantum
component (the transmitted qubits themselves). A similar partitioning
mechanism is observed in the SARG04 protocol \cite{SAR+04}. However,
in certain other protocols, such as the GV protocol \cite{GV95},
the information is distributed across two quantum components. The
security of these protocols fundamentally relies on the impossibility
of an eavesdropper (Eve) simultaneously accessing multiple information
fragments. This observation raises a fundamental question: can the
efficiency and/or secret-key rate bounds of a protocol be modified
by altering the protocol in a manner that reduces the information
present in the classical component? To explore this, we consider the
SARG04 protocol as a reference framework. In Chapter \ref{Ch3:Chapter3_QKD}, we will present
two novel QKD protocols that share similarities with the SARG04 protocol
but incorporate a reduced amount of classical information compared
to SARG04. The SARG04 protocol was originally developed to significantly
mitigate the probability of a PNS attack, as discussed by Huttner
et al. \cite{HIG+95}. However, it exhibited lower efficiency\footnote{The efficiency assessment follows Cabello's framework \cite{C2000},
wherein the transmission cost of qubits is equated to that of classical
bits. This model assumes a quantum channel with minimal noise, which
may not always align with practical scenarios, particularly in long-distance
communication with current technological constraints.} relative to various other single-photon-based QKD schemes. This limitation
motivated our investigation into an alternative approach, wherein
PNS attacks are countered by employing a greater reliance on quantum
resources rather than classical ones, thereby circumventing the current
technological constraints such as channel loss and channel noise.
Accordingly, Chapter \ref{Ch3:Chapter3_QKD} will introduce two new QKD protocols that enhance
efficiency beyond SARG04 while maintaining resilience against PNS
attacks and a range of other established attack strategies.


\section{Introductory discussion on quantum key agreement}\label{sec:Chapter1_Sec7}

Key agreement (KA) constitutes a critical area of research in cryptography
and plays a fundamental role in ensuring perfect forward secrecy.
A KA protocol enables two or more parties to collaboratively establish
a key---essentially generating a shared random number---such that
all legitimate participants contribute to the outcome. Moreover, neither
a subset of legitimate parties nor an eavesdropper should be capable
of imposing a predetermined key. This represents a relatively weaker
characterization of KA. A more rigorous definition, frequently adopted
in the literature on QKA, requires that all legitimate parties exert
equal influence over the final key. The concept of KA was first introduced
by Diffie and Hellman in 1976 through a classical two-party protocol
\cite{DH76}. Subsequent advancements extended the Diffie-Hellman
framework to multiparty scenarios, enhancing the efficiency of KA
protocols \cite{IT+82,MW+98,ST+2000,AS+2000}. Adhering to the stronger
definition of KA necessitates that each legitimate participant contributes
equally to the establishment of a common secret key. This characteristic,
which ensures that no individual or subset of participants can disproportionately
influence the final key, is commonly referred to as ``fairness''
\cite{HWL+14,HSL+17}.

The security of classical cryptographic techniques, including KA,
primarily relies on the computational complexity of certain mathematical
problems. However, the advent of quantum computing poses a significant
threat to this security paradigm, as many of these hard problems,
which form the foundation of classical cryptographic schemes, could
potentially be solved in polynomial time with scalable quantum computers.
To address this challenge, various principles of quantum mechanics---such
as collapse on measurement, the no-cloning theorem, nonlocality, Heisenberg's
uncertainty principle, and contextuality---have been leveraged. It
has been demonstrated that numerous cryptographic tasks can achieve
unconditional security through the appropriate utilization of quantum
resources. One such task is QKA, as introduced by Hahn et al. \cite{HJP20}.
In Chapter \ref{Ch4:Chapter4_QKA}, we propose two novel QKA protocols and present their
security proofs. Before delving into QKA and its multipartite extension,
it is beneficial to briefly discuss other aspects of quantum cryptography.
This discussion will highlight the distinctive nature of QKA while
providing a broader perspective on various quantum cryptographic techniques.

In 1984, Bennett and Brassard introduced the renowned BB84 protocol
\cite{BB84}, which effectively facilitates key distribution between
two distant parties under ideal conditions by leveraging quantum resources.
The security of BB84 is fundamentally rooted in the principles of
quantum mechanics. Following this breakthrough, QKD emerged as a rapidly
advancing field that garnered significant attention in cryptographic
research, progressing swiftly in both theoretical and experimental
domains\cite{E91,BBM92,B92,SAR+04,P13,SPR17,DP+23,LCQ12,CXC+14,CPC+09}.
Moreover, several other essential aspects of quantum cryptography
have been explored, including QSDC \cite{ping-pong_BF_02,DL+03},
QSS \cite{HB+99,KK+99,MSP+15,LLW+23}, QIA \cite{DHHM_1999,DP22,DP23},
quantum private comparison \cite{YW09,YC+09,CX+10,KSP18}, and quantum
digital signatures \cite{GC01,DW+14,WD+15}. In addition to these
advancements, QKA has emerged as a crucial cryptographic primitive,
enabling two or more participants to collaboratively establish a shared
key by contributing their individual key components through quantum
resources. QKA's significance lies in its diverse applications, such
as electronic auctions, secure multiparty computation, and access
control \cite{ST+2000}, as well as its role in ensuring forward
secrecy. Given its applicability in various domains, QKA has recently
been recognized as an integral area within quantum cryptography.

To the best of our understanding, the first QKA protocol was introduced
by Zhou et al. in 2004, where two participants utilized quantum teleportation
to establish a common key \cite{ZZX04}. However, in 2009, Tsai and
Hwang identified a critical flaw in Zhou et al.'s protocol, demonstrating
that one participant could unilaterally determine the shared key and
subsequently distribute it to the other, thereby compromising fairness.
Moreover, the security of Zhou et al.'s scheme was later invalidated
by Chong et al. in 2011 \cite{CT+11}. Subsequently, Shi and Zhou
proposed the first multiparty QKA (MQKA) protocol in 2013, leveraging
entanglement swapping \cite{SZ13}. However, this scheme was also
proven insecure \cite{LG+13}. In the interim, Chong et al. developed
a QKA protocol based on the BB84 framework, incorporating delay measurement
techniques along with an authenticated classical channel \cite{CH10}.
Over time, several two-party QKA protocols were extended to multiparty
settings \cite{LG+13,SA+14,XW+14,HM15}. These MQKA protocols are
generally classified into three categories \cite{LX+16}: tree-type
\cite{XW+14}, circle-type (\cite{LZ+21}, and references therein),
and complete-graph-type \cite{LG+13}. Among them, the circle-type
structure is considered the most practical for implementation due
to its greater efficiency.

A robust QKA protocol must fulfill three essential criteria: \textit{Correctness},
ensuring that all participants obtain the agreed-upon key accurately
upon the protocol's successful execution; \textit{Security}, guaranteeing
that no unauthorized entity can intercept or extract information about
the final agreement key without detection; and \textit{Fairness}, requiring
that all legitimate participants contribute equally to the key generation,
preventing any subset of them from independently determining the final
key. Based on these principles, we introduce, for the first time,
a QKA protocol incorporating a central controller, Charlie, to facilitate
key agreement between two legitimate parties without relying on quantum
memory. Our proposed scheme in Chapter \ref{Ch4:Chapter4_QKA} is specifically designed
to meet the demands of quantum cryptographic applications. Such scenarios
frequently arise in quantum cryptography, prompting the development
of various controlled quantum cryptographic schemes in recent years.
In particular, extensive research has been conducted on controlled-QKD
\cite{DGL+03,SNG+09}, controlled-quantum dialogue \cite{DXG+08,KH17},
and controlled-secure direct quantum communication \cite{WZT06,CWG+08,LLL+13,DP23}.
However, to the best of our knowledge, no existing controlled quantum
key agreement (CQKA) scheme has been proposed without relying on quantum
memory \cite{TS+20}. This motivates us to introduce a CQKA protocol
that leverages Bell states and single-qubit states as quantum resources.
Our scheme employs a one-way quantum channel to mitigate unnecessary
noise, distinguishing it from several QKA protocols that rely on two-way
quantum channels \cite{SA+14,HM15,TS+20}, which are more susceptible
to channel-induced noise. We rigorously analyze the security of the
proposed protocol and demonstrate its feasibility using current quantum
technologies. Furthermore, comparative analysis reveals that the proposed
CQKA scheme exhibits superior efficiency over many existing approaches
in Chapter \ref{Ch4:Chapter4_QKA}. The inclusion of a controller further enhances its applicability
in specific scenarios.


\section{Discussion on quantum game and Nash equilibrium}\label{sec:Chapter1_Sec8}

Game theory analyzes and models the behavior of individuals in scenarios
that require strategic reasoning and interactive decision-making.
It plays a fundamental role in optimizing decision-making processes
and evaluating potential outcomes across various domains, including
business and daily life. Situations demanding strategic analysis are
widespread in disciplines such as economics \cite{G92}, political
science \cite{O86}, biology \cite{C13,NS99}, and military strategy
\cite{D59,H97}. In such contexts, participants select from a set
of possible actions, termed strategies, with their preferences structured
within a payoff matrix. Game theory focuses on modeling these interactions
and determining optimal strategic choices. Among its fundamental principles,
the Nash equilibrium holds particular significance, as it describes
a state where no participant can enhance their payoff by unilaterally
changing their strategy, given the decisions of others \cite{N50,N51}.

Quantum mechanics remains one of the most profound and impactful theories
in scientific history. Although its emergence was met with controversy,
its theoretical predictions have been consistently validated through
precise experimental verification \cite{AGR82}. Quantum game theory
facilitates the study of interactive decision-making processes where
players employ quantum resources. This framework serves a dual function:
acting as a quantum communication protocol while also offering a more
effective means of randomizing strategic choices compared to conventional
game-theoretic approaches \cite{L11}. The formalization of quantum
game theory dates back to 1999, with pioneering contributions from
David Meyer \cite{M99} as well as Jens Eisert, Martin Wilkens, and
Maciej Lewenstein \cite{EWL99}. Their work introduced games leveraging
quantum information, illustrating instances in which quantum strategies
outperformed their classical counterparts. Since then, numerous quantum
game variants, primarily based on the foundational studies of Meyer
and Eisert et al., have been extensively analyzed. Comprehensive discussions
on these advancements can be found in review articles such as in \cite{GZK08}.
Experimentally, the quantum version of the Prisoner's Dilemma has
been realized using an NMR quantum computer \cite{DLX+02}. Vaidman
introduced in 1999 a straightforward game where players can secure
consistent victories by initially sharing a GHZ state, in contrast
to classical counterparts, whose success remains probabilistic \cite{V99}.
Quantum strategies have been leveraged to incorporate fairness mechanisms
in remote gambling scenarios \cite{GVW99} and to develop quantum
auction algorithms that offer significant security benefits \cite{P07}.
Flitney and Abbott investigated quantum variations of Parrondo's games
\cite{FA03}. These studies not only contribute to the design of
secure networks and the discovery of novel quantum algorithms but
also provide a unique perspective on defining games and protocols
\cite{DP+23,DP2023,DP+24}. Moreover, concepts such as eavesdropping
\cite{E91,NH97} and optimal cloning \cite{W98} can be interpreted
as strategic games among participants. Consequently, the intersection
of game theory and quantum mechanics has been extensively explored.

There exist two principal frameworks that delineate this interconnection.
The first framework involves utilizing quantum resources to execute
a conventional game that can also be played in the absence of such
resources; however, the incorporation of quantum mechanics introduces
certain advantages. Conversely, the second framework employs game-theoretic
principles to model and analyze quantum mechanical scenarios. Hereafter,
these approaches will be referred to as the ``quantized game''
and ``gaming the quantum'', respectively. A quantized game can
be formally characterized as a unitary transformation that maps the
Cartesian product of quantum superpositions of players' pure strategies
onto the Hilbert space governing the game, thereby preserving the
fundamental structure of the classical game under specific constraints.
In contrast, the concept of gaming the quantum, as described by Khan
and Phoenix \cite{KP+13}, pertains to the application of game-theoretic
methodologies to quantum mechanics to derive strategic solutions.
Our work in Chapter \ref{Ch5:Chapter5_QG} follows the ``gaming the quantum'' approach
by employing non-cooperative game theory within a quantum communication
protocol to illustrate how Nash equilibrium can serve as an effective
solution framework.

By leveraging the insights discussed earlier, Nash equilibrium points
in game theory can be utilized to define secure boundaries for various
quantum information parameters. This approach involves reformulating
any quantum scheme into a game-theoretic framework and analyzing its
Nash equilibrium points. These equilibrium points represent stable
configurations or offer rational probability distributions that guide
stakeholders' decision-making within a mixed-strategy game. The probabilities
derived from these equilibrium points contribute to maintaining a
stable strategic environment for all participants. This stability
allows for evaluating key cryptographic parameters, such as determining
the threshold for the QBER. In realistic scenarios where participants
make independent decisions, achieving such stability is less likely,
potentially leading to increased uncertainty. Therefore, identifying
the minimum QBER value within a stable game-theoretic setting establishes
a secure threshold, ensuring the practical feasibility of protocols
for secure quantum communication. Chapter \ref{Ch5:Chapter5_QG} focuses on analyzing the
secure threshold bound for the QBER within the framework of the DL04
protocol proposed by Deng and Long \cite{DL04}. It is essential
to contextualize this discussion within the realm of ``direct secure
quantum communication'' protocols, as the DL04 scheme is a part of
this category. These protocols can be classified into two main types
\cite{LDW+07}. The first category includes DSQC protocols \cite{ZXF+06,HHT11,DP22,DP23},
in which the recipient can only retrieve the encoded message after
receiving at least one bit of supplementary classical information
per qubit from the sender. The second category comprises QSDC protocols
\cite{BEK+02,LL02,BF02,DLL03,DL04,DBC+04,WDL+05,ZSL20,WLY+19}, which
function without necessitating any classical information exchange.
A notable QSDC scheme was proposed by Beige et al. \cite{BEK+02},
where the message remains inaccessible until additional classical
information is transmitted for each qubit. Bostr{\"o}m and Felbinger
introduced a ping-pong QSDC scheme \cite{BF02}, which ensures security
for key distribution and provides quasi-secure direct secret communication
under ideal quantum channel conditions. In 2004, Deng and Long proposed
the DL04 QSDC protocol \cite{DL04}, which operates without relying
on entangled states. A notable discovery was the capability to achieve
secure information transmission via the two-photon component, consistent
with findings in two-way QKD \cite{DL+04,LM_05,L19}. This characteristic
is particularly relevant to the specific conditions outlined in the
DL04 QSDC protocol \cite{DL04}. In Chapter \ref{Ch5:Chapter5_QG}, the primary objective
of our work is to determine the secure threshold bound of the QBER
within the DL04 protocol. This protocol is chosen due to its extensive
adoption in experimental implementations and its practicality for
secure deployment \cite{HYJ+16,ZZS+17,QSL+19,ZSN+20,PLW+20,PSL23,NZL+18}.

\section{Overview of this chapter and outline of the remaining thesis}\label{sec:Chapter1_Sec9}

This chapter provides a concise overview of the historical development
of quantum mechanics, highlighting key inventions in the field. It
also presents the fundamental postulates and mathematical formulations
relevant to this thesis. A significant portion of the chapter is dedicated
to an in-depth discussion of QIA, which is essential for ensuring
the unconditional security of quantum protocol implementations by
verifying the identities of legitimate users. The discussion on QIA
follows a chronological approach, categorizing various schemes based
on the quantum resources utilized and the computational or communication
tasks inherent in their design. A critical analysis of earlier protocols
is conducted, identifying their limitations in terms of security and
resource utilization. Additionally, potential criticisms from other
researchers are addressed. The initial and core content of this chapter
has been published in \cite{DP22}. Following the
discussion on QIA, various quantum protocols, including QKD and QKA,
are explored. The motivation for proposing novel quantum communication
schemes is also examined. The chapter concludes with an exploration
of quantum games, specifically the concept of ``gaming the quantum'',
and introduces the rationale behind establishing a security bound
on the QBER using Nash equilibrium principles.

This thesis is structured into six chapters. The current chapter provides
fundamental concepts of quantum mechanics along with a systematic
review of QIA schemes. Additionally, it outlines the motivation for
the research presented in the subsequent chapters. Chapter \ref{Ch2:Chapter2_Authentication} introduces
novel QIA schemes that utilize single photons and Bell states as quantum
resources, accompanied by a comprehensive security analysis. The design
of these new protocols is informed by the systematic review presented
in this chapter. Specifically, two QIA protocols are proposed, inspired
by controlled secure direct quantum communication. In the first approach,
Alice and Bob authenticate each other's identities with the assistance
of a third party, Charlie, using Bell states. The second set of schemes
is designed for a two-party authentication scenario that relies solely
on single-photon states. A rigorous security analysis is conducted
to demonstrate the resilience of the proposed protocols against various
attacks, including impersonation, intercept-resend, and fraudulent
impersonation attacks. Furthermore, a comparative analysis highlights
the advantages of the proposed protocols over existing QIA schemes,
emphasizing their security and efficiency. The works in Chapter \ref{Ch2:Chapter2_Authentication}
are published in Refs. \cite{DP22,DP23,DP+24}. In Chapter \ref{Ch3:Chapter3_QKD}, we
introduce two new QKD protocols that aim to be more efficient than
SARG04 while maintaining resilience against PNS attacks and a variety
of other well-known attacks. These protocols are compatible with commercially
available single-photon sources, making them practical for implementation.
We demonstrate that the proposed schemes are secure against several
types of attacks, including the intercept-resend attack and certain
collective attacks. We derive key rate bounds and show that specific
classical pre-processing techniques can increase the tolerable error
threshold. Our analysis indicates a trade-off between the quantum
resources used and the information that could potentially be accessed
by an eavesdropper. By employing slightly more quantum resources,
the new protocols achieve higher efficiency compared to similar protocols,
such as SARG04. In particular, our schemes outperform SARG04 in efficiency,
although at the cost of increased quantum resource usage. Additionally,
the proposed protocols exhibit greater critical distances under PNS
attacks than both the BB84 and SARG04 protocols in comparable scenarios.
The findings presented in this chapter have been reported in Ref.
\cite{DP+23}. In Chapter \ref{Ch4:Chapter4_QKA}, we present, for the first time, a QKA
protocol that involves a controller (Charlie) to facilitate key agreement
between two legitimate parties without relying on quantum memory.
This design meets the growing need for practical quantum cryptographic
solutions, as quantum memory is not yet available commercially. We
introduce both a controlled quantum key agreement protocol and a standard
key agreement protocol, providing a thorough security analysis. Our
security proof demonstrates the protocols' resilience against impersonation
and collective attacks, confirming their adherence to essential properties
such as fairness and correctness. Additionally, we conduct a detailed
comparison of our protocols with several existing QKA schemes, including
the controlled QKA protocol of Tang et al \cite{TS+20}. Our proposed
scheme stands out by not requiring quantum memory, unlike many existing
protocols. Furthermore, our controlled QKA protocol uses simpler quantum
resources, such as Bell states and single-photon states, instead of
the more complex GHZ states employed in Tang et al.'s protocol. The
findings presented in this chapter have been published in Ref. \cite{DP2023}.
The analysis in Chapter \ref{Ch5:Chapter5_QG} considers the sender, receiver, and eavesdropper
(Eve) as quantum players, capable of executing quantum operations.
Notably, Eve is assumed to possess the ability to perform various
quantum attacks, such as W{\'o}jcik's original attack, its symmetrized
variant, and Pavi{\v{c}}i{\'c}'s attack, as well as a classical intercept-and-resend
strategy. A game-theoretic evaluation of the DL04 protocol's security
under these conditions is conducted by examining multiple game scenarios.
The findings indicate the absence of a Pareto optimal Nash equilibrium
in these cases. Therefore, mixed strategy Nash equilibrium points
are identified and employed to derive upper and lower bounds for QBER.
Additionally, the analysis highlights the susceptibility of the DL04
protocol to Pavi{\v{c}}i{\'c}'s attack in the message transmission
mode. It is also observed that the quantum attacks by Eve are more
effective than the classical attacks, as they result in lower QBER
values and a decreased probability of detecting Eve's intrusion. The
results reported in this chapter are published in Ref. \cite{DP24}.
This Chapter \ref{Ch6:asd} summarizes the findings of the entire thesis will an emphasis
on the newly proposed quantum protocols, highlighting their advantages,
and the thesis as a whole---ends with a discussion of potential future
work suggested by these findings.

\newpage


%
%
%
%
%
%

\chapter{QUANTUM IDENTITY AUTHENTICATION SCHEMES WITH DIFFERENT QUANTUM RESOURCES}\label{Ch2:Chapter2_Authentication} 
\graphicspath{{Chapter2/Chapter2Figs/}{Chapter2/Chapter2Figs/}}


\section{Introduction}

With the proliferation of online banking, e-commerce systems, and
IoT devices, users have become familiar with identity verification
mechanisms. The frequent utilization of these technologies highlights
the critical role of authentication frameworks, which serve as structured
approaches to confirming the legitimacy of users or connected devices.
These authentication methods are fundamental to various cryptographic
operations, ensuring secure communication and computational integrity.
Quantum cryptography has fundamentally transformed security by introducing
unconditional security, an aspect unattainable in classical systems.
The groundbreaking BB84 protocol was the first unconditionally secure
QKD scheme \cite{BB84}. Unlike classical cryptographic methods that
depend on computational complexity, BB84 derives its security from
quantum mechanical principles. Various other QKD protocols (\cite{E91,B92,WSL+21,SZ22}
and additional works \cite{LCQ12,CLY+20,JJL+13,HAL+20,WYH+22}) facilitate
secure key distribution between a sender (Alice) and a receiver (Bob).
However, to prevent impersonation by an adversary (Eve), Alice and
Bob must first authenticate each other before initiating any QKD protocol.
Identity authentication is thus essential for QKD and broader quantum
cryptographic applications. Initially, classical authentication schemes,
such as the Wegman--Carter scheme \cite{WC81}, were suggested for
QKD authentication, including in the original BB84 protocol. However,
these schemes lack unconditional security. Current QKD implementations,
both commercial and experimental, still rely on classical or post-quantum
authentication schemes, meaning QKD is not yet fully quantum or unconditionally
secure. To address this, researchers have proposed various QIA protocols
\cite{CS_1995,LB_2004,WZT_2006,ZZZX_2006,ZCSL_2020,KHHYHM_2018,CXZY_2014,T.Mihara_2002,DP22,DP23,JWC+23,LZZ+22}
leveraging quantum resources for enhanced security. Some recent efforts
focus on device-independent QIA \cite{FG21} and QIA using homomorphic
encryption with qubit rotation \cite{CWJ+23}.

The first QIA scheme, introduced by Cr{\'e}peau et al. \cite{CS_1995}
in 1995, OT as its cryptographic foundation. However, Lo
and Chau \cite{LC_1997} later proved that quantum OT lacks unconditional
security in two-party settings, rendering Cr{\'e}peau et al.'s scheme
similarly insecure. Notably, OT is not the only cryptographic primitive
applicable to QIA design. Subsequent QIA protocols \cite{LB_2004,WZT_2006,ZZZX_2006,ZCSL_2020,KHHYHM_2018,CXZY_2014,T.Mihara_2002,ZW_1998,DHHM_1999,DP22}
leveraged diverse cryptographic frameworks. Some were adapted from
secure direct quantum communication, which enables message transmission
via quantum resources without key generation \cite{LL02,LM_05,P13}.
For a detailed exploration of QIA schemes and their cryptographic
underpinnings, refer to Chapter \ref{Ch1:Chapter1_Introduction}. Before summarizing existing QIA
schemes derived from modified secure direct communication methods,
it is essential to distinguish between two secure quantum direct communication
types: QSDC and DSQC. Various QSDC and DSQC protocols exist \cite{LL02,STP20,STP17,YSP14,BP12,DLL03,DP24,PMS+23},
with early versions often relying on entangled states \cite{LL02,DLL03,BF02}.
For example, Zhang et al. \cite{ZZZX_2006} adapted the entangled-state
ping-pong protocol \cite{BF02} for QSDC to develop a QIA scheme,
while Yuan et al. \cite{YLP+14} introduced a QIA protocol based on
the LM05 protocol \cite{LM_05}, a single-photon counterpart of the
ping-pong protocol. Other researchers have modified DSQC and QSDC
protocols to construct QIA schemes \cite{LB_2004,ZLG_2000,LC_2007}.
Additionally, controlled DSQC (CDSQC) schemes have been proposed \cite{STP17,P15},
where Alice, under Charlie\textquoteright s supervision, securely
communicates with Bob using quantum resources. The adaptation of CDSQC
for QIA remains underexplored. Given the importance of identity authentication,
we propose a new QIA protocol incorporating simultaneous authentication
of legitimate entities, inspired by CDSQC principles. The following
sections introduce two single photon-based and two Bell state-based
QIA protocol leveraging CDSQC concepts and analyze its robustness
against common attacks.

The remainder of this chapter is organized as follows: Section \ref{sec:Chapter2_Sec2}
presents the proposed QIA protocols utilizing single-qubit states
and evaluates their security against various attacks. Section \ref{sec:Chapter2_Sec3}
introduces a new controlled QIA scheme based on Bell states and permutation
operations, providing a comprehensive security analysis. This section
also includes a comparison of the proposed protocol with recent relevant
QIA schemes. In Section \ref{sec:Chapter2_Sec4}, another controlled
QIA scheme is proposed, employing Bell states and CNOT operations.
Its resilience against multiple attacks is demonstrated, along with
a comparison to existing protocols within the same category. Finally,
Section \ref{sec:Chapter2_Sec5} concludes the chapter with a discussion
of key insights.


\section{New protocols for QIA based on QSDC}\label{sec:Chapter2_Sec2}

In Chapter \ref{Ch1:Chapter1_Introduction}, we previously noted that secure direct quantum communication
protocols can, in principle, serve as a foundation for constructing
QIA schemes. Several QIA protocols have already been developed based
on QSDC and DSQC. However, recognizing the inherent symmetry among
these protocols enables the formulation of novel schemes. As an illustration,
we introduce four new QIA protocols, they derived from QSDC principles.
We begin with the first protocol.

\subsection{Protocol 2.1}

This QIA protocol, built upon QSDC, exhibits both distinct characteristics
and structural similarities to the frameworks proposed in \cite{CXZY_2014,YLP+14}.
Hereafter, this QSDC-inspired approach is designated as Protocol 2.1,
with its procedural steps indexed as Step $1.j$ (more generally,
Step $i.j$, where $i,j\in\{1,2,\cdots\}$, denotes the $j^{th}$
step of the $i^{th}$ proposed protocol). The protocol assumes that
Alice and Bob share a pre-shared authentication key $K=\{k_{1},k_{2},k_{3},\cdots,k_{2n}\}$.
Furthermore, the execution of the protocol relies on non-entangled
quantum states as the fundamental resource. The procedural steps are
outlined as follows.
\begin{itemize}
\item \textit{Encoding mode} 
\end{itemize}
\begin{table}[H]
\caption{The preparation method for the decoy sequence (Protocol 2.1).}\label{tab:Chapter2_Tab1}

\centering{}%
\begin{tabular*}{15.9cm}{@{\extracolsep{\fill}}@{\extracolsep{\fill}}|c|c|c|}
\hline 
$(2i)^{th}$ bit value: $k_{2i}$  & 0  & 1\tabularnewline
\hline 
$i^{th}$ qubit of decoy sequence: $K_{d_{i}}$  & $|0\rangle$~or~$|1\rangle$  & $|+\rangle$~or~$|-\rangle$\tabularnewline
\hline 
\end{tabular*}
\end{table}

\begin{description}
\item [{Step~1.1:}] Alice constructs a structured sequence of $n$-qubit
decoy states denoted as $K_{d}$, which is determined by the bit
values of a pre-shared key $K$. If $k_{2i}=0$, the $i^{th}$ qubit
of $K_{d}$ is prepared in either $\vert0\rangle$ or $\vert1\rangle$.
Otherwise, it is initialized in $\vert+\rangle$ or $\vert-\rangle$,
with Alice being the sole entity aware of the precise quantum states
assigned to each qubit in $K_{d}$.
\item [{Step~1.2:}] Alice then generates an additional qubit sequence,
$K_{a}$, designated for authentication purposes. The $i^{th}$ qubit
in $K_{a}$ is assigned the state $\vert0\rangle$ if $k_{2i-1}\oplus k_{2i}=0$,
otherwise, it is prepared in the state $\vert-\rangle$. To form an
extended sequence $K_{A}$, Alice interleaves the authentication qubits
from $K_{a}$ into $K_{d}$ based on the following rule: if $k_{2i-1}\oplus k_{2i}=0$,
the $i^{th}$ qubit of $K_{a}$ is placed immediately after the corresponding
$i^{th}$ qubit in $K_{d}$; otherwise, it is inserted before it.
Finally, Alice transmits the augmented quantum sequence $K_{A}$ to
Bob.
\end{description}
\begin{table}[H]
\caption{The preparation method of the authentication sequence (Protocol 2.1).}\label{tab:Chapter2_Tab2}
\centering{}%
\begin{tabular*}{15.9cm}{@{\extracolsep{\fill}}@{\extracolsep{\fill}}|c|c|c|}
\hline 
Additional modulo 2: $k_{2i-1}\oplus k_{2i}$  & 0  & 1\tabularnewline
\hline 
$i^{th}$ qubit of authentication sequence: $K_{a_{i}}$  & $|0\rangle$  & $|-\rangle$\tabularnewline
\hline 
\end{tabular*}
\end{table}

\begin{itemize}
\item \textit{Decoding mode} 
\end{itemize}
\begin{table}[H]
\caption{Bob employs a predetermined measurement basis to extract results
from the authentication sequence (Protocol 2.1). }\label{tab:Chapter2_Tab3}
\centering{}%
\begin{tabular*}{15.9cm}{@{\extracolsep{\fill}}@{\extracolsep{\fill}}|c|c|c|}
\hline 
Additional modulo 2: $k_{2i-1}\oplus k_{2i}$  & 0  & 1\tabularnewline
\hline 
Measurement basis: $B_{i}=\{B_{z},B_{x}\}$  & $B_{z}=\{|0\rangle,|1\rangle\}$  & $B_{x}=\{|+\rangle,|-\rangle\}$\tabularnewline
\hline 
Measurement result:  & $|0\rangle$  & $|-\rangle$\tabularnewline
\hline 
\end{tabular*}
\end{table}

\begin{description}
\item [{Step~1.3:}] Based on the pre-shared key $K$, Bob determines the
placement and measurement basis for the decoy and authentication qubits
that Alice has prepared. He then sequentially measures the decoy qubits
$K_{d}$ and authentication qubits $K_{a}$ following a predefined
rule: if $k_{2i}=0$, Bob employs the $Z$-basis ($\{\vert0\rangle,\vert1\rangle\}$)
for measuring $K_{d}$; otherwise, he utilizes the $X$-basis ($\{\vert+\rangle,\vert-\rangle\}$).
The authentication qubits $K_{a}$ are measured based on the function
\textbf{$k_{2i-1}\oplus k_{2i}$}, and the results are compared with
the expected values outlined in Table \ref{tab:Chapter2_Tab3}. To
assess the security of the channel, Bob computes the QBER using the
decoy qubit measurement results. If the QBER remains within an acceptable
threshold, the channel is deemed secure. However, if discrepancies
in the authentication sequence $K_{a}$ exceed the permissible limit,
Bob terminates the protocol. Otherwise, he verifies Alice's identity.
\item [{Step~1.4:}] Bob generates a new decoy sequence
$K_{d}^{\prime}$ using the same procedure described in Step 1.1.
Similarly, he encodes the authentication qubit sequence $K_{a}^{\prime}$
following the same encoding rules and constructing an extended sequence
$K_{B}$. He then transmits $K_{B}$ to Alice.
\item [{Step~1.5:}] Upon receiving the sequence, Alice extracts the decoy
and authentication qubits. She then verifies Bob\textquoteright s
identity using the same approach outlined in Step 1.3.
\end{description}

\subsection{Protocol 2.2}

Similar to the previous protocol, Alice and Bob are assumed to have
pre-shared an authentication key denoted as $K=\{k_{1},k_{2},k_{3},\cdots,k_{4n}\}$.
The primary distinction from the earlier scheme is that the length
of the shared key is now $4n$, which is twice that of the prior approach.
\begin{itemize}
\item \textit{Encoding mode} 
\end{itemize}
\begin{table}[H]
\caption{The preparation method of the authentication sequence (Protocol 2.2).}\label{tab:Chapter2_Tab4}
\centering{}%
\begin{tabular*}{15.9cm}{@{\extracolsep{\fill}}@{\extracolsep{\fill}}|c|c|c|}
\hline 
Additional modulo 2: $k_{2i-1}\oplus k_{2i}$  & 0  & 1\tabularnewline
\hline 
Bit value: $k_{2i}$  & %
\begin{tabular}{cc}
0  & 1\tabularnewline
\end{tabular} & %
\begin{tabular}{cc}
0  & 1\tabularnewline
\end{tabular}\tabularnewline
\hline 
Prepared authentication state: $K_{A_{i}}$  & %
\begin{tabular}{cc}
$|0\rangle$  & $|1\rangle$\tabularnewline
\end{tabular} & %
\begin{tabular}{cc}
$|+\rangle$  & $|-\rangle$\tabularnewline
\end{tabular}\tabularnewline
\hline 
\end{tabular*}
\end{table}

\begin{description}
\item [{Step~2.1:}] Alice constructs an ordered sequence of $n$ particles,
$K_{A}$, utilizing the first half of the pre-shared key (comprising
2$n$ bits). The particle states are assigned based on the following condition:
if $k_{2i-1}\oplus k_{2i}=0$, then Alice encodes the states as $\vert0\rangle$
and $\vert1\rangle$, corresponding to$k_{2i}=0$ and $k_{2i}=1$,
respectively. Conversely, if $k_{2i-1}\oplus k_{2i}=1$, Alice prepares
the states as $\vert+\rangle$ and $\vert-\rangle$, corresponding
to $k_{2i}=0$ and $k_{2i}=1$, respectively. She then transmits the
ordered particle sequence, $K_{B}$, to Bob.
\item [{Step~2.2:}] Upon receiving the sequence, Bob measures the $i^{th}$
particle of $K_{B}$ in the $Z$-basis ($\{\vert0\rangle,\vert1\rangle\}$)
if $k_{2i-1}\oplus k_{2i}=0$. Otherwise, he employs the $X$-basis
($\{\vert+\rangle,\vert-\rangle\}$) for measurement. He then constructs
a classical bit sequence, $S_{B}$, by assigning a bit value of 0
to the states $\vert0\rangle$ and $\vert+\rangle$, and a bit value
of 1 to the states $\vert1\rangle$ and $\vert-\rangle$. Bob then
compares the $i^{th}$ bit of $S_{B}$ with $k_{2i}$, which ideally
should match $K_{B}=k_{2i}$. If the observed error rate remains within
an acceptable threshold, Bob validates Alice\textquoteright s identity.
\end{description}
\begin{itemize}
\item \textit{Decoding mode} 
\end{itemize}
\begin{table}[H]
\caption{The measurement basis employed by Bob to extract results for the
authentication sequence (Protocol 2.2).}\label{tab:Chapter2_Tab5}
\centering{}%
\begin{tabular*}{15.9cm}{@{\extracolsep{\fill}}@{\extracolsep{\fill}}|c|c|c|}
\hline 
Additional modulo 2: $k_{2i-1}\oplus k_{2i}$  & 0  & 1\tabularnewline
\hline 
Bit value: $k_{2i}$  & %
\begin{tabular}{cc}
0  & 1\tabularnewline
\end{tabular} & %
\begin{tabular}{cc}
0  & 1\tabularnewline
\end{tabular}\tabularnewline
\hline 
Measurement basis: $B_{i}=\{B_{z},B_{x}\}$  & $B_{z}=\{|0\rangle,|1\rangle\}$  & $B_{z}=\{|+\rangle,|-\rangle\}$\tabularnewline
\hline 
Measurement result:  & %
\begin{tabular}{cc}
$|0\rangle$  & $|1\rangle$\tabularnewline
\end{tabular} & %
\begin{tabular}{cc}
$|+\rangle$  & $|-\rangle$\tabularnewline
\end{tabular}\tabularnewline
\hline 
\end{tabular*}
\end{table}

\begin{description}
\item [{Step~2.3:}] Bob executes a procedure analogous to Step 2.1, utilizing
the remaining half of the pre-shared authentication key, similar to
Alice\textquoteright s approach.
\item [{Step~2.4:}] Alice replicates the process in Step 2.2 to authenticate
Bob\textquoteright s identity.
\end{description}
As outlined earlier, Protocols 1 and 2 serve as illustrative examples
of protocols that can be constructed using QSDC-based schemes. In
a similar manner, other existing QSDC schemes can be adapted to develop
protocols for QIA. A detailed security analysis of these QIA schemes
will be presented in Section \ref{subsec:Chapter2_Sec2.4}.

\subsection{Comparison of the existing and the proposed protocols}\label{sec:Chapter2_Sec2.3}

Many of the existing protocols, including the seminal work of Cr{\'e}peau
et al. \cite{CS_1995}, have been criticized due to their reliance
on two-party secure multiparty computation (two-party SMC), which
is not feasible within the framework of non-relativistic quantum mechanics.
To comprehend this limitation affecting a subset of protocols, it
is essential to highlight key findings by Lo in 1997 \cite{L97} and
by Lo and Chau in 1998 \cite{LC98}. These studies demonstrated that
fundamental cryptographic primitives such as quantum bit commitment,
quantum remote coin tossing\footnote{It is important to note that only a weak form of coin tossing can
be realized using quantum resources.}, and quantum OT cannot achieve unconditional security within a two-party
quantum setting. Notably, the groundbreaking QIA protocol by Cr{\'e}peau
et al. \cite{CS_1995} was constructed on OT; however, at the time
of its publication, the results of Lo and Chau were not yet established.
Interestingly, despite the findings of Lo, Chau, and subsequent researchers
\cite{BCS12,C07}, numerous QIA protocols have continued to emerge,
leveraging computational tasks deemed infeasible for unconditional
security as per Lo-Chau and related impossibility results. For instance,
while the secure quantum private comparison is known to be unachievable
in a two-party setting, several QIA protocols have nonetheless been
designed based on this very mechanism. The question at hand is whether
the findings of Lo and Chau invalidate all proposed schemes for QIA.
The answer remains negative. Here, we present arguments supporting
this stance. Furthermore, the security vulnerabilities inherent in
Mihara\textquoteright s scheme \cite{T.Mihara_2002} have been analyzed,
along with potential countermeasures. Additionally, numerous theoretical
frameworks for QIA necessitate quantum memory, as exemplified by the
protocols outlined in Refs. \cite{ZW_1998,T.Mihara_2002,LB_2004,ZLG_2000,ZZZZ_2005,DXXN_2010,LLY_2006,WZT_2006,YTXZ_2013,TJ_2014,AHKB_2017,KHHYHM_2018,WZGZ_2019}.
Given that a fully functional quantum memory is yet to be realized,
these QIA protocols are currently impractical for real-world implementation.
While certain approaches can partially mitigate the requirement for
quantum memory through the use of time delays, this remains a constrained
solution. Another critical limitation affecting the feasibility of
QIA over long distances is the absence of quantum repeaters. Some
protocols, such as the QIA scheme proposed by Zhou et al. \cite{ZZZZ_2005},
attempt to overcome this challenge via entanglement swapping. However,
this method relies on a trusted intermediary for entanglement swapping,
which introduces potential security vulnerabilities.

\subsection{Security analysis }\label{subsec:Chapter2_Sec2.4}

Here, we evaluate the security of the proposed protocols against common
external attacks, specifically those by an adversary (Eve). Insider
attacks are outside the scope of identity authentication and are therefore
not considered in this analysis. We begin by examining impersonation
attacks and systematically assessing the security of each proposed protocol
against this threat.

\subsection*{Impersonation attack}

In this category of attacks, an external adversary, Eve, attempts
to impersonate a legitimate user, either Alice or Bob, to successfully
complete the authentication process. The following analysis demonstrates
that the proposed protocols remain secure against such impersonation
attempts by Eve.

\subsubsection{Security analysis of Protocol 2.1 against impersonation attack by
Eve}

Consider the scenario where Eve attempts to impersonate Alice. According
to the protocol, $n$ authentication particles are embedded within
an $n$-particle decoy sequence. The probability of Eve correctly identifying
the authentication state of a particle is $\frac{1}{2}$. Additionally,
she must accurately determine the position of the decoy particle,
which also occurs with a probability of $\frac{1}{2}$. Consequently,
the overall probability of Eve successfully impersonating Alice is
$(\frac{1}{4})^{n}$. Thus, the probability of detecting Eve's presence
is given by$P_{1}=1-(\frac{1}{4})^{n}$, which approaches 1 for sufficiently
large $n$ (see Figure \ref{fig:Chapter2_Fig1}). A similar analysis
applies when Eve attempts to impersonate Bob.

\subsubsection{Security analysis of Protocol 2.2 against impersonation attack by
Eve}

\begin{figure}[H]
\centering{} \includegraphics[scale=0.6]{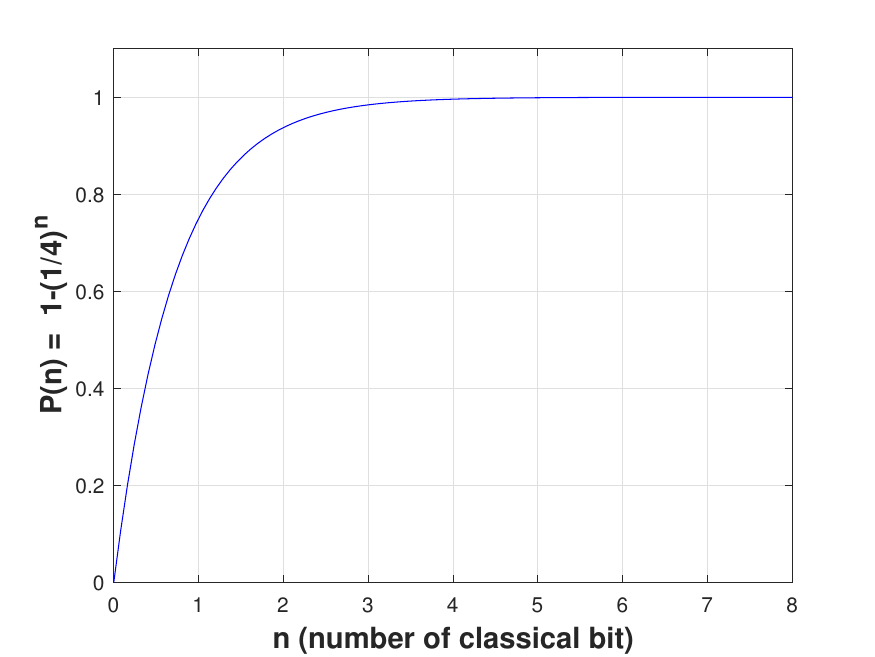}
\caption{The graph presents the correlation between the probability of detecting
Eve\textquoteright s presence, $P(n)$, and the number of classical
bits used, $n$. }\label{fig:Chapter2_Fig1}
\end{figure}

Assume Eve generates a sequence of $2n$ random classical bits in
an attempt to impersonate Alice. She then computes $k_{2i-1}\oplus k_{2i}$.
The probability that Eve's result matches Alice's is $\frac{1}{2}$.
Furthermore, the probability that Eve correctly prepares and transmits
the expected quantum states to Bob is also $\frac{1}{2}$. Consequently,
the probability of detecting Eve\textquoteright s intrusion is given
by $P_{2}=1-(\frac{1}{4})^{n}$. As $n$ increases, $P_{2}$ approaches
1, ensuring that an impersonation attempt is effectively identified.
Therefore, Protocol 2.2 maintains the same level of security against
this attack as Protocol 2.1. It follows that $P_{1}=P_{2}=P(n)=1-(\frac{1}{4})^{n}$,
where the dependence of $P(n)$ on $n$ is depicted in Figure \ref{fig:Chapter2_Fig1}.
This figure illustrates that, to reliably detect Eve\textquoteright s
presence, Protocol 2.1 requires a minimum of 6 pre-shared classical
bits, whereas Protocol 2.2 necessitates at least 10.

\subsection*{Measurement and resend attack}

\subsubsection{Security analysis of Protocol 2.1 against the measurement-resend
attack}\label{subsec:ch2_sec2.4.3}

In this attack scenario, Eve intercepts and measures the qubits transmitted
from Alice to Bob, then forwards newly prepared states to Bob based
on her measurement outcomes. To successfully impersonate Alice, Eve
would need to distinguish between decoy qubits and authentication
qubits. However, since she lacks prior knowledge of the positions
of the decoy states, any attempt to execute the measure-resend attack
inevitably affects the information qubits as well, thereby leaving
detectable traces of her interference. To illustrate this, consider
the case where Eve measures all qubits traveling through the channel
(a similar analysis holds even if she intercepts only a fraction $f$
of them). She performs measurements on each qubit randomly in either
the $\{|0\rangle,|1\rangle\}$ basis or the $\{|+\rangle,|-\rangle\}$
basis, with an equal probability of $\frac{1}{2}$ for each basis
selection. To further analyze this scenario, we compute the mutual
information between Alice and Bob in the presence of Eve, specifically
focusing on authentication particles. Let $P(B,A)$ denote the joint
probability of Bob obtaining measurement result $B$ when Alice initially
prepares the state $A$, and let $P(B)$ represent the total probability
of Bob measuring the state $B$. when measurements of $A$ and $B$ are
performed in an identical basis, yet yield distinct outcomes.

\[
P(B,A)=\left\{ \begin{array}{cc}
\frac{1}{8} & {\rm measurement}\,{\rm of}\,A\,{\rm and}\,B\,{\rm is\,performed\,with\,identical\,basis,\,with\,distinct\,outcomes}\\
\frac{3}{8} & {\rm measurement}\,{\rm of}\,A\,{\rm and}\,B\,{\rm is\,performed\,with\,identical\,basis,\,with\,distinct\,outcomes}\\
0 & {\rm when}\,B\,{\rm and}\,A\,{\rm are\,meassured\,in\,the\,different\,basis\,}
\end{array}\,\right.
\]
and 
\[
P(B=|0\rangle)=P(B=|-\rangle)=\frac{3}{8}
\]

\[
P(B=|1\rangle)=P(B=|+\rangle)=\frac{1}{8}
\]
where $A\in\left\{ |0\rangle,|-\rangle\right\} $ and $B\in\left\{ |0\rangle,|1\rangle,|+\rangle,|-\rangle\right\} .$

The conditional entropy is given by\footnote{The conditional probability is expressed as, $P(B=|i\rangle|A=|j\rangle)=\frac{P(B=|i\rangle,A=|j\rangle)}{P(A=|j\rangle)}$
and the conditional entropy is formulated as $H(B|A)=-\sum_{j}P(A=|j\rangle)\sum_{i}P(B=|i\rangle|A=|j\rangle)\log_{2}P(B=|i\rangle|A=|j\rangle)$.}, $H(B|A)=\left[2\times\frac{1}{2}\left(-\frac{3}{4}\log_{2}\frac{3}{4}-\frac{1}{4}\log_{2}\frac{1}{4}\right)\right]=0.811278$
bit, where $H(A)=1$ and $H(B)$ is computed as $H(B)=2\times\left(-\frac{3}{8}\log_{2}\frac{3}{8}-\frac{1}{8}\log_{2}\frac{1}{8}\right)=1.811278$
bit. The mutual information shared between Alice and Bob is determined
by, $I(A:B)=H(B)-H(B|A)=1.0$ bit\footnote{The mutual information shared between two parties is constrained by
the Shannon entropy of the probability distributions associated with
a single particle. Mathematically, this is expressed as $0\leq I(A:B)\leq min[H(A),H(B)]$,
where the joint entropy is defined as $H(A,B):=[max[H(A),H(B)],H(A)+H(B)]$.
Consequently, the mutual information is given by $I(A:B):=H(A)+H(B)-[max[H(A),H(B)],H(A)+H(B)]]$
which simplifies to $I(A:B):=[0,min[H(A),H(B)]]$.}.

Similarly, the probability relations for Alice and Eve follow the
same structure.

\[
P(E,A)=\left\{ \begin{array}{cc}
0 & {\rm measurement}\,{\rm of}\,E\,{\rm and}A\,{\rm is\,performed\,with\,identical\,basis,\,with\,distinct\,outcomes}\\
\frac{1}{4} & {\rm measurement}\,{\rm of}\,E\,{\rm and}A\,{\rm is\,performed\,with\,identical\,basis,\,with\,distinct\,outcomes}\\
\frac{1}{8} & {\rm when}\,E\,{\rm and}\,A\,{\rm are\,meassured\,in\,the\,different\,basis\,}
\end{array}\,\right.
\]
and 
\[
P(E=|0\rangle)=P(E=|-\rangle)=\frac{3}{8},
\]

\[
P(E=|1\rangle)=P(E=|+\rangle)=\frac{1}{8},
\]
where $A\in\left\{ |0\rangle,|-\rangle\right\} $ and $E\in\left\{ |0\rangle,|1\rangle,|+\rangle,|-\rangle\right\} $.

The conditional entropy is computed as $H(E|A)=\left[2\times\frac{1}{2}(-\frac{1}{2}\log_{2}\frac{1}{2}-\frac{1}{4}\log_{2}\frac{1}{4}-\frac{1}{4}\log_{2}\frac{1}{4}\right]=1.5$
bit. Similarly, the entropy of $E$ is given by: $H(E)=2\times\left(-\frac{3}{8}\log_{2}\frac{3}{8}-\frac{1}{8}\log_{2}\frac{1}{8}\right)=1.811278$.
The mutual information between Eve and Alice is determined as $I(A:E)=H(E)-H(E|A)=0.311278$
bit. Furthermore, it is established that the mutual information $I(A:E)$
is constrained by the Holevo bound \cite{H_73}, which in this case
is 0.600876. Clearly, $I(A:E)\le0.600876$ and $I(A:B)>I(A:E)$, indicating
that Eve's accessible information is significantly lower than Bob's.

For Eve to successfully impersonate Bob, she must target the expanded
particle sequence $K_{B}$ that Bob returns. This requirement mirrors
the conditions necessary for her to impersonate Alice.

\subsubsection{Security analysis of Protocol 2.2 against measurement-resend attack}

Eve conducts arbitrary measurements on the quantum particles transmitted
by Alice, obtaining certain outcomes. Analogous to Protocol 2.1, the
mutual information shared between Alice and Bob can be determined
as follows:

\textbf{ 
\[
P(B,A)=\frac{1}{16}+\frac{1}{8}\delta_{A,B}\,{\rm and}\,P(B=|0\rangle)=P(B=|+\rangle)=(B=|1\rangle)=P(B=|-\rangle)=\frac{1}{4},
\]
}where \textbf{$\delta_{A,B}=1$ }if Alice\textquoteright s and Bob\textquoteright s
measurement results match when using the same basis, and \textbf{$\delta_{A,B}=0$
}otherwise.

The conditional entropy is calculated as $H(B|A)=\frac{1}{4}\left[4\times\left(-\frac{3}{4}\log_{2}\frac{3}{4}\right)+4\times\left(-\frac{1}{4}\log_{2}\frac{1}{4}\right)\right]=0.81127$
bit. Additionally, the entropy values for Alice and Bob are, $H(A)=H(B)=4\times\left[-\frac{1}{4}\log_{2}\frac{1}{4}\right]=2$
bits. Consequently, the mutual information between Alice and Bob is
derived as, $I(A:B)=H(B)-H(B|A)=1.18872$ bit which exceeds 1, as
they already share identical classical information established via
a pre-shared key.

To get mutual information between Alice and Eve,
\[
P(E,A)=\left\{ \begin{array}{cc}
0 & {\rm measurement}\,{\rm of}\,E\,{\rm and}A\,{\rm is\,performed\,with\,identical\,basis,\,with\,distinct\,outcomes}\\
\frac{1}{8} & {\rm measurement}\,{\rm of}\,E\,{\rm and}A\,{\rm is\,performed\,with\,identical\,basis,\,with\,distinct\,outcomes}\\
\frac{1}{16} & {\rm when}\,E\,{\rm and}\,A\,{\rm are\,meassured\,in\,the\,different\,basis\,}
\end{array}\,\right.
\]
and 
\[
P(E=|0\rangle)=P(E=|1\rangle)=P(E=|+\rangle)=P(E=|-\rangle)=\frac{1}{4}.
\]
The conditional entropy is given by $H(E|A)=\frac{1}{4}\times[8\times(-\frac{1}{4}\log_{2}\frac{1}{4})-4\times(-\frac{1}{2}\log_{2}\frac{1}{2})]=1.5$
bits, while the entropy $H(E)$ is calculated as $H(E)=4\times[-\frac{1}{4}\log_{2}\frac{1}{4}]=2$
bits. The mutual information shared between Eve and Alice is then
determined as$I(A:E)=H(E)-H(E|A)=0.5$. Additionally, it is established
that the mutual information $I(A:E)$ is constrained by the Holevo
bound \cite{H_73}, which, in this scenario, equals 1. Consequently,
it follows that $I(A:E)\le1$ and $I(A:B)>I(A:E)$, fulfilling the
security condition that ensures Alice\textquoteright s communication
remains confidential without leaking significant information to Eve.
A similar outcome holds in cases where Eve attempts to impersonate
Bob.

\subsection*{Impersonated fraudulent attack}

\subsubsection{Security evaluation of Protocol 2.1 against impersonated fraudulent
attack via forged new qubits}

In this scenario, Eve attempts to assume Alice\textquoteright s identity
by employing a unitary transformation $U_{E}$ to establish a correlation
between Alice\textquoteright s qubit and an ancillary qubit. Once
the unitary operation is applied, the resulting quantum state can
be represented as

\[
U_{E}|0\chi\rangle_{Ae}=|\psi\rangle_{0}=a_{0}|00\rangle+b_{0}|01\rangle+c_{0}|0+\rangle+d_{0}|0-\rangle,
\]

\[
U_{E}|1\chi\rangle_{Ae}=|\psi\rangle_{1}=a_{1}|10\rangle+b_{1}|11\rangle+c_{1}|1+\rangle+d_{1}|1-\rangle,
\]

\[
U_{E}|+\chi\rangle_{Ae}=|\psi\rangle_{+}=a_{+}|+0\rangle+b_{+}|+1\rangle+c_{+}|++\rangle+d_{+}|+-\rangle,
\]

\[
U_{E}|-\chi\rangle_{Ae}=|\psi\rangle_{-}=a_{-}|-0\rangle+b_{-}|-1\rangle+c_{-}|-+\rangle+d_{-}|--\rangle.
\]
Now, we get the entire state as,

\begin{equation}
|\rho\rangle=\frac{1}{4}\left(|\psi\rangle_{00}\langle\psi|+|\psi\rangle_{11}\langle\psi|+|\psi\rangle_{++}\langle\psi|+|\psi\rangle_{--}\langle\psi|\right).\label{eq:Chapter2_Eq1}
\end{equation}
The probability of Bob obtaining the correct measurement outcome for
each authentication state and the decoy sequence is $\frac{1}{2}$.
Assuming the correct state is $|0\rangle$, the probability of detecting
Eve is given by$P_{0}=\frac{1}{2}$. Similarly, the probabilities
for other states are $P_{1}=P_{+}=P_{-}=\frac{1}{2}$. The overall
detection probability for each qubit is computed as $P_{d}=\frac{1}{4}(P_{0}+P_{1}+P_{+}+P_{-})=\frac{1}{2}$.
Based on Simmons theory \cite{S_88}, the protocol ensures unconditional
security against impersonation attacks using (\ref{eq:Chapter2_Eq1}).

\subsubsection{Security evaluation of Protocol 2.2 against impersonated fraudulent
attack via forged qubits}

In this protocol, Alice selects authentication states within the computational
and Hadamard bases with equal probability, and Bob's expected measurement
results are also evenly distributed. Consequently, the attack scenario
is the same with of Protocol 2.1. Under these conditions, Protocol 2.2
remains secure against impersonation-based fraudulent attacks.


\section{Controlled quantum identity authentication protocol using Bell pairs (Protocol 2.3)}\label{sec:Chapter2_Sec3}

In this section, we propose a novel framework for QIA, modeled similarly
to approaches derived from QSDC. Since our protocol incorporates Bell
states, it can be categorized as an entanglement-based method for
QIA. Within this framework, two authorized participants authenticate
each other with the assistance of an untrusted intermediary, utilizing
unitary transformations. Before presenting the detailed structure
of our protocol, it is essential to outline the foundational principles
that inform its design.

\subsection{Principal concept of Protocol 2.3}\label{subsec:Chapter2_Sec3.1}

This protocol is fundamentally based on Bell-state entanglement swapping
and the application of Pauli operations. The correlation between the
Bell state and the pre-shared authentication key can be articulated
as follows:

\begin{equation}
\begin{array}{lcl}
00:|\phi^{+}\rangle & = & \frac{1}{\sqrt{2}}\left(\left|00\right\rangle +\left|11\right\rangle \right),\\
01:|\phi^{-}\rangle & = & \frac{1}{\sqrt{2}}\left(\left|00\right\rangle -\left|11\right\rangle \right),\\
10:|\psi^{+}\rangle & = & \frac{1}{\sqrt{2}}\left(\left|01\right\rangle +\left|10\right\rangle \right),\\
11:|\psi^{-}\rangle & = & \frac{1}{\sqrt{2}}\left(\left|01\right\rangle -\left|10\right\rangle \right).
\end{array}\label{eq:Bell-State=000020with=000020Pre-Shared_Key}
\end{equation}
The Equations (\ref{eq:Bell-State=000020with=000020Pre-Shared_Key})
and (\ref{eq:Pauli-Operation=000020with=000020Pre-Shared_Key}) define
the relationship between the pre-shared key, the Bell state initialized
by Alice and Bob, and the Pauli operations they execute. The connection
between the corresponding Pauli operations and the pre-shared authentication
key can be expressed as follows:

\begin{equation}
\begin{array}{lcl}
00:\mathds{1}_{2} & = & \left|0\right\rangle \left\langle 0\right|+\left|1\right\rangle \left\langle 1\right|,\\
01:\sigma_{x} & = & \left|0\right\rangle \left\langle 1\right|+\left|1\right\rangle \left\langle 0\right|,\\
10:i\sigma_{y} & = & \left|0\right\rangle \left\langle 1\right|-\left|1\right\rangle \left\langle 0\right|,\\
11:\sigma_{z} & = & \left|0\right\rangle \left\langle 0\right|-\left|1\right\rangle \left\langle 1\right|.
\end{array}\label{eq:Pauli-Operation=000020with=000020Pre-Shared_Key}
\end{equation}
We present a succinct summary of the described protocol. Alice and
Bob shares a pre-established authentication key sequence comprising
$n+1$ pairs of bits, where each pair encodes two bits of information.
To illustrate, consider a specific case where the first two secret
keys in the sequence are represented as 11 and 00. Alice selects the
second key (00), typically denoted as the $\left(m+1\right)^{{\rm th}}$
key in the sequence, while Bob chooses the XOR operation between the
second and first keys, computed as $00\oplus11=11$. More generally,
this corresponds to the XOR of the $\left(m+1\right)^{{\rm th}}$
and $m^{{\rm th}}$ keys, where $m=1,2,\cdots,n.$ Subsequently, Alice
and Bob utilize their respective keys to generate Bell states according
to Equation (\ref{eq:Bell-State=000020with=000020Pre-Shared_Key}).
In this process, they prepare the quantum states $\left|\phi^{+}\right\rangle _{12}$
and $\left|\psi^{-}\right\rangle _{34}$, where subscripts $1,2$
correspond to Alice\textquoteright s particles and $3,4$ to Bob\textquoteright s
particles. By leveraging the pre-shared authentication key sequence,
they generate multiple Bell states in a similar manner. Alice retains
the sequence of particle 1 from these Bell states and transmits the
sequence associated with particle 2 to an untrusted third party, Charlie.
Charlie then applies a permutation operator $\Pi_{n}$ \cite{YSP14,TP15,SPS12,P15}
on the sequence of particle 2 while keeping the original sequence
undisclosed. After performing the permutation, Charlie forwards the
modified sequence to Bob. Meanwhile, Bob independently prepares two
sequences corresponding to particles 3 and 4, sending the sequence
associated with particle 4 directly to Alice. As a result, Alice and
Bob respectively obtain sequences comprising particles $(1,4)$ and
$(2,3)$. Given these specific secret keys, the resulting composite
system is characterized by:

\[
\begin{array}{lcl}
\left|\phi^{+}\right\rangle _{12}\left|\psi^{-}\right\rangle _{34} & = & \frac{1}{2}\left(\left|\psi^{+}\right\rangle _{14}\left|\phi^{-}\right\rangle _{23}+\left|\psi^{-}\right\rangle _{14}\left|\phi^{+}\right\rangle _{23}-\left|\phi^{+}\right\rangle _{14}\left|\psi^{-}\right\rangle _{23}-\left|\phi^{-}\right\rangle _{14}\left|\psi^{+}\right\rangle _{23}\right).\end{array}
\]
Alice (Bob) implements a Pauli $\sigma_{z}$ operation on qubit 1
(qubit 3) based on the initial key, which is designated as 11. This\footnote{For an extended pre-shared key sequence, the Pauli operation is dictated
by the$m^{{\rm th}}$ key.} operation follows the mapping defined in Equation (\ref{eq:Pauli-Operation=000020with=000020Pre-Shared_Key}).
After executing their respective operations, Alice and Bob publicly
confirm their completion. Subsequently, Charlie announces the permutation
operation $\Pi_{n}$. In response, Bob applies the inverse permutation
to his sequence of particle 2, thereby restoring the original ordering.
Following this step, Alice and Bob proceed with a Bell-state measurement
on their respective qubits (Alice on qubits 1 and 4, and Bob on qubits
2 and 3). The resultant quantum system is given by:

\begin{equation}
\begin{array}{lcl}
\left|\Psi^{11}\right\rangle  & = & \sigma_{z1}\otimes\sigma_{z3}\left(\left|\phi^{+}\right\rangle _{12}\left|\psi^{-}\right\rangle _{34}\right)\\
 & = & \frac{1}{2}\left(\left|\psi^{-}\right\rangle _{14}\left|\phi^{+}\right\rangle _{23}+\left|\psi^{+}\right\rangle _{14}\left|\phi^{-}\right\rangle _{23}-\left|\phi^{-}\right\rangle _{14}\left|\psi^{+}\right\rangle _{23}-\left|\phi^{+}\right\rangle _{14}\left|\psi^{-}\right\rangle _{23}\right).
\end{array}\label{eq:Chapter2_Eq4}
\end{equation}
Alice (or Bob) encodes the result of her (his) Bell-state measurement
into classical bits following Equation (\ref{eq:Bell-State=000020with=000020Pre-Shared_Key}),
while keeping the measurement outcome undisclosed. In this framework,
Alice\textquoteright s measurement outcomes, each occurring with a
probability of $\frac{1}{4}$, correspond to the Bell states $\left|\psi^{-}\right\rangle ,$
$\left|\psi^{+}\right\rangle ,$ $\left|\phi^{-}\right\rangle $ and
$\left|\phi^{+}\right\rangle $, which are mapped to classical bit
values 11, 10, 01, and 00, respectively. Conversely, Bob\textquoteright s
measurement outcomes align with $\left|\phi^{+}\right\rangle ,$ $\left|\phi^{-}\right\rangle ,$
$\left|\psi^{+}\right\rangle $ and $\left|\psi^{-}\right\rangle $,
associated with classical bit values 00, 01, 10, and 11, respectively.
Within this setting, Alice assumes the verifier role, given that two
secret keys are involved in the scheme. If Bob announces his bit value
as 00 via an unjammable classical communication channel, Alice performs
an XOR operation between Bob\textquoteright s transmitted bit and
her classical bit value (11) derived from her measurement. In an ideal
scenario with no quantum channel disturbances such as noise or eavesdropping,
the computed XOR value should correspond to the first key from the
pre-shared sequence, which is 11. More generally, this XOR result
would match the $m^{{\rm th}}$ key within the pre-established key
sequence. The complete set of potential measurement outcomes and their
correlations is presented in Table \ref{tab:Chapter2_Tab6}, assuming
the initial pre-shared keys are 11 and 00.

\begin{table}[H]
\caption{This outlines the possible measurement results for the legitimate
participants within the proposed QIA scheme.}\label{tab:Chapter2_Tab6}

\centering{}%
\begin{tabularx}{15.9cm}{|>{\centering\arraybackslash}X|>{\centering\arraybackslash}X|}
\hline 
Alice and Bob's possible measurement outcomes  & Additional modulo 2 \tabularnewline
\hline 
$\left|\psi^{-}\right\rangle _{14}\otimes\left|\phi^{+}\right\rangle _{23}$  & $11\oplus00=11$\tabularnewline
\hline 
$\left|\psi^{+}\right\rangle _{14}\otimes\left|\phi^{-}\right\rangle _{23}$  & $10\oplus01=11$\tabularnewline
\hline 
$\left|\phi^{-}\right\rangle _{14}\otimes\left|\psi^{+}\right\rangle _{23}$  & $01\oplus10=11$\tabularnewline
\hline 
$\left|\phi^{+}\right\rangle _{14}\otimes\left|\psi^{-}\right\rangle _{23}$  & $00\oplus11=11$\tabularnewline
\hline 
\end{tabularx}
\end{table}

We examine three additional cases to validate the robustness of our
conclusion across the remaining possible secret keys positioned at
the $m^{{\rm th}}$ place in the pre-shared key sequence, specifically
\{00\}, \{01\} and \{10\}. This allows us to illustrate the final
quantum state shared between Alice and Bob. The resultant composite
systems, where the secret keys at the $m^{{\rm th}}$ and $\left(m+1\right)^{{\rm th}}$
positions in the sequence take values $\left\{ 00,01\right\} ,$ $\left\{ 01,10\right\} $
and $\left\{ 10,11\right\} $, respectively, are formulated as follows:

\begin{equation}
\begin{array}{lcl}
\left|\Psi^{00}\right\rangle  & = & \mathds{1}_{1}\otimes\mathds{1}_{3}\left(\left|\phi^{-}\right\rangle _{12}\left|\phi^{-}\right\rangle _{34}\right)\\
 & = & \frac{1}{2}\left(\left|\phi^{+}\right\rangle _{14}\left|\phi^{+}\right\rangle _{23}+\left|\phi^{-}\right\rangle _{14}\left|\phi^{-}\right\rangle _{23}-\left|\psi^{+}\right\rangle _{14}\left|\psi^{+}\right\rangle _{23}-\left|\psi^{-}\right\rangle _{14}\left|\psi^{-}\right\rangle _{23}\right)
\end{array},\label{eq:Chapter2_Eq5}
\end{equation}

\begin{equation}
\begin{array}{lcl}
\left|\Psi^{01}\right\rangle  & = & \sigma_{x1}\otimes\sigma_{x3}\left(\left|\psi^{+}\right\rangle _{12}\left|\psi^{-}\right\rangle _{34}\right)\\
 & = & \frac{1}{2}\left(-\left|\phi^{+}\right\rangle _{14}\left|\phi^{-}\right\rangle _{23}-\left|\phi^{-}\right\rangle _{14}\left|\phi^{+}\right\rangle _{23}+\left|\psi^{+}\right\rangle _{14}\left|\psi^{-}\right\rangle _{23}+\left|\psi^{-}\right\rangle _{14}\left|\psi^{+}\right\rangle _{23}\right)
\end{array},\label{eq:Chapter2_Eq6}
\end{equation}
and 
\begin{equation}
\begin{array}{lcl}
\left|\Psi^{10}\right\rangle  & = & i\sigma_{y1}\otimes i\sigma_{y3}\left(\left|\psi^{-}\right\rangle _{12}\left|\phi^{-}\right\rangle _{34}\right)\\
 & = & \frac{1}{2}\left(\left|\psi^{+}\right\rangle _{14}\left|\phi^{+}\right\rangle _{23}+\left|\psi^{-}\right\rangle _{14}\left|\phi^{-}\right\rangle _{23}+\left|\phi^{+}\right\rangle _{14}\left|\psi^{+}\right\rangle _{23}+\left|\phi^{-}\right\rangle _{14}\left|\psi^{-}\right\rangle _{23}\right)
\end{array},\label{eq:Chapter2_Eq7}
\end{equation}
respectively. In Equations (\ref{eq:Chapter2_Eq4} - \ref{eq:Chapter2_Eq7}),
we evaluate four distinct pairwise combinations of the $m^{{\rm th}}$
and $\left(m+1\right)^{{\rm th}}$ keys. However, there are twelve
additional key pairings that yield identical measurement outcomes,
thereby reinforcing authentication between legitimate users.

\subsection{Description of Protocol 2.3}

In the scenario examined in this part, two legitimate parties, Alice
and Bob, aim to authenticate themselves with assistance from an untrusted
intermediary, Charlie. Within this authentication framework, Alice
and Bob share a pre-established classical secret key, denoted as $K_{AB}=\left\{ k_{1}^{1}k_{2}^{1},k_{1}^{2}k_{2}^{2},k_{1}^{3}k_{2}^{3},\cdots,k_{1}^{m}k_{2}^{m},\cdots k_{1}^{n+1}k_{2}^{n+1}\right\} $.
This secret key sequence, $K_{AB}$, consists of two-bit key elements
distributed uniformly, represented as $k_{1}^{m}k_{2}^{m}$, where
$k_{1}^{m}k_{2}^{m}\in\{00,01,10,11\}$. Before detailing the authentication
protocol's execution, it is essential to outline the role of the decoy
states in establishing secure communication between two entities \cite{LMC05}.
In this approach, one party randomly integrates decoy qubits---ideally
matching the original information sequence in quantity---into the
transmitted qubit sequence. This expanded sequence is then forwarded
to the receiving party. Upon reception, the receiving party requests
information regarding the positions and encoding of the decoy qubits
via a secure public channel. The sender publicly discloses this data,
enabling the recipient to validate it by measuring the decoy qubits.
If the detected error rate surpasses an acceptable threshold, both
parties discard the entire qubit sequence. The implementation of decoy
states strengthens channel security, ensuring reliable information
exchange between communicating entities.

\begin{figure}
\begin{centering}
\includegraphics[scale=0.45]{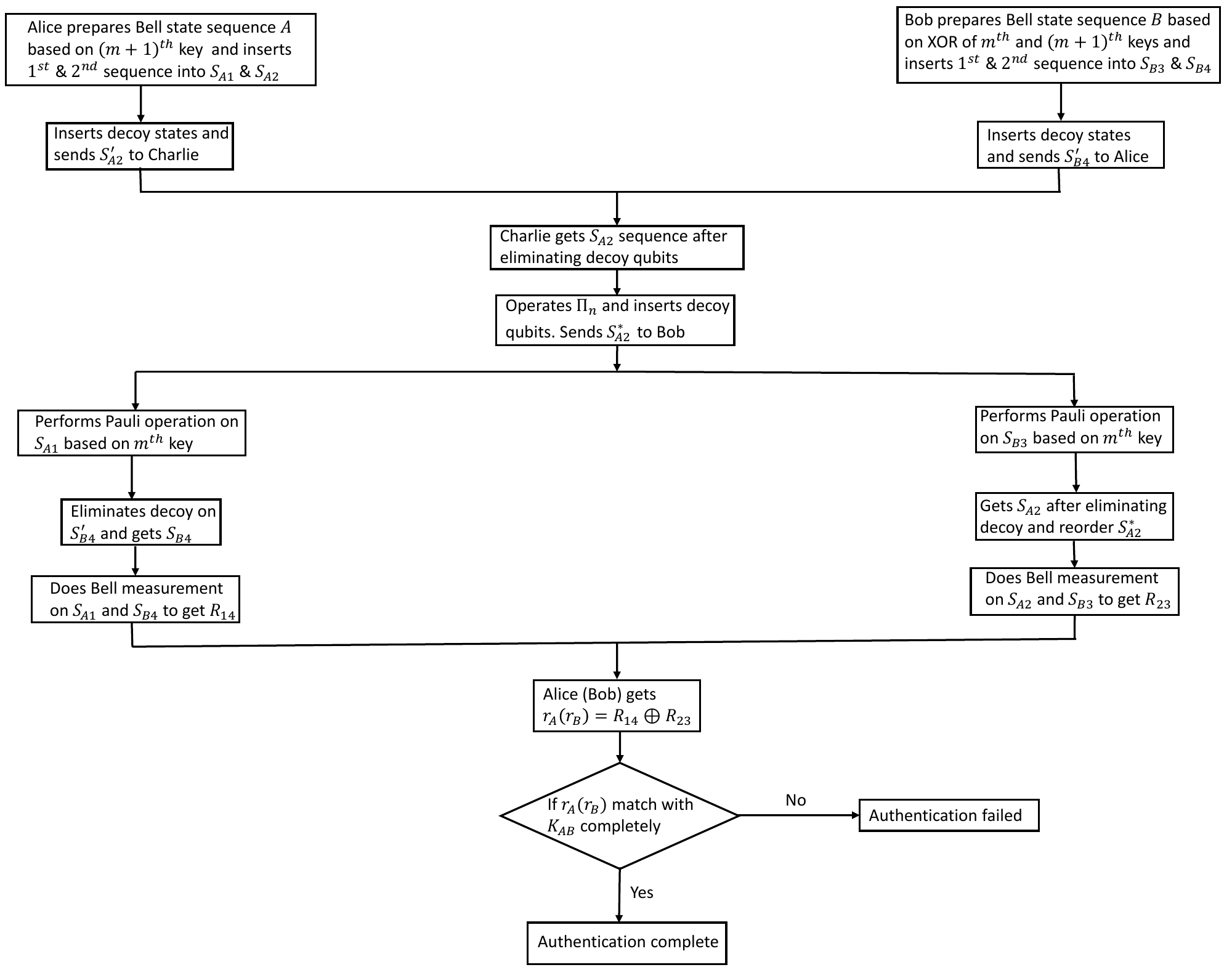} 
\par\end{centering}
\caption{Flowchart depicting the operational workflow of the proposed QIA
protocol.}\label{fig:Chapter2_Fig2}
\end{figure}

The subsequent part elaborates on the stepwise procedure involved
in the authentication process.
\begin{description}
\item [{Step~3.1}] Alice and Bob construct Bell state sequences $A_{12}$
and $B_{34}$ by utilizing the $\left(m+1\right)^{{\rm th}}$ key
and performing an XOR operation on the $m^{{\rm th}}$ and $\left(m+1\right)^{{\rm th}}$
keys from the entire sequence $K_{AB}$, as prescribed by the protocol
described in Equation (\ref{eq:Bell-State=000020with=000020Pre-Shared_Key}),
where $m=1,2,\cdots,n$.\\
 
\[
\begin{array}{lcl}
A & = & \left\{ |A\rangle_{12}^{1},|A\rangle_{12}^{2},|A\rangle_{12}^{3},\cdots,|A\rangle_{12}^{m},\cdots,|A\rangle_{12}^{n}\right\} ,\\
B & = & \left\{ |B\rangle_{34}^{1},|B\rangle_{34}^{2},|B\rangle_{34}^{3},\cdots,|B\rangle_{34}^{m},\cdots,|B\rangle_{34}^{n}\right\} .
\end{array}
\]
Here, the subscripts $1,2$ and $3,4$ denote the respective particles
assigned to Alice and Bob. Ideally, both sets of states should be
identical.
\item [{Step~3.2}] Alice and Bob divide the states of their Bell pairs
into two separate sequences, each containing $n$ particles. Within
these sequences, the first particle of each Bell pair forms one sequence,
while the second particle constitutes the other. Consequently, Alice
and Bob hold the sequences $S_{A1},\,S_{A2}$ and $S_{B3},\,S_{B4}$,
respectively.\\
 
\[
\begin{array}{lcl}
S_{A1} & = & \left\{ s_{1}^{1},s_{1}^{2},s_{1}^{3},\cdots,s_{1}^{m},\cdots,s_{1}^{n}\right\} ,\\
S_{A2} & = & \left\{ s_{2}^{1},s_{2}^{2},s_{2}^{3},\cdots,s_{2}^{m},\cdots,s_{2}^{n}\right\} ,\\
S_{B3} & = & \left\{ s_{3}^{1},s_{3}^{2},s_{3}^{3},\cdots,s_{3}^{m},\cdots,s_{3}^{n}\right\} ,\\
S_{B4} & = & \left\{ s_{4}^{1},s_{4}^{2},s_{4}^{3},\cdots,s_{4}^{m},\cdots,s_{4}^{n}\right\} .
\end{array}
\]
Specifically, $S_{A1}$ ($S_{A2}$) consists of the first (second)
particles of all Bell pairs in sequence $A$, while $S_{B3}$ ($S_{B4}$)
contains the first (second) particles of all Bell pairs in sequence
$B$. Alice and Bob retain the sequences $S_{A1}$ and $S_{B3}$ with
themselves, respectively. Additionally, Alice (Bob) embeds decoy particles
$D_{A}$ $\left(D_{B}\right)$ into $S_{A2}$ ($S_{B4}$), generating
the modified sequence $S_{A2}^{\prime}$ ($S_{B4}^{\prime}$). Alice
(Bob) then transmits $S_{A2}^{\prime}$ ($S_{B4}^{\prime}$) through
a quantum channel to Charlie (Alice).
\item [{Step~3.3}] Upon receiving $S_{A2}^{\prime}$, Charlie performs
security verification using the decoy particles. Once the security
tests are completed, Charlie discards $D_{A}$ to reconstruct the
original sequence $S_{A2}$. After obtaining $S_{A2}$,
Charlie applies a permutation operation $\Pi_{n}$ and introduces
additional decoy states, forming a new sequence $S_{A2}^{*}$. Finally,
Charlie forwards $S_{A2}^{*}$ to Bob.
\item [{Step~3.4}] Alice and Bob execute a security validation using decoy
particles. Once the security test is completed, Alice
discards the decoy states, labeled as$D_{B}$, from the expanded sequence
$S_{B4}^{'}$, thereby reverting it to the original sequence $S_{B4}$
as initially prepared by Bob. Following this, Charlie publicly announces
the exact arrangement of the sequence$S_{A2}^{*}$, which corresponds
to the inverse operation of the $\Pi_{n}$ operator. Bob then reorders
his sequence accordingly to reconstruct its initial configuration,
denoted as $S_{A2}$.
\item [{Step~3.5}] Alice (Bob) applies Pauli operations to each $m^{{\rm th}}$
particle in the sequence $S_{A1}$ ($S_{B3}$) based on the corresponding
$m^{{\rm th}}$key from their pre-shared key sequence, in accordance
with Equation \ref{eq:Pauli-Operation=000020with=000020Pre-Shared_Key}).
Afterward, Alice (Bob) retains the sequences $S_{A1},\,S_{B4}$ ($S_{A2},\,S_{B3}$)
and proceeds to perform Bell measurements on the paired particles
within these sequences\footnote{A Bell measurement is conducted on the $m^{{\rm th}}$ particle of
sequence $S_{A1}$($S_{A2}$) along with the corresponding $m^{{\rm th}}$
particle of sequence $S_{B4}$ ($S_{B3}$).}. The results from these Bell state measurements are then recorded
by Alice and Bob as classical key sequences, denoted as $R_{14}=\left\{ r_{14}^{1},r_{14}^{2},r_{14}^{3},\cdots,r_{14}^{m},\cdots,r_{14}^{n}\right\} $
and $R_{23}=\left\{ r_{23}^{1},r_{23}^{2},r_{23}^{3},\cdots,r_{23}^{m},\cdots,r_{23}^{n}\right\} $,
following Equation (\ref{eq:Bell-State=000020with=000020Pre-Shared_Key}).
\item [{Step~3.6}] Alice publicly reveals the positions and values of
$\frac{n}{2}$ keys from the sequence $R_{14}$, while Bob discloses
$\frac{n}{2}$ keys from the sequence $R_{23}$ at the corresponding
remaining positions within $R_{14}$. Alice (Bob) performs XOR operations
between the key announced by Bob (Alice) and the corresponding key
from her (his) sequence $R_{14}$ ($R_{23}$), thereby generating
a new two-bit key sequence $r_{A}$ ($r_{B}$) of length $n$.
\item [{Step~3.7}] In the absence of errors, the elements of $r_{A}$
and $r_{B}$ should correspond to the respective key elements of the
initial sequence $K_{AB}$\footnote{In this process, the $n$ elements of the sequences $r_{A}$ and $r_{B}$
must align with the corresponding $n$ elements of $K_{AB}$. It is
important to note that $K_{AB}$ consists of $n+1$ pre-shared keys.}. If this condition is satisfied, mutual authentication between Alice
and Bob is successfully established. This authentication procedure
is conducted simultaneously by both legitimate parties. \\
The proposed QIA protocol is visually represented in the flowchart
shown in Figure \ref{fig:Chapter2_Fig2}.
\end{description}

\subsection{Assessment of the security aspects in Protocol 2.3}\label{sec:Chapter2_Sec3.3}

This section assesses the security resilience of the proposed protocol
against several well-established attacks that an eavesdropper, commonly
referred to as Eve, might attempt. Notably, insider threats originating
from Alice and Bob are excluded from this discussion, as they fall
outside the scope of QIA. The evaluation begins with the impersonation
attack, where an untrusted entity, Charlie, seeks unauthorized access
to sensitive information while adhering to all protocol steps. Additionally,
an impersonation-based fraudulent attack involves an external adversary,
such as Eve, attempting to masquerade as a legitimate participant---Alice
or Bob---to successfully navigates the authentication mechanism. Among
the various threats analyzed in this section, the intercept-resend
attack and impersonation-based fraudulent attack specifically exploit
vulnerabilities within the quantum channel.

\subsubsection{Security evaluation of Protocol 2.3 against an impersonation attack
by Eve}

To analyze the security of the protocol, consider the impersonation
scenario discussed in Section \ref{subsec:Chapter2_Sec3.1}. In this
case, Eve assumes Alice\textquoteright s identity and selects the
state $|\phi^{+}\rangle_{12}$, subsequently transmitting the particle
sequence $S_{e2}$ (corresponding to particle 2) to Charlie\footnote{The term ``sequence'' is used here for generalization purposes,
though the security assessment follows a specific example from Section
\ref{subsec:Chapter2_Sec3.1}.}. Charlie proceeds according to the protocol, performing the designated
permutation operation before forwarding the modified sequence $S_{e2}^{\prime}$
to Bob. Once Charlie publicly discloses the correct ordering of $S_{e2}^{\prime}$,
Bob reconstructs the original sequence $S_{e2}$. Additionally, Bob
transmits the particle sequence $S_{B4}$ directly to Eve. Eve, impersonating
Alice, executes all protocol steps as required, applying the Pauli
operation ($\sigma_{z1}$) on Particle 1. In parallel, Bob applies
the Pauli operation ($\sigma_{z3}$) on his own Particle 3. Following
this, both Bob and Eve conduct Bell-state measurements. The outcomes
of these measurements, as detailed in Table \ref{tab:Chapter2_Tab6},
expose Eve\textquoteright s attempt at impersonation.

\begin{figure}
\begin{centering}
\includegraphics[scale=0.5]{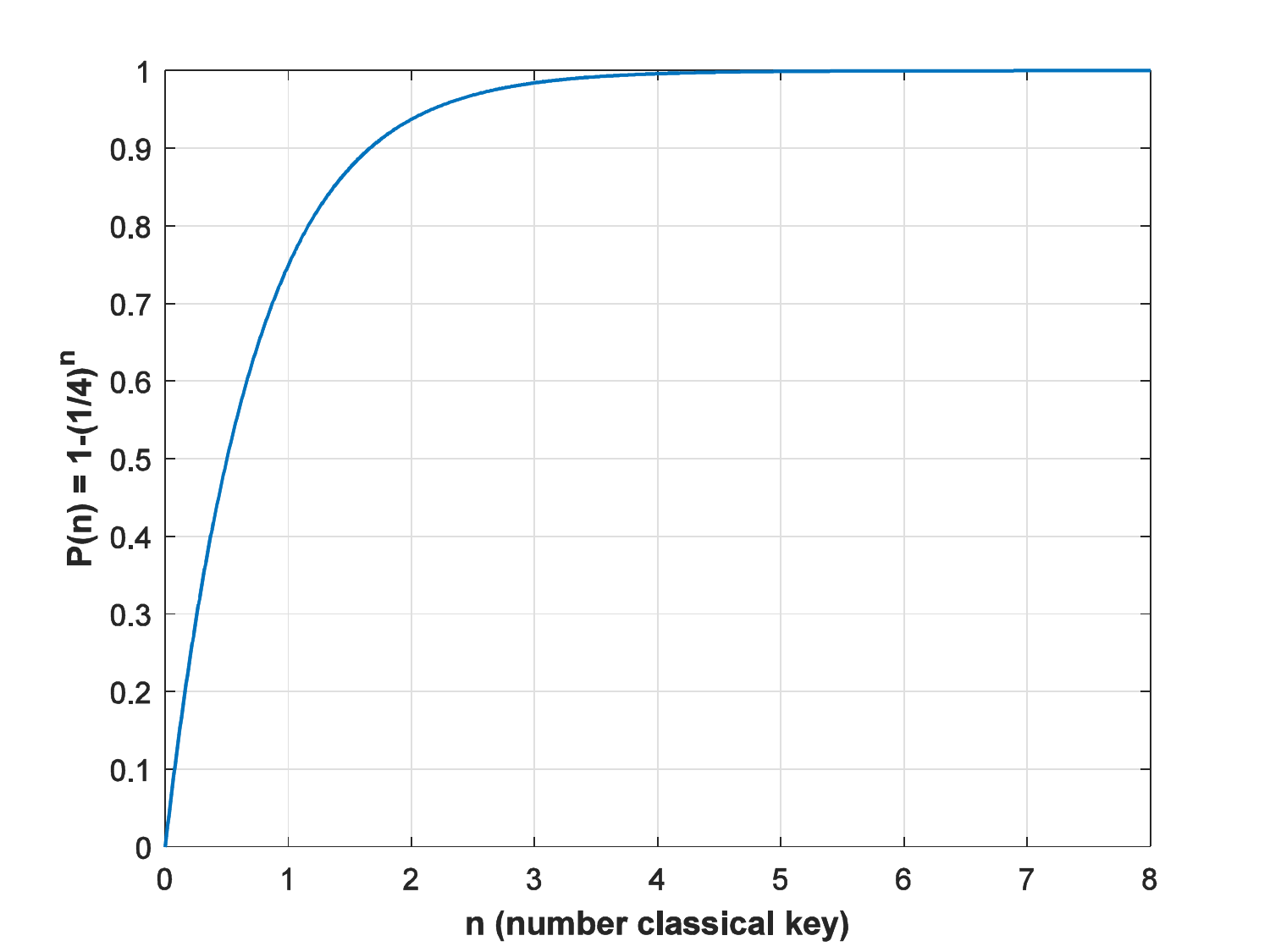} 
\par\end{centering}
\caption{The probability $P\left(n\right)$ of detecting Eve's presence is
correlated with the number of classical authentication keys $n$ that
are pre-shared.}\label{fig:Chapter2_Fig3}
\end{figure}

Since Eve lacks access to the legitimate key pair $K_{AB}$ (e.g.,
11 and 00) and the permutation operation $\Pi_{n}$ performed by Charlie,
her probability of correctly determining the Bell state is $\frac{1}{4}$.
For Eve to remain undetected, she must correctly guess all $n$ Bell
states. Consequently, the probability of a successful impersonation
attempt is $\left(\frac{1}{4}\right)^{n}$. As $n$ increases, this
probability approaches zero, making successful impersonation virtually
impossible. Therefore, the probability $P\left(n\right)$ of detecting
Eve\textquoteright s presence is given by $1-\left(\frac{1}{4}\right)^{n}$.
For sufficiently large $n$, $P\left(n\right)$ approaches 1, ensuring
the detection of an impersonation attack with near certainty. The
dependence of $P\left(n\right)$ on $n$ is depicted in Figure \ref{fig:Chapter2_Fig3},
demonstrating that a minimum of six pre-shared keys is required to
reliably identify Eve\textquoteright s intrusion.

\subsubsection{Security evaluation of Protocol 2.3 against intercept and resend
attack}

\begin{figure}
\begin{centering}
\includegraphics[scale=0.5]{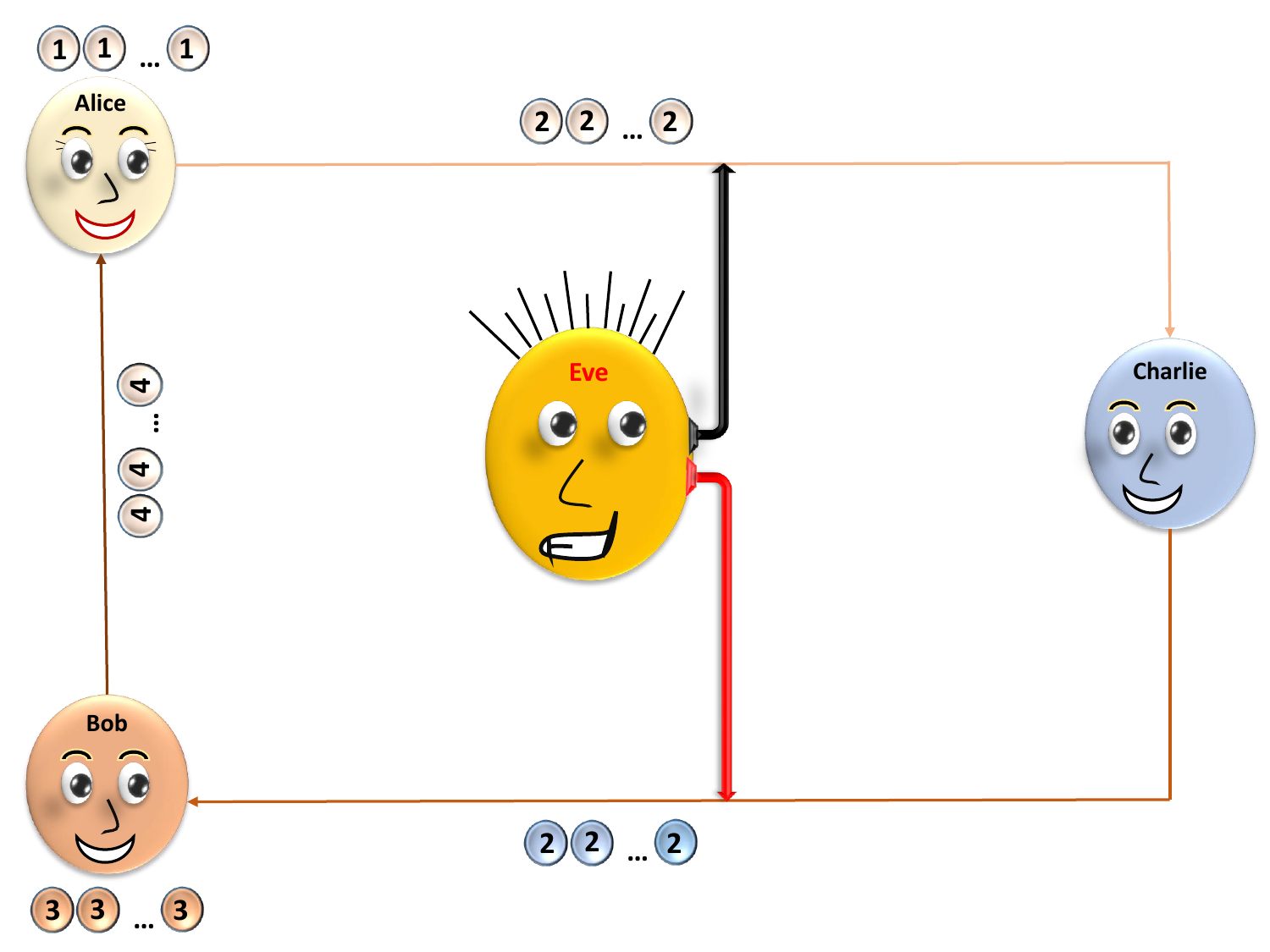} 
\par\end{centering}
\caption{The IR attack strategy utilized by Eve is structured such that qubits
labeled 1, 2, 3, and 4 correspond to sequences $S_{A1}$, $S_{A2}$,
$S_{B3}$ and $S_{B4}$, respectively.}\label{fig:Chappter2_Fig4}
\end{figure}

In this quantum protocol, Alice and Bob utilize their pre-shared secret
key to generate Bell states. Rather than transmitting their complete
quantum states, they selectively send Particle 2 and Particle 4 via
the quantum channel. This method confines Eve's attack opportunities
to these specific particles at any given time. For clarity, we assume
that Eve attempts to intercept Particle 2 during its transmission
from Alice to Charlie and Particle 4 when sent from Bob to Alice.
The amount of information Eve can extract from the quantum channel
is fundamentally restricted by the Holevo bound or Holevo quantity
\cite{H_73}, given by

\begin{equation}
\begin{array}{lcl}
\chi\left(\rho\right) & = & S\left(\rho\right)-\underset{i}{\sum}{\rm p}_{i}S\left(\rho_{i}\right),\end{array}\label{eq:Chapter2_Eq8}
\end{equation}
which defines the upper limit of accessible information. Here, $S\left(\rho\right)=-{\rm Tr}\left(\rho\,{\rm log_{2}}\rho\right)$
represents the von Neumann entropy, where $\rho_{i}$ denotes an individual
component of the mixed state $\rho$ with probability ${\rm p}_{i}$.
The density matrix $\rho$ is formulated as
\[
\rho=\frac{1}{4}\left[\left|\Psi^{00}\right\rangle +\left|\Psi^{01}\right\rangle +\left|\Psi^{10}\right\rangle +\left|\Psi^{11}\right\rangle \right]_{1234}=\underset{i}{\sum\,}{\rm p}_{i}\rho_{i},
\]
where $\rho_{{\rm i}}$ represents the density matrix of the states
defined in Equations (\ref{eq:Chapter2_Eq4}), (\ref{eq:Chapter2_Eq5}),
(\ref{eq:Chapter2_Eq6}) and (\ref{eq:Chapter2_Eq7}). As previously
established, Eve's objective is to target Particles 2 and 4. We aim to demonstrate a scenario where she can simultaneously intercept
both. However, in such a case, the amount of useful information Eve
can acquire remains constrained by the Holevo quantity. To ensure
consistency with this constraint, we modify Equation (\ref{eq:Chapter2_Eq8})
accordingly.

\begin{equation}
\begin{array}{lcl}
\chi\left(\rho^{24}\right) & = & S\left(\rho^{24}\right)-\underset{i}{\sum}{\rm p}_{i}S\left(\rho_{i}^{24}\right),\end{array}\label{eq:Chapter2_Eq9}
\end{equation}
In this context, $\rho^{24}$ and $\rho_{i}^{24}$ denote the reduced
density matrices derived from $\rho$ and $\rho_{i}$, respectively,
by executing a partial trace over particles 1 and 3. Through direct
computation, we find that $\rho^{24}={\rm Tr}_{13}\left(\rho\right)=\frac{1}{4}\left(|11\rangle\langle11|+|10\rangle\langle10|+|01\rangle\langle01|+|00\rangle\langle00|\right)=\frac{1}{4}\mathds{1}_{4}$,
where $\mathds{1}_{4}$ represents the $4\times4$ identity matrix.
Likewise, the von Neumann entropy for each component of the mixed
state $\rho$ after the partial trace, expressed as $\rho_{i}^{24}={\rm Tr}_{13}\left(\rho_{i}\right)=\frac{1}{4}\mathds{1}_{4}$,
follows the same structure. By substituting $\rho^{24}$ and $\rho_{i}^{24}$
into Equation (\ref{eq:Chapter2_Eq9}), it follows that $\chi\left(\rho^{24}\right)=0$.
This result implies that Eve is unable to extract any key information
through direct interception of the transmitted particles. The intercept-resend
(IR) attack strategy adopted by Eve is illustrated in Figure \ref{fig:Chappter2_Fig4}.
To mitigate or nullify Eve's information gain from the communication
channel in an IR attack, the maximally mixed state is formulated with
an equal probability distribution among the Bell states, as described
in Equation (\ref{eq:Chapter2_Eq9}). However, if these probabilities
are asymmetric, the system's security against IR attacks will depend
on the specific probability values of the Bell states within the mixed
state. The protocol remains secure against the IR attack as long as
the mutual information between Alice and Bob exceeds the Holevo bound,
i.e., $I(A|B)>\chi\left(\rho\right)$ \cite{CRE_04}.

\subsubsection{Security evaluation of Protocol 2.3 against impersonated fraudulent
attack}

To analyze the resilience of our protocol against impersonation-based
fraudulent attacks, we consider two pre-shared secret keys, namely
11 and 00. Initially, Alice and Bob generate the Bell states $|\phi^{+}\rangle_{12}$
and $|\psi^{-}\rangle_{34}$, respectively. Simultaneously, Eve fabricates
two single-qubit states, $|\chi\rangle_{5}$ and $|\chi\rangle_{6}$,
which serve as counterfeit (fake) states. These states are defined
as $a|0\rangle+b|1\rangle$ and $c|0\rangle+d|1\rangle$, respectively,
where the normalization conditions $\left|a\right|^{2}+\left|b\right|^{2}=1$
and $\left|c\right|^{2}+\left|d\right|^{2}=1$ hold. To execute the
attack, Eve applies a CNOT gate, using qubit 2 (4) as the control
and qubit 5 (6) as the target. She retains qubits 2 and 4 while transmitting
qubits 5 and 6 to Bob and Alice, respectively. The resultant composite
state after Eve's manipulation through the CNOT operation is given
by:

\begin{equation}
\begin{array}{lcl}
 &  & {\rm CNOT_{2(4)\rightarrow5(6)}|\phi^{+}\rangle_{12}\otimes|\psi^{-}\rangle_{34}\otimes|\chi\rangle_{5}\otimes|\chi\rangle_{6}}\\
 & = & \frac{1}{2}\left[|0001\rangle\left(a|0\rangle+b|1\rangle\right)\left(c|1\rangle+d|0\rangle\right)-|0010\rangle\left(a|0\rangle+b|1\rangle\right)\left(c|0\rangle+d|1\rangle\right)\right.\\
 & + & \left.|1101\rangle\left(a|1\rangle+b|0\rangle\right)\left(c|1\rangle+d|0\rangle\right)-|1110\rangle\left(a|1\rangle+b|0\rangle\right)\left(c|0\rangle+d|1\rangle\right)\right]_{123456}\\
 & = & \frac{1}{2}\left[ac|000101\rangle+ad|000100\rangle+bc|000111\rangle+bd|000110\rangle\right.\\
 & - & ac|001000\rangle-ad|001001\rangle-bc|001010\rangle-bd|001011\rangle\\
 & + & ac|110111\rangle-ad|110110\rangle-bc|110101\rangle-bd|110100\rangle\\
 & - & \left.ac|111010\rangle-ad|111011\rangle-bc|111000\rangle-bd|111001\rangle\right]_{123456}
\end{array}.\label{eq:Chapter2_Eq10}
\end{equation}
According to the protocol, Alice and Bob each perform a $\sigma_{z}$
operation on their respective qubits 1 and 3. Following this, they
conduct Bell-state measurements, with Alice measuring the entangled
pair ($1,6$) and Bob measuring ($5,3$). Meanwhile, Eve retains qubits
2 and 4. The overall quantum system encompassing Alice, Bob, and Eve,
post-Pauli operations and Bell measurements as per Equation (\ref{eq:Chapter2_Eq10}),
is represented as: 

\begin{equation}
\begin{array}{lcl}
|\Psi\rangle & = & \frac{1}{2\sqrt{2}}\left[ac\left(|\psi^{+}\rangle|\phi^{+}\rangle|\psi^{-}\rangle+|\psi^{+}\rangle|\phi^{-}\rangle|\psi^{+}\rangle+|\psi^{-}\rangle|\phi^{+}\rangle|\psi^{+}\rangle+|\psi^{-}\rangle|\phi^{-}\rangle|\psi^{+}\rangle\right.\right.\\
 & + & \left.|\phi^{+}\rangle|\psi^{+}\rangle|\phi^{-}\rangle+|\phi^{+}\rangle|\psi^{-}\rangle|\phi^{+}\rangle+|\phi^{-}\rangle|\psi^{+}\rangle|\phi^{+}\rangle+|\phi^{-}\rangle|\psi^{-}\rangle|\phi^{-}\rangle\right)\\
 & + & ad\left(|\phi^{+}\rangle|\phi^{+}\rangle|\psi^{-}\rangle+|\phi^{+}\rangle|\phi^{-}\rangle|\psi^{+}\rangle+|\phi^{-}\rangle|\phi^{+}\rangle|\psi^{+}\rangle+|\phi^{-}\rangle|\phi^{-}\rangle|\psi^{-}\rangle\right.\\
 & + & \left.|\psi^{+}\rangle|\psi^{+}\rangle|\phi^{-}\rangle+|\psi^{+}\rangle|\psi^{-}\rangle|\phi^{+}\rangle+|\psi^{-}\rangle|\psi^{+}\rangle|\phi^{+}\rangle+|\psi^{-}\rangle|\psi^{-}\rangle|\phi^{-}\rangle\right)\\
 & + & bc\left(|\psi^{+}\rangle|\psi^{+}\rangle|\psi^{-}\rangle-|\psi^{+}\rangle|\psi^{-}\rangle|\psi^{+}\rangle+|\psi^{-}\rangle|\psi^{+}\rangle|\psi^{+}\rangle-|\psi^{-}\rangle|\psi^{-}\rangle|\psi^{-}\rangle\right.\\
 & + & \left.|\phi^{+}\rangle|\phi^{+}\rangle|\phi^{-}\rangle-|\phi^{+}\rangle|\phi^{-}\rangle|\phi^{+}\rangle+|\phi^{-}\rangle|\phi^{+}\rangle|\phi^{+}\rangle-|\phi^{-}\rangle|\phi^{-}\rangle|\phi^{-}\rangle\right)\\
 &  & bd\left(|\phi^{+}\rangle|\psi^{+}\rangle|\psi^{-}\rangle-|\phi^{+}\rangle|\psi^{-}\rangle|\psi^{+}\rangle+|\phi^{-}\rangle|\psi^{+}\rangle|\psi^{+}\rangle-|\phi^{-}\rangle|\psi^{-}\rangle|\psi^{-}\rangle\right.\\
 & + & \left.\left.|\psi^{+}\rangle|\phi^{+}\rangle|\phi^{-}\rangle-|\psi^{+}\rangle|\phi^{-}\rangle|\phi^{+}\rangle+|\psi^{-}\rangle|\phi^{+}\rangle|\phi^{+}\rangle-|\psi^{-}\rangle|\phi^{-}\rangle|\phi^{-}\rangle\right)\right]_{165324}
\end{array}.\label{eq:Final_Composite_State_Single_Qubit}
\end{equation}
When Alice and Bob receive Bell pairs in the states $\left|\psi^{+}\right\rangle _{16}\left|\phi^{-}\right\rangle _{53},\left|\psi^{-}\right\rangle _{16}\left|\phi^{+}\right\rangle _{53},\left|\phi^{+}\right\rangle _{16}\left|\psi^{-}\right\rangle _{53}\,$
and $\left|\phi^{-}\right\rangle _{16}\left|\psi^{+}\right\rangle _{53}$
following measurement, the probability of detecting an eavesdropper
becomes zero. By evaluating the density operator of the final composite
state, $|\Psi\rangle\langle\Psi|$, as given in Equation (\ref{eq:Final_Composite_State_Single_Qubit}),
the probability of Eve remaining undetected, denoted as ${\rm P_{nd}}$,
is given by $\frac{1}{2}\left(\left|ac\right|^{2}+\left|bd\right|^{2}\right)$.
Consequently, the probability of detecting Eve, ${\rm P_{d}}$, is
expressed as 
\[
{\rm P_{d}=1-}\frac{1}{2}\left(\left|ac\right|^{2}+\left|bd\right|^{2}\right).
\]
To minimize her detection probability, Eve can choose $\left|a\right|=\left|b\right|=\left|c\right|=\left|d\right|=\frac{1}{\sqrt{2}}$,
leading to her single-qubit states being $|+\rangle_{5}=\frac{1}{\sqrt{2}}\left(|0\rangle+|1\rangle\right)_{5}$
and $|+\rangle_{6}=\frac{1}{\sqrt{2}}\left(|0\rangle+|1\rangle\right)_{6}$.
Under this condition, the minimum detection probability of Eve's presence
is ${\rm P_{d}}=\frac{3}{4}$, when she employs a single-qubit fake
state.

Next, we examine the same scenario with the only distinction being
that Eve uses an entangled fake state to mimic a legitimate participant.
Retaining the same pre-shared key pairs, 11 and 00, Eve's fake state
is given by $|\chi^{\prime}\rangle_{56}=\left(a^{\prime}|00\rangle+b^{\prime}|01\rangle+c^{\prime}|10\rangle+d^{\prime}|11\rangle\right)_{56}$,
where the coefficients satisfy the normalization condition $\left|a^{\prime}\right|^{2}+\left|b^{\prime}\right|^{2}+\left|c^{\prime}\right|^{2}+\left|d^{\prime}\right|^{2}=1$.
The rest of the procedure remains unchanged from the previous case.
Ultimately, the final composite system of Alice, Bob, and Eve after
their Bell measurement is,

\begin{equation}
\begin{array}{lcl}
|\Psi^{\prime}\rangle & = & \frac{1}{2\sqrt{2}}\left[a^{\prime}\left(|\psi^{+}\rangle|\phi^{+}\rangle|\psi^{-}\rangle+|\psi^{+}\rangle|\phi^{-}\rangle|\psi^{+}\rangle+|\psi^{-}\rangle|\phi^{+}\rangle|\psi^{+}\rangle+|\psi^{-}\rangle|\phi^{-}\rangle|\psi^{+}\rangle\right.\right.\\
 & + & \left.|\phi^{+}\rangle|\psi^{+}\rangle|\phi^{-}\rangle+|\phi^{+}\rangle|\psi^{-}\rangle|\phi^{+}\rangle+|\phi^{-}\rangle|\psi^{+}\rangle|\phi^{+}\rangle+|\phi^{-}\rangle|\psi^{-}\rangle|\phi^{-}\rangle\right)\\
 & + & b^{\prime}\left(|\phi^{+}\rangle|\phi^{+}\rangle|\psi^{-}\rangle+|\phi^{+}\rangle|\phi^{-}\rangle|\psi^{+}\rangle+|\phi^{-}\rangle|\phi^{+}\rangle|\psi^{+}\rangle+|\phi^{-}\rangle|\phi^{-}\rangle|\psi^{-}\rangle\right.\\
 & + & \left.|\psi^{+}\rangle|\psi^{+}\rangle|\phi^{-}\rangle+|\psi^{+}\rangle|\psi^{-}\rangle|\phi^{+}\rangle+|\psi^{-}\rangle|\psi^{+}\rangle|\phi^{+}\rangle+|\psi^{-}\rangle|\psi^{-}\rangle|\phi^{-}\rangle\right)\\
 & + & c^{\prime}\left(|\psi^{+}\rangle|\psi^{+}\rangle|\psi^{-}\rangle-|\psi^{+}\rangle|\psi^{-}\rangle|\psi^{+}\rangle+|\psi^{-}\rangle|\psi^{+}\rangle|\psi^{+}\rangle-|\psi^{-}\rangle|\psi^{-}\rangle|\psi^{-}\rangle\right.\\
 & + & \left.|\phi^{+}\rangle|\phi^{+}\rangle|\phi^{-}\rangle-|\phi^{+}\rangle|\phi^{-}\rangle|\phi^{+}\rangle+|\phi^{-}\rangle|\phi^{+}\rangle|\phi^{+}\rangle-|\phi^{-}\rangle|\phi^{-}\rangle|\phi^{-}\rangle\right)\\
 &  & d^{\prime}\left(|\phi^{+}\rangle|\psi^{+}\rangle|\psi^{-}\rangle-|\phi^{+}\rangle|\psi^{-}\rangle|\psi^{+}\rangle+|\phi^{-}\rangle|\psi^{+}\rangle|\psi^{+}\rangle-|\phi^{-}\rangle|\psi^{-}\rangle|\psi^{-}\rangle\right.\\
 & + & \left.\left.|\psi^{+}\rangle|\phi^{+}\rangle|\phi^{-}\rangle-|\psi^{+}\rangle|\phi^{-}\rangle|\phi^{+}\rangle+|\psi^{-}\rangle|\phi^{+}\rangle|\phi^{+}\rangle-|\psi^{-}\rangle|\phi^{-}\rangle|\phi^{-}\rangle\right)\right]_{165324}
\end{array}.\label{eq:Chapter2_Eq12}
\end{equation}
In the given scenario, the probability of detecting Eve's presence
is zero when Alice and Bob obtain measurement results corresponding
to specific Bell pairs: $\left|\psi^{+}\right\rangle _{16}\left|\phi^{-}\right\rangle _{53},\left|\psi^{-}\right\rangle _{16}\left|\phi^{+}\right\rangle _{53},\left|\phi^{+}\right\rangle _{16}\left|\psi^{-}\right\rangle _{53}\,$
and $\left|\phi^{-}\right\rangle _{16}\left|\psi^{+}\right\rangle _{53}$.
The probability of Eve remaining undetected is determined using the
density matrix $|\Psi^{\prime}\rangle\langle\Psi^{\prime}|$ from
Equation (\ref{eq:Chapter2_Eq12}), expressed as ${\rm P_{nd}}=\frac{1}{2}\left(\left|a^{\prime}\right|^{2}+\left|d^{\prime}\right|^{2}\right)$.
Consequently, the probability of detecting Eve is given by ${\rm P_{d}}=1-\frac{1}{2}\left(\left|a^{\prime}\right|^{2}+\left|d^{\prime}\right|^{2}\right)$.
To minimize her probability of detection, Eve selects $\left|a^{\prime}\right|=\left|d^{\prime}\right|=\frac{1}{\sqrt{2}}$
and $\left|b^{\prime}\right|=\left|c^{\prime}\right|=0$. The entangled
state utilized by Eve is given by $|\phi^{+}\rangle_{56}=\frac{1}{\sqrt{2}}\left(|00\rangle+|11\rangle\right)_{56}$,
which results in a detection probability of ${\rm P_{d}}=\frac{1}{2}$.
This indicates that employing an entangled state rather than a single
qubit state reduces the probability of Eve\textquoteright s detection.
Consequently, the minimal detection probabilities for single qubit
and entangled states are $\frac{3}{4}$ and $\frac{1}{2}$, respectively.
Based on this analysis, it can be concluded that the protocol remains
secure against an impersonation attack by Eve, particularly under
her optimal strategy.

\subsection{Collective noise analysis of Protocol 2.3}\label{sec:Chapter2_Sec3.4}

The previously discussed QIA protocol was examined under the assumption
of an ideal quantum channel. However, in practical scenarios, quantum
channels introduce noise, which can perturb the transmitted particles'
states. Moreover, an adversary may exploit this noise to conceal an
attack, making it challenging to distinguish between errors originating
from environmental noise and those introduced by malicious interference.
Mitigating the effects of collective noise in quantum communication
remains a significant challenge\cite{CGL+18,HM16,HWL++14}. Walton
et al. \cite{WAS+03} proposed the concept of a decoherence-free subspace
(DFS) \cite{SQM+24,QMN24}, which serves as a means to counteract
collective noise by preserving the states of quantum particles within
these protected subspaces. Subsequent research has further explored
quantum communication under the influence of collective noise \cite{HM17,GCQ18}.
Collective noise generally encompasses both collective-dephasing and
collective-rotation noise \cite{LDZ08}. The following analysis examines
the impact of these noise types on the proposed scheme.

\subsubsection{Collective-dephasing noise}

Collective-dephasing noise can be described as \cite{LDZ08,GCQ18}

\[
\begin{array}{lcl}
U_{dp}|0\rangle=|0\rangle & , & U_{dp}|1\rangle=e^{i\phi}|1\rangle\end{array}.
\]
This type of noise is associated with a time-dependent parameter $\phi$.
A logical qubit encoded in the product states of two physical qubits,
given by
\[
\begin{array}{lclc}
|0\rangle_{L}=|01\rangle & , & |1\rangle_{L}=|10\rangle & ,\end{array}
\]
is inherently resilient to collective-dephasing noise, as both logical
qubit components acquire an identical phase factor $e^{i\phi}$.

\begin{figure}[h]
\begin{centering}
\includegraphics[scale=0.5]{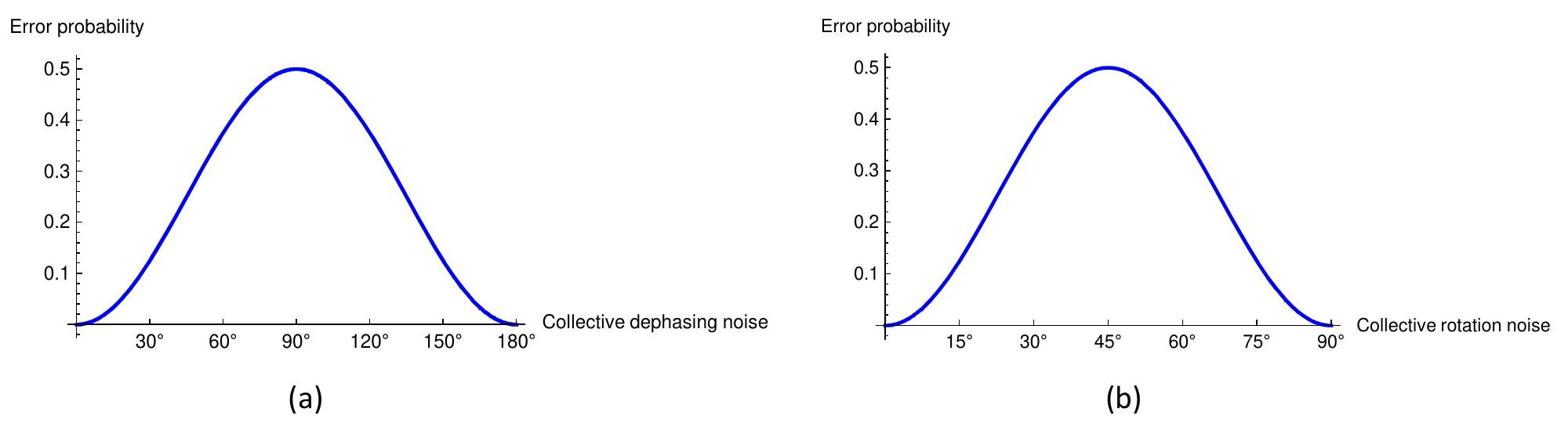}
\par\end{centering}
\caption{The collective error probability associated with the noise parameter
can be classified into: (a) Collective dephasing error probability,
characterized by the noise parameter $\phi$, and (b) Collective rotation
error probability, defined by the noise parameter $\theta$.}\label{fig:Collective-error-probability}
\end{figure}

Here, the subscript $L$ designates a logical qubit, whereas 0 and
1 correspond to horizontal and vertical polarization states, respectively---these
being the eigenstates of the Pauli operator$\sigma_{z}$ ($Z$ basis).
The logical states $|+_{L}\rangle$ and $|-_{L}\rangle$ are defined
as follows:

\[
\begin{array}{ccccc}
|+_{L}\rangle & = & \frac{1}{\sqrt{2}}\left(|0_{L}\rangle+|1_{L}\rangle\right) & = & \frac{1}{\sqrt{2}}\left(|01\rangle+|10\rangle\right)\\
|-_{L}\rangle & = & \frac{1}{\sqrt{2}}\left(|0_{L}\rangle-|1_{L}\rangle\right) & = & \frac{1}{\sqrt{2}}\left(|01\rangle+|10\rangle\right)
\end{array}.
\]
Based on the equations presented in Section \ref{subsec:Chapter2_Sec3.1},
particularly Equations (\ref{eq:Chapter2_Eq4}) and (\ref{eq:Chapter2_Eq7}),
it is evident that particles 2 and 4 within the composite states serve
as carriers of quantum information through the quantum channel. For
instance, considering Equation (\ref{eq:Chapter2_Eq4}):

\[
\begin{array}{lcl}
\left|\Psi^{11}\right\rangle  & = & \sigma_{z1}\otimes\sigma_{z3}\left(\left|\phi^{+}\right\rangle _{12}\left|\psi^{-}\right\rangle _{34}\right)\\
 & = & \frac{1}{2}\sigma_{z1}\otimes\sigma_{z3}\left(|0001\rangle-|0010\rangle+|1101\rangle-|1110\rangle\right)_{1234}\\
 & = & \frac{1}{2}\left(|0001\rangle+|0010\rangle-|1101\rangle-|1110\rangle\right)_{1234}
\end{array}.
\]
Upon introducing collective dephasing noise to particles 2 and 4,
the system evolves as follows:

\[
\begin{array}{lcl}
\left|\Psi^{11}\right\rangle _{dp} & = & \frac{1}{2}\left(e^{i\phi}|0001\rangle+|0010\rangle-e^{2i\phi}|1101\rangle-e^{i\phi}|1110\rangle\right)_{1234}\\
 & = & \frac{1}{2}\left(e^{i\phi}|01\rangle|00\rangle+|00\rangle|01\rangle-e^{2i\phi}|11\rangle|10\rangle-e^{i\phi}|10\rangle|11\rangle\right)_{1423}\\
 & = & \frac{1}{4}\left[2e^{i\phi}\left(|\psi^{+}\rangle|\phi^{-}\rangle+|\psi^{-}\rangle|\phi^{+}\rangle\right)+\left(1-e^{2i\phi}\right)\left(|\phi^{+}\rangle|\psi^{+}\rangle+|\phi^{-}\rangle|\psi^{-}\rangle\right)_{1423}\right..\\
 & + & \left.\left(1+e^{2i\phi}\right)\left(|\phi^{+}\rangle|\psi^{-}\rangle+|\phi^{-}\rangle|\psi^{+}\rangle\right)\right]_{1423}
\end{array}
\]
In an ideal scenario where no errors occur in the channel, the Bell
pairs shared between Alice and Bob belong to the set $\left\{ |\psi^{-}\rangle|\phi^{+}\rangle,|\psi^{+}\rangle|\phi^{-}\rangle,|\phi^{-}\rangle|\psi^{+}\rangle,|\phi^{+}\rangle|\psi^{-}\rangle\right\} $.
However, the presence of collective dephasing noise can alter the
output pairs to $\left\{ |\phi^{+}\rangle|\psi^{+}\rangle,|\phi^{-}\rangle|\psi^{-}\rangle\right\} $,
thereby introducing errors in the transmission. The probability of
such errors occurring due to this noise is given by $\frac{1}{4}\left(1-\cosh(2i\phi)\right)$,
where $\phi$ denotes the noise parameter. Figure \ref{fig:Collective-error-probability}
illustrates how this probability varies with respect to the noise
parameter. As shown in Figure \ref{fig:Collective-error-probability}.(a),
the error probability follows a characteristic curve, reaching a peak
value of 0.5 at $\phi=90^{\circ}$ and approaching zero at $\phi=0^{\circ}\,{\rm and}\,180^{\circ}$.
To mitigate or eliminate errors induced by this noise, the quantum
channel must be chosen such that the error remains within an acceptable
threshold. Given a tolerable error limit $p$, the solutions to $p=\frac{1}{4}\left(1-\cosh(2i\phi)\right)$
within the range $0^{\circ}<\phi<180^{\circ}$ must be determined.
If two solutions, $\phi_{1}$ and $\phi_{2}$, are found such that
$\phi_{1}<\phi_{2}$, the protocol remains secure when operating in
a channel where $\phi<\phi_{1}$ or $\phi>\phi_{2}$. To verify this
condition in practical implementations, it is essential to conduct
channel characterization prior to executing the protocol.

\subsubsection{Collective-rotation noise}

The collective rotation noise is mathematically expressed as:

\[
\begin{array}{lcl}
U_{r}|0\rangle=\cos\theta|0\rangle+\sin\theta|1\rangle & , & U_{r}|1\rangle=-\sin\theta|0\rangle+\cos\theta|1\rangle\end{array},
\]
Here, the parameter $\theta$ represents the collective rotation noise,
which varies dynamically within the quantum channel. For instance,
when this noise affects particles 2 and 4 as per Equation (\ref{eq:Chapter2_Eq4}),
the resulting expression is obtained after extensive calculations,

\[
\begin{array}{lcl}
U_{r2}\otimes U_{r4}\left|\Psi^{11}\right\rangle  & = & \frac{1}{2}\left[-\sin\theta\cos\theta|0000\rangle-\sin^{2}\theta|0010\rangle+\cos^{2}\theta|0100\rangle+\sin\theta\cos\theta|0110\rangle\right.\\
 & + & \sin\theta\cos\theta|1001\rangle-\cos^{2}\theta|1011\rangle+\sin^{2}\theta|1101\rangle-\sin\theta\cos\theta|1111\rangle\\
 & - & \left(\cos^{2}\theta|0001\rangle+\sin\theta\cos\theta|0011\rangle+\sin\theta\cos\theta|0101\rangle+\sin^{2}\theta|0111\rangle\right.\\
 & - & \left.\left.\sin^{2}\theta|1000\rangle+\sin\theta\cos\theta|1010\rangle+\sin\theta\cos\theta|1100\rangle-\cos^{2}\theta|1110\rangle\right)\right]_{1423}.\\
 & = & \frac{1}{2}\left[2\sin\theta\cos\theta\left(-|\phi^{+}\rangle|\phi^{+}\rangle-|\psi^{-}\rangle|\psi^{-}\rangle\right)\right.\\
 & + & \sin^{2}\theta\left(|\phi^{+}\rangle|\psi^{-}\rangle-|\phi^{-}\rangle|\psi^{+}\rangle+|\psi^{+}\rangle|\phi^{-}\rangle-|\psi^{-}\rangle|\phi^{+}\rangle\right)\\
 & + & \left.\cos^{2}\theta\left(|\psi^{+}\rangle|\phi^{-}\rangle+|\psi^{-}\rangle|\phi^{+}\rangle-|\phi^{+}\rangle|\psi^{-}\rangle-|\phi^{-}\rangle|\psi^{+}\rangle\right)\right]_{1423}
\end{array}
\]
 From this configuration, the probability of an error induced by collective
rotation noise is determined as $2\sin^{2}\theta\cos^{2}\theta$,
which is contingent on the noise parameter $\theta$. Likewise, when
collective rotation noise is applied to particles 2 and 4 as referenced
in Equation (\ref{eq:Chapter2_Eq5}), the corresponding expression
follows,
\[
\begin{array}{lcl}
U_{r2}\otimes U_{r4}\left|\Psi^{00}\right\rangle  & = & \frac{1}{2}\left[\cos^{2}\theta|0000\rangle+\sin\theta\cos\theta|0010\rangle+\sin\theta\cos\theta|0100\rangle+\sin^{2}\theta|0110\rangle\right.\\
 & + & \sin^{2}\theta|1001\rangle-\sin\theta\cos\theta|1011\rangle-\sin\theta\cos\theta|1101\rangle+\cos^{2}\theta|1111\rangle\\
 & - & \left(-\sin\theta\cos\theta|0001\rangle-\sin^{2}\theta|0011\rangle+\cos^{2}\theta|0101\rangle+\sin\theta\cos\theta|0111\rangle\right.\\
 & - & \left.\left.-\sin\theta\cos\theta|1000\rangle+\cos^{2}\theta|1010\rangle-\sin^{2}\theta|1100\rangle+-\sin\theta\cos\theta|1110\rangle\right)\right]_{1423}.\\
 & = & \frac{1}{2}\left[2\sin\theta\cos\theta\left(|\phi^{-}\rangle|\psi^{+}\rangle+|\psi^{+}\rangle|\phi^{-}\rangle\right)\right.\\
 & + & \sin^{2}\theta\left(|\psi^{+}\rangle|\psi^{+}\rangle-|\psi^{-}\rangle|\psi^{-}\rangle+|\phi^{+}\rangle|\phi^{+}\rangle-|\phi^{-}\rangle|\phi^{-}\rangle\right)\\
 & + & \left.\cos^{2}\theta\left(|\phi^{+}\rangle|\phi^{+}+|\phi^{-}\rangle|\phi^{-}\rangle+|\psi^{+}\rangle|\psi^{+}\rangle+|\psi^{-}\rangle|\psi^{-}\rangle\right)\right]_{1423}
\end{array}
\]
For the state $|\Psi^{00}\rangle$, errors do not occur if the final
Bell pairs shared between Alice and Bob are among the set $\left\{ |\phi^{+}\rangle|\phi^{+}\rangle,|\phi^{-}\rangle|\phi^{-}\rangle,|\psi^{+}\rangle|\psi^{+}\rangle,|\psi^{-}\rangle|\psi^{-}\rangle\right\} $.
Analyzing this outcome reveals that the probability of error due to
collective rotation noise exhibits the same functional dependence,
$2\sin^{2}\theta\cos^{2}\theta$, as observed for $|\Psi^{11}\rangle$.
The variation of error probability with respect to $\theta$ is illustrated
in Figure \ref{fig:Collective-error-probability}.(b). From this figure,
it is evident that the probability of a rotation-induced error reaches
its peak value of 0.5 when $\theta=45^{\circ}$ and reduces to zero
at $\theta=0^{\circ}$ and $\theta=90^{\circ}$. To mitigate or entirely
suppress errors arising from this noise, it is crucial to choose a
communication channel where the error probability remains within an
acceptable threshold. Denoting this threshold as $p^{\prime}$, one
must solve the equation $p^{\prime}=2\sin^{2}\theta\cos^{2}\theta$
for $0^{\circ}<\theta<90^{\circ}$. If the solutions to this equation
are $\theta_{1}$ and $\theta_{2}$ with $\theta_{1}<\theta_{2}$,
the protocol will operate securely in a channel where $\theta<\theta_{1}$
or $\theta>\theta_{2}$. To ensure this condition is met in real-world
applications, channel characterization must be performed before implementing
the protocol. Recent developments in quantum communication have demonstrated
intrinsic robustness against collective noise \cite{TFI+16,WCS22,LWZ+16}.
Similar approaches can be incorporated into this scheme to further
enhance its resilience against collective noise.

\subsection{Comparison with the existing protocols}\label{sec:Chapter2_Sec3.5}

This section presents a concise comparative evaluation of the proposed
protocol against a selection of existing QIA schemes. The analysis
considers key factors, including the quantum resource requirements,
the role of third-party entities (distinguishing between honest, semi-honest,
and untrusted models), the minimum number of pre-shared authentication
keys, and whether the scheme supports bidirectional mutual authentication
or restricts verification to a single legitimate party. Given the
extensive range of QIA schemes, a representative subset has been carefully
selected, with emphasis on those grounded in secure direct quantum
communication protocols, similar to the proposed approach, which is
influenced by CDSQC.

We initiate the comparative analysis by evaluating Zhang et al.'s
QIA scheme \cite{ZZZX_2006}, which utilizes the ping-pong protocol
for QSDC. Within this framework, Alice functions as the trusted certification
authority, while Bob is the general user whose identity must be authenticated
by Alice. Similarly, Yuan et al.'s scheme \cite{YLP+14}, which is
based on the LM05 protocol for single-photon QSDC, follows the same
structural approach, with Alice serving as the certification authority.
Both of these QSDC-driven QIA protocols \cite{ZZZX_2006,YLP+14} operate
in a unidirectional manner, in contrast to the proposed bidirectional
scheme, thereby demonstrating an advantage of the latter. This superiority
extends to various unidirectional QIA protocols, including the one
introduced by Hong et al. \cite{HCJ+17}. However, this bidirectional
feature is not exclusive, as similar capabilities are incorporated
in Kang et al.'s schemes \cite{KHHYHM_2018,KHH+20} and Zhang et al.'s
2020 study \cite{ZCSL_2020}. Notably, all these protocols \cite{ZCSL_2020,KHHYHM_2018,KHH+20}
require a minimum of six pre-established keys for authentication,
whereas the proposed approach necessitates only four, making it more
resource-efficient given the high cost of quantum resources. Furthermore,
the studies conducted by Kang et al. \cite{KHHYHM_2018,KHH+20} incorporate
GHZ-like states, which are tripartite entangled states. These states
present greater challenges in both preparation and stability compared
to the Bell states employed in our protocol. Additionally, the 2020
study by Zhang et al. assumes a semi-honest third party, whereas our
proposed framework operates under the assumption that the third party
is untrusted, thereby improving security. Jiang et al. introduced
a semi-quantum QIA protocol analogous to the ping-pong protocol \cite{JZH21},
where Bob possesses only classical computational abilities with restricted
quantum interaction, whereas Alice has extensive quantum capabilities,
enabling her to prepare and measure Bell states along with other quantum
operations. However, this approach increases noise probability due
to the two-way communication and allows Eve complete access to the
Bell state within the quantum channel. In contrast, our proposed scheme
employs a one-way quantum channel, effectively preventing Eve from
obtaining full access to the Bell state. Another approach integrating
QIA with QKA, utilizing Bell and GHZ states under the assumption of
a semi-honest third party, was introduced by Wu et al. \cite{WCG+21}.
However, their protocol necessitates substantial quantum resources
and presents maintenance difficulties in comparison to our scheme.
Moreover, their reliance on hash functions for authentication may
not fully comply with quantum security principles. Similarly, Li et
al. proposed a simultaneous QIA scheme based on GHZ states \cite{LZZ+22},
but sustaining these states proves to be more complex than managing
Bell states. Thus, our proposed protocol achieves the essential QIA
properties while optimizing resource efficiency, particularly when
evaluated against other QIA schemes reliant on entangled states. The
comparative analysis provided aims to highlight the significance and
benefits of our scheme. A summarized comparison is presented in Table
\ref{tab:Chapter2_Tab7} for enhanced clarity.

\begin{table}
\caption{A comprehensive comparison of Protocol 2.3 is conducted with various
existing QIA protocols. The column labeled ``Minimum secret key required''
presents multiple values corresponding to different protocols discussed
within the same reference. The following abbreviations are employed
in this table: TC=two-way channel, IR=intercept-resend, IF=impersonated
fraudulent, EM=entangled measure.}\label{tab:Chapter2_Tab7}

\centering{}%
\begin{tabular*}{15.9cm}{@{\extracolsep{\fill}}@{\extracolsep{\fill}}|>{\centering}p{1.5cm}|>{\centering}p{1.5cm}|>{\centering}p{2cm}|c|>{\centering}p{1.5cm}|>{\centering}p{3.0cm}|}
\hline 
Protocol  & Quantum resources  & Minimum secret key required  & Way of authentication  & Nature of the third party  & Secure against attacks \tabularnewline
\hline 
Zhang et al. \cite{ZZZX_2006}  & Bell state  & 3  & Unidirectional  & No  & IF, direct measurement on channel particles, attack on TC\tabularnewline
\hline 
Yuan et al. \cite{YLP+14}  & Single photon  & 1  & Unidirectional  & No  & IR, EM on TC\tabularnewline
\hline 
Hong et al. \cite{HCJ+17}  & Single photon  & 1  & Unidirectional  & No  & IF, IR, EM\tabularnewline
\hline 
Kang et al. \cite{KHHYHM_2018,KHH+20}  & GHZ-like  & 6  & Bidirectional  & Untrusted  & $-$\tabularnewline
\hline 
Zhang et al. \cite{ZCSL_2020}  & Bell state  & 6  & Bidirectional  & Semi-honest  & IF, EM, IR\tabularnewline
\hline 
Jiang et al. \cite{JZH21}  & Bell state  & $18,35$  & Bidirectional  & No  & IF, IR, EM\tabularnewline
\hline 
Wu et al. \cite{WCG+21}  & Bell state, GHZ state  & 6  & Unidirectional  & Semi-honest  & External, dishonest participants', IF\tabularnewline
\hline 
Li et al. \cite{LZZ+22}  & GHZ state  & $10,3$  & Bidirectional  & Trusted  & IF, EM, IR\tabularnewline
\hline 
Our protocol  & Bell state  & 4  & Bidirectional  & Untrusted  & IF, IR, EM\tabularnewline
\hline 
\end{tabular*}
\end{table}


\section{New protocol for quantum identity authentication (Protocol 2.4)}\label{sec:Chapter2_Sec4}

In this work, we introduce a novel QIA scheme, which can be interpreted
as a secure direct quantum communication-inspired approach for QIA.
Since the protocol utilizes Bell states, it can also be classified
as an entangled-state-based QIA protocol. Prior to detailing the proposed
scheme, we outline the fundamental concept underlying its design.

\subsection{Core concept}

The protocol leverages the properties of Bell states and entanglement
swapping. The relationships between Bell states and the pre-shared
authentication key can be considered as Equation (\ref{eq:Bell-State=000020with=000020Pre-Shared_Key}).
Additionally, the corresponding mappings between Pauli operations
and the pre-shared authentication key can be defined as Equation (\ref{eq:Pauli-Operation=000020with=000020Pre-Shared_Key}).
To summarize the underlying mechanism of this protocol, consider that
Alice and Bob have pre-established an authentication key represented
as a bit string of a specific length. Suppose this bit string is 10;
accordingly, Alice and Bob generate Bell states based on Equation
(\ref{eq:Bell-State=000020with=000020Pre-Shared_Key}).

\begin{equation}
\begin{array}{lcl}
\vert\psi^{+}\rangle & = & \frac{1}{\sqrt{2}}(\vert01\rangle_{12(34)}+\vert10\rangle_{12(34)}).\end{array}\label{eq:Chapter2_Eq15}
\end{equation}
Here, subscripts $1,2$ and $3,4$ denote the particles associated
with Alice and Bob, respectively. Alice (Bob) transmits particle 2
(4) to a semi-honest third party, Charlie, who serves as the authentication
verifier. Charlie then prepares one of the following quantum states

\begin{equation}
\begin{array}{c}
\vert\pm\rangle=\frac{1}{\sqrt{2}}(\vert0\rangle_{5}\pm\vert1\rangle_{5}).\end{array}\label{eq:Chapter2_Eq16}
\end{equation}
Consider the scenario where Charlie selects the $\vert-\rangle$ state.
The collective quantum state of Alice, Bob, and Charlie can be represented
as follows:

\[
\begin{array}{lcl}
\vert\psi^{+}\rangle_{12}\otimes\vert\psi^{+}\rangle_{34}\otimes\vert-\rangle_{5} & = & \frac{1}{2\sqrt{2}}[(\vert01\rangle+\vert10\rangle)_{12}\otimes(\vert01\rangle+\vert10\rangle)_{34}\otimes(\vert0\rangle-\vert1\rangle)_{5}].\end{array}
\]
Charlie executes a CNOT operation, where Particle 5 serves as the
control qubit, and the target qubit is randomly chosen between Particle
2 or Particle 4. Following this, Charlie transmits Particle 2 to Bob
and Particle 4 to Alice. Concurrently, Alice and Bob apply the $i\sigma_{y}$
operator to their respective particles (Particles 1 and 3) based on
their shared secret key $S_{AB}=\{10\}$, as indicated in (\ref{eq:Pauli-Operation=000020with=000020Pre-Shared_Key}).
Assuming Particle 2 is designated as the target qubit for the CNOT
operation, the resulting joint quantum state after the application
of both the CNOT and Pauli operations can be formulated as follows:

\begin{equation}
\begin{array}{lcl}
|\Psi^{10}\rangle_{12345} & = & \frac{1}{2\sqrt{2}}[\vert11110\rangle-\vert10111\rangle-\vert11000\rangle+\vert10001\rangle\\
 & - & \vert00110\rangle+\vert01111\rangle+\vert00000\rangle-\vert01001\rangle]_{12345}.
\end{array}\label{eq:Chapter2_Eq17}
\end{equation}
Rearranging this expression, it can be rewritten in an equivalent
form

\begin{equation}
\begin{array}{lcl}
|\Psi^{10}\rangle_{14235} & = & \frac{1}{2\sqrt{2}}[(\vert\phi^{+}\rangle_{14}\otimes|\phi^{+}\rangle_{23}+\vert\phi^{-}\rangle_{14}\otimes|\phi^{-}\rangle_{23}\\
 & - & \vert\psi^{+}\rangle_{14}\otimes|\psi^{+}\rangle_{23}-\vert\psi^{-}\rangle_{14}\otimes|\psi^{-}\rangle_{23})\otimes|0\rangle_{5}\\
 & + & (-\vert\phi^{+}\rangle_{14}\otimes|\psi^{+}\rangle_{23}+\vert\phi^{-}\rangle_{14}\otimes|\psi^{-}\rangle_{23}\\
 & + & \vert\psi^{+}\rangle_{14}\otimes|\phi^{+}\rangle_{23}-\vert\psi^{-}\rangle_{14}\otimes|\phi^{-}\rangle_{23})\otimes|1\rangle_{5}].
\end{array}\label{eq:Chapter2_Eq18}
\end{equation}
Subsequently, Alice and Bob each conduct Bell measurements on their
respective particles---Alice on Particles 1 and 4, and Bob on Particles
2 and 3. They then communicate their classical measurement outcomes
to Charlie, following the procedure detailed in (\ref{eq:Bell-State=000020with=000020Pre-Shared_Key}).
Charlie proceeds to measure Particle 5 in the computational basis
$(\{\vert0\rangle,|1\rangle\})$, obtaining either $|0\rangle$ or
$|1\rangle$ with equal probability ($\frac{1}{2}$). After receiving
the classical bits from Alice and Bob, Charlie applies an XOR operation
on their results. The set of possible outcomes corresponding to the
measurements performed by Alice, Bob, and Charlie in this specific
instance is summarized in Table \ref{tab:Chapter2_Tab8}.

\begin{table}[h]
\caption{This outlines the potential measurement results observed by all participants
within Protocol 2.4 for QIA.}\label{tab:Chapter2_Tab8}

\centering{}%
\begin{tabular*}{15.9cm}{@{\extracolsep{\fill}}@{\extracolsep{\fill}}|c|c|c|}
\hline 
Alice and Bob's possible combination result  & Charlie's result  & Additional modulo 2 \tabularnewline
\hline 
$|\phi^{+}\rangle_{14}\oplus|\phi^{+}\rangle_{23}$  & $|0\rangle_{5}$  & $00\oplus00=00$\tabularnewline
\hline 
$\vert\phi^{-}\rangle_{14}\oplus|\phi^{-}\rangle_{23}$  & $|0\rangle_{5}$  & $01\oplus01=00$\tabularnewline
\hline 
$\vert\psi^{+}\rangle_{14}\oplus|\psi^{+}\rangle_{23}$  & $|0\rangle_{5}$  & $10\oplus10=00$\tabularnewline
\hline 
$\vert\psi^{-}\rangle_{14}\oplus|\psi^{-}\rangle_{23}$  & $|0\rangle_{5}$  & $11\oplus11=00$\tabularnewline
\hline 
$\vert\phi^{+}\rangle_{14}\oplus|\psi^{+}\rangle_{23}$  & $|1\rangle_{5}$  & $00\oplus10=10$\tabularnewline
\hline 
$\vert\phi^{-}\rangle_{14}\oplus|\psi^{-}\rangle_{23}$  & $|1\rangle_{5}$  & $01\oplus11=10$\tabularnewline
\hline 
$\vert\psi^{+}\rangle_{14}\oplus|\phi^{+}\rangle_{23}$  & $|1\rangle_{5}$  & $10\oplus00=10$\tabularnewline
\hline 
$\vert\psi^{-}\rangle_{14}\oplus|\phi^{-}\rangle_{23}$  & $|1\rangle_{5}$  & $11\oplus01=10$\tabularnewline
\hline 
\end{tabular*}
\end{table}

From Table \ref{tab:Chapter2_Tab8}, it is evident that if Charlie
receives the state $|0\rangle$ , the XOR of the classical bits transmitted
by Alice and Bob consistently results in 00. Conversely, if Charlie
obtains the state $|1\rangle$, the XOR operation always yields 10.
Any deviation from these expected results would indicate the presence
of an eavesdropper. A similar methodology can be applied to derive
the same conclusion for alternative secret keys. For instance, consider
the secret key $S_{AB}=\{11\}$. In this scenario, both Alice and
Bob select the Bell state $|\psi^{-}\rangle$, with Charlie designating
his particle $|+\rangle$ as the control qubit. The collective quantum
state of the three parties can be represented as follows:

\[
\begin{array}{lcl}
\vert\psi^{-}\rangle_{12}\otimes\vert\psi^{-}\rangle_{34}\otimes\vert+\rangle & = & \frac{1}{2\sqrt{2}}[(\vert01\rangle-\vert10\rangle)_{12}\otimes(\vert01\rangle-\vert10\rangle)_{34}\otimes(\vert0\rangle+\vert1\rangle)_{5}].\end{array}
\]
Alice and Bob send their respective particles 2 and 4 to Charlie and
apply the $\sigma_{z}$ operator on their respective particles 1 and
3. Charlie then executes a CNOT operation, using his particle as the
control qubit and particle 4 as the target qubit. Following this,
he returns particle 2 (4) to Bob (Alice). After completing these operations,
the quantum state evolves into

\begin{equation}
\begin{array}{lcl}
|\Psi^{11}\rangle_{12345} & = & \frac{1}{2\sqrt{2}}[\vert01010\rangle-\vert01001\rangle-\vert01100\rangle+\vert01111\rangle\\
 & - & \vert10010\rangle+\vert10001\rangle+\vert10100\rangle-\vert10111\rangle]_{12345}.
\end{array}\label{eq:Chapter2_Eq19}
\end{equation}
Alice and Bob subsequently perform Bell measurements on their respective
particle pairs (1 and 4 for Alice, 2 and 3 for Bob), obtaining outcomes
correlated with Charlie\textquoteright s measurement results, as previously
discussed. For clarity, the above expression can be further simplified
to

\begin{equation}
{\begin{array}{lcl}
|\Psi^{11}\rangle_{14235} & = & \frac{1}{2\sqrt{2}}[(\vert\phi^{+}\rangle_{14}\otimes|\phi^{+}\rangle_{23}-\vert\phi^{-}\rangle_{14}\otimes|\phi^{-}\rangle_{23}\\
 & + & \vert\psi^{+}\rangle_{14}\otimes|\psi^{+}\rangle_{23}-\vert\psi^{-}\rangle_{14}\otimes|\psi^{-}\rangle_{23})\otimes|0\rangle_{5}\\
 & + & (\vert\phi^{+}\rangle_{14}\otimes|\psi^{+}\rangle_{23}-\vert\phi^{-}\rangle_{14}\otimes|\psi^{-}\rangle_{23}\\
 & + & \vert\psi^{+}\rangle_{14}\otimes|\phi^{+}\rangle_{23}-\vert\psi^{-}\rangle_{14}\otimes|\phi^{-}\rangle_{23})\otimes|1\rangle_{5}].
\end{array}}\label{eq:Chapter2_Eq20}
\end{equation}
Analyzing Equation (\ref{eq:Chapter2_Eq20}) confirms the same conclusions
outlined in Table \ref{tab:Chapter2_Tab8}. Furthermore, to establish
that this conclusion holds for the remaining secret keys \{01\} and
\{10\}, the final quantum state shared among Alice, Bob, and Charlie
can be expressed as follows:

\begin{equation}
\begin{array}{lcl}
|\Psi^{01}\rangle_{14235} & = & \frac{1}{2\sqrt{2}}[(|01100\rangle+|00101\rangle+|00110\rangle+|01111\rangle\\
 & + & |11000\rangle+|10001\rangle+|10010\rangle+|11011\rangle]_{14235}\\
 & = & \frac{1}{2\sqrt{2}}[(|\psi^{+}\rangle_{14}\otimes|\psi^{+}\rangle_{23}-|\psi^{-}\rangle_{14}\otimes|\psi^{-}\rangle_{23}\\
 & + & |\phi^{+}\rangle_{14}\otimes|\phi^{+}\rangle_{23}-|\phi^{-}\rangle_{14}\otimes|\phi^{-}\rangle_{23})\otimes|0\rangle_{5}\\
 & + & (|\phi^{+}\rangle_{14}\otimes|\psi^{+}\rangle_{23}-|\phi^{-}\rangle_{14}\otimes|\psi^{-}\rangle_{23}\\
 & + & |\psi^{+}\rangle_{14}\otimes|\phi^{+}\rangle_{23}-|\psi^{-}\rangle_{14}\otimes|\phi^{-}\rangle_{23})\otimes|1\rangle_{5}],
\end{array}\label{eq:Chapter2_Eq21}
\end{equation}
and

\begin{equation}
\begin{array}{lcl}
|\Psi^{00}\rangle_{14235} & = & \frac{1}{2\sqrt{2}}[(|10010\rangle+|11011\rangle-|11000\rangle-|10001\rangle\\
 & - & |00110\rangle-|01111\rangle+|01100\rangle+|00101\rangle]_{14235}\\
 & = & \frac{1}{2\sqrt{2}}[(|\psi^{+}\rangle_{14}\otimes|\psi^{+}\rangle_{23}-|\psi^{-}\rangle_{14}\otimes|\psi^{-}\rangle_{23}\\
 & - & |\phi^{+}\rangle_{14}\otimes|\phi^{+}\rangle_{23}+|\phi^{-}\rangle_{14}\otimes|\phi^{-}\rangle_{23})\otimes|0\rangle_{5}\\
 & + & (|\phi^{+}\rangle_{14}\otimes|\psi^{+}\rangle_{23}-|\phi^{-}\rangle_{14}\otimes|\psi^{-}\rangle_{23}\\
 & - & |\psi^{+}\rangle_{14}\otimes|\phi^{+}\rangle_{23}+|\psi^{-}\rangle_{14}\otimes|\phi^{-}\rangle_{23})\otimes|1\rangle_{5}].
\end{array}\label{eq:Chapter2_Eq22}
\end{equation}
Equations (\ref{eq:Chapter2_Eq21}) and (\ref{eq:Chapter2_Eq22})
align with the authentication criteria specified in Table \ref{tab:Chapter2_Tab8}
for the secret keys $S_{AB}=\{01\}$ and $\{00\}$, respectively.
In this protocol, the controller does not require any permutation
operation to manage authentication; instead, quantum mechanical principles
facilitate the process.

\subsection{Description of Protocol 2.4}

\begin{figure}
\begin{centering}
\includegraphics[scale=0.4]{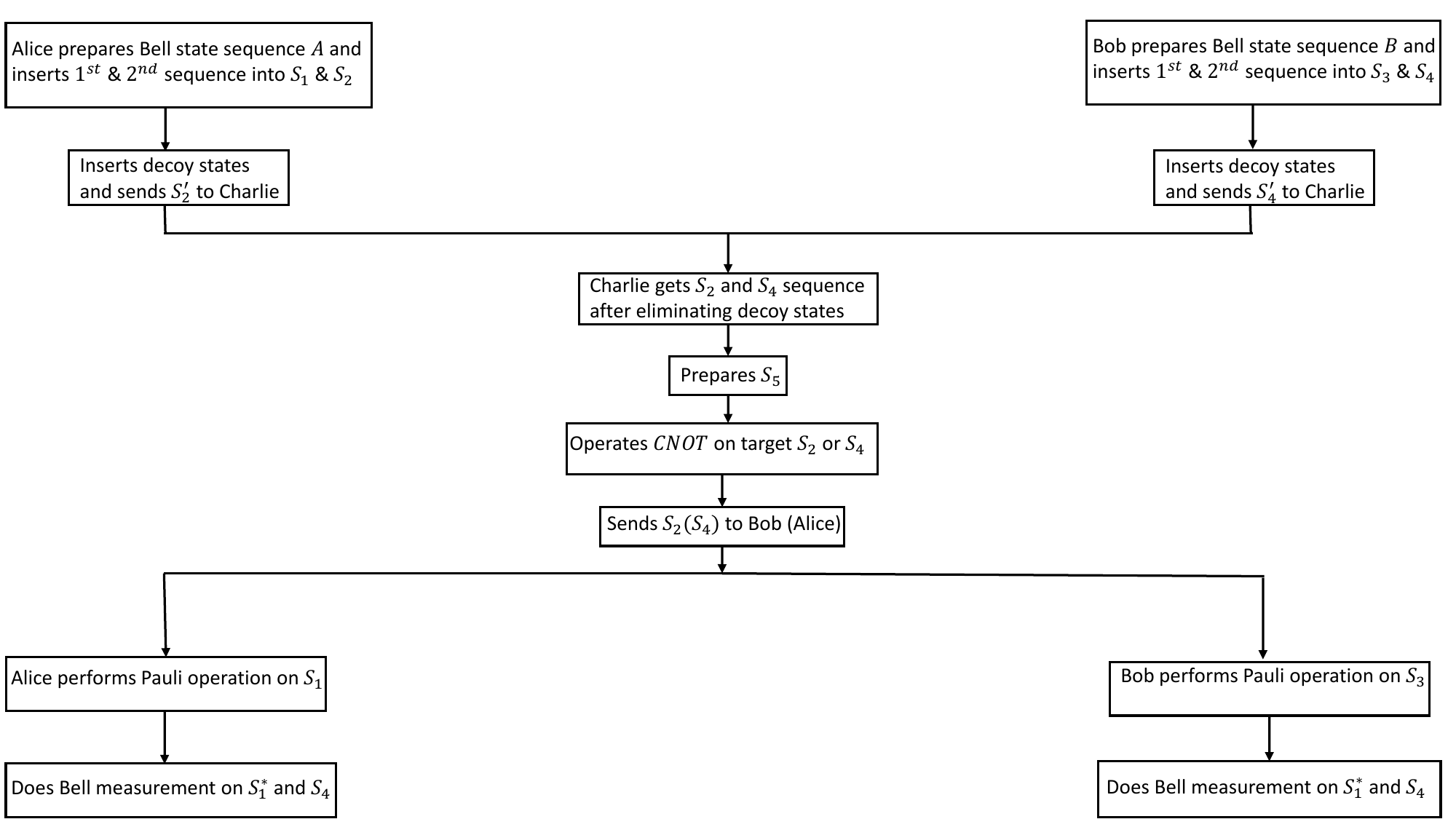} 
\par\end{centering}
\begin{centering}
\includegraphics[scale=0.4]{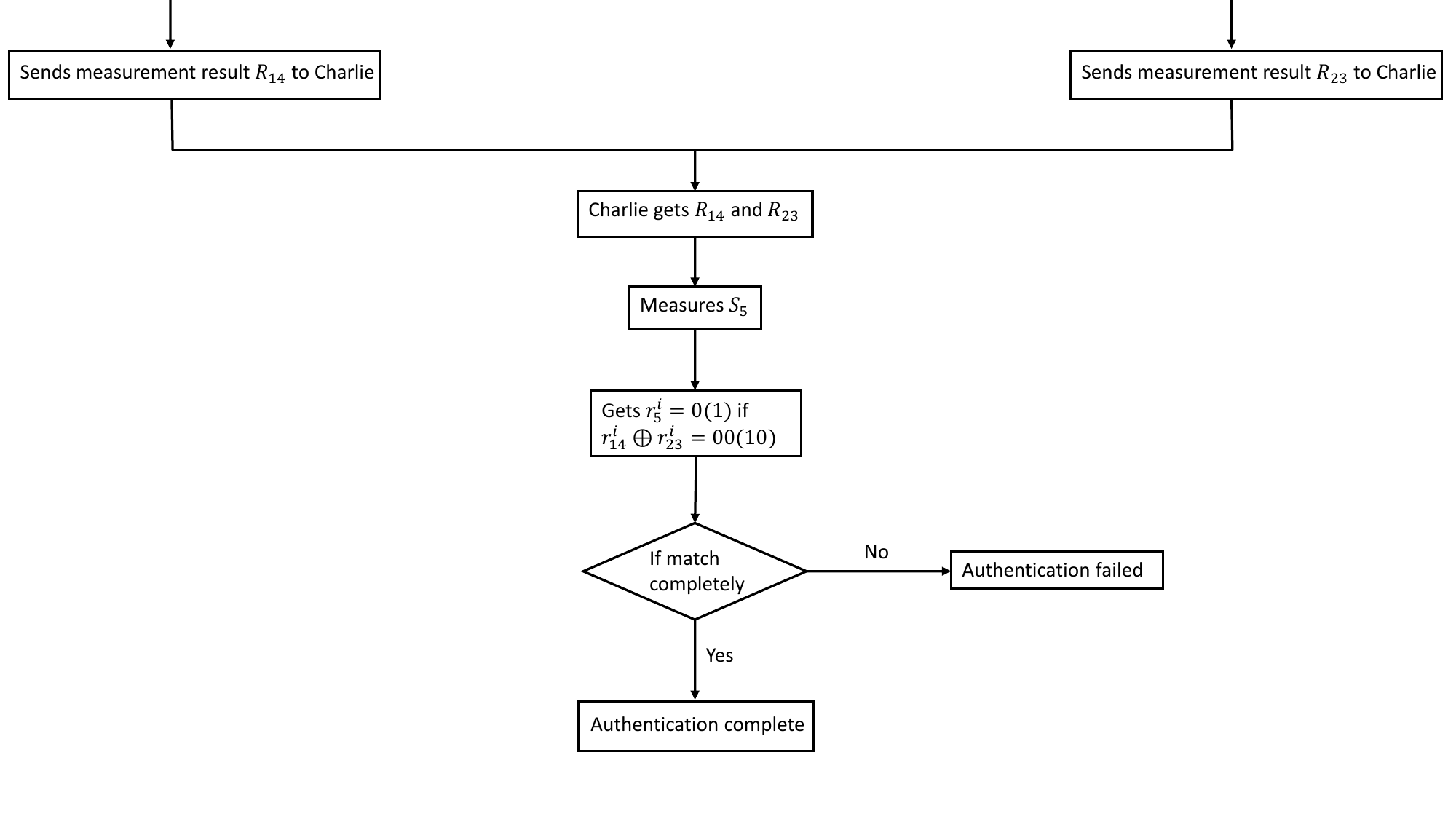} 
\par\end{centering}
\caption{A flowchart visually represents the operational framework of the
proposed QIA protocol.}\label{fig:Chapter2_Fig5}
\end{figure}

Alice and Bob aim to authenticate themselves as legitimate users with
the assistance of a third party, Charlie. Within this protocol, Alice
and Bob share a pre-established classical secret key sequence, denoted
as $K_{AB}=\{k_{1}^{1}k_{2}^{1},k_{1}^{2}k_{2}^{2},k_{1}^{3}k_{2}^{3},\cdots,k_{1}^{i}k_{2}^{i},\cdots,k_{1}^{n}k_{2}^{n},\}$.
Before detailing the protocol in a stepwise manner, it is essential
to briefly discuss the role of decoy states in establishing a secure
communication channel between any two entities \cite{LMC05,STP+16}.
In general, one party introduces decoy qubits at random positions
within the sequence of information qubits and transmits this extended
sequence to the other party. Upon receiving it, the second party requests
the positional and encoding details of the decoy qubits via an unjammable
public channel. The first party then publicly discloses this information,
allowing the second party to verify it by measuring the decoy qubits.
If the detected error rate exceeds an acceptable threshold, both parties
discard the entire qubit sequence. This decoy-state mechanism enhances
channel security, ensuring the secure transmission of information
between parties. The authentication process consists of the following
steps:
\begin{description}
\item [{Step~4.1:}] Alice and Bob independently generate a series of $n$
Bell states, denoted as $A_{12}$ and $B_{34}$, respectively. The
preparation follows the pre-shared key $K_{AB}$ and adheres to the
rule specified in (\ref{eq:Bell-State=000020with=000020Pre-Shared_Key}).
The sequences are represented as follows:
\begin{equation}
\begin{array}{c}
A=\{\vert A\rangle_{12}^{1},|A\rangle_{12}^{2},|A\rangle_{12}^{3},\cdots,|A\rangle_{12}^{i},\cdots,|A\rangle_{12}^{n}\},\\
B=\{\vert B\rangle_{34}^{1},|B\rangle_{34}^{2},|B\rangle_{34}^{3},\cdots,|B\rangle_{34}^{i},\cdots,|B\rangle_{34}^{n}\}.
\end{array}
\end{equation}
The subscripts 1, 2, 3, and 4 are used to distinguish the particles
in the sequences. Ideally, $A=B$. Here, the subscripts (1, 2, 3,
4) are used to distinguish individual particles in each sequence.
Ideally, the prepared sequences satisfy $A=B$.
\item [{Step~4.2:}] Each Bell state is then separated into two ordered
sequences of $n$ particles. The first particle of each Bell state
forms one sequence, while the second particle forms another.
\end{description}
\begin{equation}
\begin{array}{c}
S_{1}=\{s_{1}^{1},s_{1}^{2},s_{1}^{3},\cdots,s_{1}^{i},\cdots,s_{1}^{n}\},\\
S_{2}=\{s_{2}^{1},s_{2}^{2},s_{2}^{3},\cdots,s_{2}^{i},\cdots,s_{2}^{n}\},\\
S_{3}=\{s_{3}^{1},s_{3}^{2},s_{3}^{3},\cdots,s_{3}^{i},\cdots,s_{3}^{n}\},\\
S_{4}=\{s_{4}^{1},s_{4}^{2},s_{4}^{3},\cdots,s_{4}^{i},\cdots,s_{4}^{n}\}.
\end{array}
\end{equation}
The sequences $S_{1}$ and $S_{2}$ consist of the first and second
particles of all Bell states in $A$, respectively. Likewise, $S_{3}$
and $S_{4}$ are composed of the first and second particles of all
Bell states in $B$. Alice and Bob retain sequences $S_{1}$ and $S_{3}$
while proceeding with further operations. To enhance security, Alice
and Bob each randomly introduce decoy particles, $d_{a}$ and $d_{a}$,
into sequences $S_{2}$ and $S_{4}$, respectively. This results in
expanded sequences $S_{2}^{\prime}$ and $S_{4}^{\prime}$. Alice
then transmits $S_{2}^{\prime}$ to Charlie via a quantum channel,
while Bob also sends $S_{4}^{\prime}$ to Charlie.
\begin{description}
\item [{Step~4.3:}] Upon receiving $S_{2}^{\prime}$ and $S_{4}^{\prime}$,
Charlie performs security verification using the embedded decoy particles.
If the security check is successful, Charlie removes the decoy particles
to recover the original sequences $S_{2}$ and $S_{4}$. Next, Charlie
generates a sequence of $n$ qubits:
\[
S_{5}=\{s_{5}^{1},s_{5}^{2},s_{5}^{3},\cdots,s_{5}^{i},\cdots,s_{5}^{n}\}
\]
\\
Each qubit in $S_{5}$ is prepared in either the $\vert+\rangle$
or $\vert-\rangle$ state at random. Charlie then applies a CNOT operation,
using the qubits in $S_{5}$ as the control qubits and those in $S_{2}$
or $S_{4}$ as the target qubits, selected randomly.
\item [{Step~4.4:}] Charlie generates decoy particles, denoted as $d_{c}$
and $d_{d}$, and randomly integrates them into the sequences $S_{2}$
and $S_{4}$, thereby forming the extended sequences $S_{2}^{\prime}$
and $S_{4}^{\prime}$. These modified sequences, $S_{2}^{\prime}$
and $S_{4}^{\prime}$, are then transmitted to Bob and Alice, respectively.
Following successful security validation using the decoy particles
and their subsequent removal, Alice retains the sequences $S_{1}$
and $S_{4}$, while Bob holds $S_{2}$ and $S_{3}$.
\item [{Step~4.5:}] Alice (Bob) applies the Pauli operation on the particles
in sequence $S_{1}$ ($S_{3}$) based on a pre-established classical
key, as specified in (\ref{eq:Pauli-Operation=000020with=000020Pre-Shared_Key}).
This operation transforms the sequences into $S_{1}^{*}$ and $S_{3}^{*}$.
Alice then conducts a Bell-state measurement on the particles in $S_{1}^{*}$
and $S_{4}$, recording the measurement outcomes as a classical sequence
$R_{14}=\{r_{14}^{1},r_{14}^{2},\cdots,r_{14}^{i},\cdots,r_{14}^{n}\}$,
following (\ref{eq:Bell-State=000020with=000020Pre-Shared_Key}).
Similarly, Bob performs a Bell-state measurement on the particles
in $S_{2}$ and $S_{3}^{*}$, obtaining the classical sequence\\ 
$R_{23}=\{r_{23}^{1},r_{23}^{2},\cdots,r_{23}^{i},\cdots,r_{23}^{n}\}$. Both Alice and Bob forward their respective measurement results, $R_{14}$
and $R_{23}$, to Charlie. Concurrently, Charlie measures the sequence
$S_{5}$ using the computational basis \textbf{$\{\vert0\rangle,|1\rangle\}$},
producing the sequence $R_{5}=\{r_{5}^{1},r_{5}^{2},\cdots,r_{5}^{i},\cdots,r_{5}^{n}\}$,
where each outcome $r_{5}^{i}$ takes a value of either 0 or 1, corresponding
to the states $|0\rangle$ and $|1\rangle$, respectively.
\item [{Step~4.6:}] At this stage, the third party, Charlie, possesses
three classical sequences: $R_{14}=\{r_{14}^{1},r_{14}^{2},\cdots,r_{14}^{i},\cdots,r_{14}^{n}\}$,
$R_{23}=\{r_{23}^{1},r_{23}^{2},\cdots,r_{23}^{i},\cdots,r_{23}^{n}\}$
and $R_{5}=\{r_{5}^{1},r_{5}^{2},\cdots,r_{5}^{i},\cdots,r_{5}^{n}\}$.
Charlie executes an XOR operation on the bit strings $R_{14}$ and
$R_{23}$ at corresponding positions. If $r_{5}^{i}$ is 0 and the
XOR result is$r_{14}^{i}\oplus r_{23}^{i}=00$, or if $r_{5}^{i}$
is 1 and the XOR result is $r_{14}^{i}\oplus r_{23}^{i}=10$, then
Charlie confirms the authentication of Alice and Bob.
\item [{Step~4.7:}] Charlie then publicly verifies and announces the authentication
status of Alice and Bob via an unjammable public channel.
\end{description}
The QIA protocol elaborated above is also illustrated through a flowchart,
as depicted in Figure \ref{fig:Chapter2_Fig5}.

\subsection{Security analysis of Protocol 2.4 }\label{sec:Chapter2_Sec4.3}

This section aims to evaluate the security of the proposed protocol
against several well-known attacks that an eavesdropper (Eve) could
potentially execute. It is important to note that insider attacks
are not considered in this analysis, as they are irrelevant in the
context of QIA. The following discussion will demonstrate the security
of the protocol against specific attack scenarios. Initially, we assess
its resilience against impersonation attacks, where an external adversary,
Eve attempts to mimic a legitimate user, Alice or Bob, to bypass
the authentication mechanism.

\subsubsection{Security evaluation of Protocol 2.4 against impersonation attack
by Eve}

\begin{figure}[h]
\centering{} \includegraphics[scale=0.6]{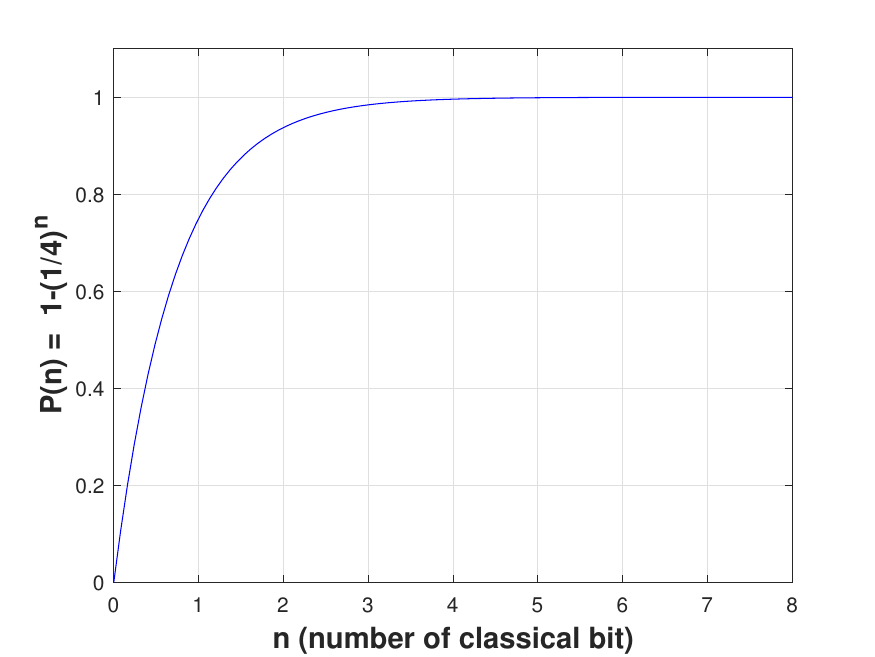}
\caption{The probability of detecting Eve\textquoteright s interference denoted
as $P(n)$, is correlated with the number of classical bits, $n$,
utilized as the pre-shared authentication key.}\label{fig:Chapter2_Fig6}
\end{figure}

To analyze the security of the protocol, consider the scenario outlined
in Section \ref{sec:Chapter2_Sec4}, where an adversary, Eve, attempts
to impersonate Alice. Eve initiates the attack by selecting the Bell
state $|\phi^{+}\rangle_{12}$ and transmits the sequence $S_{e2}$
(comprising particle 2) to Charlie. Following the protocol, Charlie
processes the sequence and returns $S_{2e}$ (denoted as $S_{4}$)
to Bob, who in this case is Eve. Then Eve executes the protocol steps
as Alice would, applying the Pauli identity operation ($I$) on Particle
1. Subsequently, Bob and Eve perform Bell state measurements. The
results, presented in Table \ref{tab:Chapter2_Tab9}, indicate scenarios
that reveal Eve\textquoteright s presence. Since Eve lacks knowledge
of the legitimate key $K_{AB}$, her probability of selecting the
correct Bell state at each step is $\frac{1}{4}$. To bypass detection,
she must correctly guess all $n$ Bell states, making the overall
probability of a successful impersonation attack $(\frac{1}{4})^{n}$.
As $n$ increases significantly, this probability approaches zero.
Therefore, the probability $P(n)$ of detecting Eve is given by, $P(n)=1-(\frac{1}{4})^{n}$.
For sufficiently large $n$, $P(n)$ approaches 1, ensuring the detection
of an impersonation attempt. Figure \ref{fig:Chapter2_Fig6} illustrates
the dependence of $P(n)$ on $n$, demonstrating that at least six
bits of pre-shared classical information are required to reliably
detect Eve\textquoteright s presence.

\begin{table}[h]
\begin{centering}
\captionsetup{justification=justified, singlelinecheck=false}\caption{Possible results of measurements.}\label{tab:Chapter2_Tab9}
\par\end{centering}
\centering{}%
\begin{tabular*}{15.9cm}{@{\extracolsep{\fill}}@{\extracolsep{\fill}}ccc}
\toprule 
Eve and Bob's possible combined result  & Charlie's result  & Modulo 2 addition\tabularnewline
\midrule 
$|\psi^{+}\rangle_{14}\otimes|\psi^{-}\rangle_{23}$  & $|0\rangle_{5}$  & $10\oplus11=01$\tabularnewline
$\vert\psi^{-}\rangle_{14}\otimes|\psi^{+}\rangle_{23}$  & $|0\rangle_{5}$  & $11\oplus10=01$\tabularnewline
$\vert\phi^{+}\rangle_{14}\otimes|\phi^{-}\rangle_{23}$  & $|0\rangle_{5}$  & $00\oplus01=01$\tabularnewline
$\vert\phi^{-}\rangle_{14}\otimes|\phi^{+}\rangle_{23}$  & $|0\rangle_{5}$  & $01\oplus00=01$\tabularnewline
$\vert\psi^{+}\rangle_{14}\otimes|\phi^{-}\rangle_{23}$  & $|1\rangle_{5}$  & $10\oplus01=11$\tabularnewline
$\vert\psi^{-}\rangle_{14}\otimes|\phi^{+}\rangle_{23}$  & $|1\rangle_{5}$  & $11\oplus00=11$\tabularnewline
$\vert\phi^{+}\rangle_{14}\otimes|\psi^{-}\rangle_{23}$  & $|1\rangle_{5}$  & $00\oplus11=11$\tabularnewline
$\vert\phi^{-}\rangle_{14}\otimes|\psi^{+}\rangle_{23}$  & $|1\rangle_{5}$  & $01\oplus10=11$\tabularnewline
\bottomrule
\end{tabular*}
\end{table}

\subsubsection{Security analysis of Protocol 2.4 against intercept and resend attack}\label{subsec:Chapter2_Sec4.3.2}

\begin{figure}
\begin{centering}
\includegraphics[scale=0.4]{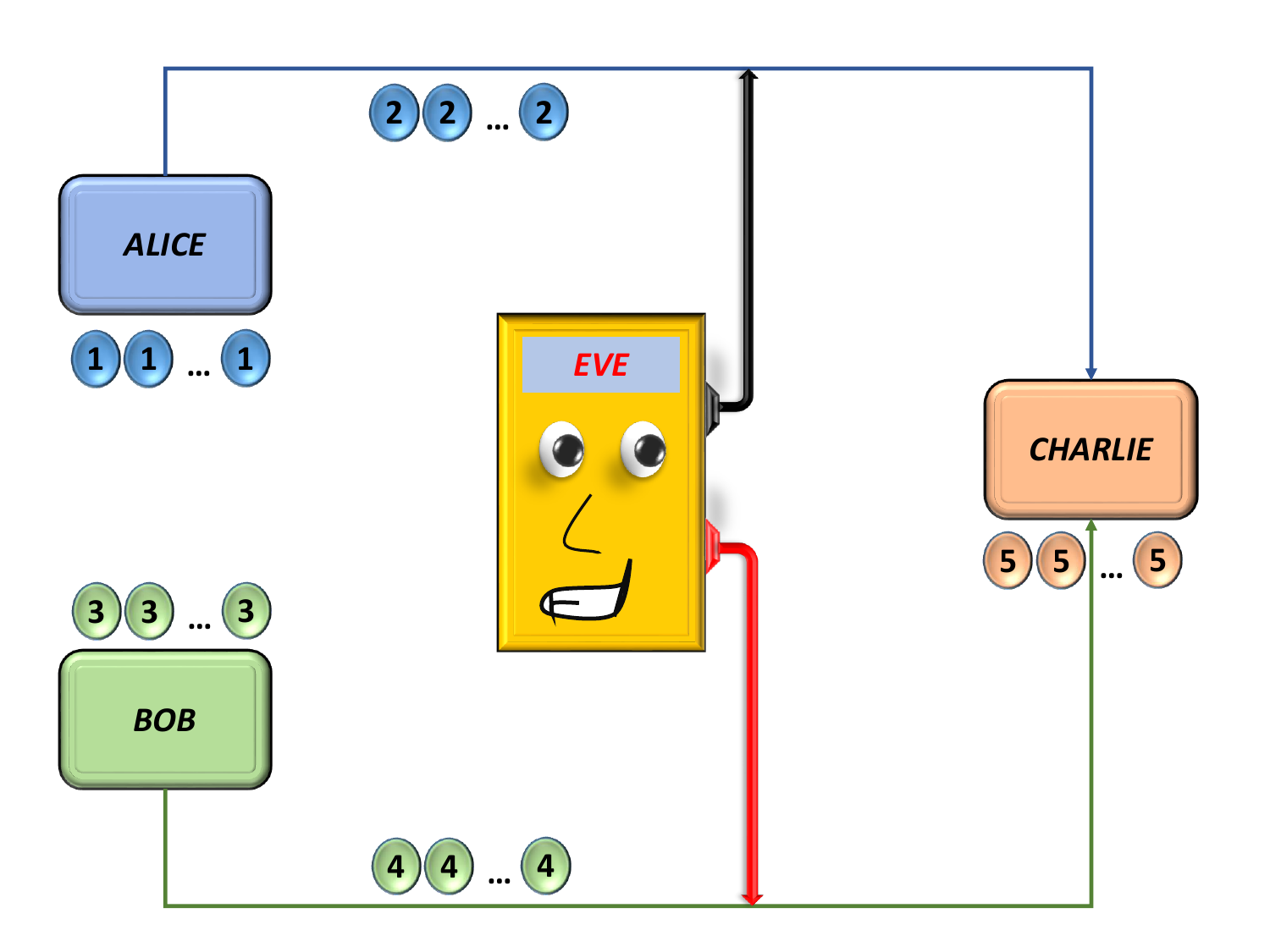} 
\par\end{centering}
\caption{Eve employs an intercept-and-resend strategy. In this context, the
qubits labeled as 1, 2, 3, 4, and 5 correspond to the sequences $S_{1}$,
$S_{2}$, $S_{3}$, $S_{4}$, and $S_{5}$, respectively.}\label{fig:Chapter2_Fig7}
\end{figure}

In this protocol, Alice and Bob generate Bell states based on their
pre-shared secret key. Neither Alice nor Bob transmits their quantum
states in their entirety, forcing Eve to intercept only one particle
at a time. Without loss of generality, we assume Eve targets Particle
2 (Particle 4) when Alice sends it to Charlie (or when Charlie sends
it to Alice). The amount of information Eve can extract from the quantum
channel is fundamentally constrained by the ``Holevo bound or
``Holevo quantity'' \cite{H_73}.

\begin{equation}
\chi(\rho)=S(\rho)-\sum_{i}p_{i}S(\rho_{i}),\label{eq:Chapter2_Eq25}
\end{equation}
Considering the von Neumann entropy, defined as $S(\rho)=-Tr(\rho\log_{2}\rho)$,
where $\rho_{i}$ represents a component of the mixed state $\rho$
with probability $p_{i}$, the density matrix can be expressed as:
$\rho=\sum_{i}p_{i}\rho_{i}=$$\frac{1}{4}\left[|\Psi^{00}\rangle\langle\Psi^{00}|+|\Psi^{01}\rangle\langle\Psi^{01}|+|\Psi^{10}\rangle\langle\Psi^{10}|+|\Psi^{11}\rangle\langle\Psi^{11}|\right]_{14235}$.
Here, $\rho_{i}$ corresponds to the density matrices of the states
described in (\ref{eq:Chapter2_Eq17}), (\ref{eq:Chapter2_Eq19}),
(\ref{eq:Chapter2_Eq21}), and (\ref{eq:Chapter2_Eq22}). Since Eve
targets particle 2 (or 4), Equation (\ref{eq:Chapter2_Eq25}) can
be reformulated accordingly.

\begin{equation}
\chi(\rho_{2(4)})=S(\rho_{2(4)})-\sum_{i}p_{i}S(\rho_{2(4)_{i}}).\label{eq:Chapter2_Eq26}
\end{equation}
Here, the reduced density matrices $\rho_{2(4)}$ and $\rho_{2(4)_{i}}$
correspond to $\rho$ and $\rho_{i}$, respectively, after performing
a partial trace over particles 1, 4(2), 3, and 5. The resultant reduced
density matrix for particle 2 (or 4) consistently takes the form:

\[
\begin{array}{lcl}
\rho_{2(4)} & = & Tr_{14(2)35}(\rho)\\
 & = & \frac{1}{8}\times\left(|0\rangle\langle0|+|1\rangle\langle1|\right)\times4\\
 & = & \frac{1}{2}I_{2}.
\end{array}
\]
Furthermore, it follows that $\rho_{2(4)_{i}}=\frac{1}{2}I$. By substituting
$\rho_{2(4)}$ and $\rho_{2(4)_{i}}$ into Equation (\ref{eq:Chapter2_Eq26}),
we obtain $\chi(\rho_{2(4)})=0$. This result indicates that an adversary,
Eve would be unable to extract any useful information through a direct
measurement attack on particle 2 (or 4). Beyond this scenario, we
must also consider an alternative attack strategy where Eve attempts
to gain information from both particles 2 and 4 simultaneously. Employing
the same reasoning, we determine that the Holevo bound constraints
Eve from accessing any meaningful data. To align with our conditions,
we reformulate Equation (\ref{eq:Chapter2_Eq25}) in the following
manner:

\begin{equation}
\chi(\rho_{24})=S(\rho_{24})-\sum_{i}p_{i}S(\rho_{24_{i}}).\label{eq:Chapter2_Eq27}
\end{equation}
Here, $\rho_{24}$ and $\rho_{24_{i}}$ represent the reduced density
matrices of $\rho$ and $\rho_{i}$, respectively, after tracing out
particles 1, 3, and 5. Since the overall conditions remain unchanged,
a straightforward calculation yields:

\[
\rho_{24}=Tr_{135}(\rho)=\frac{1}{8}\times\left(|00\rangle\langle00|+|01\rangle\langle01|+|10\rangle\langle10|+|11\rangle\langle11|\right)\times2=\frac{1}{4}I_{4},
\]
where $I_{4}$ denotes the $4\times4$ identity matrix. The von Neumann
entropy of each component of the mixed state $\rho$ after partial
tracing, specifically $\rho_{24_{i}}$, is given by $\rho_{24_{i}}=\frac{1}{4}I_{4}$.
Substituting $\rho_{24}$ and $\rho_{24_{i}}$ into Equation (\ref{eq:Chapter2_Eq27})
results in $\chi(\rho_{2})=0$. Consequently, it can be concluded
that Eve is unable to acquire any key information through a direct
intercept attack on the transmitted particles. A schematic representation
of Eve\textquoteright s intercept-resend attack strategy is illustrated
in Figure \ref{fig:Chapter2_Fig7}.

\subsubsection{Security analysis of Protocol 2.4 against impersonated fraudulent
attack}

To successfully impersonate Alice, Eve's optimal approach involves
interacting with the traveling particle (qubit 2) that Alice transmits,
using an ancillary state she prepares. After applying a general operation
to the traveling qubit, the resulting state is expressed as follows:

\begin{equation}
U_{E}|1\chi\rangle_{2e}=\left(a_{0}|10\rangle+b_{0}|11\rangle+c_{0}|00\rangle+d_{0}|01\rangle\right)_{2e},
\end{equation}

\begin{equation}
U_{E}|0\chi\rangle_{2e}=\left(a_{1}|10\rangle+b_{1}|11\rangle+c_{1}|00\rangle+d_{1}|01\rangle\right)_{2e}.
\end{equation}
Here, $|\chi\rangle_{e}$ represents Eve\textquoteright s ancillary
state, where the subscript $e$ signifies that it is prepared by Eve,
and 2 denotes Alice\textquoteright s traveling qubit. The normalization
condition is maintained as $|a_{1}^{2}|+|b_{1}^{2}|+|c_{1}^{2}|+|d_{1}^{2}|=|a_{0}^{2}|+|b_{0}^{2}|+|c_{0}^{2}|+|d_{0}^{2}|=1$.
Through her operation, Eve generates the state:
\begin{equation}
|\Psi^{\prime}\rangle_{12e}=\frac{1}{\sqrt{2}}\left(a_{0}|010\rangle+b_{0}|011\rangle+c_{0}|000\rangle+d_{0}|001\rangle+a_{1}|110\rangle+b_{1}|111\rangle+c_{1}|100\rangle+d_{1}|101\rangle\right).\label{eq:17}
\end{equation}
For simplicity, Eve applies a general operation on the traveling qubit
while retaining her own qubit. She then forwards Alice\textquoteright s
qubit to Charlie, who processes it according to the prescribed protocol
before returning it. The objective is to determine the final composite
state, factoring in Eve\textquoteright s attack within the scenario
outlined in Section \ref{sec:Chapter2_Sec4}.

\begin{equation}
\begin{array}{lcl}
|\Psi^{\prime\prime}\rangle_{12e345} & = & \frac{1}{2\sqrt{2}}[a_{0}|110110\rangle-a_{0}|100111\rangle-a_{0}|110000\rangle+a_{0}|100001\rangle\\
 & + & b_{0}|111110\rangle-b_{0}|101111\rangle-b_{0}|111000\rangle+b_{0}|101001\rangle\\
 & - & c_{0}|100110\rangle-c_{0}|110111\rangle-c_{0}|100000\rangle+c_{0}|110001\rangle\\
 & + & d_{0}|101110\rangle-d_{0}|111111\rangle-d_{0}|101000\rangle+d_{0}|111001\rangle\\
 & - & a_{1}|010110\rangle+a_{1}|000111\rangle+a_{1}|010000\rangle+a_{1}|000001\rangle\\
 & - & b_{1}|011110\rangle+b_{1}|001111\rangle+b_{1}|011000\rangle-b_{1}|001001\rangle\\
 & - & c_{1}|000110\rangle+c_{1}|010111\rangle+c_{1}|000000\rangle-c_{1}|010001\rangle\\
 & - & d_{1}|001110\rangle+d_{1}|011111\rangle+d_{1}|001000\rangle-d_{1}|011001\rangle].
\end{array}\label{eq:CNOT=000020operation}
\end{equation}
Upon receiving particle 4 from Charlie, Eve and Bob conduct a Bell
measurement on their respective particles and transmit the classical
sequence of measurement outcomes to Charlie. By enforcing the authentication
condition, the probabilities of Eve successfully passing authentication
and failing authentication can be determined as $\frac{1}{4}(|b_{0}|^{2}+|c_{0}|^{2})+\frac{1}{8}$
and $\frac{1}{4}(|a_{0}|^{2}+|d_{0}|^{2}+1)+\frac{3}{8}$, respectively.
If Eve prepares the ancillary state to maximize $|b_{0}|^{2}+|c_{0}|^{2}$
while minimizing $|a_{0}|^{2}+|d_{0}|^{2}$, she can achieve the highest
possible deception probability as an authenticated participant, given
by $P_{3}^{2}=\frac{3}{8}$. For all other scenarios, the probability
satisfies $P_{3}^{2}\leq\frac{3}{8}$. This analysis confirms that
the protocol remains secure under an acceptable error threshold even
in the most favorable conditions for a collective attack by Eve.

\subsection{Comparison of Protocol 2.4 with a set of existing protocols }\label{sec:Chapter2_Sec4.4}

In this section, we provide a concise comparison of the proposed protocol
with previously introduced QIA schemes. The evaluation considers key
aspects such as quantum resource utilization, the nature of the third
party (whether it is assumed to be honest, semi-honest, or untrusted),
the minimum number of qubits required for authentication (i.e., the
total number of qubits needed to execute the protocol when the pre-shared
authentication key size is 1 bit), and whether the protocol supports
bidirectional authentication (mutual verification) or only allows
one party to authenticate the other (unidirectional). Given the extensive
range of QIA schemes, a representative subset has been selected for
comparison, with a particular focus on those derived from secure direct
quantum communication protocols, as the proposed scheme shares a similar
foundation. Specifically, the protocol presented here is inspired
by CDSQC.

To initiate the comparison, we analyze Zhang et al.'s QIA scheme \cite{ZZZX_2006}
from 2006, which employs the ping-pong protocol for QSDC. In this
protocol, Alice functions as the trusted ``certification authority'',
while Bob represents a regular user whose identity requires validation
by Alice. A similar framework is observed in Yuan et al.'s protocol
\cite{YLP+14}, which is based on the single-photon QSDC protocol
known as the LM05 protocol, wherein Alice also serves as the certification
authority. Notably, both of these QSDC-based QIA protocols \cite{ZZZX_2006,YLP+14}
operate in a unidirectional manner, unlike the proposed scheme, which
enables bidirectional authentication. This characteristic provides
a distinct advantage of the proposed protocol over Zhang et al.'s
scheme. Furthermore, this advantage extends beyond Zhang et al.'s
protocol to a broader category of unidirectional QIA protocols, such
as Hong et al.'s scheme \cite{HCJ+17}. However, this advantage is
not exclusive, as similar bidirectional capabilities are also demonstrated
in Kang et al.'s protocols \cite{KHHYHM_2018,KHH+20} and Zhang et
al.'s 2020 work \cite{ZCSL_2020}. A key distinction is that these
protocols \cite{KHHYHM_2018,KHH+20,ZCSL_2020} necessitate at least
six qubits for authentication, whereas the proposed scheme requires
a minimum of only five qubits. Given the high cost of quantum resources,
this reduction can be considered beneficial. Additionally, Kang et
al.'s protocols \cite{KHHYHM_2018,KHH+20} utilize GHZ-like states,
which are tripartite entangled states that are comparatively more
complex to generate and maintain than the Bell states adopted in the
proposed scheme. Moreover, in Zhang et al.'s 2020 protocol, the third
party is assumed to be semi-honest, whereas in the proposed scheme,
the third party is treated as untrusted, thereby enhancing security.
Consequently, the proposed scheme efficiently achieves the desired
QIA features while optimizing resource utilization. The comparison
conducted to establish the significance of the proposed protocol,
particularly concerning entangled-state-based QIA schemes, is summarized
in Table \ref{tab:Chapter2_Tab10}.

\begin{table}[h]
\caption{Detailed comparison of Protocol 2.4 with previous protocols.}\label{tab:Chapter2_Tab10}

\centering{}%
\begin{tabular*}{15.9cm}{@{\extracolsep{\fill}}@{\extracolsep{\fill}}|>{\centering}p{2.5cm}|>{\centering}p{3cm}|>{\centering}p{2.2cm}|>{\centering}p{3cm}|>{\centering}p{2.8cm}|}
\hline 
Protocol  & Quantum Resources  & Minimum number of qubits required  & Way of authentication  & Nature of the third party\tabularnewline
\hline 
Zhang et al. \cite{ZZZX_2006}  & Bell state  & 3  & unidirectional  & no\tabularnewline
\hline 
Yuan et al.\cite{YLP+14}  & Single photon  & 1  & unidirectional  & no\tabularnewline
\hline 
Hong et al. \cite{HCJ+17}  & Single photon  & 1  & unidirectional  & no\tabularnewline
\hline 
Kang et al. \cite{KHHYHM_2018,KHH+20}  & GHZ-like  & 6  & bidirectional  & untrusted\tabularnewline
\hline 
Zhang et al. \cite{ZCSL_2020}  & Bell state  & 6  & bidirectional  & semi-honest\tabularnewline
\hline 
Proposed Protocol 2.4  & Bell state  & 5  & bidirectional  & untrusted\tabularnewline
\hline 
\end{tabular*}
\end{table}

\section{Conclusions}\label{sec:Chapter2_Sec5}

We have examined the existing protocols for QIA in Chapter \ref{Ch1:Chapter1_Introduction} to discern
their inherent symmetry. This analysis has identified symmetries across
various protocols, facilitating their classification and enabling
the formulation of straightforward strategies to adapt these protocols
for different quantum computing and communication applications, ultimately
leading to novel QIA schemes. To substantiate this approach, four
new QIA protocols have been introduced and demonstrated to be secure
against specific potential attacks. Notably, many of the discussed
protocols, including several existing ones, are not practically implementable
with current technology. For instance, certain protocols require a
user to retain a photon until the travel photon(s) return, such as
in ping-pong-based QIA schemes \cite{ZZZX_2006} or authentication
mechanisms outlined in \cite{T.Mihara_2002,LB_2004,ZLG_2000,ZZZZ_2005,WZT_2006,TJ_2014,KHHYHM_2018,WZGZ_2019}.
The feasibility of such QIA protocols is hindered by the necessity
of quantum memory, which remains commercially unviable. In particular,
most entangled-state-based QSDC protocols for QIA, as well as entangled-state
protocols relying on quantum private comparison and those involving
a third party (e.g., Trent) responsible for storing a photon, encounter
similar limitations. At present, it is preferable to prioritize QIA
protocols that do not rely on quantum memory. Notably, the protocols
introduced in this work do not require quantum memory, making them
feasible with existing technology. While numerous approaches have
been proposed for constructing quantum memory, and such memory-dependent
schemes may become beneficial in the future, direct teleportation-based
QIA schemes (such as those in \cite{ZZZZ_2005,TJ_2014}) are unlikely
to have widespread applications, even in the long term. This limitation
arises because teleportation can only facilitate secure quantum communication
in the absence of noise in the quantum channel. Additionally, any
protocol dependent on shared entanglement (e.g., those discussed in
\cite{CS_01,CSPF_02,ZLG_2000,T.Mihara_2002,LB_2004,ZZZZ_2005}) may
necessitate entanglement purification or concentration, as shared
entanglement degrades due to decoherence. These processes could introduce
undesirable and potentially insecure interactions between Alice and
Bob. Despite these challenges and technological constraints, research
in QIA is advancing rapidly, largely due to the fact that the purported
unconditional security of quantum cryptographic protocols hinges on
the robustness of the identity authentication mechanisms employed.

The current chapter, in conjunction with our research references \cite{DP23,DP+24},
highlights the potential for developing a diverse range of novel QIA
schemes. A thorough investigation in this domain could be instrumental
in identifying the most effective QIA scheme that can be realized
with existing technology. The proposed QIA protocol is an entangled-state-based
approach that employs Bell states. Drawing inspiration from secure
direct quantum communication, it enables Alice to transmit information
securely to Bob, contingent upon authorization from controller Charlie,
without requiring pre-shared keys. While the protocol exclusively
utilizes Bell states, its implementation demands quantum memory, which
is not yet commercially accessible. However, this constraint is not
specific to the proposed protocol but is a common challenge across
various QIA, QSDC, DSQC, and quantum dialogue protocols. With increasing
research efforts and advancements in quantum memory technology, along
with recent proposals for its development, commercial availability
is expected in the near future \cite{LLY+24}. Meanwhile, a delay
mechanism can function as an interim solution to replace quantum memory
in implementing the proposed QIA protocol, thereby improving the robustness
of secure quantum communication protocols. The other two proposed
protocols in this chapter utilize a single-photon source as the qubit.
These protocols are viable for implementation using current technology
to facilitate bi-directional identity authentication.

Furthermore, to provide a comprehensive discussion, it is important
to acknowledge the numerous post-quantum authentication schemes introduced
in recent years \cite{WZW+21,MGB+20}. However, these schemes have
not been covered here, as they are fundamentally classical and only
conditionally secure. A key underlying assumption of these schemes
is that quantum computers will be unable to efficiently solve problems
beyond the bounded-error ``quantum polynomial time'' complexity
class. This assumption lacks formal proof and is essentially equivalent
to presume that if an efficient algorithm for a given computational
problem is not currently known, it will remain undiscoverable in the
future.

\newpage



\chapter{QUANTUM KEY DISTRIBUTION WITH TIGHT SECURITY BOUNDS}\label{Ch3:Chapter3_QKD}
\graphicspath{{Chapter3/Chapter3Figs/}{Chapter3/Chapter3Figs/}}

\section{Introduction}

Cryptography has been a fundamental tool for securing information
since ancient times. Traditionally, cryptographic techniques concealed
secret data, but advancements in cryptanalysis often led to their
decryption. A major breakthrough occurred in the 1970s with the advent of the schemes for
public-key cryptography, including RSA \cite{RSA78} and Diffie-Hellman (DH)
\cite{DH76} schemes, whose security relies on the computational complexity
of the tasks like factorization of odd square-free bi-primes and discrete
logarithm problems \cite{P13}, respectively. The security of these schemes is essentially 
obtained from the fact that the computational problems (i.e., factorization of odd square-free bi-primes and discrete
logarithm problems) that are used to construct these schemes for cryptography belong to the NP complexity class, and thus these problems cannot be solved efficiently, i.e., in a polynomial number of steps. In contrast to these classical results, in 1994, Peter W. Shor \cite{S94}
demonstrated that quantum computers can efficiently (i.e., in polynomial
time) solve these two problems. This established that classical cryptographic schemes like RSA and DH are vulnerable against attacks using quantum computers. Thus, a challenge arose from quantum algorithms outperforming classical counterparts
in solving complex tasks. However, QKD offers a solution by ensuring
security through fundamental physical laws rather than computational
complexity. Notably, the first QKD protocol was introduced a decade
before Shor\textquoteright s breakthrough by Bennett and Brassard
in 1984 \cite{BB84}. Principles such as the no-cloning theorem \cite{WZ82},
measurement collapse, and Heisenberg's uncertainty principle underpin
QKD security, enabling implementations with polarization-encoded single
photons and other photonic qubits. While QKD ideally detects eavesdropping,
practical limitations in devices may allow undetected breaches. The
BB84 protocol paved the way for various QKD protocols \cite{B92,E91,GV95,STP20}
and other cryptographic applications \cite{YSP14,SKB+13,LXS_18,BST+17,STP17,thapliyal2018orthogonal,TP15,STP20,DP+23}
(for a review, see \cite{SPR17,GRT+02}). Each protocol offers distinct
advantages and limitations. While most are unconditionally secure
in an ideal setting, practical implementations suffer from device
imperfections, leading to vulnerabilities via side-channel attacks.
Quantum identity authentication \cite{CS_01,DP22,DP23} is essential
before executing a QKD protocol to ensure secure communication. However,
real-world QKD implementations face challenges, such as the requirement
for a true single-photon source. Protocols like BB84 and B92 \cite{B92}
ideally depend on single-photon states sent from Alice to Bob. Although
significant experimental progress has been made toward reliable single-photon
sources \cite{LP21,TS21}, most commercial systems use WCP by attenuating
laser outputs as an approximation. The quantum state of an attenuated
WCP can be described as:
\begin{equation}
|\alpha\rangle=|\sqrt{\mu}\exp(i\theta)\rangle=\sum_{n=0}^{\infty}\left(\frac{e^{-\mu}\mu^{n}}{n!}\right)^{\frac{1}{2}}\exp(in\theta)|n\rangle.\label{eq:Chapter3_Eq1}
\end{equation}
A Fock state $|n\rangle$ represents an $n$-photon state, with Alice
generating a quantum state as a superposition of Fock states following
a Poissonian photon number distribution: $p(n,\mu)=\frac{e^{-\mu}\mu^{n}}{n!}$,
where the mean photon number is $\mu=|\alpha|^{2}\ll1$. Using such
a source, Alice emits the desired one-photon state with probability
$p(1,\mu)$, while multi-photon pulses occur with probability $1-p(0,\mu)-p(1,\mu)$.
This enables Eve to exploit a PNS attack \cite{HIG+95}. In long-distance
communication, channel loss further facilitates eavesdropping, as
an adversary with advanced technology can replace the lossy channel
with a transparent one and intercept signals undetected \cite{BLM+2000}.
To mitigate PNS attacks, Scarani et al. introduced the SARG04 QKD
protocol in 2004 \cite{SAR+04}. Here, we propose two novel QKD protocols
designed to counter not only PNS attacks but also a broader range
of attacks while offering specific advantages over SARG04 and similar
existing protocols.

Information splitting is a fundamental aspect of every QKD protocol.
In BB84 and B92, information is divided into a classical component
(basis choice) and a quantum component (transmitted qubits), whereas
in SARG04, a similar division occurs. However, protocols like GV protocol
\cite{GV95} split information into two quantum parts. The security
of these protocols stems from Eve's inability to access both pieces
simultaneously. This raises an important question: Can protocol efficiency
and secret-key rate bounds be modified by reducing the classical information
content? To explore this, we use SARG04 as a test case and introduce
two new QKD protocols that contain less classical information than
SARG04. The SARG04 protocol was designed to mitigate the PNS attack
but is less efficient than other single-photon QKD schemes. Efficiency
is assessed using Cabello\textquoteright s definition \cite{C2000},
assuming equal costs for qubit and classical bit transmission and
a low-noise quantum channel---conditions that are often unrealistic
for long-distance communication. Motivated by these limitations, we
investigate whether increasing quantum resource utilization can counteract
the PNS attack while reducing dependence on classical resources. Our
goal is to develop two new QKD protocols that enhance efficiency compared
to SARG04 while maintaining resilience against PNS and other standard
attacks.

The structure of the chapter is outlined as follows. Section \ref{sec:Chapter3_Sec2}
introduces a novel QKD protocol based on single-photon transmissions,
which eliminates the necessity for an ideal single-photon source.
This protocol, designated as Protocol 3.1, is initially presented
in a generalized framework, followed by a stepwise explanation. A
minor modification in the sifting subprotocol of Protocol 3.1 results
in an improved protocol (Protocol 3.2) with enhanced efficiency. Section
\ref{sec:Chapter3_Sec3} provides a comprehensive security analysis,
utilizing a depolarizing channel to model errors introduced by either
Eve or the communication channel itself. This approach enables the
determination of the tolerable error threshold for the initial quantum
particle sequence generated by Bob. Additionally, security is evaluated
against various collective attack strategies. Section \ref{sec:Chapter3_Sec4}
examines the PNS attack on both Protocol 3.1 and Protocol 3.2, deriving
the critical distance and demonstrating the benefits of employing
a relatively higher quantum resource allocation. Finally, Section
\ref{sec:Chapter3_Sec5} presents the conclusion of the chapter.

\section{Proposed quantum key distribution protocols (Protocol 3.1 and 3.2)}\label{sec:Chapter3_Sec2}

As previously discussed, numerous QKD protocols that rely on non-orthogonal
state sequences require the partitioning of information into quantum
and classical components. This segmentation ensures that any eavesdropping
attempt by Eve leaves detectable traces through measurement disturbances.
In all such QKD frameworks, Alice and Bob must eventually compare
the initial states (or bases) used for encoding with those measured
at the receiving end. This comparison helps identify correlations
that could indicate the presence of an eavesdropper. After performing
this verification, Alice and Bob retain only the states that satisfy
predefined conditions, forming the basis for the final key generation.
This process constitutes a specific stage within the protocol, often
referred to as the classical key-sifting subprotocol. In this chapter,
we employ a bi-directional quantum channel for the secure distribution
of quantum information using single-photon states. The objective is
to establish a secret key between Alice and Bob after executing the
key-sifting subprotocol. Here, Alice possesses prior knowledge of
the quantum states in the initial sequence that she prepares for Bob.
This prior information facilitates agreement on the positions of the
sifted key following the information reconciliation process. To formalize
our approach, we define the following notation: When encoding a bit
value $x$, Alice prepares the quantum state $\psi_{J}^{x}$, utilizing
mutually unbiased bases (MUBs) in a Hilbert space $\mathcal{H}$ of
dimension\footnote{Consider two orthonormal basis sets in a $d$-dimensional Hilbert
space, represented as $\psi_{j_{1}}:=\{\psi_{1},\psi_{2,}\ldots,\psi_{d}\}$
and $\psi_{j_{2}}:=\{\psi_{1}^{'},\psi_{2,}^{'}\ldots,\psi_{d}^{'}\},$
These bases are mutually unbiased if the squared modulus of the inner
product between any two different basis vectors equals $\frac{1}{d}$,
mathematically expressed as $\left|\langle\psi_{a}|\psi_{b}^{'}\rangle\right|^{2}=\frac{1}{d}$,
$\forall a,b\in\{1,2,\ldots,d\}.$ When a quantum system is prepared
in one of these mutually unbiased bases (MUBs), measuring it in another
basis results in a completely random outcome, maximizing measurement
uncertainty.} $d$. The bit value $x$ represents the encoded information, while
$J$ specifies the basis used for encoding. Without loss of generality,
we define $J\coloneqq\{Z,X\}$, where the basis sets $Z$ and $X$
correspond to $\left\{ |0\rangle,|1\rangle\right\} $ and $\left\{ |+\rangle,|-\rangle\right\} $,
respectively. These basis sets are commonly referred to as the computational
and diagonal bases. For ease of classical key-sifting, we designate
$J=0$ for the $Z$-basis and $J=1$ for the $X$-basis. Building
upon this notation, we now outline the generalized structure of our
proposed protocol.

(1) \textit{ State generation, transmission, and measurement:} Alice
prepares and transmits a sequence of qubits, denoted as $S_{A}$,
to Bob. Each qubit is encoded in one of the four quantum states$\psi_{J}^{x}\coloneqq\{\text{\ensuremath{\psi_{Z}^{x},}}\psi_{X}^{x}\}$,
where the bit value $x$ is randomly selected from $\{0,1\}$. Bob
then measures each qubit using either the computational or diagonal
basis at random, resulting in a sequence of quantum states $\psi_{J}^{y}\coloneqq\{\text{\ensuremath{\psi_{Z}^{x},}}\psi_{X}^{x/x^{\perp}}\}$,
where $\psi_{J}^{x^{\perp}}$ represents the state orthogonal to $\psi_{J}^{x}$,
and $x,y\in\{0,1\}$. It is assumed that no decoherence occurs during
qubit transmission. Additionally, Alice does not disclose her chosen
measurement basis to Bob. Up to this stage, the procedure is analogous
to the BB84 QKD protocol \cite{BB84}.

Based on his measurement results, Bob constructs a sequence of quantum
states, denoted as $S_{B1}$, and transmits it back to Alice. Upon
receiving $S_{B1}$, Alice measures each qubit using the same basis
originally used in $S_{A}$ and records the results. Given the assumption
of a long sequence and a noise-free quantum channel, Alice will observe
the state $\psi_{J}^{x^{\perp}}$ with a probability of $\frac{1}{4}$.
If this probability falls within the predefined threshold around $\frac{1}{4}$,
Alice publicly instructs Bob to send the next sequence of qubits,
denoted as $S_{B2}$.

(2) \textit{Preparation and measurement of the second sequence ($S_{B2}$):}
Upon receiving Alice\textquoteright s request, Bob prepares the sequence
$S_{B2}$ using the mutually unbiased basis (MUB) alternative to the
one previously used for the $n$th qubit in sequence $S_{B1}$. Specifically,
if the $Z$ (or $X$) basis was employed for a particular qubit in
$S_{B1}$, then the $X$ (or $Z$) basis is chosen for the corresponding
qubit in $S_{B2}$, ensuring the same bit values at equivalent positions
as in sequence $S_{B1}$ ($\psi_{J}^{y}$). Bob then transmits $S_{B2}$
to Alice. Upon receiving it, Alice performs measurements based on
the following rule: If her measurement of $S_{B1}$ yields the state
$\psi_{J}^{x}$, she utilizes the alternate MUB (the second basis).
However, if she obtains the orthogonal state $\psi_{J}^{x^{\perp}}$
(relative to the corresponding elements of her initial sequence $S_{A}$),
she continues using the same basis.

(3) \textit{Condition for key-sifting:} To enhance the retention of
raw key bits post-sifting while reducing the amount of disclosed classical
information compared to the SARG04 protocol, a classical subprotocol
is introduced. Alice reveals only the positions of qubits for which
Bob\textquoteright s measurement outcomes of $S_{A}$ contribute to
key generation, based on the following conditions: (i) If Alice\textquoteright s
measurement of SB1SB1 results in an orthogonal
state $\psi_{Z}^{x^{\perp}}$ with respect to her initial sequence
$S_{A}$, and the corresponding measurement outcome for $S_{B2}$
is either $\psi_{Z}^{x}$ or $\psi_{Z}^{x^{\perp}}$, Alice infers
that Bob\textquoteright s measurement result for $S_{A}$ was either
$\psi_{X}^{x}$ or $\psi_{X}^{x^{\perp}}$. (ii) If Alice measures
$S_{B1}$ and obtains $\psi_{Z}^{x}$, and the corresponding measurement
outcome of $S_{B2}$ from Bob is $\psi_{X}^{x}$, then Alice deduces
that Bob\textquoteright s measurement outcome for $S_{A}$ was $\psi_{Z}^{x}$.
However, this holds only if Bob's announced $J$ value for each measurement
matches the corresponding $J$ value for elements in Alice's initial
sequence $S_{A}$. The $J$ value is disclosed selectively for a subset
of qubits---specifically, for those where Alice's measurement of
$S_{B1}$ aligns with the respective elements of $S_{A}$, given that
the corresponding qubits in $S_{B2}$ maintain bitwise consistency
in a distinct basis relative to $S_{B1}$.

In the subsequent sections, we provide a comprehensive step-by-step
breakdown of our primary protocol, designated as Protocol 3.1. Following
this, we illustrate how a minor refinement to the key-sifting subprotocol
within Protocol 3.1 can enhance the efficiency of our proposed QKD
protocol. The optimized version will be referred to as Protocol 3.2
(see Table \ref{tab:Chapter3_Tab1} for further details).

\begin{longtable}[H]{|>{\centering}p{0.60cm}|>{\centering}p{0.60cm}|>{\centering}p{0.60cm}|>{\centering}p{2.1cm}|>{\centering}p{2.1cm}|>{\centering}p{1.8cm}|>{\centering}p{0.75cm}|>{\centering}p{1.30cm}|>{\centering}p{0.75cm}|>{\centering}p{1.30cm}|}
\caption{ This table outlines the encoding and decoding principles for both
Protocol 3.1 and Protocol 3.2, while also presenting the measurement
results obtained after the classical sifting subprotocol. }\label{tab:Chapter3_Tab1} \\
\hline 
$S_{A}$  & $S_{B1}$  & $S_{B2}$  & Measurement result of $S_{B1}$ by Alice  & Measurement result of $S_{B2}$ by Alice  & Probability  & $J$ value for P1  & Result determine by P1  & $M$ value for P2  & Result determine by P2\tabularnewline
\hline 
 & $|0\rangle$  & $|+\rangle$  & $|0\rangle$  & $|+\rangle$  & $\nicefrac{1}{8}$  & 0  & $|0\rangle$  & 0  & $|0\rangle$\tabularnewline
\cline{2-10}
 &  &  & $|0\rangle$  & $|+\rangle$  & $\nicefrac{1}{64}$  & 1  & $-$  & 0  & $|0\rangle$\tabularnewline
 & $|+\rangle$  & $|0\rangle$  & $|0\rangle$  & $|-\rangle$  & $\nicefrac{1}{64}$  & $-$  & $-$  & 0  & $|+\rangle$\tabularnewline
$|0\rangle$  &  &  & $|1\rangle$  & $|0\rangle$  & $\nicefrac{1}{32}$  & $-$  & $|+\rangle$  & $-$  & $|+\rangle$\tabularnewline
\cline{2-10}
 &  &  & $|0\rangle$  & $|+\rangle$  & $\nicefrac{1}{64}$  & 1  & $-$  & 1  & $|-\rangle$\tabularnewline
 & $|-\rangle$  & $|1\rangle$  & $|0\rangle$  & $|-\rangle$  & $\nicefrac{1}{64}$  & $-$  & $-$  & 1  & $|-\rangle$\tabularnewline
 &  &  & $|1\rangle$  & $|1\rangle$  & $\nicefrac{1}{32}$  & $-$  & $|-\rangle$  & $-$  & $|-\rangle$\tabularnewline
\hline 
 & $|1\rangle$  & $|-\rangle$  & $|1\rangle$  & $|-\rangle$  & $\nicefrac{1}{8}$  & 0  & $|1\rangle$  & 1  & $|1\rangle$\tabularnewline
\cline{2-10}
 &  &  & $|1\rangle$  & $|+\rangle$  & $\nicefrac{1}{64}$  & $-$  & $-$  & 0  & $|+\rangle$\tabularnewline
 & $|+\rangle$  & $|0\rangle$  & $|1\rangle$  & $|-\rangle$  & $\nicefrac{1}{64}$  & 1  & $-$  & 0  & $|+\rangle$\tabularnewline
$|1\rangle$  &  &  & $|0\rangle$  & $|0\rangle$  & $\nicefrac{1}{32}$  & $-$  & $|+\rangle$  & $-$  & $|+\rangle$\tabularnewline
\cline{2-10}
 &  &  & $|1\rangle$  & $|+\rangle$  & $\nicefrac{1}{64}$  & $-$  & $-$  & 1  & $|-\rangle$\tabularnewline
 & $|-\rangle$  & $|1\rangle$  & $|1\rangle$  & $|-\rangle$  & $\nicefrac{1}{64}$  & 1  & $-$  & 1  & $|1\rangle$\tabularnewline
 &  &  & $|0\rangle$  & $|1\rangle$  & $\nicefrac{1}{32}$  & $-$  & $|-\rangle$  & $-$  & $|-\rangle$\tabularnewline
\hline 
 & $|+\rangle$  & $|0\rangle$  & $|+\rangle$  & $|0\rangle$  & $\nicefrac{1}{8}$  & 1  & $|+\rangle$  & 0  & $|+\rangle$\tabularnewline
\cline{2-10}
 &  &  & $|+\rangle$  & $|0\rangle$  & $\nicefrac{1}{64}$  & 0  & $-$  & 0  & $|+\rangle$\tabularnewline
 & $|0\rangle$  & $|+\rangle$  & $|+\rangle$  & $|1\rangle$  & $\nicefrac{1}{64}$  & $-$  & $-$  & 0  & $|0\rangle$\tabularnewline
$|+\rangle$  &  &  & $|-\rangle$  & $|+\rangle$  & $\nicefrac{1}{32}$  & $-$  & $|0\rangle$  & $-$  & $|0\rangle$\tabularnewline
\cline{2-10}
 &  &  & $|+\rangle$  & $|0\rangle$  & $\nicefrac{1}{64}$  & 0  & $-$  & 1  & $|1\rangle$\tabularnewline
 & $|1\rangle$  & $|-\rangle$  & $|+\rangle$  & $|1\rangle$  & $\nicefrac{1}{64}$  & $-$  & $-$  & 1  & $|1\rangle$\tabularnewline
 &  &  & $|-\rangle$  & $|-\rangle$  & $\nicefrac{1}{32}$  & $-$  & $|1\rangle$  & $-$  & $|1\rangle$\tabularnewline
\hline 
 & $|-\rangle$  & $|1\rangle$  & $|-\rangle$  & $|1\rangle$  & $\nicefrac{1}{8}$  & 1  & $|-\rangle$  & 1  & $|-\rangle$\tabularnewline
\cline{2-10}
 &  &  & $|-\rangle$  & $|0\rangle$  & $\nicefrac{1}{64}$  & $-$  & $-$  & 0  & $|0\rangle$\tabularnewline
 & $|0\rangle$  & $|+\rangle$  & $|-\rangle$  & $|1\rangle$  & $\nicefrac{1}{64}$  & 0  & $-$  & 0  & $|0\rangle$\tabularnewline
$|-\rangle$  &  &  & $|+\rangle$  & $|+\rangle$  & $\nicefrac{1}{32}$  & $-$  & $|0\rangle$  & $-$  & $|0\rangle$\tabularnewline
\cline{2-10}
 &  &  & $|-\rangle$  & $|0\rangle$  & $\nicefrac{1}{64}$  & $-$  & $-$  & 1  & $|1\rangle$\tabularnewline
 & $|1\rangle$  & $|-\rangle$  & $|-\rangle$  & $|1\rangle$  & $\nicefrac{1}{64}$  & 0  & $-$  & 1  & $|-\rangle$\tabularnewline
 &  &  & $|+\rangle$  & $|-\rangle$  & $\nicefrac{1}{32}$  & $-$  & $|1\rangle$  & $-$  & $|1\rangle$\tabularnewline
\hline 
\end{longtable}

\subsection*{Protocol 3.1}

To define these protocols, we employ elements from the $Z$ and $X$
bases, utilizing a notation where basis states are expressed as $|+z\rangle/|-z\rangle(|+x\rangle/|-x\rangle):=|0\rangle/|1\rangle(|+\rangle/|-\rangle).$.
The $Z$ and $X$ basis elements are further defined as:

\begin{equation}
\begin{array}{lcl}
|+x\rangle=\frac{1}{\sqrt{2}}\left(|0\rangle+|1\rangle\right) & , & |-x\rangle=\frac{1}{\sqrt{2}}\left(|0\rangle-|1\rangle\right)\\
|+z\rangle=\frac{1}{\sqrt{2}}\left(|+x\rangle+|-x\rangle\right) & , & |-z\rangle=\frac{1}{\sqrt{2}}\left(|+x\rangle-|-x\rangle\right).
\end{array}\label{eq:define=000020the=000020bases=000020elements}
\end{equation}

\begin{description}
\item [{Step~1}] Alice randomly prepares a sequence of single qubits,
$S_{A}$, in either the $Z$ or $X$ basis, keeping the basis choice
confidential, and transmits the sequence to Bob.
\item [{Step~2}] Bob performs measurements on the qubits in $S_{A}$,
choosing randomly between the $Z$ and $X$ bases, and records the
outcomes. He then generates a new qubit sequence, $S_{B1}$, with
states corresponding to his measurement results and forwards it back
to Alice.
\item [{Step~3}] Alice measures each qubit in $S_{B1}$ using the same
basis initially chosen to prepare the corresponding qubit in $S_{A}$.
Specifically, if the $i^{th}$ qubit in $S_{A}$ was prepared in the
$Z$ basis ($X$ basis), Alice measures the $i^{th}$ qubit of $S_{B1}$
using the $Z$ basis ($X$ basis). She records the outcomes and instructs
Bob to proceed only if the measurement results remain within an acceptable
deviation from the expected probability distribution.
\item [{Step~4}] Bob generates a second qubit sequence, $S_{B2}$, which
mirrors the bit values of $S_{B1}$ but is prepared using the complementary
basis. Specifically, if the $i^{th}$ qubit in $S_{B1}$ is in the
state $|\pm z\rangle(|\pm x\rangle)$, then the corresponding qubit
in $S_{B2}$ is prepared in the state $\text{\ensuremath{|\pm x\rangle(|\pm z\rangle)}}$
before being transmitted to Alice.
\item [{Step~5}] Alice measures each qubit in $S_{B2}$, choosing the
measurement basis based on her results from $S_{B1}$. If the measurement
outcome for a qubit in $S_{B1}$ is $|\pm z\rangle$ or $|\pm x\rangle$,
she uses the $X$ or $Z$ basis accordingly. Conversely, if the outcome
is the orthogonal state ($|\mp z\rangle$or $|\mp x\rangle$), she
switches the basis for measurement on $S_{B2}$. The measurement strategy
is aligned with the initial sequence $S_{A}$.
\item [{Step~6}] Alice extracts the conclusive measurement results---those
that definitively determine Bob's corresponding measurements---by
analyzing the outcomes from both $S_{B1}$ and $S_{B2}$. If she originally
prepared the $i^{th}$ qubit of $S_{A}$ in $|\pm z\rangle(|\pm x\rangle)$
and observes measurement results of $|\mp z\rangle(|\mp x\rangle)$
from $S_{B1}$ and $|\pm z\rangle(|\pm x\rangle)$ or $|\mp z\rangle(|\mp x\rangle)$
from $S_{B2}$, she deduces Bob's measurement result for $S_{A}$
as either $|\pm x\rangle(|\pm z\rangle)$ or $|\mp x\rangle(|\mp z\rangle)$.
This step directly corresponds to the point (a) of the ``condition for key-sifting''
and facilitates key generation without publicly revealing the value
of $J$.
\item [{Step~7}] Alice retains only those bits as part of the sifted key
where the $J$ value matches for both parties. For instance, if Alice
prepares $S_{A}$ in $|\pm z\rangle(|\pm x\rangle)$ and obtains measurement
results of $|\pm z\rangle(|\pm x\rangle)$ from $S_{B1}$ and $|\pm x\rangle(|\pm z\rangle)$
from $S_{B2}$, then she determines Bob's corresponding result as
$|\pm z\rangle(|\pm x\rangle)$ only if the preparation and measurement
bases align (i.e., the same $J$ value is used). This step aligns
with point (b) of ``condition for key-sifting'' and finalizes the
sifted key using $J$ (refer to Table \ref{tab:Chapter3_Tab2}). Notably,
the $J$ value is revealed only for qubits where Alice's $S_{B1}$
measurement matches $S_{A}$, and the corresponding qubits in $S_{B2}$
maintain the same bit values under an alternate basis relative to
$S_{B1}$.\\
This process ensures that the key-sifting mechanism is executed efficiently,
establishing a shared secret key between Alice and Bob.
\end{description}
\begin{table}[H]
\caption{Table mapping between measurement outcomes and Alice's determined
results in Protocol 3.1.}\label{tab:Chapter3_Tab2}

\centering{}%
\begin{tabular*}{15.9cm}{@{\extracolsep{\fill}}|@{\extracolsep{\fill}}|>{\centering}p{1.5cm}|>{\centering}p{4.5cm}|>{\centering}p{4.0cm}|>{\centering}p{4.0cm}|}
\hline 
$S_{A}$ & Measurement result of $S_{B1}$, $S_{B2}$ by Alice & Result determined without $J$ value & Result determined with same $J$ value\tabularnewline
\hline 
$|\pm z\rangle$ & $|\pm z\rangle$,$|\pm x\rangle$ & $-$ & $|\pm z\rangle$\tabularnewline
 & $|\mp z\rangle,$$|\pm z\rangle$ or $|\mp z\rangle,$$|\pm z\rangle$ & $|\pm x\rangle$ or $|\mp x\rangle$ & $-$\tabularnewline
\hline 
$|\pm x\rangle$ & $|\pm x\rangle$,$|\pm z\rangle$ & $-$ & $|\pm x\rangle$\tabularnewline
 & $|\mp x\rangle,$$|\pm x\rangle$ or $|\mp x\rangle,$$|\mp x\rangle$ & $|\pm z\rangle$ or $|\mp z\rangle$ & $-$\tabularnewline
\hline 
\end{tabular*}
\end{table}

\subsection*{Protocol 3.2}

To facilitate the interpretation of Bob's measurement results for
the sequence $S_{A}$, we introduce a new variable $M\in\{0,1\}$.
This variable is essential for the classical key-sifting process and
is defined as follows: $M=0$ corresponds to the states $\{|+z\rangle,|+x\rangle\}$,
while $M=1$ corresponds to $\{|-z\rangle,|-x\rangle\}$. Protocol
3.2 follows the same steps (1 to 6) as Protocol 3.1, but with certain
modifications in the classical post-processing, specifically in Step
7. By employing this modified sifting process, the resulting sifted
key retains a maximum intrinsic error probability of $\nicefrac{1}{16}$,
yet it enhances efficiency compared to Protocol 3.1. A detailed analysis
of this trade-off will be presented later.
\begin{description}
\item [{Step~7}] If Alice prepares the sequence $S_{A}$ using the states
$|+z\rangle/|+x\rangle$ or $|-z\rangle/|-x\rangle$, the following
rules apply: If Bob publicly announces $M=1(0)$, Alice assigns Bob\textquoteright s
measurement outcome for $S_{A}$ as $|-x\rangle/|-z\rangle$ (or $|+x\rangle/|+z$),
regardless of the measurement outcomes of $S_{B1}$ and $S_{B2}$.
If Bob announces $M=0(1)$, Alice determines Bob\textquoteright s
measurement result as (i) $|+x\rangle/|+z\rangle$ (or $|-x\rangle/|-z\rangle$)
if the corresponding measurement outcomes of $S_{B1}$ and $S_{B2}$
are $|+z\rangle/|+x\rangle$ (or $|-z\rangle/|-x\rangle$) and $|-x\rangle/|-z\rangle$
(or $|+x\rangle/|+z\rangle$), respectively. (ii) $|+z\rangle/|+x\rangle$
(or $|-z\rangle/|-x\rangle$) if the measurement outcomes of $S_{B1}$
and $S_{B2}$ are $|+z\rangle/|+x\rangle$ (or $|-z\rangle/|-x\rangle$)
and $|+x\rangle/|+z\rangle$ (or $|-x\rangle/|-z\rangle$), respectively
(see Table \ref{tab:Chapter3_Tab3}). The value of $M$ is disclosed
only for a subset of qubits where Alice's measurements of $S_{B1}$
match the corresponding elements of $S_{A}$, ensuring that the qubits
in $S_{B1}$ are measured in a basis distinct from their corresponding
qubits in $S_{B2}$. Once Bob reveals the values of $M$ for specific
qubits, the encryption criteria transition from the BB84 protocol
to the SARG04 protocol. Specifically, in Protocol 3.2, the encryption
rule is defined such that the states $|\pm z\rangle$ encode bit value
0, while $|\pm x\rangle$ encode bit value 1. Consequently, revealing
the values of $M$ do not provide sufficient information to reconstruct
the final key.
\end{description}
Assuming an inherent error probability of $\frac{1}{16}$, Protocol
3.2 achieves a higher key rate than Protocol 3.1 in an environment
free of an eavesdropper. In particular, the efficiencies of Protocol
3.2 and Protocol 3.1 are 0.192 and 0.2069, respectively. However, unlike
Protocol 3.2, and Protocol 3.1 do not introduce intrinsic errors. A
comprehensive analysis of these results and the related trade-offs
is provided in Section \ref{sec:Chapter3_Sec5}.

\begin{table}
\caption{Table mapping between measurement results and Alice's determined
outcomes in Protocol 3.2.}\label{tab:Chapter3_Tab3}

\centering{}%
\begin{tabular*}{15.75cm}{@{\extracolsep{\fill}}|@{\extracolsep{\fill}}|>{\centering}p{2.25cm}|>{\centering}p{2.00cm}|>{\centering}p{3.00cm}|>{\centering}p{3.00cm}|>{\centering}p{3.25cm}|}
\hline 
$S_{A}$ & Value of $M$ & Measurement result of $S_{B1}$ by Alice & Measurement result of $S_{B2}$ by Alice & Result determined\tabularnewline
\hline 
 & 1 & $-$ & $-$ & $|-x\rangle/|-z\rangle$\tabularnewline
\cline{2-5}
$|+z\rangle/|+x\rangle$ & 0 & $|+z\rangle/|+x\rangle$ & $|-x\rangle/|-z\rangle$ & $|+x\rangle/|+z\rangle$\tabularnewline
 & 0 & $|+z\rangle/|+x\rangle$ & $|+x\rangle/|+z\rangle$ & $|+z\rangle/|+x\rangle$\tabularnewline
\hline 
 & 0 & $-$ & $-$ & $|+x\rangle/|+z\rangle$\tabularnewline
\cline{2-5}
$|-z\rangle/|-x\rangle$ & 1 & $|-z\rangle/|-x\rangle$ & $|+x\rangle/|+z\rangle$ & $|-x\rangle/|-z\rangle$\tabularnewline
 & 1 & $|-z\rangle/|-x\rangle$ & $|-x\rangle/|-z\rangle$ & $|-z\rangle/|-x\rangle$\tabularnewline
\hline 
\end{tabular*}
\end{table}

\section{Security evaluation of the proposed protocols}\label{sec:Chapter3_Sec3}

As previously discussed, Bob must await Alice\textquoteright s approval
of the sequence $S_{B1}$ before proceeding with the transmission
of $S_{B2}$. Once Alice confirms $S_{B1}$, Bob transmits the second
sequence, $S_{B2}$. The protocol concludes with Alice and Bob establishing
a shared secret key, contingent upon the computed error rate remaining
within the acceptable threshold. The primary focus of our security
analysis is to determine the maximum tolerable error rate in the presence
of collective attacks. To assess Eve\textquoteright s potential attack
strategies, we adopt an approach inspired by Ref. \cite{KGR_05},
which utilizes a depolarizing map to transform any arbitrary two-qubit
state into a Bell-diagonal state. In order to rigorously analyze the
security of the proposed QKD protocols in accordance with the principles
outlined in Ref. \cite{KGR_05}, we must reformulate them into equivalent
entanglement-based versions. A corresponding entanglement-based representation
of Protocol 3.1/2 can be conceptualized as follows: Alice generates
a set of $n$ entangled two-qubit states (such as Bell states) and
applies her encoding operation to the first qubit of each pair, while
the second qubit is transmitted to Bob. Specifically, if Alice initially
prepares the state $|\Phi^{+}\rangle$, she applies the transformation
$A_{j}\otimes I_{2}|\Phi^{+}\rangle$ and sends the second qubit to
Bob. Here, $|\Phi^{\pm}\rangle=\frac{1}{\sqrt{2}}(|00\rangle\pm|11\rangle)$,
where $A_{j}$ represents Alice's encoding operation, and $I_{2}$
denotes the identity operator in a two-dimensional Hilbert space.
Upon receiving the qubits, Bob independently and randomly applies
one of his encoding operations, $B_{j}$, to each qubit he receives.
The $2n$-qubit state shared between Alice and Bob can be represented
as $\tilde{\rho}_{AB}^{n}$. Subsequently, Alice and Bob perform measurements
on their respective qubits of $\tilde{\rho}_{AB}^{n}$ using randomly
chosen $X$ and $Z$ bases, assigning the outcomes to binary values
(0 or 1). To formalize this process, two completely positive maps
(CPMs), denoted as $\mathcal{O}_{1}$ and $\mathcal{O}_{2}$, are
introduced. The first map, $\mathcal{O}_{1}$, is entirely dictated
by the protocol, while $\mathcal{O}_{2}$ remains independent of it.
Specifically, these maps are defined as:

\[
\begin{array}{lcl}
\mathcal{O}_{1}(\rho) & = & \frac{1}{N}\sum_{j}p_{j}A_{j}\otimes B_{j}(\rho)A_{j}^{\dagger}\otimes B_{j}^{\dagger}\\
\\\mathcal{O}_{2}(\rho) & = & \sum_{l}M_{l}\otimes M_{l}(\rho)M_{l}^{\dagger}\otimes M_{l}^{\dagger}
\end{array}.
\]
Here, $p_{j}\ge0$ represents the probability that Alice and Bob retain
a bit value in the sifting subprotocol, and $N$ is a normalization
constant. The operator $M_{l}$ characterizes a quantum operation,
where $M_{l}\in I_{2},\sigma_{x},\sigma_{y},\sigma_{z}$, with

\[
\begin{array}{lcl}
I_{2}=|0\rangle\langle0|+|1\rangle\langle1|, &  & \sigma_{x}=|0\rangle\langle1|+|1\rangle\langle0|,\\
\\i\sigma_{y}=|0\rangle\langle1|+|1\rangle\langle0|, &  & \sigma_{z}=|0\rangle\langle0|-|1\rangle\langle1|.
\end{array}
\]
Since $\mathcal{O}_{2}(\rho)$ applies identical operators to both
qubits, the possible two-qubit operations are given by $M_{l}\otimes M_{l}\in\left\{ I\otimes I,\,\sigma_{x}\otimes\sigma_{x},\,\sigma_{y}\otimes\sigma_{y},\,\sigma_{z}\otimes\sigma_{z}\right\} $.
These transformations occur randomly with uniform probability. Interestingly,
this random application effectively simulates the behavior of a depolarizing
channel, which maps any two-qubit state to a Bell-diagonal state.
If Alice and Bob implement a unitary transformation\footnote{The encoding and decoding operations can be expressed as $A_{j}=|0\rangle\langle(\phi_{j}^{0})^{*}|+|1\rangle\langle(\phi_{j}^{1})^{*}|$
and $B_{j}=|0\rangle\langle(\phi_{j}^{1})^{\perp}|+|1\rangle\langle(\phi_{j}^{0})^{\perp}|$
where $|(\phi_{j}^{i})^{*}\rangle$ denotes the complex conjugate
of $|\phi_{j}^{i}\rangle$ and $|(\phi_{j}^{i})^{\perp}\rangle$ represents
its orthogonal complement in the computational basis. The index $j\in\{1,\cdots,m\}$
corresponds to the states encoding the bit values $i=0,1$ \cite{RGK_05,KGR_05}.} $A_{j}\otimes B_{j}$, upon completing the sifting phase, Alice and
Bob obtain their sifted key, with the normalization constraint $\sum_{j}p_{j}=1$.
The normalized two-qubit density operator from Equation (1) of Ref.
\cite{RGK_05} is employed for $n=1$ (see \cite{KGR_05} for further
details). Finally, the notation $P_{|\Phi\rangle}=|\Phi\rangle\langle\Phi|$
represents a state projection operator, which projects a quantum state
of the corresponding dimension onto $|\Phi\rangle$. Now,

\begin{equation}
\rho^{1}[\boldsymbol{\mathcal{\mu}}]=\mu_{1}P_{|\Phi^{+}\rangle}+\mu_{2}P_{|\Phi^{-}\rangle}+\mu_{3}P_{|\Psi^{+}\rangle}+\mu_{4}P_{|\Psi^{-}\rangle}.\label{eq:Chapter3_Eq3}
\end{equation}
The operators $P_{|\Phi^{\pm}\rangle}$ and $P_{|\Psi^{\pm}\rangle}$
represent the state projections onto the Bell states $|\Phi^{\pm}\rangle\text{=\ensuremath{\frac{1}{\sqrt{2}}(|00\rangle\pm|11\rangle})}$
and $|\Psi^{\pm}\rangle$$=\ensuremath{\frac{1}{\sqrt{2}}(|01\rangle\pm|10\rangle})$.
The parameters $\mu_{1/2}$ and $\mu_{3/4}$ denote the corresponding
probabilities of obtaining these Bell states when a system undergoes
depolarization. To evaluate the security of the sequence $S_{B1}$,
the following key rate equation is employed:

\begin{equation}
r:=I(A:B)-\underset{\rho\in\mathscr{\mathcal{R}}}{max}S(\rho),\label{eq:Chapter3_Eq4}
\end{equation}
where $A$ and $B$ represent the quantum states acquired by Alice
and Bob after measurement. Here, $I(A:B)$ quantifies the mutual information
between Alice and Bob, while $S(\rho)$ denotes the von Neumann entropy
of the joint quantum state $\rho$. The set $\mathcal{R}$ represents
the density range\footnote{Let $\mathcal{S}(\mathcal{H})$ represent the collection of density
operators on the Hilbert space$\mathcal{H}\equiv H_{\textrm{A}}\otimes\mathcal{H}_{\textrm{B}}$.
Consider a density operator $\rho^{\prime}$ defined on $\mathcal{H}^{\otimes n}$,
meaning $\rho^{\prime}\in\mathcal{S}(\mathcal{H}^{\otimes n})$, with
an associated density range $\mathcal{R}\subseteq\mathcal{S}(\mathcal{H})$.
The set $\mathcal{R}$, known as the density range, consists of reduced
density operators corresponding to individual subsystems obtained
from $\rho^{\prime}$. More specifically, $\mathcal{R}:=\mathcal{R}(a,b)$
defines the set of density operators on $\mathcal{H_{\textrm{A}}\otimes\mathcal{H}_{\textrm{B}}}$,
ensuring that the measurement statistics for any $\rho\in\mathcal{R}$
align with the measurement operations performed by Alice and Bob \cite{CRE_04}.} of $\rho$, as formally defined in Definition 3.16 of \cite{CRE_04}.
This equation, initially introduced in Ref. \cite{CRE_04}, Equation
(\ref{eq:Chapter3_Eq4}) does not directly determine the key rate
for the protocols under consideration. Instead, it establishes the
threshold for secure errors in $S_{B1}$. Since $S_{B1}$ is transmitted
through a one-way quantum channel from Bob to Alice, this key rate
equation is applicable. Bob discloses partial information about his
measurement results for sequence $S_{A}$ (or the states prepared
for $S_{B1}$), enabling both parties to validate security through
correlation checks following this classical disclosure. Given that
Alice possesses knowledge of her initial sequence $S_{A}$, she can
utilize the key rate Equation (\ref{eq:Chapter3_Eq4}) to determine
the tolerable error threshold for $S_{B1}$. Assume that the QBER
is $\mathcal{E}\in[0,1]$ for measurements conducted in both the $X$
and $Z$ bases. The outcome of projective measurements on the quantum
state $\rho$ can be described by a random variable $V$. Since measurements
in the $Z$ and $X$ bases yield four possible outcomes, we associate
four distinct probabilities with these outcomes. Specifically, the
probabilities of obtaining each measurement outcome in the respective
bases can be expressed as $\mu_{i}$ for $i\in\{1,2,3,4\}$, corresponding
to different values of $V$. The entropy associated with the variable
$V$ is given by $H(V)=-\sum_{V\in\mu_{i}}V\log_{2}V\ge S(\rho)$.
The probabilities $\mu_{i}$ can be efficiently determined by computing
the expectation values of $\rho$ with respect to the relevant states.
Specifically, in this scenario, we have $\mu_{1}=\langle\Phi^{+}|\rho|\Phi^{+}\rangle,\,\mu_{2}=\langle\Phi^{-}|\rho|\Phi^{-}\rangle,\,\mu_{3}=\langle|\Psi^{+}|\rho|\Psi^{+}\rangle$
and $\mu_{4}=\langle|\Psi^{-}|\rho|\Psi^{-}\rangle$. A systematic
yet lengthy derivation establishes relationships among these probabilities
for the system described in Equation (\ref{eq:Chapter3_Eq3}), yielding
the following equations: $\mu_{3}+\mu_{4}=\mathcal{E}$, $\mu_{2}+\mu_{4}=\mathcal{E}$,
$\mu_{1}+\mu_{2}=1-\mathcal{E}$, and $\mu_{1}+\mu_{3}=1-\mathcal{E}$.
However, these four equations are not linearly independent. Instead,
only three of them form an independent set, which can be chosen in
different ways---for instance, by selecting (i) the first three equations
or (ii) the first two along with the last one. Due to this dependence,
direct resolution of the system is not feasible. Instead, one probability
can be treated as a free parameter, with the remaining ones expressed
in terms of it. Here, $\mu_{4}$ is selected as the free parameter,
leading to the expressions $\mu_{1}=1-2\mathcal{E}+\mu_{4}$ and $\mu_{2}=\mu_{3}=\mathcal{E}-\mu_{4}$.
The range of $\mu_{4}$ is constrained by $0\le\mu_{4}\le\mathcal{E}$,
as probability values must lie within $[0,1]$, and the condition
$\mu_{2}+\mu_{4}=\mathcal{E}$ holds (see the mathematical analysis
in the next part). The proposed framework prioritizes maximizing the
quantum component while minimizing the classical contribution. Our
analysis primarily explores the information-theoretic security limits
of each quantum sequence. If the QBER remains within the acceptable
secure threshold, both parties continue with the next steps of the
protocol. This security assessment applies to both newly introduced
protocols since all operations before the classical sub-protocol (i.e.,
the quantum phase) are identical in both cases. For simplicity, the
security evaluation is confined to the quantum aspect of the proposed
approach.

As discussed in the previous text, the Bell states are defined as:
$|\Phi^{\pm}\rangle\text{=\ensuremath{\frac{1}{\sqrt{2}}(|00\rangle\pm|11\rangle})}$
and $|\Psi^{\pm}\rangle$$=\ensuremath{\frac{1}{\sqrt{2}}(|01\rangle\pm|10\rangle})$.
These Bell states can also be represented in the diagonal basis as
$|\Phi^{+}\rangle=\frac{1}{\sqrt{2}}\left(|++\rangle+|--\rangle\right)$,
$|\Phi^{-}\rangle=\frac{1}{\sqrt{2}}\left(|+-\rangle+|-+\rangle\right)$,
$|\Psi^{+}\rangle=\frac{1}{\sqrt{2}}\left(|++\rangle-|--\rangle\right)$,
$|\Psi^{-}\rangle=\frac{1}{\sqrt{2}}\left(|-+\rangle-|+-\rangle\right)$.
From Equation (\ref{eq:Chapter3_Eq3}),

\begin{equation}
\begin{array}{lcl}
\mu_{2} & = & \langle\Phi^{-}|\rho|\Phi^{-}\rangle\\
 & = & \frac{1}{2}\left(\langle+-|\rho|+-\rangle+\langle+-|\rho|-+\rangle+\langle-+|\rho|+-\rangle+\langle-+|\rho|-+\rangle\right),
\end{array}\label{eq:Chapter3_AEq1}
\end{equation}

\begin{equation}
\begin{array}{lcl}
\mu_{4} & = & \langle\Psi^{-}|\rho|\Psi^{-}\rangle\\
 & = & \frac{1}{2}\left(\langle+-|\rho|+-\rangle-\langle+-|\rho|-+\rangle-\langle-+|\rho|+-\rangle+\langle-+|\rho|-+\rangle\right),
\end{array}\label{eq:Chapter3_AEq2}
\end{equation}

\begin{equation}
\begin{array}{lcl}
\mu_{3} & = & \langle\Psi^{+}|\rho|\Psi^{+}\rangle\\
 & = & \frac{1}{2}\left(\langle01|\rho|01\rangle+\langle01|\rho|10\rangle+\langle10|\rho|01\rangle+\langle10|\rho|10\rangle\right),
\end{array}\label{eq:Chapter3_AEq3}
\end{equation}
and

\begin{equation}
\begin{array}{lcl}
\mu_{4} & = & \langle\Psi^{-}|\rho|\Psi^{-}\rangle\\
 & = & \frac{1}{2}\left(\langle01|\rho|01\rangle-\langle01|\rho|10\rangle-\langle10|\rho|01\rangle+\langle10|\rho|10\rangle\right).
\end{array}\label{eq:Chapter3_AEq4}
\end{equation}
We define $\mathcal{E}$ as a symmetric error, ensuring that the following
conditions hold:

\begin{equation}
\begin{array}{lcl}
\langle00|\rho|00\rangle+\langle11|\rho|11\rangle & = & 1-\mathcal{E},\\
\langle++|\rho|++\rangle+\langle--|\rho|--\rangle & = & 1-\mathcal{E},\\
\langle01|\rho|01\rangle+\langle10|\rho|10\rangle & = & \mathcal{E},\\
\langle+-|\rho|+-\rangle+\langle-+|\rho|-+\rangle & = & \mathcal{E}.
\end{array}\label{eq:Chapter3_AEq5}
\end{equation}
Using these Equations (\ref{eq:Chapter3_AEq1}), (\ref{eq:Chapter3_AEq2}),
(\ref{eq:Chapter3_AEq3}), and (\ref{eq:Chapter3_AEq4}) in Equation
(\ref{eq:Chapter3_AEq5}), we obtain

\[
\begin{array}{lcl}
\mu_{2}+\mu_{4} & = & \left(\langle+-|\rho|+-\rangle+\langle-+|\rho|-+\rangle\right)=\mathcal{E}\\
\mu_{2} & = & \mathcal{E}-\mu_{4}
\end{array},
\]
and

\textbf{ 
\[
\begin{array}{lcl}
\mu_{3}+\mu_{4} & = & \left(\langle01|\rho|01\rangle+\langle10|\rho|10\rangle\right)=\mathcal{E}\\
\mu_{3} & = & \mathcal{E}-\mu_{4}.
\end{array}.
\]
}The total probability must adhere to the condition $\mu_{1}+\mu_{2}+\mu_{3}+\mu_{4}=1$.
Leveraging the established relationship with preceding results, we
derive $\mu_{1}=1-2\mathcal{E}+\mu_{4}$ and $\mu_{2}=\mu_{3}=\mathcal{E}-\mu_{4}$.

\begin{figure}
\begin{centering}
\includegraphics[scale=0.5]{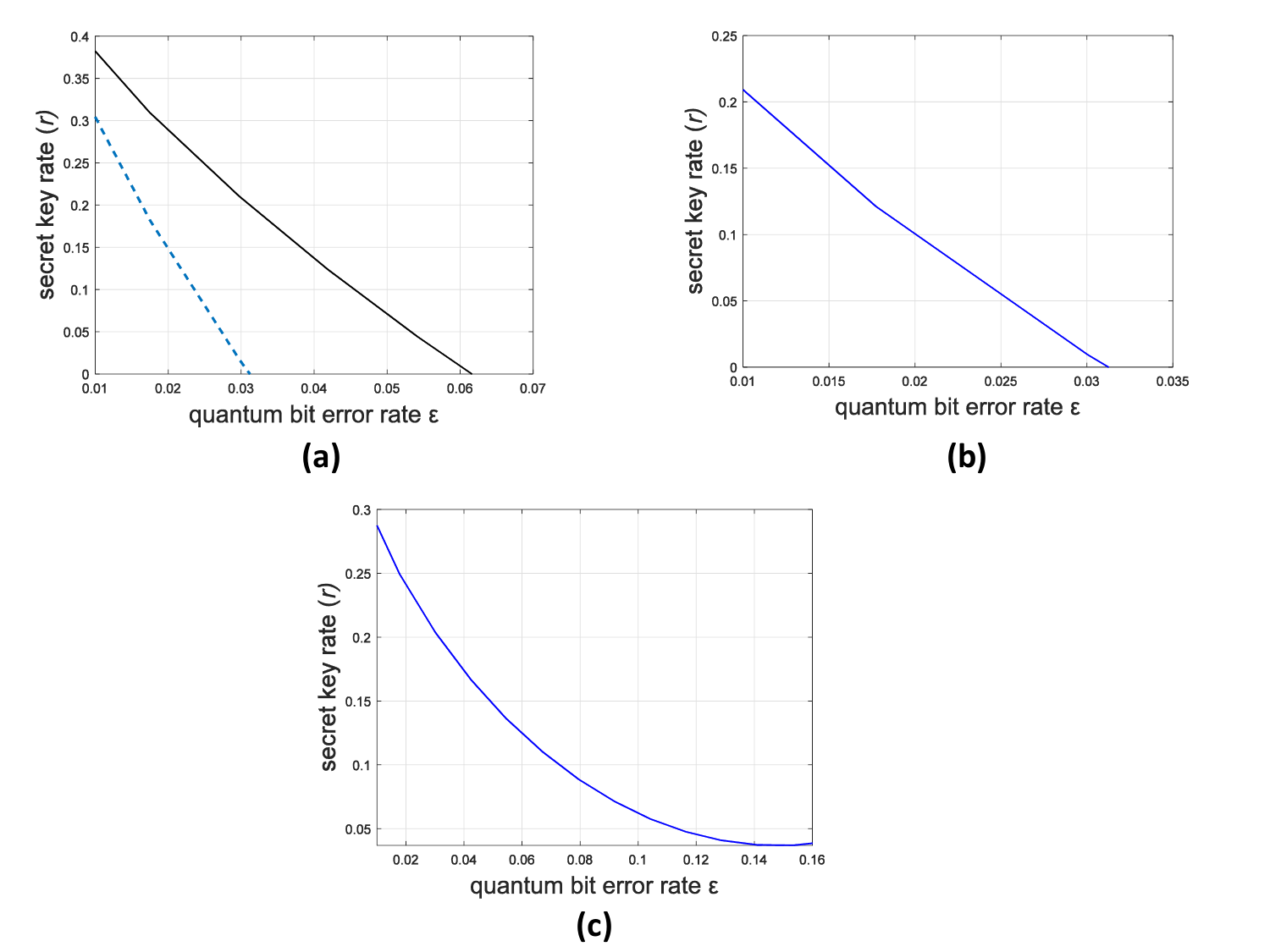} 
\par\end{centering}
\caption{The secret key rate is plotted as a function of the quantum bit error
rate ($\mathcal{E}$): $(a)$ a plot assessing the maximum tolerable
error threshold (security limit) for sequence $S_{B1}$, where the
solid (black) line and dashed (blue) line represent scenarios with
and without the incorporation of the new variable $\mathcal{Y}$,
respectively; (b) a plot evaluating the maximum tolerable error threshold
(security limit) for sequence $S_{B2}$ without introducing the new
variable $\mathcal{X}$; and $(c)$ a plot assessing the maximum tolerable
error threshold (security limit) for sequence $S_{B2}$ with the inclusion
of the new variable $\mathcal{X}$.}\label{fig:Chapter3_Fig1}
\end{figure}

This analysis establishes the secure error threshold for the sequence
$S_{B1}$ both with and without classical pre-processing \cite{CRE_04}.
To optimize the entropy of the random variable $V$, we solve the
equation $\frac{d\left(H\left(V\right)\right)}{d\mu_{4}}=0$, which
results in $\mu_{4}=\mathcal{E}^{2}$. Consequently, the entropy $H(V)$
is given by $2h(\mathcal{E})$, where $h(\mathcal{E})=-\mathcal{E}\log_{2}\mathcal{E}-(1-\mathcal{E})\log_{2}(1-\mathcal{E})$
represents the binary entropy function. Bob's measurement results
yield an entropy of $H(B)=2$, while the conditional entropy of $B$
given $A$ is expressed as $H(B|A)=1-\frac{1-\mathcal{E}}{2}\log_{2}\frac{1-\mathcal{E}}{2}$$-\frac{\mathcal{E}}{2}\log_{2}\frac{\mathcal{E}}{2}$.
The security threshold, or the highest permissible error rate, corresponds
to the maximum value of $\mathcal{E}$ for which Equation (\ref{eq:Chapter3_Eq4})
remains positive for $S_{B1}$. Under these constraints, solving $1+\frac{1-\mathcal{E}}{2}\log_{2}\frac{1-\mathcal{E}}{2}+\frac{\mathcal{E}}{2}\log_{2}\frac{\mathcal{E}}{2}-2h(\mathcal{E})=0$
yields $\mathcal{E}\approx0.0314$ (i.e., a QBER of $3.14\%$). Further
details can be found in this analysis,

These expressions are utilized to determine the key rate. The conditional
probability is given by, $Pr\left(B=|i\rangle\left|A=|j\rangle\right.\right)=\frac{Pr(B=|i\rangle,A=|j\rangle)}{Pr(A=|j\rangle)}$
and alongside the corresponding conditional entropy,

\begin{equation}
H(B|A)=-\sum_{j}Pr(A=|j\rangle)\sum_{i}Pr(B=|i\rangle|A=|j\rangle)\log_{2}Pr(B=|i\rangle|A=|j\rangle),\label{eq:Chapter3_BEq1}
\end{equation}

\[
\begin{array}{lcl}
Pr\left(B=|0\rangle|A=|0\rangle\right)=Pr\left(B=|1\rangle|A=|1\rangle\right) & = & \frac{\mu_{1}+\mu_{2}}{2},\\
Pr\left(B=|1\rangle|A=|0\rangle\right)=Pr\left(B=|0\rangle|A=|1\rangle\right) & = & \frac{\mu_{3}+\mu_{4}}{2},\\
Pr\left(B=|+\rangle|A=|+\rangle\right)=Pr\left(B=|-\rangle|A=|-\rangle\right) & = & \frac{\mu_{1}+\mu_{3}}{2},\\
Pr\left(B=|-\rangle|A=|+\rangle\right)=Pr\left(B=|+\rangle|A=|-\rangle\right) & = & \frac{\mu_{2}+\mu_{4}}{2},
\end{array}
\]

\[
\begin{array}{lcl}
Pr\left(B=|0\rangle|A=|0\rangle\right)=Pr\left(B=|1\rangle|A=|1\rangle\right) & = & \frac{\mu_{1}+\mu_{2}}{2},\\
Pr\left(B=|1\rangle|A=|0\rangle\right)=Pr\left(B=|0\rangle|A=|1\rangle\right) & = & \frac{\mu_{3}+\mu_{4}}{2},\\
Pr\left(B=|+\rangle|A=|+\rangle\right)=Pr\left(B=|-\rangle|A=|-\rangle\right) & = & \frac{\mu_{1}+\mu_{3}}{2},\\
Pr\left(B=|-\rangle|A=|+\rangle\right)=Pr\left(B=|+\rangle|A=|-\rangle\right) & = & \frac{\mu_{2}+\mu_{4}}{2},
\end{array}
\]

\[
Pr\left(B=|m\rangle|A=|n\rangle\right)=Pr\left(B=|n\rangle|A=|m\rangle\right)=\frac{1}{4},
\]

\[
Pr\left(B=|m\rangle\right)=Pr\left(B=|n\rangle\right)=\frac{1}{4},
\]

\[
Pr\left(A=|m\rangle\right)=Pr\left(A=|n\rangle\right)=\frac{1}{4}.
\]
For indices $m\neq n$, where $m\in\{|0\rangle,|1\rangle\}$ and $n\in\{|+\rangle,|-\rangle,\}$,
we can derive relevant expressions using Equation (\ref{eq:Chapter3_BEq1}),

\[
\begin{array}{lcl}
H(B|A) & = & -\frac{\mu_{1}+\mu_{2}}{4}\log_{2}\frac{\mu_{1}+\mu_{2}}{2}-\frac{\mu_{3}+\mu_{4}}{4}\log_{2}\frac{\mu_{3}+\mu_{4}}{2}\\
 & - & \frac{\mu_{1}+\mu_{3}}{4}\log_{2}\frac{\mu_{1}+\mu_{3}}{2}-\frac{\mu_{2}+\mu_{4}}{4}\log_{2}\frac{\mu_{2}+\mu_{4}}{2}-\frac{1}{2}\log_{2}\frac{1}{4}\\
 & = & -\frac{1-\mathcal{E}}{2}\log_{2}\frac{1-\mathcal{E}}{2}-\frac{\mathcal{E}}{2}\log_{2}\frac{\mathcal{E}}{2}+1,
\end{array}
\]
and 
\[
\begin{array}{lcl}
H(B) & = & -4\times\frac{1}{4}\log_{2}\frac{1}{4}\\
 & = & 2
\end{array}.
\]
Therefore,

\[
\begin{array}{lcl}
I(A:B) & = & H(B)-H(B|A)\\
 & = & 1+\frac{1-\mathcal{E}}{2}\log_{2}\frac{1-\mathcal{E}}{2}+\frac{\mathcal{E}}{2}\log_{2}\frac{\mathcal{E}}{2},
\end{array}
\]
utilizing the secret key rate we have,

\[
\begin{array}{lcl}
r & = & I(A:B)-H(V)\\
 & = & 1+\frac{1-\mathcal{E}}{2}\log_{2}\frac{1-\mathcal{E}}{2}+\frac{\mathcal{E}}{2}\log_{2}\frac{\mathcal{E}}{2}-2h(\mathcal{E}).
\end{array}
\]
This result indicates that, in the absence of classical pre-processing,
the maximum theoretical error tolerance for $S_{B1}$ is $3.14\%$
following Bob's classical announcement and correlation evaluation
for security verification. To enhance this security threshold, we
define a new variable\footnote{This framework can be interpreted as a classical pre-processing approach
aimed at enhancing the key rate while simultaneously increasing the
maximum permissible error threshold \cite{CRE_04}. Additionally,
$\mathcal{X}$ serves a similar role in the context of the sequence
$S_{B2}$.} $\mathscr{\mathcal{Y}}=j_{A}\oplus j_{B}$, where $j_{A}$ and $j_{B}$
denote the basis choices made by Alice and Bob when measuring particles
within $S_{B1}$ and $S_{A}$, respectively. Here, the basis indices
satisfy $j_{A},j_{B}\in\{0,1\}$. Solving the equation:

\[
\begin{array}{lcl}
1+\frac{1-\mathcal{E}}{2}\log_{2}\frac{1-\mathcal{E}}{2}+\frac{\mathcal{E}}{2}\log_{2}\frac{\mathcal{E}}{2}-h(\mathcal{E}) & = & 0\end{array},
\]
yields $\mathcal{E}\approx0.0617$ (equivalent to a $6.17\%$ bit
error rate; refer to Figure \ref{fig:Chapter3_Fig1}. a. This outcome
indicates that when $\mathcal{Y}$ is disclosed, the highest permissible
error threshold for the measurement of $S_{A}$ and $S_{B1}$ rises
to $6.17\%$. Consequently, under classical pre-processing, the theoretical
maximum tolerable error rate for $S_{B1}$ is established at $6.17\%$.
As discussed in Section \ref{sec:Chapter3_Sec2}, the probability
of obtaining the expected outcomes from $S_{A}$ is $\frac{1}{2}$
($\text{\ensuremath{\psi_{Z}^{x}}}$) when using the same basis and
$\frac{1}{4}$ ($\psi_{X}^{x/x^{\perp}}$) when different bases are
employed. For $S_{B1}$, the probability is given by $\frac{1}{4}$
($\psi_{J}^{x^{\perp}}$). The maximum allowable deviations from these
probability values correspond to an error margin of $3.14\%$ when
$\mathcal{Y}$ is not announced, increasing to $6.17\%$ upon its
disclosure. The introduction of new variables $\mathcal{Y}$ (alongside
$\mathcal{X}$, introduced subsequently) follows the approach proposed
by Christandl et al. \cite{CRE_04}. The methodology for determining
these variables is rooted in the concepts of information reconciliation
and privacy amplification against quantum adversaries. Within information
reconciliation, hash and guessing functions are employed to establish
these variables. Privacy amplification, on the other hand, involves
the use of hash functions and a transformation based on the measurement
outcomes of quantum states relative to an arbitrary POVM, which is
also influenced by the hash function (refer to Sections 4.2, 4.3,
and 5.1 in \cite{CRE_04}).

Alice begins by verifying that the security threshold for the sequence
$S_{B1}$ remains within the designated limits. She then measures
the second sequence, $S_{B2}$, which is sent by Bob, thereby completing
the sifting subprotocol. Following this, both Alice and Bob assess
whether the QBER is below the established security threshold. The
method used to determine this threshold value is identical to the
previously described approach. The entropy associated with Bob\textquoteright s
final bit string, $b$, after executing the sifting subprotocol is
given by $H(b)=1$, while the conditional entropy of $b$ given Alice\textquoteright s
bit string, $a$, is expressed as $H(b|a)=h(\frac{1}{6}+\frac{2\mathcal{E}}{3})$,
where $a,b\in\{0,1\}$. The detailed mathematical derivation, along
with the key rate equation, is subsequently determined by analyzing
the key-sifting subprotocol performed by both parties, ensuring that
only probabilities meeting the necessary conditions are considered.

\[
\begin{array}{lcl}
Pr\left(b=0|a=0\right)=Pr\left(b=1|a=1\right) & = & \frac{1}{6}+\frac{2\mu_{1}+\mu_{2}+\mu_{3}}{3},\\
Pr\left(b=0|a=1\right)=Pr\left(b=1|a=0\right) & = & \frac{1}{6}+\frac{\mu_{2}+\mu_{3}+2\mu_{4}}{3},
\end{array}
\]

\[
\begin{array}{lcl}
Pr\left(b=0\right)=Pr\left(b=1\right) & = & \frac{1}{2}.\end{array}
\]
Applying a similar methodology as in Equation (\ref{eq:Chapter3_BEq1}),
we obtain the final expression for the secret key rate.
\[
\begin{array}{lcl}
H(b|a) & = & -\frac{1+4\mu_{1}+2\mu_{2}+2\mu_{3}}{6}\log_{2}\frac{1+4\mu_{1}+2\mu_{2}+2\mu_{3}}{6}-\frac{1+2\mu_{2}+2\mu_{3}+4\mu_{4}}{6}\log_{2}\frac{1+2\mu_{2}+2\mu_{3}+4\mu_{4}}{6}\\
 & = & -\frac{5-4\mathcal{E}}{6}\log_{2}\frac{5-4\mathcal{E}}{6}-\frac{1+4\mathcal{E}}{6}\log_{2}\frac{1+4\mathcal{E}}{6}\\
 & = & h\left(\frac{1}{6}+\frac{2\mathcal{E}}{3}\right),
\end{array}
\]

\[
\begin{array}{lcl}
I(a|b) & = & H(b)-H(b|a)\\
 & = & -\frac{1}{2}\log_{2}\frac{1}{2}-h(b|a)\\
 & = & 1-h\left(\frac{1}{6}+\frac{2\mathcal{E}}{3}\right),
\end{array}
\]
the secret key rate,

\[
\begin{array}{lcl}
r & = & I(a|b)-H(V)\\
 & = & 1-h\left(\frac{1}{6}+\frac{2\mathcal{E}}{3}\right)-2h(\mathcal{E}).
\end{array}
\]
Using a similar approach, the equation governing the positive key
rate for a one-way quantum channel is derived to determine the tolerable
QBER for $S_{B2}$. By solving the equation

\begin{equation}
1-h(\frac{1}{6}+\frac{2\mathcal{E}}{3})-2h(\mathcal{E})=0.\label{eq:Chapter3_Eq5}
\end{equation}
Under the condition $r=0$, the security threshold is calculated as
$\mathcal{E}\approx0.0316$, corresponding to a $3.16\%$ QBER (see
Figure \ref{fig:Chapter3_Fig1}. b). Enhancing the security threshold
can be achieved by introducing a random variable $\mathcal{X}=a\oplus b$,
which carries information regarding error positions. This inclusion
reduces the quantum contribution (final term) in Equation (\ref{eq:Chapter3_Eq4}),
while preserving the minimum entropy of $b$ (refer to Sec. 5.1 of
Ref. \cite{CRE_04} for more details). To check the effect of this
variable mathematically, we formulate it as follows,

Alice and Bob assess the security threshold of particle sequences,
denoted as $S_{A}$ and $S_{B1}$, under the condition that they use
the same basis for state preparation or measurement. The presence
or absence of errors is determined based on their measurement of the
states $|\Phi^{\pm}\rangle$ (or $|\Psi^{\pm}\rangle$) within the
computational basis. Likewise, error-free or erroneous outcomes arise
when both parties measure $|\Phi^{+}\rangle$ and $|\Psi^{+}\rangle$
(or $|\Phi^{-}\rangle$ and $|\Psi^{-}\rangle$) using the diagonal
basis. This results in four distinct scenarios to analyze. Furthermore,
it follows that the quantum state $\rho$ can be measured with equal
probability by both Alice and Bob, regardless of whether they employ
the computational or diagonal basis. To illustrate, consider a case
where both parties measure the Bell states $|\Phi^{+}\rangle$ and
$|\Psi^{+}\rangle$ using the computational basis. The corresponding
probabilities of obtaining an error-free outcome and an erroneous
outcome are $\frac{\mu_{1}}{2}$ and $\frac{\mu_{3}}{2}$, respectively\footnote{For simplicity, we assume that both participants measure all qubits
in the same basis---either both using the computational basis or
both using the diagonal basis---during the error verification stage.}. Consequently, the probabilities of obtaining an error-free result
and an error when measuring $|\Phi^{+}\rangle$ and $|\Psi^{+}\rangle$
are given by $\frac{1-\mathcal{E}}{2}$ and $\frac{\mathcal{E}}{2}$,
respectively\footnote{In this framework, the factor of 2 arises due to the equal distribution
of the total error-checking qubits between computational and diagonal
basis measurements.}. Considering the total number of qubits, the probabilities of encountering
an error-free or erroneous outcome when measuring $|\Phi^{+}\rangle$
and $|\Psi^{+}\rangle$ in the computational basis are given by $\frac{\frac{\mu_{1}}{2}}{\frac{1-\mathcal{E}}{2}}$
and $\frac{\frac{\mu_{3}}{2}}{\frac{\mathcal{E}}{2}}$, which simplify
to $\frac{\mu_{1}}{1-\mathcal{E}}$ and $\frac{\mu_{3}}{\mathcal{E}}$,
respectively. Similarly, when measuring $|\Phi^{-}\rangle$ and $|\Psi^{-}\rangle$
within the computational basis, the corresponding probabilities are
$\frac{\mu_{2}}{1-\mathcal{E}}$ and $\frac{\mu_{4}}{\mathcal{E}}$.
For measurements in the diagonal basis, the probabilities of obtaining
an error-free outcome and an error when measuring $|\Phi^{+}\rangle$
and $|\Psi^{+}\rangle$ are given by $\frac{\mu_{1}}{1-\mathcal{E}}$
and $\frac{\mu_{3}}{1-\mathcal{E}}$, respectively. Likewise, for
the states $|\Phi^{-}\rangle$ and \textbar$\Psi^{-}\rangle$, the
probabilities are $\frac{\mu_{2}}{\mathcal{E}}$ and $\frac{\mu_{4}}{\mathcal{E}}$,
respectively. Finally, in a scenario where no errors are present,
the entropy can be computed as follows:

\[
\begin{array}{lcl}
{\rm H}_{{\rm no-error}} & = & -\frac{1}{2}\left[\frac{\mu_{1}}{1-\mathcal{E}}\log_{2}\frac{\mu_{1}}{1-\mathcal{E}}+\frac{\mu_{2}}{1-\mathcal{E}}\log_{2}\frac{\mu_{2}}{1-\mathcal{E}}+\frac{\mu_{1}}{1-\mathcal{E}}\log_{2}\frac{\mu_{1}}{1-\mathcal{E}}+\frac{\mu_{3}}{1-\mathcal{E}}\log_{2}\frac{\mu_{3}}{1-\mathcal{E}}\right]\\
 & = & -\frac{1}{2}\left[\frac{2\mu_{1}}{1-\mathcal{E}}\log_{2}\frac{\mu_{1}}{1-\mathcal{E}}+\frac{2\mu_{2}}{1-\mathcal{E}}\log_{2}\frac{\mu_{2}}{1-\mathcal{E}}\right]\,\,\,\,{\rm as}\,\,\mu_{2}=\mu_{3}\\
 & = & -\left[\frac{\left(1-2\mathcal{E}+\mu_{4}\right)}{1-\mathcal{E}}\log_{2}\frac{\left(1-2\mathcal{E}+\mu_{4}\right)}{1-\mathcal{E}}+\frac{\left(\mathcal{E}-\mu_{4}\right)}{1-\mathcal{E}}\log_{2}\frac{\left(\mathcal{E}-\mu_{4}\right)}{1-\mathcal{E}}\right]\,\,\,\,({\rm using\,Eqs\,(\ref{eq:Chapter3_AEq1})\,-\,(\ref{eq:Chapter3_AEq5}}))\\
 & = & h\left(\frac{1-2\mathcal{E}+\mu_{4}}{1-\mathcal{E}}\right).
\end{array}
\]
In the presence of errors, entropy can be determined as follows:

\[
\begin{array}{lcl}
{\rm H_{error}} & = & -\frac{1}{2}\left[\frac{\mu_{3}}{\mathcal{E}}\log_{2}\frac{\mu_{3}}{\mathcal{E}}+\frac{\mu_{4}}{\mathcal{E}}\log_{2}\frac{\mu_{4}}{\mathcal{E}}+\frac{\mu_{2}}{\mathcal{E}}\log_{2}\frac{\mu_{2}}{\mathcal{E}}+\frac{\mu_{4}}{\mathcal{E}}\log_{2}\frac{\mu_{4}}{\mathcal{E}}\right]\\
 & = & -\frac{1}{2}\left[\frac{2\mu_{2}}{\mathcal{E}}\log_{2}\frac{2\mu_{2}}{\mathcal{E}}+\frac{2\mu_{4}}{\mathcal{E}}\log_{2}\frac{2\mu_{4}}{\mathcal{E}}\right]\,\,{\rm as}\,\mu_{2}=\mu_{3}\\
 & = & -\left[\frac{\left(\mathcal{E}-\mu_{4}\right)}{\mathcal{E}}\log_{2}\frac{\left(\mathcal{E}-\mu_{4}\right)}{\mathcal{E}}+\frac{\mu_{4}}{\mathcal{E}}\log_{2}\frac{\mu_{4}}{\mathcal{E}}\right]\,\,({\rm using\,Eqs\,(\ref{eq:Chapter3_AEq1})\,-\,(\ref{eq:Chapter3_AEq5}}))\\
 & = & h\left(\frac{\mathcal{E}-\mu_{4}}{\mathcal{E}}\right).
\end{array}
\]
By performing statistical averaging over both error-free and erroneous
cases, we derive:

\[
\begin{array}{l}
\left(1-\mathcal{E}\right){\rm H}_{{\rm no-error}}+\mathcal{E}H_{error}\\
=\left(1-\mathcal{E}\right)h\left(\frac{1-2\mathcal{E}+\mu_{4}}{1-\mathcal{E}}\right)+\mathcal{E}h\left(\frac{\mathcal{E}-\mu_{4}}{\mathcal{E}}\right)\\
=-\left(1-\mathcal{E}\right)\left[\frac{\left(1-2\mathcal{E}+\mu_{4}\right)}{1-\mathcal{E}}\log_{2}\frac{\left(1-2\mathcal{E}+\mu_{4}\right)}{1-\mathcal{E}}+\frac{\left(\mathcal{E}-\mu_{4}\right)}{1-\mathcal{E}}\log_{2}\frac{\left(\mathcal{E}-\mu_{4}\right)}{1-\mathcal{E}}\right]\\
-\mathcal{E}\left[\frac{\left(\mathcal{E}-\mu_{4}\right)}{\mathcal{E}}\log_{2}\frac{\left(\mathcal{E}-\mu_{4}\right)}{\mathcal{E}}+\frac{\mu_{4}}{\mathcal{E}}\log_{2}\frac{\mu_{4}}{\mathcal{E}}\right]\\
=-\left[\left(1-2\mathcal{E}+\mu_{4}\right)\log_{2}\left(1-2\mathcal{E}+\mu_{4}\right)+\left(\mathcal{E}-\mu_{4}\right)\log_{2}\left(\mathcal{E}-\mu_{4}\right)\right.\\
\left.-\left(1-2\mathcal{E}+\mu_{4}\right)\log_{2}\left(1-\mathcal{E}\right)-\left(\mathcal{E}-\mu_{4}\right)\log_{2}\left(1-\mathcal{E}\right)\right]\\
-\left[\left(\mathcal{E}-\mu_{4}\right)\log_{2}\left(\mathcal{E}-\mu_{4}\right)+\mu_{4}\log_{2}\mu_{4}-\left(\mathcal{E}-\mu_{4}\right)\log_{2}\mathcal{E}-\mu_{4}\log_{2}\mathcal{E}\right]\\
=-\left[\left(1-2\mathcal{E}+\mu_{4}\right)\log_{2}\left(1-2\mathcal{E}+\mu_{4}\right)+2\left(\mathcal{E}-\mu_{4}\right)\log_{2}\left(\mathcal{E}-\mu_{4}\right)+\mu_{4}\log_{2}\mu_{4}\right]\\
+\left(1-2\mathcal{E}+\mu_{4}+\mathcal{E}-\mu_{4}\right)\log_{2}\left(1-\mathcal{E}\right)+\left(\mathcal{E}-\mu_{4}+\mu_{4}\right)\log_{2}\mathcal{E}\\
=H(V)+\left(1-\mathcal{E}\right)\log_{2}\left(1-\mathcal{E}\right)+\mathcal{E}\log_{2}\mathcal{E}\\
=H(V)-h(\mathcal{E}).
\end{array}
\]
To elucidate this further, the quantum system can be decomposed into
four distinct subsystems, where two correspond to error states and
the other two to error-free states for each basis. Within the $Z(X)$
basis, errors and non-errors account for fractions of $\frac{\mathcal{E}}{2}$
and $\frac{1-\mathcal{E}}{2}$, respectively, of the total qubits.
By computing the entropy for these four subsystems in both error and
no-error cases, one derives $h\left(\frac{\mathcal{E}-\mu_{4}}{\mathcal{E}}\right)$
and $h\left(\frac{1-2\mathcal{E}+\mu_{4}}{1-\mathcal{E}}\right)$
as the respective entropy values. Subsequently, performing statistical
averaging over all four subsystems yields the final result.

\begin{equation}
(1-\mathcal{E})h\left(\frac{1-2\mathcal{E}+\mu_{4}}{1-\mathcal{E}}\right)+\mathcal{E}h\left(\frac{\mathcal{E}-\mu_{4}}{\mathcal{E}}\right)=H(V)-h(\mathcal{E})\label{eq:Chapter3_Eq6}
\end{equation}
By substituting the reconditioned entropy for variable $V$ in Equation
(\ref{eq:Chapter3_Eq5}), we derive a modified key rate equation for
a one-way quantum channel, establishing the tolerable QBER as $S_{B2}$.

\begin{equation}
r=1-h(\frac{1}{6}+\frac{2\mathcal{E}}{3})-h(\mathcal{E}).\label{eq:Chapter3_Eq7}
\end{equation}
In this context, the equation's solution for a positive $r$ determines
the security threshold, yielding $\mathcal{E}\approx0.15$. Consequently,
the corresponding bit error rate is approximately $15\%$ (see Figure
\ref{fig:Chapter3_Fig1} c).

Next, we evaluate the secret-key rate under the assumption that the
protocol remains secure against collective attacks by Eve. To begin,
we characterize the initial quantum state $\rho_{AB}^{n}$, which
depends on the threshold QBER beyond which the protocol prematurely
terminates. Ideally, $\rho_{AB}^{n}$ represents a quantum state shared
exclusively between Alice and Bob, though Eve may access it and potentially
launch collective attacks. Let $\Gamma$ denote the set of all two-qubit
states $\sigma_{AB}$ that could arise from Eve's collective attack
on $\rho_{AB}^{n}$. The effectiveness of Eve's attack is determined
by whether it remains undetectable. If the attack leaves no observable
traces, the resulting state satisfies $\sigma_{AB}^{\otimes n}=\rho_{AB}^{n}$.
However, unsuccessful attacks introduce detectable disturbances, leading
to protocol termination. Our focus is on scenarios where the protocol
continues to operate. To account for such cases, we assume the existence
of an operation that allows Eve to construct the state $\sigma_{AB}^{\otimes n}=\rho_{AB}^{n}$
using ancillary qubits and the portion of the initial state accessible
via the quantum channel. Following the framework established in Ref.
\cite{RGK_05}, we define $\Gamma_{QBER}$ as a subset of $\Gamma$
containing all states $\sigma_{AB}$ for which the protocol does not
terminate. That is, if $\sigma_{AB}\in\Gamma_{QBER}$, then the protocol
is capable of generating a secure key. As demonstrated by Renner et
al. in \cite{RGK_05}, under these conditions, both lower and upper
bounds on the secret-key rate can be rigorously determined for any
one-way post-processing protocol.

\begin{equation}
r\ge\underset{c\leftarrow a}{sup}\underset{\sigma_{AB}\in\Gamma_{QBER}}{inf}\left(S(c|E)-H(c|b)\right).\label{eq:Chapter3_Eq8}
\end{equation}
In this context, $r_{c\leftarrow a}$ represents the achievable rate
when the channel $c\leftarrow a$ is employed for pre-processing\footnote{The channel $c\leftarrow a$ can be interpreted as a probabilistic
mixture of quantum transitions, formally described as $q(|1\rangle_{ca}\langle0|+|0\rangle_{ca}\langle1|)+(1-q)(|0\rangle_{ca}\langle0|+|1\rangle_{ca}\langle1|)$.
Here, $a$ denotes Alice\textquoteright s register containing the
classical measurement outcomes, while $c$ represents the register
storing the noisy counterpart of $a$.}. The quantity $S(c|E)$ corresponds to the von Neumann entropy of
$c$, conditioned on Eve\textquoteright s initial state, mathematically
defined as $S(c|E)=S(\sigma_{cE})-S(\sigma_{E})$. The bipartite state
$\sigma_{cE}$ is derived from the two-qubit state $\sigma_{AB}$
by considering a purification $\sigma_{ABE}$ of its Bell diagonal
form, denoted as $\sigma_{AB}^{diag}:=\mathcal{O}_{2}(\sigma_{AB})$,
where $\sigma_{AB}^{diag}$ retains the same diagonal elements as
$\sigma_{AB}$ in the Bell basis. The symbols $a$, $b$ and $e$
represent the measurement outcomes corresponding to Alice, Bob, and
Eve, respectively, following the application of measurements to the
first, second, and third subsystems of $\sigma_{ABE}$.

To determine an upper bound on the achievable rate, it is sufficient
to restrict our consideration to collective attacks. The joint quantum
system involving Alice, Bob, and Eve adheres to a product structure,
represented as $\rho_{ABE}^{n}:=\sigma_{ABE}^{\otimes n}$, where
$\sigma_{ABE}$ denotes a tripartite quantum state. The $n$-fold
tensor product state $\sigma_{abE}^{n}$ describes the scenario in
which the single-copy state $\sigma_{abE}$ arises from Alice and
Bob\textquoteright s measurements performed on $\sigma_{ABE}$ (for
a rigorous proof, refer to Section IV of Ref. \cite{RGK_05}). Consequently,
the upper bound on the secret key rate is given by:

\begin{equation}
r(a,b,e)=\underset{c\leftarrow a}{sup}\left(H(c|e)-H(c|b)\right).\label{eq:Chapter3_Eq9}
\end{equation}
This result indicates that when the supremum is taken over all possible
channels---including both quantum and classical, $c\leftarrow a$
establishes an upper limit on the secret key rate. Next, we analyze
the protocol in terms of both lower and upper bounds on the secret
key rate. As before, we set $n=1$ and define $\sigma_{AB}=\rho^{1}[\mu]$.
It is necessary to consider a purification $|\Psi\rangle_{ABE}$ of
the Bell diagonal state $\mathcal{O}_{2}(\sigma_{AB})$, which originates
from $\sigma_{AB}$ and is expressed as:

\begin{equation}
|\Psi\rangle_{ABE}:=\sum_{i=1}^{4}\sqrt{\mu_{i}}|\varphi_{i}\rangle_{AB}\otimes|\varepsilon_{i}\rangle_{E}.\label{eq:Chapter3_Eq10}
\end{equation}
In this context, $|\varphi_{i}\rangle_{AB}$ represents the Bell states
associated with the joint quantum system shared between Alice and
Bob\footnote{Alice performs a measurement on her qubit using the $Z$ basis, whereas
Bob measures his qubit using either the $Z$ or $X$ basis with equal
probability ($\frac{1}{2}$).}. Meanwhile, $|\varepsilon_{i}\rangle_{E}$ corresponds to a set of
mutually orthogonal states forming a basis in Eve\textquoteright s
system, denoted as $\varepsilon_{E}\in\left\{ |\varepsilon_{1}\rangle_{E},\ldots,|\varepsilon_{4}\rangle_{E}\right\} $.
It is straightforward to verify that when Alice measures her qubit
in the $Z$ (or $X$) basis and Bob performs measurements in either
the $Z$ or $X$ basis with equal probability, their respective measurement
outcomes are represented as $|\mathcal{A\rangle}$ and $|\mathcal{B\rangle}$.
For illustration, we assume $|\mathcal{A\rangle}\in\left\{ |0\rangle,|1\rangle\right\} $
and $|\mathcal{B\rangle}\in\left\{ |0\rangle,|1\rangle,|+\rangle,|-\rangle\right\} $.
Under this framework, Eve's quantum state is denoted as $|\phi^{\mathcal{A},\mathcal{B}}\rangle$.
We have,

\begin{equation}
\begin{array}{lcl}
|\phi^{0,0}\rangle & = & \frac{1}{\sqrt{2}}\left(\sqrt{\mu_{1}}|\varepsilon_{1}\rangle_{E}+\sqrt{\mu_{2}}|\varepsilon_{2}\rangle_{E}\right),\\
|\phi^{1,1}\rangle & = & \frac{1}{\sqrt{2}}\left(\sqrt{\mu_{1}}|\varepsilon_{1}\rangle_{E}-\sqrt{\mu_{2}}|\varepsilon_{2}\rangle_{E}\right),\\
|\phi^{0,1}\rangle & = & \frac{1}{\sqrt{2}}\left(\sqrt{\mu_{3}}|\varepsilon_{3}\rangle_{E}+\sqrt{\mu_{4}}|\varepsilon_{4}\rangle_{E}\right),\\
|\phi^{1,0}\rangle & = & \frac{1}{\sqrt{2}}\left(\sqrt{\mu_{3}}|\varepsilon_{3}\rangle_{E}-\sqrt{\mu_{4}}|\varepsilon_{4}\rangle_{E}\right),\\
|\phi^{0,+}\rangle & = & \frac{1}{2}\left(\sqrt{\mu_{1}}|\varepsilon_{1}\rangle_{E}+\sqrt{\mu_{2}}|\varepsilon_{2}\rangle_{E}+\sqrt{\mu_{3}}|\varepsilon_{3}\rangle_{E}+\sqrt{\mu_{4}}|\varepsilon_{4}\rangle_{E}\right),\\
|\phi^{0,-}\rangle & = & \frac{1}{2}\left(\sqrt{\mu_{1}}|\varepsilon_{1}\rangle_{E}+\sqrt{\mu_{2}}|\varepsilon_{2}\rangle_{E}-\sqrt{\mu_{3}}|\varepsilon_{3}\rangle_{E}-\sqrt{\mu_{4}}|\varepsilon_{4}\rangle_{E}\right),\\
|\phi^{1,+}\rangle & = & \frac{1}{2}\left(\sqrt{\mu_{1}}|\varepsilon_{1}\rangle_{E}-\sqrt{\mu_{2}}|\varepsilon_{2}\rangle_{E}+\sqrt{\mu_{3}}|\varepsilon_{3}\rangle_{E}-\sqrt{\mu_{4}}|\varepsilon_{4}\rangle_{E}\right),\\
|\phi^{1,-}\rangle & = & \frac{1}{2}\left(-\sqrt{\mu_{1}}|\varepsilon_{1}\rangle_{E}+\sqrt{\mu_{2}}|\varepsilon_{2}\rangle_{E}+\sqrt{\mu_{3}}|\varepsilon_{3}\rangle_{E}-\sqrt{\mu_{4}}|\varepsilon_{4}\rangle_{E}\right).
\end{array}
\end{equation}
A comprehensive explanation of the mathematical procedures leading
from Equation (\ref{eq:Chapter3_Eq10}) to Equation (\ref{eq:Chapter3_Eq12})
is provided here. Initially, we analyze a scenario where both Alice
and Bob measure their qubits in the $Z$ basis (a similar result holds
when employing the $X$ basis).

\[
\begin{array}{lcl}
|\Psi\rangle_{ABE} & := & \sum_{i=1}^{4}\sqrt{\mu_{i}}|\varphi_{i}\rangle_{AB}\otimes|\varepsilon_{i}\rangle_{E}\\
 & = & \frac{1}{2\sqrt{2}}\left[\sqrt{\mu_{1}}\left(|00\rangle+|11\rangle\right)\otimes|\varepsilon_{1}\rangle+\sqrt{\mu_{2}}\left(|00\rangle-|11\rangle\right)\otimes|\varepsilon_{2}\rangle\right.\\
 & + & \left.\sqrt{\mu_{3}}\left(|01\rangle+|10\rangle\right)\otimes|\varepsilon_{3}\rangle+\sqrt{\mu_{4}}\left(|01\rangle-|10\rangle\right)\otimes|\varepsilon_{4}\rangle\right]_{ABE}\\
 & = & \frac{1}{2\sqrt{2}}\left[|00\rangle\left(\sqrt{\mu_{1}}|\varepsilon_{1}\rangle+\sqrt{\mu_{2}}|\varepsilon_{2}\rangle\right)+|11\rangle\left(\sqrt{\mu_{1}}|\varepsilon_{1}\rangle-\sqrt{\mu_{2}}|\varepsilon_{2}\rangle\right)\right.\\
 & + & \left.|01\rangle\left(\sqrt{\mu_{3}}|\varepsilon_{3}\rangle+\sqrt{\mu_{4}}|\varepsilon_{4}\rangle\right)+|10\rangle\left(\sqrt{\mu_{3}}|\varepsilon_{3}\rangle-\sqrt{\mu_{4}}|\varepsilon_{4}\rangle\right)\right]_{ABE}\\
 & = & \frac{1}{2}\left[|00\rangle\otimes|\phi^{0,0}\rangle+|11\rangle\otimes|\phi^{1,1}\rangle+|01\rangle\otimes|\phi^{0,1}\rangle+|10\rangle\otimes|\phi^{1,0}\rangle\right]_{ABE}
\end{array}.
\]
Next, we examine the case where Alice and Bob perform measurements
using the $Z$ and $X$ bases, respectively\footnote{It is crucial to highlight that the same results arise when Alice
and Bob interchange their measurement bases to $X$ and $Z$.},

\[
\begin{array}{lcl}
|\Psi\rangle_{ABE} & := & \sum_{i=1}^{4}\sqrt{\mu_{i}}|\varphi_{i}\rangle_{AB}\otimes|\varepsilon_{i}\rangle_{E}\\
 & = & \frac{1}{2\sqrt{2}}\left[\sqrt{\mu_{1}}\left(|00\rangle+|11\rangle\right)\otimes|\varepsilon_{1}\rangle+\sqrt{\mu_{2}}\left(|00\rangle-|11\rangle\right)\otimes|\varepsilon_{2}\rangle\right.\\
 & + & \left.\sqrt{\mu_{3}}\left(|01\rangle+|10\rangle\right)\otimes|\varepsilon_{3}\rangle+\sqrt{\mu_{4}}\left(|01\rangle-|10\rangle\right)\otimes|\varepsilon_{4}\rangle\right]_{ABE}\\
 & = & \frac{1}{4}\left[\sqrt{\mu_{1}}\left\{ |0\rangle\left(|+\rangle+|-\rangle\right)+|1\rangle\left(|+\rangle-|-\rangle\right)\right\} |\varepsilon_{1}\rangle\right.\\
 & + & \sqrt{\mu_{2}}\left\{ |0\rangle\left(|+\rangle+|-\rangle\right)-|1\rangle\left(|+\rangle-|-\rangle\right)\right\} |\varepsilon_{2}\rangle\\
 & + & \sqrt{\mu_{3}}\left\{ |0\rangle\left(|+\rangle-|-\rangle\right)+|1\rangle\left(|+\rangle+|-\rangle\right)\right\} |\varepsilon_{3}\rangle\\
 & + & \left.\sqrt{\mu_{4}}\left\{ |0\rangle\left(|+\rangle-|-\rangle\right)-|1\rangle\left(|+\rangle+|-\rangle\right)\right\} |\varepsilon_{4}\rangle\right]_{ABE}\\
 & = & \frac{1}{4}\left[|0+\rangle\left\{ \sqrt{\mu_{1}}|\varepsilon_{1}\rangle+\sqrt{\mu_{2}}|\varepsilon_{2}\rangle+\sqrt{\mu_{3}}|\varepsilon_{3}\rangle+\sqrt{\mu_{4}}|\varepsilon_{4}\rangle\right\} \right.\\
 & + & |0-\rangle\left\{ \sqrt{\mu_{1}}|\varepsilon_{1}\rangle+\sqrt{\mu_{2}}|\varepsilon_{2}\rangle-\sqrt{\mu_{3}}|\varepsilon_{3}\rangle-\sqrt{\mu_{4}}|\varepsilon_{4}\rangle\right\} \\
 & + & |1+\rangle\left\{ \sqrt{\mu_{1}}|\varepsilon_{1}\rangle-\sqrt{\mu_{2}}|\varepsilon_{2}\rangle+\sqrt{\mu_{3}}|\varepsilon_{3}\rangle-\sqrt{\mu_{4}}|\varepsilon_{4}\rangle\right\} \\
 & + & \left.|1-\rangle\left\{ -\sqrt{\mu_{1}}|\varepsilon_{1}\rangle+\sqrt{\mu_{2}}|\varepsilon_{2}\rangle+\sqrt{\mu_{3}}|\varepsilon_{3}\rangle-\sqrt{\mu_{4}}|\varepsilon_{4}\rangle\right\} \right]_{ABE}\\
 & = & \frac{1}{2}\left[|0+\rangle|\phi^{0,+}\rangle+|0-\rangle|\phi^{0,-}\rangle+|1+\rangle|\phi^{1,+}\rangle+|1-\rangle|\phi^{1,-}\rangle\right]_{ABE}
\end{array},
\]
In the primary text, we have outlined the details of Eve's initial
state, represented as $\varepsilon_{E}$, along with Eve's state $|\phi^{\mathcal{A},\mathcal{B}}\rangle$
after Alice and Bob\textquoteright s measurement process. Our focus
remains exclusively on the instances accepted by Alice and Bob following
classical pre-processing. After normalization, we analyze the density
operator of Eve\textquoteright s system under the condition that Alice's
measurement yields an outcome of 0,

\[
\begin{array}{lcl}
\sigma_{E}^{0} & = & \frac{1}{2}\left[|\phi^{0,0}\rangle\langle\phi^{0,0}|+|\phi^{0,1}\rangle\langle\phi^{0,1}|\right]+\frac{1}{2}\left[|\phi^{0,+}\rangle\langle\phi^{0,+}|+|\phi^{0,-}\rangle\langle\phi^{0,-}|\right]\\
 & = & \frac{1}{2}\left(P_{|\phi^{0,0}\rangle}+P_{|\phi^{0,1}\rangle}\right)+\frac{1}{2}\left(P_{|\phi^{0,+}\rangle}+P_{|\phi^{0,-}\rangle}\right)
\end{array},
\]
and similarly, when Alice\textquoteright s measurement results in
an outcome of 1,

\[
\begin{array}{lcl}
\sigma_{E}^{1} & = & \frac{1}{2}\left[|\phi^{1,0}\rangle\langle\phi^{1,0}|+|\phi^{1,1}\rangle\langle\phi^{1,1}|\right]+\frac{1}{2}\left[|\phi^{1,+}\rangle\langle\phi^{1,+}|+|\phi^{1,-}\rangle\langle\phi^{1,-}|\right]\\
 & = & \frac{1}{2}\left(P_{|\phi^{1,0}\rangle}+P_{|\phi^{1,1}\rangle}\right)+\frac{1}{2}\left(P_{|\phi^{1,+}\rangle}+P_{|\phi^{1,-}\rangle}\right)
\end{array}.
\]
Subsequently, we express Eve\textquoteright s state in terms of the
orthonormal basis $|\varepsilon_{i}\rangle_{E}$, where $i\in\{1,\cdots,4\}$.

\begin{equation}
\sigma_{E}^{k}=\left(\begin{array}{cccc}
\mu_{1} & (-1)^{k}\sqrt{\mu_{1}\mu_{2}} & 0 & 0\\
(-1)^{k}\sqrt{\mu_{1}\mu_{2}} & \mu_{2} & 0 & 0\\
0 & 0 & \mu_{3} & (-1)^{k}\sqrt{\mu_{1}\mu_{2}}\\
0 & 0 & (-1)^{k}\sqrt{\mu_{1}\mu_{2}} & \mu_{4}
\end{array}\right),\label{eq:Chapter3_Eq12}
\end{equation}
where $k\in\left\{ 0,1\right\} .$

Additionally, we consider a communication channel where $c$ is a
noisy version of $a$, represented by the transition $c\leftarrow a$.
In this model, Alice applies a bit-flip operation with probability
$q$, leading to the conditional probabilities $p_{c|a=0}(1)=p_{c|a=1}(0)=q$.
Using standard information-theoretic relations, we simplify the right-hand
side of Equation (\ref{eq:Chapter3_Eq8}),

\begin{equation}
\begin{array}{lcl}
S(c|E) & = & S(cE)-S(E)\\
 & = & \left[H(c)+S(E|c)-S(E)\right],
\end{array}\label{eq:Chapter3_Eq13}
\end{equation}
and 
\begin{equation}
\begin{array}{lcl}
H(c|b) & = & H(cb)-H(b)\\
 & = & \left[H(c)+H(b|c)-H(b)\right].
\end{array}\label{eq:Chapter3_Eq14}
\end{equation}
By substituting Equation (\ref{eq:Chapter3_Eq13}) and Equation (\ref{eq:Chapter3_Eq14})
into Equation (\ref{eq:Chapter3_Eq8}), we express the entropy difference
in the following form:

\begin{equation}
S(c|E)-H(c|b)=S(E|c)-S(E)-\left(H(b|c)-H(b)\right).\label{eq:Chapter3_Eq15}
\end{equation}
The aforementioned substitution alters Equation (\ref{eq:Chapter3_Eq8})
in a way that facilitates the computation of the lower bound for the
secret key rate in our protocol.

\begin{figure}
\begin{centering}
\includegraphics[scale=0.5]{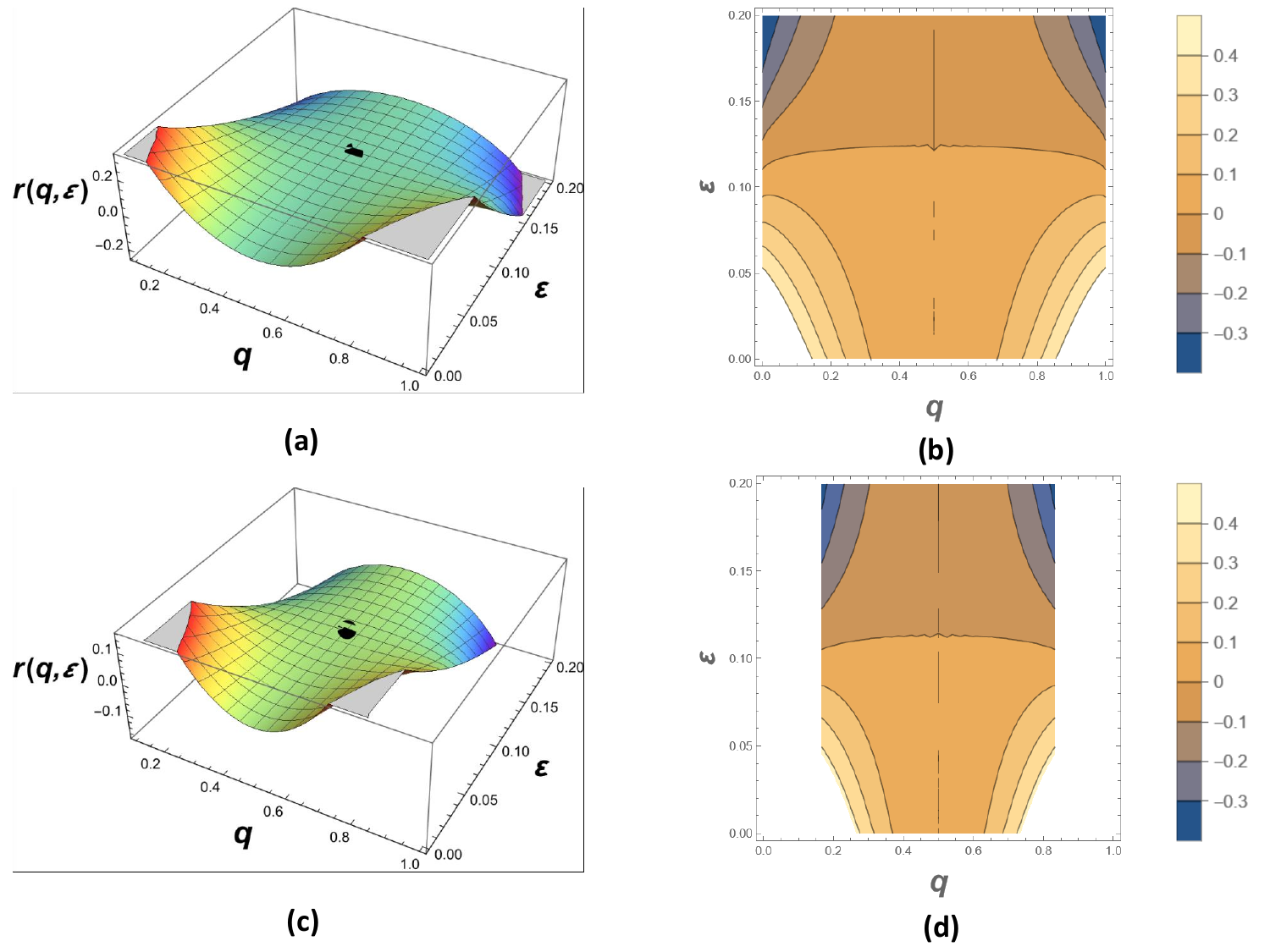} 
\par\end{centering}
\caption{The variation of the secret key rate with respect to bit-flip probability
($q$) and QBER ($\mathcal{E}$) is illustrated: $(a)$ the lower
bound on the secret key rate as a function of bit-flip probability
and QBER, $(b)$ a contour plot depicting the lower bound error limit,
correlating QBER with bit-flip probability, $(c)$ the upper bound
on the secret key rate as a function of bit-flip probability and QBER,
and $(d)$ a contour plot illustrating the upper bound error limit
with respect to QBER and bit-flip probability.}\label{fig:Chapter3_Fig2}
\end{figure}

If entropy is computed solely based on Eve's system, there exist only
two possible scenarios where Alice\textquoteright s bit value can
be either 0 or 1. Furthermore, determining the entropy of Eve\textquoteright s
system, conditioned on the value $c$ (announced by Alice), depends
on the bit flip probability. This leads to the expression:

\[
S(E|c)=\frac{1}{2}S\left((1-q)\sigma_{E}^{0}+q\sigma_{E}^{1}\right)+\frac{1}{2}S\left(q\sigma_{E}^{0}+(1-q)\sigma_{E}^{1}\right),
\]
and

\[
S(E)=S\left(\frac{1}{2}\sigma_{E}^{0}+\frac{1}{2}\sigma_{E}^{1}\right).
\]
Next, we examine Bob\textquoteright s bit string, which is derived
from the measurement outcome of his particle (system $B$) in the
state $|\Psi_{ABE}\rangle$. Conceptually, if only Bob's bit string
is taken into account, there should be an equal probability of obtaining
bit values 0 and 1. Moreover, when evaluating the conditional entropy
of Bob\textquoteright s bit string given Alice\textquoteright s noisy
bit string (value $c$), both error and no-error probabilities must
be incorporated. Consequently, we obtain:

\[
H(b)=1
\]
and 
\[
H(b|c)=h[q(1-\mathscr{\mathcal{E}})+(1-q)\mathcal{E}].
\]
By utilizing these relations and optimizing the parameter $q$, a
positive secret key can be established if $\mathcal{E}\le0.124$ (refer
to Figure \ref{fig:Chapter3_Fig2}. a). This threshold represents
the maximum tolerable error rate under classical pre-processing, which
accounts for the noise introduced by Alice.

Now, let us determine the upper bound of the secret key rate using
Equation (\ref{eq:Chapter3_Eq9}). In this scenario, the states of
Eve, corresponding to Alice and Bob obtaining the results $(0,0),(0,0),(1,1)$,
and $(1,1),$ are given by $|\phi^{0,0}\rangle$, $|\phi^{0,+}\rangle$,
$|\phi^{1,1}\rangle$, and $|\phi^{1,-}\rangle$, respectively. To
analyze the worst-case scenario for an adversary, we assume Eve performs
a von Neumann measurement using the projectors: $\frac{1}{\sqrt{2}}(|\phi^{0,0}\rangle+|\phi^{1,1}\rangle),\frac{1}{\sqrt{2}}(|\phi^{0,0}\rangle-|\phi^{1,1}\rangle),\frac{1}{\sqrt{2}}(|\phi^{0,+}\rangle+|\phi^{0,-}\rangle),$
and $\frac{1}{\sqrt{2}}(|\phi^{0,+}\rangle-|\phi^{0,-}\rangle)$\footnote{The probabilities of executing the last two projector operations are
half of those for the first two. Specifically, the probability of
applying the last two projectors is $\frac{1}{6}$, while for the
first two, it is $\frac{1}{3}$ each.} to obtain an outcome $e$. Applying this condition appropriately
modifies Equation (\ref{eq:Chapter3_Eq9}) as follows:

\begin{equation}
\begin{array}{lcl}
r(a,b,e) & = & H(c|e)-H(c|b)\\
 & = & H(e|c)-H(e)-[H(b|c)-H(b)]\\
 & \le & -\chi(E)-[H(b|c)-H(b)],
\end{array}\label{eq:Chapter3_Eq16}
\end{equation}

here

\[
\begin{array}{lcl}
\chi(E) & = & S\left[\frac{1}{3}\left(P_{|\phi^{0,0}\rangle}+P_{|\phi^{1,1}\rangle}\right)+\frac{1}{6}\left(P_{|\phi^{0,+}\rangle}+P_{|\phi^{1,-}\rangle}\right)\right]\\
 & - & \frac{1}{2}S\left[\left(1-q\right)\left(\frac{2}{3}P_{|\phi^{0,0}\rangle}+\frac{1}{3}P_{|\phi^{0,+}\rangle}\right)+q\left(\frac{2}{3}P_{|\phi^{1,1}\rangle}+\frac{1}{3}P_{|\phi^{1,-}\rangle}\right)\right]\\
 & - & \frac{1}{2}S\left[q\left(\frac{2}{3}P_{|\phi^{0,0}\rangle}+\frac{1}{3}P_{|\phi^{0,+}\rangle}\right)+\left(1-q\right)\left(\frac{2}{3}P_{|\phi^{1,1}\rangle}+\frac{1}{3}P_{|\phi^{1,-}\rangle}\right)\right]
\end{array}.
\]
The above expression is the Holevo quantity \cite{H_73} establishes
the upper limit of mutual information between $e$ and $c$ across
all conceivable measurement strategies that Eve might employ. This
enables the computation of an upper bound on the key rate by solving
$r(a,b,e)=0$. The resulting solution provides an upper bound of $\mathcal{E}\ge0.114$,
assuming the optimal selection of $q$ (refer to Figure \ref{fig:Chapter3_Fig2}.
c). In the BB84 protocol, a single transmission of the qubit sequence
occurs over a one-way quantum channel. Conversely, the proposed scheme
necessitates three quantum channel transmissions, prioritizing a more
substantial quantum component while reducing reliance on the classical
component. These three transmissions correspond to distinct sequences
generated by Alice and Bob. Although the BB84 protocol incorporates
one-way classical post-processing, the proposed scheme retains this
feature. Given the three separate quantum transmissions, security
validation for each sequence is imperative through information reconciliation.
Since the sequences differ, their security is assessed independently.
Our approach employs an information-theoretic framework, utilizing
two-qubit density operators to establish the upper and lower bounds
of the secret key rate based on the QBER, following a methodology
similar to that used in BB84. For the BB84 protocol, the lower bound
on the tolerable QBER with classical pre-processing is $\mathcal{E}\le0.124$,
while the upper bound for the key rate is $\mathcal{E}\ge0.146$.
In contrast, the proposed scheme maintains the same lower bound at
$\mathcal{E}\le0.124$ but exhibits a reduced upper bound at $\mathcal{E}\ge0.114$.

\section{Analysis of photon number splitting attack}\label{sec:Chapter3_Sec4}

\begin{figure}
\begin{centering}
\includegraphics[scale=0.4]{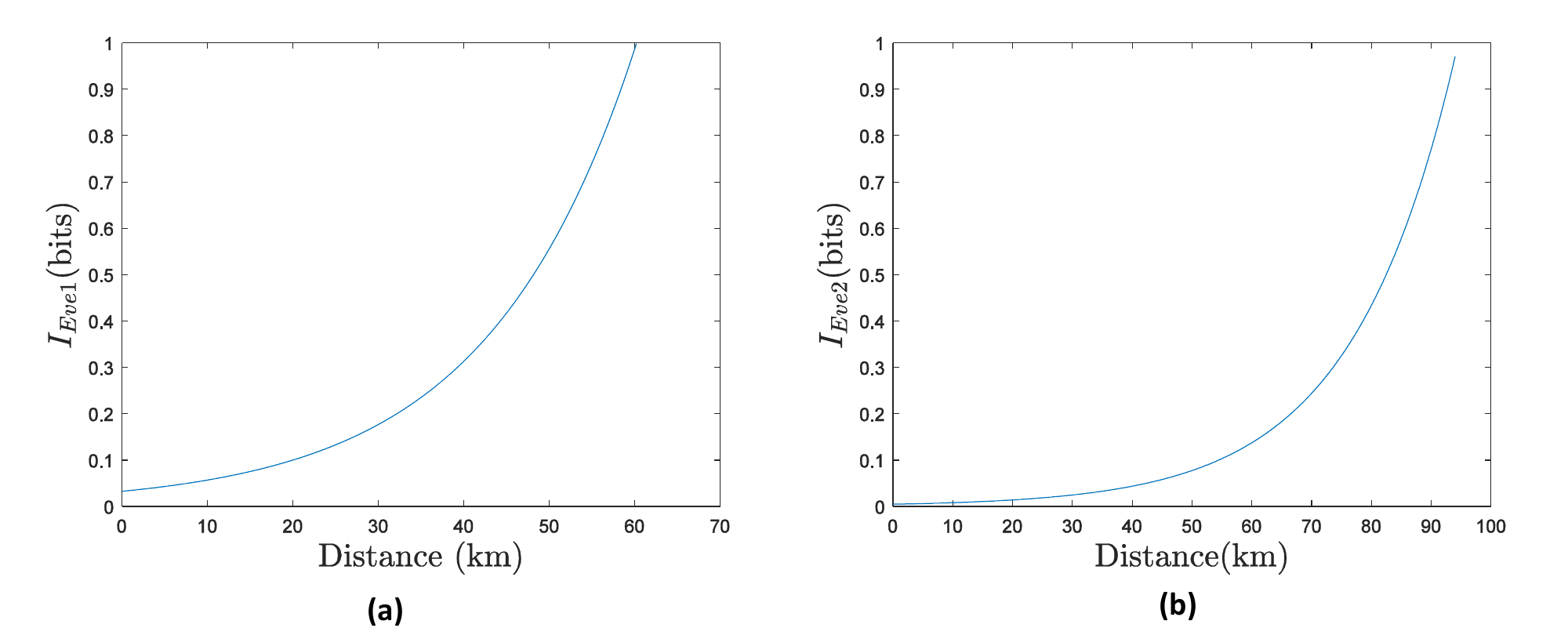} 
\par\end{centering}
\caption{The variation in Eve's information as a function of distance is analyzed
to determine the critical distance ($l_{c}$): $(a)$ Eve's information
plotted against distance to estimate the critical threshold where
the attacker extracts maximum key information via a PNS attack on
Protocol 3.1, and $(b)$ Eve's information plotted against distance
to estimate the critical threshold where the attacker gains maximum
key information using an IRUD attack on Protocol 3.2.}\label{fig:Chapter3_Fig3}
\end{figure}

As previously discussed, our proposed schemes can be implemented using
sources emitting WCPs (refer to Equation (\ref{eq:Chapter3_Eq1})).
However, cryptographic protocols relying on WCP may encounter security
vulnerabilities due to the potential exploitation of PNS and related
attacks by an adversary. Therefore, it is essential to analyze and
ensure the robustness of our schemes against various forms of PNS
attacks that an eavesdropper could execute under the assumption of
a vanishing QBER (i.e., ${\rm QBER}=0$). The sub-protocols incorporated
in our framework necessitate Bob to disclose basis information or
a specific set of non-orthogonal state parameters. With this in mind,
the following observations are noteworthy:
\begin{enumerate}
\item In Protocol 3.1 for QKD, Eve carries out a PNS attack under the assumption
of having unlimited technological resources, constrained only by the
fundamental principles of physics. The PNS attack, inherently dependent
on quantum memory, is also referred to as a quantum storage attack.
Here, we consider an optimal scenario for Eve's strategy. The probability
of an $n$-photon state occurring follows the distribution $p(n,\mu)=\frac{e^{-\mu}\mu^{n}}{n!}$,
where $\mu$ represents the mean photon number. Both Alice and Bob
possess knowledge of the channel transmittance ($\eta$) and the mean
photon number ($\mu$). In an ideal case without eavesdropping, the
expected probability of detecting at least one photon is given by
$\underset{n\ge1}{\sum}p(n,\mu\eta)$, commonly referred to as the
raw detection rate per pulse. Within this framework, sequence $S_{B2}$
is generated independently but encodes the same bit information as
sequence $S_{B1}$, albeit in a distinct basis. Eve strategically
targets only $S_{B1}$ to maximize her probability of extracting information.
Specifically, we consider a scenario where Eve initially performs
a photon-number QND measurement on $S_{B1}$ to determine the number
of photons present. She subsequently discards single-photon pulses
while selectively retaining one photon from multi-photon pulses. The
remaining photons are then transmitted to Alice through an ideal lossless
channel\footnote{The transmission efficiency in an optical fiber of length ll is given
by $\eta=10^{-\frac{\delta}{10}}$, where $\delta$ is defined as
$\delta=\text{\ensuremath{\alpha l}{\rm [dB]}}$. Here, $\alpha$
represents the fiber loss in dB/km. Since both $\alpha$ and $l$
are non-negative, $\delta$ also remains non-negative. The boundary
conditions for $\delta$, specifically $\delta=0$ and $\delta\rightarrow\infty$,
yield $\eta=1$ and $\eta=0$, respectively. Consequently, the range
$0\leq\eta\leq1$ characterizes the attenuation in the transmission
channel, where $\eta=1$ corresponds to an ideal, lossless channel
allowing full transmission, and $\eta=0$ represents a completely
opaque channel with no transmission.} ($\eta=1$). However, Eve cannot arbitrarily discard multi-photon
pulses, as maintaining a QBER of zero imposes constraints on such
actions. The extent to which Eve can filter multi-photon pulses is
fundamentally dependent on the channel transmittance $\eta$ between
Alice and Bob. For this attack to be feasible, the following condition
must hold \cite{AGS04}:

\[
\underset{n\ge1}{\sum}p(n,\mu\eta)=t_{1}p(1,\mu)+\underset{n\ge2}{\sum}p(n,\mu).
\]

In this context, $t_{1}$ represents the proportion of single-photon
pulses that successfully reach Bob. When faced with a highly capable
eavesdropper, losses can be so significant that $t_{1}=0$. Under
these circumstances, Eve exclusively intercepts multi-photon pulses,
occurring with a probability given by $\underset{n\ge2}{\sum}p(n,\mu)$.
The amount of information Eve acquires is expressed as:

\[
I_{{\rm Eve1}}=\frac{0.5\times\underset{n\ge2}{\sum}p(n,\mu)}{0.75\times\underset{n\ge1}{\sum}p(n,\mu\eta)},
\]

The factor of 0.5 in the numerator originates from the classical information
disclosed by Bob during the generation of the sifted key\footnote{For Protocol 3.1, a total of $0.625$ bits of information
is revealed to Eve, out of which only $0.5$ bits are
useful for generating the sifted key.}, while the factor of 0.75 in the denominator arises from Alice\textquoteright s
information related to non-empty pulses. Specifically, 0.5 accounts
for Bob\textquoteright s revealed classical information, and the remaining
0.25 corresponds to Alice\textquoteright s information gain in the
absence of Bob\textquoteright s announcement. Figure \ref{fig:Chapter3_Fig3}.
a depicts the variation of $I_{{\rm Eve}1}$ with distance $l$ for
an attenuation of $\alpha=0.25$ dB/km and a mean photon number $\mu=0.1$,
ensuring a fair comparison with the BB84 protocol ($\mu=0.1$). The
estimated critical attenuation is $\delta_{c}=15.05$ dB, corresponding
to a critical distance of $l_{c}=60.2$ km, beyond which the attacker
acquires full bit information under the PNS attack. Comparatively,
the critical distance for the BB84 protocol under similar conditions
is 52 km \cite{AGS04}, which is shorter than that of our protocol.
Additionally, the figure indicates that Eve gains almost no information
up to a distance of 30 km. This advantage stems from the utilization
of a higher amount of quantum signal or equivalently a higher amount
of quantum resources in the communication process. Furthermore, as
this protocol reveals less classical information than BB84, Eve's
ability to extract information after the classical announcement in
a collective attack scenario is reduced, thereby enhancing the security
of the final key.
\item In Protocol 3.2 for QKD, Bob discloses information about the non-orthogonal
states. Given this setup, Eve can employ a PNS attack variant known
as the intercept-resend unambiguous discrimination (IRUD) attack.
This strategy begins with Eve conducting a quantum non-demolition
(QND) measurement to ascertain the photon number within each pulse.
She discards all pulses containing fewer than three photons and proceeds
to measure the remaining ones (those with at least three photons)
using a specific measurement process\footnote{The measurement $\mathcal{\mathscr{\mathcal{M}}}$ represents a von
Neumann measurement capable of distinguishing the following four quantum
states: $|\Phi_{1}\rangle=\frac{1}{\sqrt{2}}\left(|000\rangle-|011\rangle\right),|\Phi_{2}\rangle=\frac{1}{2}\left(|101\rangle+|010\rangle+|100\rangle+|110\rangle\right),|\Phi_{3}\rangle=\frac{1}{\sqrt{2}}\left(|111\rangle-|001\rangle\right)$,
and $|\Phi_{4}\rangle=\frac{1}{2}\left(|101\rangle-|010\rangle-|100\rangle+|110\rangle\right)$\cite{SAR+04}.}, denoted as $\mathcal{\mathscr{\mathcal{M}}}$. Upon obtaining a
definite outcome from $\mathcal{\mathscr{\mathcal{M}}}$, she reconstructs
a new photon state and transmits it to Bob. This attack assumes that
Eve operates in a lossless channel ($\eta=1$) and possesses quantum
memory. The attack is exclusively executed on sequence $S_{B1}$,
following the same principle as the PNS attack in Protocol 3.1 for
QKD. However, unlike the standard PNS attack, Eve cannot indiscriminately
discard pulses with fewer than three photons, as maintaining a QBER
of zero is essential. For the attack to be effective, the following
condition must be met: 
\[
\underset{n\ge1}{\sum}p(n,\mu\eta)=t_{1}p(1,\mu)+t_{2}p(2,\mu)+\underset{n\ge3}{\sum}p(n,\mu),
\]
where $t_{1}$ and $t_{2}$ represent the proportions of single-photon
and two-photon pulses, respectively, that reach Bob. Since the probability
of pulses containing more than three photons is negligible, it can
be approximated as $\underset{n\ge3}{\sum}p(n,\mu)\approx p(3,\mu)$.
For an adversary with maximal capabilities, channel losses ensure
that $t_{1}=0$ and $t_{2}=0$, meaning Eve exclusively targets three-photon
pulses, which occur with a probability of $p(3,\mu)$. The amount
of information Eve extracts in this scenario is given by: 
\[
I_{{\rm Eve2}}=\frac{I(3,\chi)p(3,\mu)}{\underset{n\ge1}{\sum}p(n,\mu\eta)},
\]
where $I(n,\chi)$ represents the maximum information Eve can extract
using $n$ photons in a single pulse. $\chi$ is the overlap of two
states within each set of non-orthogonal states announced by Bob\footnote{$I(n,\chi)=1-h(P,1-P)$ with $h(P,1-P)$ being the binary entropy
function and $P=\frac{1}{2}\left(1+\sqrt{1-\chi^{2n}}\right)$\cite{P97}.
In our case, the overlap, $\chi=\frac{1}{\sqrt{2}}.$}. The denominator represents the raw detection rate per pulse in the
absence of an eavesdropper, given a channel transmittance $\eta$
between Alice and Bob. For a fair comparison with the SARG04 protocol
($\mu=0.2$), we also set $\mu=0.2$. Assuming an attenuation factor
of $\alpha=0.25$ dB/km and $\mu=0.2$, we analyze the variation of
Eve's information $(I_{Eve2})$ with distance to determine the critical
attenuation (see Figure \ref{fig:Chapter3_Fig3}. b). From Figure
\ref{fig:Chapter3_Fig3}. b, the critical attenuation is found to
be $\delta_{c}=23.75$ dB, corresponding to a critical distance of
$l_{c}=95$ km. In comparison, under similar conditions, the critical
distance for the SARG04 protocol ranges from approximately 50 km to
100 km \cite{SAR+04}, which is comparable to the critical distance
achieved with our protocol. For this attack, Eve gains almost no information
up to 60 km. Additionally, our protocol reveals less classical information,
thereby offering better protection against Eve's collective attack,
which relies on the information disclosed in the classical subprotocol.
\end{enumerate}

\section{Conclusion}\label{sec:Chapter3_Sec5}

This chapter introduces a novel protocol for QKD along with a variant
of the same. While the proposed protocols require greater quantum
resources compared to SARG04 and similar schemes, they minimize the
amount of classical information transmitted over the public channel,
thereby reducing the probability of certain side-channel attacks.
A comprehensive security analysis has been conducted for the proposed
protocols. Additionally, the tolerable error thresholds for the upper
and lower bounds of the secret key rate under collective attacks have
been determined. The findings indicate that the implementation of
specific classical pre-processing techniques enhances the tolerable
error limit, a result that is further illustrated through graphical
representations. Before concluding, it is important to highlight key
observations from the analysis. In \cite{RGK_05}, the density operators
of Eve\textquoteright s final state were derived for the six-state
QKD protocol. Interestingly, despite structural differences, the density
operators characterizing Eve\textquoteright s final system in the
proposed protocols are found to be identical to those in \cite{RGK_05}.
Notably, in Protocol 3.2 (Protocol 3.1), Alice and Bob (Alice) do
not disclose their (her) measurement outcomes when different bases
are used for state preparation and measurement. The reason behind
the equivalence of density operators lies in the cancellation of terms
in the density matrix in cases of basis mismatch. Furthermore, it
is explicitly shown that for the proposed protocols, the tolerable
QBER is constrained within $\mathcal{E}\le0.124$ for the lower bound
and $\mathcal{E}\ge0.114$ for the upper bound of the secret key rate
when classical pre-processing is applied. In the absence of such pre-processing,
a reduction in these tolerable error limits is expected.

In practical cryptographic implementations, various types of errors
can arise during qubit transmission. When the QBER exceeds zero, an
eavesdropper may attempt to exploit partial cloning machines \cite{CI00,NG98,C00}.
Ac\'{i}n et al. demonstrated that legitimate users of the SARG04
protocol can tolerate an error rate of up to 15\% when Eve employs
the most effective known partial cloning strategy \cite{AGS04}. Additionally,
they established that this tolerable error threshold surpasses that
of the BB84 protocol. In our analysis, for ${\rm QBER}>0$, we similarly
determine the tolerable error limit to be 15\% (refer to Section \ref{sec:Chapter3_Sec3}),
which exceeds the performance of BB84 and its variants. Furthermore,
we conduct a security-efficiency trade-off evaluation for the proposed
schemes, benchmarking their efficiency against the SARG04 protocol.

\textit{Security-efficiency trade-off for our protocol:} In 2000,
Cabello \cite{C2000} introduced a measure of the efficiency of quantum
communication protocols as $\eta=\frac{b_{s}}{q_{t}+b_{t}}$, where
$b_{s}$ is the number of secret bits exchanged by the protocol, $q_{t}$
is the number of qubits interchanged via the quantum channel in each
step of the protocol, and $b_{t}$ is the classical bit information
exchanged between Alice and Bob via the classical channel\footnote{The classical bit which is used for detecting eavesdropping is neglected
here.}. As far as Protocol 3.2 is concerned, an inherent error arises during
its execution. If Alice's initial state is $|0\rangle$, and Bob's
measurement outcomes are $|0\rangle$ or $|-\rangle$, no error occurs.
However, if Bob's outcome is $|+\rangle$, an inherent error is introduced.
The probabilities of Bob obtaining $|0\rangle$, $|-\rangle$, and
$|+\rangle$ are $\frac{1}{8}$, $\frac{1}{16}$, and $\frac{1}{16}$,
respectively (see Table \ref{tab:Chapter3_Tab1}). In this case, when
the measurement result is determined as $|0\rangle$, the error probability
corresponding to Bob\textquoteright s measurement outcomes $|0\rangle$
and $|+\rangle$ is calculated as $\left(\frac{1}{64}\right)/\left(\frac{1}{64}+\frac{1}{8}\right)=\frac{1}{9}$.
A similar scenario applies when Alice\textquoteright s initial state
is $|1\rangle$, $|+\rangle$ or $|-\rangle$. For instance, if Alice\textquoteright s
initial state is $|1\rangle$, no error occurs when Bob\textquoteright s
measurement results are $\left\{ |1\rangle,|+\rangle\right\} $, whereas
an error arises when the outcome is $|-\rangle$. The same probability
calculations hold for the remaining cases of Alice\textquoteright s
initial states. Finally, the secret key bits ($b_{s}$) are transformed
as follows:

\[
\begin{array}{lcl}
b_{s} & = & \left[\left(\frac{1}{16}+\frac{1}{32}+\frac{1}{64}\right)+\left(\frac{1}{8}+\frac{1}{64}\right)\left(1-h(\frac{1}{9})\right)\right]\times4\\
 & = & 0.72
\end{array}.
\]
For the sifting subprotocol of the second QKD protocol (Protocol 3.2),
the values of the essential parameters are $b_{s}=0.72$, $q_{t}=3$
and $b_{t}=0.75$, resulting in an efficiency of $\eta=0.192$. For
the sifting subprotocol of our Protocol 3.1, the values of the essential
parameter are $b_{s}=0.75$, $q_{t}=3$ and $b_{t}=0.625$, giving
an efficiency of $\eta=0.2069$ with no inherent error. In this specific
sifting condition, the basis information will be revealed at the end
of the protocol, which may increase the chance of a PNS attack by
a powerful Eve. To apply the more efficient Protocol 3.2, one must
consider the inherent error probability value of $0.0625$, and the
exchange of classical information is also more than Protocol 3.1.
We want to stress that the Protocol 3.2 is more robust against PNS
attack because Alice and Bob do not reveal the basis information;
instead, two non-orthogonal state information values are announced
for the sifting process ($M$ value). Our two protocols are more efficient
than the SARG04 protocol\footnote{The values of essential parameters for SARG04 protocol are $b_{s}=0.25$,
$q_{t}=1$ and $b_{t}=1$, resulting in an efficiency of $\eta=0.125$.}($\eta=0.125$). It is worth noting that for both protocols, the amount
of classical information disclosed during the classical sifting phase
is lower than that of the SARG04 protocol. This reduction decreases
the probability of an information gain by Eve using the announced
classical information. One can use either of our two protocols as
needed for the necessary task.

\newpage



\chapter{CONTROLLED QUANTUM KEY AGREEMENT WITHOUT USING QUANTUM MEMORY}\label{Ch4:Chapter4_QKA}
\graphicspath{{Chapter4/Chapter4Figs/}{Chapter4/Chapter4Figs/}}

\section{Introduction}\label{sec:Chapter4_Sec1}

Key agreement is a vital cryptographic domain, essential for achieving
perfect forward secrecy. It enables two or more parties to collaboratively
generate a shared key, ensuring all legitimate participants influence
the outcome while preventing any subset or eavesdropper from dictating
the key. A stronger KA definition, commonly adopted in QKA, mandates
equal influence from all legitimate users. The foundational two-party
KA protocol was introduced by Diffie and Hellman in 1976 \cite{DH76},
with later extensions to multiparty settings for enhanced efficiency
\cite{IT+82,MW+98,ST+2000,AS+2000}. This stronger definition enforces
``fairness'', ensuring no individual or group can manipulate the
final key \cite{HWL+14,HSL+17}. Classical KA security relies on computational
complexity, but advancements in quantum computing threaten this foundation,
as many underlying problems could be efficiently solved with scalable
quantum computers. To counteract this, cryptographic schemes leverage
quantum mechanics principles---such as collapse on measurement, the
no-cloning theorem, and Heisenberg\textquoteright s uncertainty principle---enabling
cryptographic tasks with unconditional security. QKA, the quantum
counterpart of KA, benefits from these properties \cite{HJP20}. This
work introduces two novel QKA schemes and their security proofs. Before
discussing QKA and its multipartite extensions, a brief overview of
quantum cryptography will highlight its unique attributes and broader
applications.

QKA has emerged as a fundamental cryptographic primitive, enabling
two or more parties to establish an identical key with equal contribution
using quantum resources. Its significance lies in applications such
as electronic auctions, multiparty secure computing, and access control
\cite{ST+2000}, along with its role in ensuring forward secrecy.
Due to its broad applicability, QKA has recently been recognized as
a distinct domain within quantum cryptography. The first QKA protocol,
introduced by Zhou et al. in 2004 \cite{ZZX04}, employed quantum
teleportation for key agreement between two users. However, Tsai and
Hwang (2009) later identified a fairness issue \cite{TH09}, where
one party could unilaterally determine the key and distribute it undetected.
Subsequent research confirmed the protocol\textquoteright s insecurity
\cite{CT+11}. In 2013, Shi and Zhou proposed the first MQKA scheme
using entanglement swapping \cite{SZ13}, which was later found to
be insecure \cite{LG+13}. Meanwhile, Chong et al. (2010) developed
a QKA protocol based on BB84, integrating delay measurement and authenticated
classical channels \cite{CH10}. Over time, various two-party QKA
protocols have been extended to MQKA \cite{LG+13,SA+14,XW+14,HM15},
classified into tree-type \cite{XW+14}, complete-graph-type \cite{LG+13},
and circle-type \cite{LZ+21}. The circle-type structure offers better
feasibility and efficiency. Research in quantum conference key agreement
and MQKA continues to grow, leveraging methods such as differential
phase reference \cite{IH24} and discrete variable entangled quantum
states. Experimental advancements in QCKA have been demonstrated by
Proietti et al. \cite{PHG+21}, while Wang et al. \cite{WZL24} have
explored new schemes for mutually authenticated QKA.

A secure QKA protocol must ensure three key properties: \textit{Correctness},
where all participants derive the agreed key accurately; \textit{Security},
preventing unauthorized eavesdropping without detection; and \textit{Fairness},
ensuring no subset of participants can solely determine the final
key. Leveraging these principles, we introduce a novel CQKA protocol
with a controller, Charlie, enabling key agreement between two legitimate
parties without requiring quantum memory. Controlled quantum cryptography
has gained significant attention, with numerous schemes proposed for
controlled-QKD \cite{DGL+03,SNG+09}, controlled-quantum dialogue
\cite{DXG+08,KH17}, and controlled-secure direct communication \cite{WZT06,CWG+08,LLL+13,DP23}.
However, to the best of our knowledge, no existing CQKA protocol achieves
this without quantum memory \cite{TS+20}. Addressing this gap, our
scheme employs Bell states and single-qubit states, using a one-way
quantum channel to minimize noise, unlike two-way QKA protocols \cite{SA+14,HM15,TS+20}
that suffer from increased channel noise. We further analyze the impact
of Markovian and non-Markovian noise on our protocol, examining fidelity
as a function of the noise parameter. The security of the proposed
CQKA scheme is rigorously proven, demonstrating feasibility with current
quantum technologies. Additionally, we establish its efficiency over
many existing schemes, with the controller mechanism providing distinct
advantages in specific scenarios.

The structure of this chapter is outlined as follows: Section \ref{sec:Chapter4_Sec2}
introduces the fundamental concepts necessary for articulating our
schemes, followed by a step-by-step explanation of our proposed protocol.
Section \ref{sec:Chapter4_Sec3} presents a security evaluation, addressing
both impersonation-based fraudulent attacks and broader (more generalized)
threats, such as collective attacks. The influence of noise on the
protocol's performance is examined in Section \ref{sec:Chapter4_Sec4}.
A comparative analysis between our protocol and existing approaches
is conducted in Section \ref{sec:Chapter4_Sec5}. Section \ref{sec:Chapter4_Sec6}
provides an in-depth discussion of the results along with the details
of the experimental setup. Lastly, the conclusions of this chapter
are drawn in Section \ref{sec:Chapter4_Sec7}.

\section{New protocol for CQKA (Protocol 4.1)}\label{sec:Chapter4_Sec2}

In the previous section, we highlighted the significance of correctness
as a fundamental requirement for a QKA protocol. This section provides
a detailed explanation of our proposed CQKA protocol, with a specific
focus on ensuring ``correctness''. To illustrate this, we consider
a particular example where Alice, Bob, and Charlie select the $i^{th}$
key bit to prepare the corresponding $i^{th}$ quantum state in their
particle sequence for subsequent protocol steps. The first part of
this section outlines the steps taken to satisfy the correctness condition,
while the second part presents the protocol in a structured, step-by-step
manner for a more generalized understanding.

\subsection*{Basis concept of the proposed Protocol 4.1}

Our proposed CQKA protocol is formulated based on core principles
of quantum mechanics. To define the process, we establish the relationship
between classical bit values and quantum states. Two distinct mappings
are utilized: one for public announcement and another for the final key
agreement between Alice and Bob \cite{SHW16}, which remains undisclosed
to uphold privacy constraints. During protocol execution, Charlie
and Alice publicly declare their respective classical bit sequences,
$k_{C}$ and $k_{A}$, based on the quantum states they generate,
as specified in Equations (\ref{eq:Chapter4_Eq1}) and (\ref{eq:Chapter4_Eq2}).
Meanwhile, Bob determines his classical bit sequence $k_{B}$ from
measurement outcomes, following the mapping defined in Equation (\ref{eq:Chapter4_Eq3}).
The association between these classical sequences and their corresponding
quantum states is detailed in the subsequent discussion.

\begin{equation}
\begin{array}{lcl}
0 & : & |\phi^{+}\rangle=\frac{1}{\sqrt{2}}\left(\vert00\rangle+\vert11\rangle\right),\\
1 & : & \vert\phi^{-}\rangle=\frac{1}{\sqrt{2}}\left(\vert00\rangle-\vert11\rangle\right).
\end{array}\label{eq:Chapter4_Eq1}
\end{equation}
In the proposed CQKA scheme, if Charlie prepares the Bell state $|\phi^{+}\rangle$
$(\vert\phi^{-}\rangle)$, the corresponding bit value in the sequence
$k_{C}$ is assigned as 0 (1). The mapping process utilized by Alice
and Bob follows a similar approach, as outlined below.

\begin{equation}
\begin{array}{lcl}
0 & : & |0\rangle,\\
1 & : & |1\rangle,
\end{array}\label{eq:Chapter4_Eq2}
\end{equation}
and 
\begin{equation}
\begin{array}{ccccc}
0 & : & \vert\phi^{+}\rangle=\frac{1}{\sqrt{2}}\left(\vert00\rangle+\vert11\rangle\right) & {\rm or} & \vert\psi^{-}\rangle=\frac{1}{\sqrt{2}}\left(\vert01\rangle-\vert10\rangle\right),\\
1 & : & \vert\phi^{-}\rangle=\frac{1}{\sqrt{2}}\left(\vert00\rangle-\vert11\rangle\right) & {\rm or} & \vert\psi^{+}\rangle=\frac{1}{\sqrt{2}}\left(\vert01\rangle+\vert10\rangle\right).
\end{array}\label{eq:Chapter4_Eq3}
\end{equation}
Alice and Bob employ a mapping technique that associates two-bit classical
information with their final measurement outcomes after performing
Bell measurements. This mapping is structured as follows:

\begin{equation}
\begin{array}{c}
00:\vert\phi^{+}\rangle=\frac{1}{\sqrt{2}}\left(\vert00\rangle+\vert11\rangle\right),\\
01:\vert\phi^{-}\rangle=\frac{1}{\sqrt{2}}\left(\vert00\rangle-\vert11\rangle\right),\\
10:\vert\psi^{+}\rangle=\frac{1}{\sqrt{2}}\left(\vert01\rangle+\vert10\rangle\right),\\
11:\vert\psi^{-}\rangle=\frac{1}{\sqrt{2}}\left(\vert01\rangle-\vert10\rangle\right).
\end{array}\label{eq:Chapter4_Eq4}
\end{equation}
The first row of the mapping table indicates that if Alice's (Bob's)
Bell measurement results in $|\phi^{+}\rangle$, they store the bit
sequence 00 privately for later use in the protocol. Similar mappings
apply to other cases.

Now, we outline the fundamental concept behind our CQKA protocol.
Initially, Alice and Bob request Charlie to initiate the protocol.
In response, Charlie generates a sequence of Bell states, choosing
between $|\phi^{+}\rangle_{{\rm C_{1}C_{2}}}$ and $|\phi^{-}\rangle_{{\rm C_{1}C_{2}}}$.
For instance, if Charlie selects $|\phi^{+}\rangle_{{\rm C_{1}C_{2}}}$,
where subscripts ${\rm C_{1}}$ and ${\rm C_{2}}$ denote the first
and second qubits of Charlie\textquoteright s Bell state, he records
the classical sequence $k_{C}$ (in this case, 0) to store information
about the prepared state. Charlie then transmits qubit ${\rm C_{1}}$
to Alice and qubit ${\rm C_{2}}$ to Bob while keeping $k_{C}$ confidential.
Upon receiving qubit ${\rm C_{1}}$ (${\rm C_{2}}$) from Charlie,
Alice (Bob) prepares their respective qubits in the computational
basis ($Z\in\left\{ |0\rangle,|1\rangle\right\} $), choosing $|0\rangle$
($|1\rangle$) as an example. Alice records her classical sequence
$k_{A}$ (here, 0) and keeps it private. Alice\textquoteright s and
Bob\textquoteright s prepared qubits are denoted as ${\rm A}$ and
${\rm B}$, respectively. Next, Alice (Bob) applies a CNOT operation,
where ${\rm C_{1}}$ (${\rm C_{2}}$) serves as the control qubit
and ${\rm A}$ (${\rm B}$) as the target qubit. The resulting quantum
state after the CNOT operation is represented by the following equation:

\begin{equation}
\begin{array}{lcl}
|\psi_{2}\rangle & = & \frac{1}{\sqrt{2}}{\rm CNOT_{C_{1}\rightarrow A}CNOT_{C_{2}\rightarrow B}}\left(|0\rangle_{{\rm A}}|\phi^{+}\rangle_{{\rm C_{1}C_{2}}}|1\rangle_{{\rm B}}\right)\\
 & = & \frac{1}{\sqrt{2}}\left(|00\rangle_{{\rm AC_{1}}}|01\rangle_{{\rm C_{2}B}}+|11\rangle_{{\rm AC_{1}}}|10\rangle_{{\rm C_{2}B}}\right).
\end{array}\label{eq:Chapter4_Eq5}
\end{equation}
In this scenario, Alice and Bob each possess two qubits, denoted as
${\rm C_{1},A}$ for Alice and ${\rm C_{2},B}$ for Bob. Both parties
perform Bell basis measurements, resulting in the following quantum
state:

\begin{equation}
\begin{array}{lcl}
|\psi_{2}\rangle & = & \frac{1}{\sqrt{2}}\left(|\phi^{+}\rangle_{{\rm AC_{1}}}|\psi^{+}\rangle_{{\rm C_{2}B}}+|\phi^{-}\rangle_{{\rm AC_{1}}}|\psi^{-}\rangle_{{\rm C_{2}B}}\right).\end{array}\label{eq:Chapter4_Eq6}
\end{equation}
Upon obtaining their measurement outcomes, they request Charlie to
disclose his bit sequence $k_{C}$ (set to 0 in this case), corresponding
to his initially prepared states based on the mapping in Equation
(\ref{eq:Chapter4_Eq1}). Additionally, Alice and Bob announce their
respective bit sequences $k_{A}$ and $k_{B}$, where $k_{A}$ is
determined by Alice\textquoteright s prepared state and $k_{B}$ is
derived from Bob\textquoteright s measurement on qubits 3 and 4 after
performing a Bell measurement, as defined in Equation $(\ref{eq:Chapter4_Eq2})$
and Equation $(\ref{eq:Chapter4_Eq3})$, respectively. In this instance,
Alice announces 0, while Bob announces either 0 or 1 with equal probability
($\frac{1}{2}$), as he may obtain the measurement outcomes $|\psi^{-}\rangle_{34}$
or $|\psi^{+}\rangle_{34}$ with equal probability (refer to Equation
(\ref{eq:Chapter4_Eq6})). The relationship between the announced
values of Charlie, Alice, and Bob, along with their undisclosed measurement
outcomes and the hidden measurement results of the other party, is
detailed in Table \ref{tab:Chapter4_Tab1}. For instance, if Charlie,
Alice, and Bob announce the bit values 0, 0, and 1, respectively,
Alice and Bob can infer their own two-bit classical measurement results
using Equation $(\ref{eq:Chapter4_Eq4})$. Based on Table \ref{tab:Chapter4_Tab1},
Alice deduces Bob\textquoteright s measurement outcome as 10, while
Bob infers Alice\textquoteright s outcome as 00. To derive the final
agreement key, Alice and Bob perform modulo-2 addition on their respective
measurement results, as described in Equation (\ref{eq:Chapter4_Eq4}).
This produces a two-bit result, which is then further processed by
applying modulo-2 addition between the first and second bits. In this
example, the computation follows: $10\oplus00=10$ Thus, the final
key is $1\oplus0=1$.

There exist eight possible state combinations resulting from the independent
choices of Alice, Bob, and Charlie. These states can be represented
as $|\psi_{i}\rangle$, where $i\in\left\{ 1,2,\cdots,8\right\} $.
Since $|\psi_{2}\rangle$ has already been discussed, the remaining
state combinations are defined as follows.

\begin{equation}
\begin{array}{lcl}
|\psi_{1}\rangle & = & \frac{1}{\sqrt{2}}{\rm CNOT}{}_{{\rm C_{1}\rightarrow{\rm A}}}{\rm CNOT_{C_{2}\rightarrow B}}\left(|0\rangle_{{\rm A}}|\phi^{+}\rangle_{{\rm C_{1}C_{2}}}|0\rangle_{{\rm B}}\right)\\
 & = & \frac{1}{\sqrt{2}}\left(|\phi^{+}\rangle_{{\rm AC_{1}}}|\phi^{+}\rangle_{{\rm C_{2}B}}+|\phi^{-}\rangle_{{\rm AC_{1}}}|\phi^{-}\rangle_{{\rm C_{2}B}}\right),
\end{array}\label{eq:Chapter4_Eq7}
\end{equation}

\begin{equation}
\begin{array}{lcl}
|\psi_{3}\rangle & = & \frac{1}{\sqrt{2}}{\rm CNOT}{}_{{\rm C_{1}\rightarrow{\rm A}}}{\rm CNOT_{C_{2}\rightarrow B}}\left(|1\rangle_{{\rm A}}|\phi^{+}\rangle_{{\rm C_{1}C_{2}}}|0\rangle_{{\rm B}}\right)\\
 & = & \frac{1}{\sqrt{2}}\left(|\psi^{+}\rangle_{{\rm AC_{1}}}|\phi^{+}\rangle_{{\rm C_{2}B}}-|\psi^{-}\rangle_{{\rm AC_{1}}}|\phi^{-}\rangle_{{\rm C_{2}B}}\right),
\end{array}\label{eq:Chapter4_Eq8}
\end{equation}

\begin{equation}
\begin{array}{lcl}
|\psi_{4}\rangle & = & \frac{1}{\sqrt{2}}{\rm CNOT}{}_{{\rm C_{1}\rightarrow{\rm A}}}{\rm CNOT_{C_{2}\rightarrow B}}\left(|1\rangle_{{\rm A}}|\phi^{+}\rangle_{{\rm C_{1}C_{2}}}|1\rangle_{{\rm B}}\right)\\
 & = & \frac{1}{\sqrt{2}}\left(|\psi^{+}\rangle_{{\rm AC_{1}}}|\psi^{+}\rangle_{{\rm C_{2}B}}-|\psi^{-}\rangle_{{\rm AC_{1}}}|\psi^{-}\rangle_{{\rm C_{2}B}}\right),
\end{array}\label{eq:Chapter4_Eq9}
\end{equation}

\begin{equation}
\begin{array}{lcl}
|\psi_{5}\rangle & = & {\rm \frac{1}{\sqrt{2}}CNOT{}_{C_{1}\rightarrow{\rm A}}CNOT_{C_{2}\rightarrow B}}\left(|0\rangle_{{\rm A}}|\phi^{-}\rangle_{{\rm C_{1}C_{2}}}|0\rangle_{{\rm B}}\right)\\
 & = & \frac{1}{\sqrt{2}}\left(|\phi^{+}\rangle_{{\rm AC_{1}}}|\phi^{-}\rangle_{{\rm C_{2}B}}+|\phi^{-}\rangle_{{\rm AC_{1}}}|\phi^{+}\rangle_{{\rm C_{2}B}}\right),
\end{array}\label{eq:Chapter4_Eq10}
\end{equation}

\begin{equation}
\begin{array}{lcl}
|\psi_{6}\rangle & = & \frac{1}{\sqrt{2}}{\rm CNOT_{C_{1}\rightarrow{\rm A}}CNOT_{C_{2}\rightarrow B}}\left(|0\rangle_{{\rm A}}|\phi^{-}\rangle_{{\rm C_{1}C_{2}}}|1\rangle_{{\rm B}}\right)\\
 & = & \frac{1}{\sqrt{2}}\left(|\phi^{+}\rangle_{{\rm AC_{1}}}|\psi^{-}\rangle_{{\rm C_{2}B}}+|\phi^{-}\rangle_{{\rm AC_{1}}}|\psi^{+}\rangle_{{\rm C_{2}B}}\right),
\end{array}\label{eq:Chapter4_Eq11}
\end{equation}

\begin{equation}
\begin{array}{lcl}
|\psi_{7}\rangle & = & \frac{1}{\sqrt{2}}{\rm CNOT_{C_{1}\rightarrow{\rm A}}CNOT_{C_{2}\rightarrow B}}\left(|1\rangle_{{\rm A}}|\phi^{-}\rangle_{{\rm C_{1}C_{2}}}|0\rangle_{{\rm B}}\right)\\
 & = & \frac{1}{\sqrt{2}}\left(|\psi^{+}\rangle_{{\rm AC_{1}}}|\phi^{-}\rangle_{{\rm C_{2}B}}-|\psi^{-}\rangle_{{\rm AC_{1}}}|\phi^{+}\rangle_{{\rm C_{2}B}}\right),
\end{array}\label{eq:Chapter4_Eq12}
\end{equation}

\begin{equation}
\begin{array}{lcl}
|\psi_{8}\rangle & = & \frac{1}{\sqrt{2}}{\rm CNOT_{C_{1}\rightarrow{\rm A}}CNOT_{C_{2}\rightarrow B}}\left(|1\rangle_{{\rm A}}|\phi^{-}\rangle_{{\rm C_{1}C_{2}}}|1\rangle_{{\rm B}}\right)\\
 & = & \frac{1}{\sqrt{2}}\left(|\psi^{+}\rangle_{{\rm AC_{1}}}|\psi^{-}\rangle_{{\rm C_{2}B}}-|\psi^{-}\rangle_{{\rm AC_{1}}}|\psi^{+}\rangle_{{\rm C_{2}B}}\right).
\end{array}\label{eq:Chapter4_Eq13}
\end{equation}

\begin{center}
\begin{table}
\caption{The table below illustrates the relationship between the announcements
made by Charlie, Alice, and Bob, and the final key.}\label{tab:Chapter4_Tab1}

\centering{}%
\begin{tabular*}{15.9cm}{@{\extracolsep{\fill}}@{\extracolsep{\fill}}|>{\raggedright}p{1.8cm}|>{\raggedright}p{1.75cm}|>{\raggedright}p{1.75cm}|>{\raggedright}p{2.25cm}|>{\raggedright}p{2.25cm}|>{\raggedright}p{1.4cm}|>{\raggedright}p{1.15cm}|}
\hline 
\centering{}Announced bit by Charlie $(k_{C})$  & \centering{}Alice's Announced bit $(k_{A})$  & \centering{}Bob's Announced bit $(k_{B})$  & \centering{}Bob's guess of Alice's measurement outcome $(r_{A})$  & \centering{}Alice's guess of Bob's measurement outcome $(r_{B})$  & \begin{centering}
Value of 
\par\end{centering}
\centering{}$r_{A}\oplus r_{B}$  & \centering{}Final key after QKA $(K)$\tabularnewline
\hline 
\centering{}0  & \centering{}0  & \centering{}0  & \centering{}00  & \centering{}00  & \centering{}00  & \centering{}0\tabularnewline
\centering{}  & \centering{}0  & \centering{}1  & \centering{}01  & \centering{}01  & \centering{}00  & \centering{}0\tabularnewline
\centering{}  & \centering{}0  & \centering{}1  & \centering{}00  & \centering{}10  & \centering{}10  & \centering{}1\tabularnewline
\centering{}  & \centering{}0  & \centering{}0  & \centering{}01  & \centering{}11  & \centering{}10  & \centering{}1\tabularnewline
\hline 
\centering{}  & \centering{}1  & \centering{}0  & \centering{}10  & \centering{}00  & \centering{}10  & \centering{}1\tabularnewline
\centering{}  & \centering{}1  & \centering{}1  & \centering{}11  & \centering{}01  & \centering{}10  & \centering{}1\tabularnewline
\centering{}  & \centering{}1  & \centering{}1  & \centering{}10  & \centering{}10  & \centering{}00  & \centering{}0\tabularnewline
\centering{}  & \centering{}1  & \centering{}0  & \centering{}11  & \centering{}11  & \centering{}00  & \centering{}0\tabularnewline
\hline 
\centering{}1  & \centering{}0  & \centering{}1  & \centering{}00  & \centering{}01  & \centering{}01  & \centering{}1\tabularnewline
\centering{}  & \centering{}0  & \centering{}0  & \centering{}01  & \centering{}00  & \centering{}01  & \centering{}1\tabularnewline
\centering{}  & \centering{}0  & \centering{}0  & \centering{}00  & \centering{}11  & \centering{}11  & \centering{}0\tabularnewline
\centering{}  & \centering{}0  & \centering{}1  & \centering{}01  & \centering{}10  & \centering{}11  & \centering{}0\tabularnewline
\hline 
\centering{}  & \centering{}1  & \centering{}1  & \centering{}10  & \centering{}01  & \centering{}11  & \centering{}0\tabularnewline
\centering{}  & \centering{}1  & \centering{}0  & \centering{}11  & \centering{}00  & \centering{}11  & \centering{}0\tabularnewline
\centering{}  & \centering{}1  & \centering{}0  & \centering{}10  & \centering{}11  & \centering{}01  & \centering{}1\tabularnewline
\centering{}  & \centering{}1  & \centering{}1  & \centering{}11  & \centering{}10  & \centering{}01  & \centering{}1\tabularnewline
\hline 
\end{tabular*}
\end{table}
\par\end{center}

\subsection*{Description of proposed Protocol 4.1}

\begin{figure}
\begin{centering}
\includegraphics[scale=0.4]{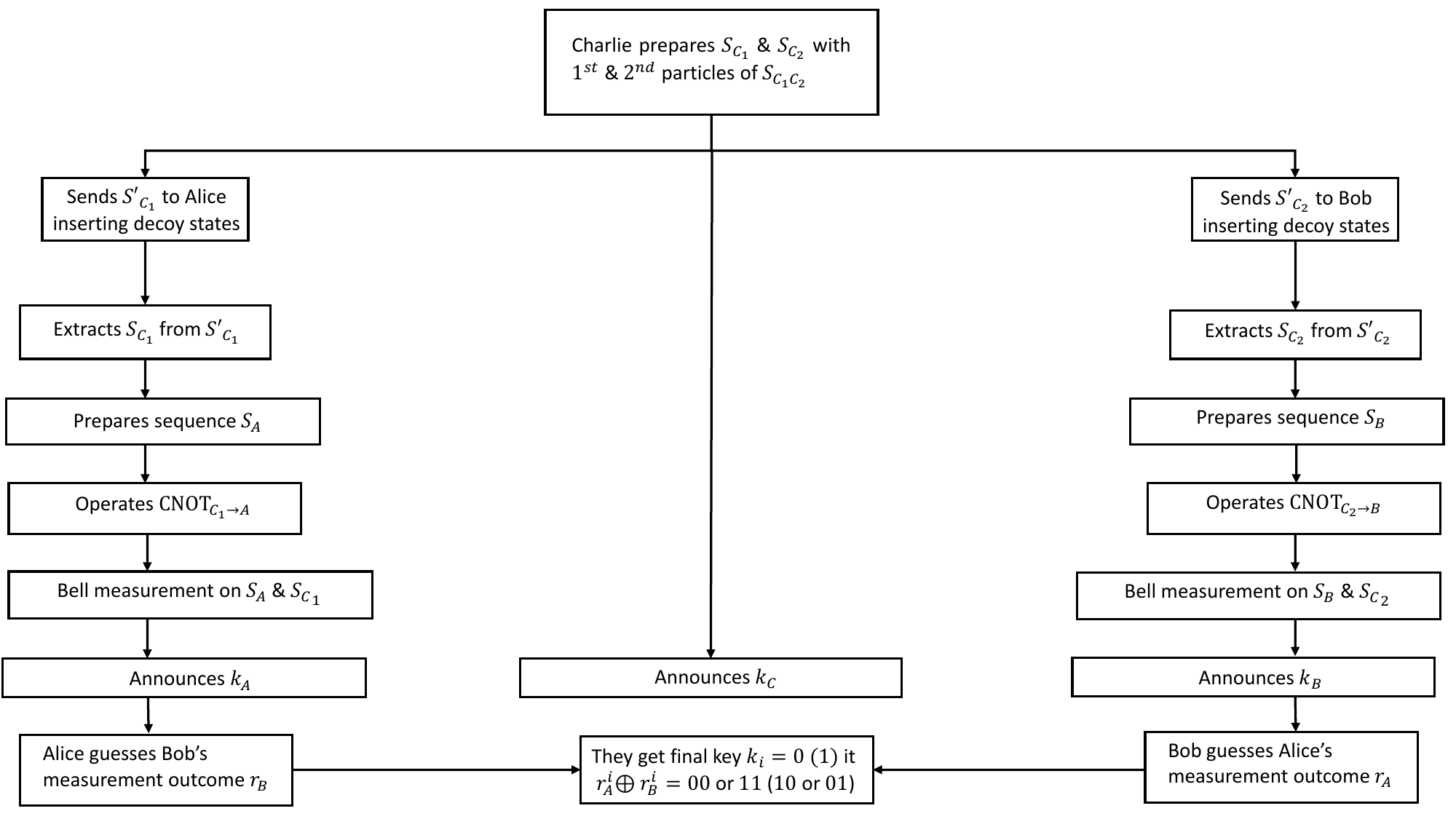} 
\par\end{centering}
\caption{The flowchart illustrates the operational framework of the proposed
CQKA protocol.}\label{fig:Chapter4_Fig1}
\end{figure}

The new CQKA protocol is outlined in a stepwise manner:
\begin{description}
\item [{Step~1}] Charlie generates a sequence of $n$ Bell states, denoted
as $\left\{ S_{C_{1}C_{2}}\right\} $, and partitions it into two
sequences: $S_{C_{1}}$, containing the first qubits, and $S_{C_{2}}$,
containing the second qubits of the Bell states.
\[
\begin{array}{c}
S_{C_{1}}=\left\{ \vert s\rangle_{C_{1}}^{1},|s\rangle_{C_{1}}^{2},|s\rangle_{C_{1}}^{3},\cdots,|s\rangle_{C_{1}}^{i},\cdots,|s\rangle_{C_{1}}^{n}\right\} ,\\
S_{C_{2}}=\left\{ \vert s\rangle_{C_{2}}^{1},|s\rangle_{C_{2}}^{2},|s\rangle_{C_{2}}^{3},\cdots,|s\rangle_{C_{2}}^{i},\cdots,|s\rangle_{C_{2}}^{n}\right\} .
\end{array}
\]
The subscripts ${\rm C_{1}}$ and ${\rm C_{2}}$ denote the first
and second qubits of Charlie\textquoteright s Bell state sequence
$S_{C_{1}C_{2}}$. Each Bell state in this sequence is either $|\phi^{+}\rangle$
or $|\phi^{-}\rangle$, while the superscript (e.g., $\left\{ 1,2,3,\cdots,i,\cdots,n\right\} $)
represents the positional index of the qubits in $S_{C_{1}}$ and
$S_{C_{2}}$.
\item [{Step~2}] Charlie randomly generates $2p$ decoy qubits using either
the computational basis ($Z$-basis: $\left\{ |0\rangle,|1\rangle\right\} $)
or the diagonal basis ($X$-basis: $\left\{ |+\rangle,|-\rangle\right\} $).
He inserts $p$ decoy qubits at arbitrary positions within $S_{C_{1}}$
and another $p$ within $S_{C_{2}}$, expanding the sequences to $S_{C_{1}}^{\prime}$
and $S_{C_{2}}^{\prime}$, respectively. Charlie then transmits $S_{C_{1}}^{\prime}$
to Alice and $S_{C_{2}}^{\prime}$ to Bob while retaining a classical
bit sequence $k_{C}$, which encodes the original Bell states in $S_{C_{1}C_{2}}$.\\
The correlation between $k_{C}$ and $S_{C_{1}C_{2}}$ adheres to
a predefined mathematical relation (Equation (\ref{eq:Chapter4_Eq1})).
\item [{Step~3}] Upon receiving sequences $S_{C_{1}}^{\prime}$ and $S_{C_{2}}^{\prime}$
from Charlie, Alice, and Bob conduct security verification using the
decoy qubits, extracting $S_{C_{1}}$ and $S_{C_{2}}$.\\
Charlie then publicly discloses the positions and encoding basis of
the decoy states, enabling Alice and Bob to assess channel security
by evaluating the error rate. If the observed error rate remains within
the acceptable threshold, they proceed.
\item [{Step~4}] Alice and Bob prepare their respective single-qubit sequences,
$S_{A}$ and $S_{B}$, each of length $n$, using the $Z$-basis.
Alice records the classical bit sequence$k_{A}$, which determines
$S_{A}$ based on a predefined relation (Equation (\ref{eq:Chapter4_Eq2})).
The sequences $S_{A}$ and $S_{B}$ can be visualized as follows:
\[
\begin{array}{lcl}
S_{A} & = & \left\{ |s\rangle_{A}^{1},|s\rangle_{A}^{2},|s\rangle_{A}^{3},\cdots,|s\rangle_{A}^{i},\cdots,|s\rangle_{A}^{n}\right\} ,\\
S_{B} & = & \left\{ |s\rangle_{B}^{1},|s\rangle_{B}^{2},|s\rangle_{B}^{3},\cdots,|s\rangle_{B}^{i},\cdots,|s\rangle_{B}^{n}\right\} .
\end{array}
\]
\item [{Step~5}] Alice and Bob each apply a CNOT operation, where the
qubits in $S_{A}$ and $S_{B}$ act as target qubits, while the corresponding
qubits in $S_{C_{1}}$ and $S_{C_{2}}$ serve as control qubits, respectively.
This operation establishes a correlated two-qubit system, forming
the pairs $({\rm C_{1},A)}$ for Alice and ${\rm (C_{2},B})$ for
Bob.
\item [{Step~6}] Both Alice and Bob conduct Bell-state measurements on
their respective two-qubit systems, obtaining results $r_{A}$ and
$r_{B}$, as determined by Equation (\ref{eq:Chapter4_Eq4}).\\
These results remain private. Additionally, Bob generates a classical
bit sequence $k_{B}$ based on his measurement outcomes, utilizing
the correlation rule specified in Equation (\ref{eq:Chapter4_Eq3}).
\item [{Step~7}] Alice and Bob then request Charlie to disclose his classical
bit sequence $k_{C}$ publicly.
\item [{Step~8}] Following Charlie\textquoteright s announcement, Alice
and Bob reveal their respective classical bit sequences $k_{A}$ and
$k_{B}$ publicly.
\item [{Step~9}] Using the mapping outlined in Table \ref{tab:Chapter4_Tab1},
Alice predicts Bob\textquoteright s bit sequence $r_{B}$, while Bob
predicts Alice\textquoteright s bit sequence $r_{A}$. They then compute
the modulo-2 addition of the corresponding elements in $r_{A}$ and
$r_{B}$. If the result is either 00 or 11, the final key bit is $K_{i}=0$;
if it is 01 or 10, then $K_{i}=1$.
\end{description}
To verify key consistency, Alice and Bob compare a subset of the final
key $K_{i}$. If the mismatch probability or error rate exceeds the
acceptable threshold, the protocol is aborted. Otherwise, the final
key is accepted for secure communication after post-processing steps,
including error correction and privacy amplification. A flowchart
representation of the CQKA protocol is provided in Figure \ref{fig:Chapter4_Fig1}.

In this section, we introduce an alternative approach to our CQKA
protocol by removing the controller\textquoteright s role, thereby
transforming it into a two-party QKA protocol. This modification enhances
feasibility and reduces noise compared to the original scheme. A detailed
comparison of these two protocols, analyzing their advantages and
limitations alongside previously proposed schemes, is provided in
Section \ref{sec:Chapter4_Sec5}. The choice between these protocols
depends on the specific requirements of different scenarios. To simplify
the description, the modified two-party QKA protocol is outlined in
the following steps without incorporating the decoy state. The mapping
of Alice and Bob\textquoteright s announcements and the final outcome
are presented in Table \ref{tab:Chapter4_Tab2}.

\subsection*{ Description of proposed QKA protocol (Protocol 4.2)}
\begin{description}
\item [{Step~1}] Alice begins by generating $n$ Bell states, denoted
as $\left\{ S_{C_{1}C_{2}}\right\} $, and divides them into two sequences:
$S_{C_{1}}$, containing the first qubits of the Bell states, and
$S_{C_{2}}$, containing the second qubits. Additionally, she prepares
a sequence of $n$ single qubits, $S_{A})$, in the $Z$-basis, ensuring
that the bit values correspond to $k_{C}$ and $k_{A}$, maintaining
their equivalence \footnote{The relations established in Equations (\ref{eq:Chapter4_Eq1},\ref{eq:Chapter4_Eq2}
and \ref{eq:Chapter4_Eq3}) are also applicable to proposed QKA protocol.}. The sequences prepared by Alice are as follows:
\[
\begin{array}{c}
S_{C_{1}}=\left\{ \vert s\rangle_{C_{1}}^{1},|s\rangle_{C_{1}}^{2},|s\rangle_{C_{1}}^{3},\cdots,|s\rangle_{C_{1}}^{i},\cdots,|s\rangle_{C_{1}}^{n}\right\} ,\\
S_{C_{2}}=\left\{ \vert s\rangle_{C_{2}}^{1},|s\rangle_{C_{2}}^{2},|s\rangle_{C_{2}}^{3},\cdots,|s\rangle_{C_{2}}^{i},\cdots,|s\rangle_{C_{2}}^{n}\right\} ,\\
S_{A}=\left\{ |s\rangle_{A}^{1},|s\rangle_{A}^{2},|s\rangle_{A}^{3},\cdots,|s\rangle_{A}^{i},\cdots,|s\rangle_{A}^{n}\right\} .
\end{array}
\]
\item [{Step~2}] Alice retains the $S_{C_{1}}$ sequence and transmits
$S_{C_{2}}$ to Bob.
\item [{Step~3}] Simultaneously, Bob independently generates a sequence
of $n$ single qubits, $S_{B}$, in the $Z$-basis, represented as:
\[
S_{B}=\left\{ |s\rangle_{B}^{1},|s\rangle_{B}^{2},|s\rangle_{B}^{3},\cdots,|s\rangle_{B}^{i},\cdots,|s\rangle_{B}^{n}\right\} .
\]
\item [{Step~4}] This step follows Step 5 of CQKA protocol.
\item [{Step~5}] This step follows Step 6 of CQKA protocol.
\item [{Step~6}] Since$k_{A}$ and $k_{C}$ are identical, Alice and Bob
publicly disclose their respective sequences, $k_{A}$ and $k_{B}$.
\item [{Step~7}] This step follows Step 9 of CQKA protocol.
\end{description}
\begin{center}
\begin{table}
\caption{The table below illustrates the relationship between the announcements
made by Alice, Bob, and the final key.}\label{tab:Chapter4_Tab2}

\centering{}%
\begin{tabular*}{15.9cm}{@{\extracolsep{\fill}}@{\extracolsep{\fill}}|>{\raggedright}p{2.5cm}|>{\raggedright}p{2.5cm}|>{\raggedright}p{2.25cm}|>{\raggedright}p{2.25cm}|>{\raggedright}p{1.75cm}|>{\raggedright}p{1.75cm}|}
\hline 
\centering{} Alice's Announced bit $(k_{A})$  & \centering{}Bob's Announced bit $(k_{B})$  & \centering{}Bob's guess of Alice's measurement outcome $(r_{A})$  & \centering{} Alice's guess of Bob's measurement outcome $(r_{B})$  & \begin{centering}
Value of 
\par\end{centering}
\centering{}$r_{A}\oplus r_{B}$  & \centering{}Final key after QKA $(K)$\tabularnewline
\hline 
\centering{}0  & \centering{}0  & \centering{}00  & \centering{}00  & \centering{}00  & \centering{}0\tabularnewline
\centering{}0  & \centering{}1  & \centering{}01  & \centering{}01  & \centering{}00  & \centering{}0\tabularnewline
\centering{}0  & \centering{}1  & \centering{}00  & \centering{}10  & \centering{}10  & \centering{}1\tabularnewline
\centering{}0  & \centering{}0  & \centering{}01  & \centering{}11  & \centering{}10  & \centering{}1\tabularnewline
\hline 
\centering{}1  & \centering{}1  & \centering{}10  & \centering{}01  & \centering{}11  & \centering{}0\tabularnewline
\centering{}1  & \centering{}0  & \centering{}11  & \centering{}00  & \centering{}11  & \centering{}0\tabularnewline
\centering{}1  & \centering{}0  & \centering{}10  & \centering{}11  & \centering{}01  & \centering{}1\tabularnewline
\centering{}1  & \centering{}1  & \centering{}11  & \centering{}10  & \centering{}01  & \centering{}1\tabularnewline
\hline 
\end{tabular*}
\end{table}
\par\end{center}

\section{Security analysis of the proposed protocols}\label{sec:Chapter4_Sec3}

Security plays a crucial role not only in QKA protocols but in all
quantum communication protocols. This section examines the security
of the proposed CQKA protocol against impersonation attacks and general
collective attacks. The same security analysis is also applicable
to the proposed QKA protocol. Additionally, we demonstrate that our protocols
remain secure even with a relatively small final agreement key length
$({\rm n=6})$.

\subsection{Security analysis against impersonation fraudulent attack}

To evaluate the resilience of our protocol against impersonation attempts,
we consider a scenario where an adversary, Eve, attempts to pose as
Charlie by employing a generalized two-qubit system. For simplicity,
we disregard the security enhancements provided by decoy qubits. In
Step 2 of our protocol, an impersonating Eve transmits the first and
second qubits of her generated two-qubit system to Alice and Bob,
mimicking Charlie\textquoteright s role. We first analyze the case
where Alice and Bob select the same state $|0\rangle$ as an element
of their respective sequences, $S_{A}$ and $S_{B}$, and proceed
with the subsequent protocol steps. After both parties perform the
CNOT operation, the resulting quantum state can be expressed as follows:

\begin{equation}
\begin{array}{lcl}
|\Phi_{1}\rangle & = & {\rm CNOT_{C_{1}\rightarrow A}CNOT_{C_{2}\rightarrow B}}\left(|0\rangle_{{\rm A}}\left({\rm {\rm a}|00\rangle+b|01\rangle+c|10\rangle+d|11\rangle}\right)_{{\rm C_{1}C_{2}}}|0\rangle_{{\rm B}}\right)\\
 & = & \left({\rm a|0\rangle|00\rangle|0\rangle+b|0\rangle|01\rangle|1\rangle+c|1\rangle|10\rangle|0\rangle+d|1\rangle|11\rangle|1\rangle}\right)_{{\rm AC_{1}C_{2}B}}.
\end{array}\label{eq:Chapter4_Eq14}
\end{equation}
Let ${\rm a,b,c}$, and ${\rm d}$ be the probability amplitudes associated
with the states $|00\rangle,|01\rangle,|10\rangle$, and $|11\rangle$
in Eve's two-qubit system\footnote{One can examine the well-established two-qubit states generated by
Eve and assess their resilience against an impersonation attack by
assigning specific values to the nonzero parameters ${\rm a,b,c}$,
and ${\rm d}$.}, constrained by the normalization condition ${\rm |a|^{2}+|b|^{2}+|c|^{2}+|d|^{2}=1}$.
Alice and Bob perform measurements on their respective two-qubit subsystems,
${\rm A,C_{1}}$ and ${\rm C_{2},B}$ in the Bell basis. To facilitate
the analysis of post-measurement outcomes, the above expression can
be rewritten in the Bell basis as follows:

\begin{equation}
\begin{array}{lcl}
|\Phi_{1}\rangle & = & \frac{1}{2}\left[{\rm a}\left(|\phi^{+}\rangle|\phi^{+}\rangle+|\phi^{+}\rangle|\phi^{-}\rangle+|\phi^{-}\rangle|\phi^{+}\rangle+|\phi^{-}\rangle|\phi^{-}\rangle\right)\right.\\
 & + & {\rm b\left(|\phi^{+}\rangle|\phi^{+}\rangle-|\phi^{+}\rangle|\phi^{-}\rangle+|\phi^{-}\rangle|\phi^{+}\rangle-|\phi^{-}\rangle|\phi^{-}\rangle\right)}\\
 & + & {\rm c}\left(|\phi^{+}\rangle|\phi^{+}\rangle+|\phi^{+}\rangle|\phi^{-}\rangle-|\phi^{-}\rangle|\phi^{+}\rangle-|\phi^{-}\rangle|\phi^{-}\rangle\right)\\
 & + & {\rm \left.d\left(|\phi^{+}\rangle|\phi^{+}\rangle-|\phi^{+}\rangle|\phi^{-}\rangle-|\phi^{-}\rangle|\phi^{+}\rangle+|\phi^{-}\rangle|\phi^{-}\rangle\right)\right]_{AC_{1}C_{2}B}}
\end{array}.\label{eq:Chapter4_Eq15}
\end{equation}
Equation (\ref{eq:Chapter4_Eq15}) represents the composite quantum
system involving Alice, Bob, and Eve, following an impersonation attack
where Eve generates a two-qubit state. The probability of detecting
Eve\textquoteright s presence can be determined by comparing with
Equation (\ref{eq:Chapter4_Eq7}), which describes the expected results
for Alice and Bob. After performing the necessary calculations, the
detection probability of Eve is derived as:

\begin{equation}
\begin{array}{lcl}
{\rm P_{|0\rangle|\phi^{+}\rangle|0\rangle}} & = & \frac{1}{2}\left[{\rm \left(|a|^{2}+|b|^{2}+|c|^{2}+|d|^{2}\right)-\left(a^{*}b+b^{*}c+c^{*}b+d^{*}a\right)}\right]\\
 & = & \frac{1}{2}\left[{\rm 1-\left(a^{*}b+b^{*}c+c^{*}b+d^{*}a\right)}\right],
\end{array}\label{eq:Chapter4_Eq116}
\end{equation}
and similarly,

\begin{equation}
\begin{array}{lcl}
{\rm P_{|0\rangle|\phi^{-}\rangle|0\rangle}} & = & \frac{1}{2}\left[{\rm \left(|a|^{2}+|b|^{2}+|c|^{2}+|d|^{2}\right)+\left(a^{*}b+b^{*}c+c^{*}b+d^{*}a\right)}\right]\\
 & = & \frac{1}{2}\left[{\rm 1+\left(a^{*}b+b^{*}c+c^{*}b+d^{*}a\right)}\right].
\end{array}\label{eq:Chapter4_Eq17}
\end{equation}
Here, the subscripts $|0\rangle,|\phi^{+}\rangle$, and $|0\rangle$
correspond to quantum states prepared by Alice, Bob, and Charlie,
respectively. Evaluating additional scenarios---where Alice and Bob
generate alternative state combinations such as $|0\rangle_{{\rm A}},|1\rangle_{{\rm B}};|1\rangle_{{\rm A}},|0\rangle_{{\rm B}}$;
and $|1\rangle_{{\rm A}},|1\rangle_{{\rm B}}$---yields the corresponding
detection probabilities:

\begin{equation}
\begin{array}{lcl}
P_{|0\rangle|\phi^{-}\rangle|0\rangle}=P_{|0\rangle|\phi^{+}\rangle|1\rangle} & = & P_{|1\rangle|\phi^{+}\rangle|0\rangle}={\rm P_{|1\rangle|\phi^{+}\rangle|1\rangle},}\\
{\rm P_{|0\rangle|\phi^{-}\rangle|0\rangle}=P_{|0\rangle|\phi^{-}\rangle|1\rangle}} & = & P_{|1\rangle|\phi^{-}\rangle|0\rangle}={\rm P_{|1\rangle|\phi^{-}\rangle|1\rangle}.}
\end{array}\label{eq:Chapter4_Eq18}
\end{equation}
Thus, the probability of detecting Eve\textquoteright s intervention
for each transmission instance is given by:

\begin{equation}
\begin{array}{lcl}
{\rm P_{d}} & = & \frac{1}{8}\left[{\rm P_{|0\rangle|\phi^{+}\rangle|0\rangle}}+{\rm P_{|0\rangle|\phi^{+}\rangle|1\rangle}+{\rm P_{|1\rangle|\phi^{+}\rangle|0\rangle}+{\rm P_{|1\rangle|\phi^{+}\rangle|1\rangle}}}}\right.\\
 & + & \left.{\rm P_{|0\rangle|\phi^{-}\rangle|0\rangle}}+{\rm P_{|0\rangle|\phi^{-}\rangle|1\rangle}+{\rm P_{|1\rangle|\phi^{-}\rangle|0\rangle}+{\rm P_{|1\rangle|\phi^{-}\rangle|1\rangle}}}}\right]\\
 & = & \frac{1}{2}\left({\rm |a|^{2}+|b|^{2}+|c|^{2}+|d|^{2}}\right)\\
 & = & \frac{1}{2}.
\end{array}\label{eq:Chapter4_Eq19}
\end{equation}
A detection probability of $\frac{1}{2}$ signifies the highest level
of statistical randomness attainable for identifying an adversary,
thereby ensuring protocol security. Mathematically, this means that
in each message transmission, there is a $0.5$ probability of detecting
an attacker (Eve) attempting to manipulate or forge a message. As
established by Simmons\textquoteright{} theorem \cite{S_88}, a secure
protocol ensures that an adversary cannot decrease the detection probability
below the randomness threshold of $\frac{1}{2}$ without being noticed.
This threshold corresponds to the maximum entropy in a binary detection
scenario (i.e., detected or undetected), preventing Eve from deterministically
avoiding detection. If Eve engages in $n$ independent tampering attempts
across $n$ transmissions, the probability of evading detection is
$\left(\frac{1}{2}\right)^{n}$, which declines exponentially as $n$
grows, making detection almost inevitable with repeated attempts.
Consequently, the inherent randomness of a $\frac{1}{2}$ detection
probability ensures unpredictability for the adversary and satisfies
Simmons's security criteria, providing a strong defense against message
tampering and forgery. Utilizing Simmons\textquoteright s theory \cite{S_88},
the proposed protocol achieves unconditional security against impersonation-based
fraudulent attacks, as demonstrated in Equation (\ref{eq:Chapter4_Eq19}).

\subsection{Security analysis against collective attack strategies in two-way
channel}

Secure quantum communication schemes are susceptible to three primary
classes of attacks by an eavesdropper: (i) Individual attacks, (ii)
Collective attacks, and (iii) Coherent attacks. Among these, individual
attacks are the least effective. In this scenario, Eve interacts separately
with each of Alice\textquoteright s signal systems by coupling them
with an auxiliary system and applying a fixed unitary transformation.
She then measures each system independently immediately after the
sifting step, before Alice and Bob engage in classical post-processing.
Collective attacks \cite{BM97,BM+97} share similarities with individual
attacks but allow Eve to postpone her measurement until the protocol
concludes. In this case, Eve\textquoteright s measurement strategy
can be influenced by the messages exchanged during error correction
and privacy amplification, and she can conduct joint measurements
on all auxiliary systems. This capability grants her greater power
compared to an eavesdropper restricted to individual attacks. Assessing
a protocol\textquoteright s resilience against collective attacks
is a crucial step toward proving security against coherent attacks,
which represent the most general and advanced adversarial strategy.
This section examines the protocol's security under collective attacks
to determine the tolerable error threshold within the security bounds.
We first outline Eve\textquoteright s attack methodology in a two-way
quantum channel, incorporating non-orthogonal ancilla states, and
computing the probability of detecting her presence. Additionally, we
evaluate Eve\textquoteright s probability of gaining information by
considering key parameters such as her detection probability and the
final key information. To establish a secure bound, we employ the
asymptotic secret key rate formulated by Devetak and Winter \cite{DW05}
for collective attacks, referencing secure bounds from \cite{RGK_05}.

\begin{figure}[h]
\begin{centering}
\includegraphics[scale=0.5]{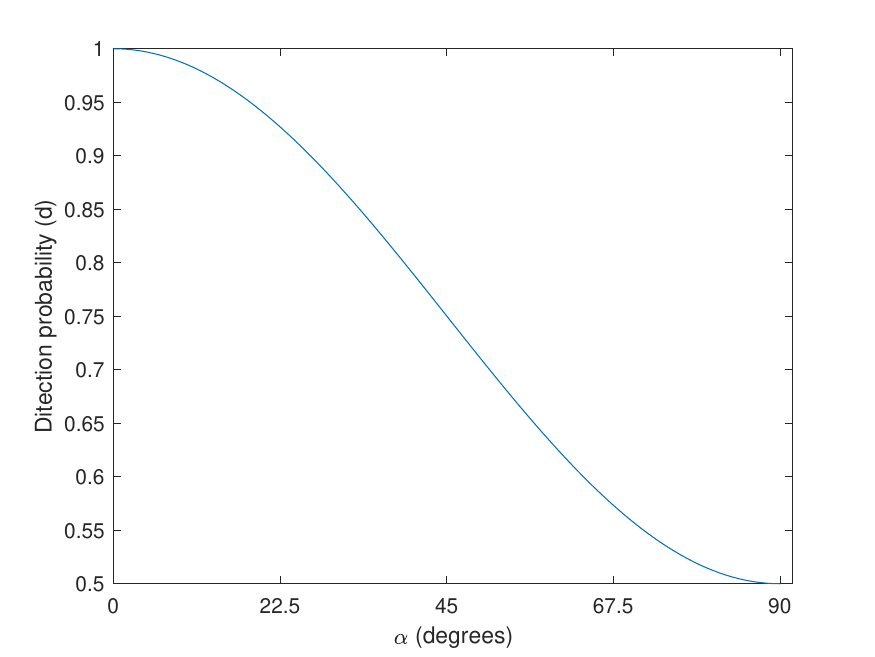} 
\par\end{centering}
\caption{The correlation between Eve\textquoteright s detection probability
$({\rm d})$ and the parameter $\alpha$, which characterizes non-orthogonality.}\label{fig:Chapter4_Fig2}
\end{figure}

In the proposed CQKA protocol, two quantum channels are established:
the Charlie-Alice (CA) channel and the Charlie-Bob (CB) channel. Suppose
an adversary, Eve, attempts to exploit these channels to extract the
final key information while minimizing the probability of detection.
To execute this attack, Eve prepares two ancillary qubits, denoted
as $|\zeta\rangle$ and $|\eta\rangle$, which together form a two-qubit
system specifically designed to target the CA and CB channels. Eve
begins by intercepting a qubit, ${\rm C_{1}}$, from the CA channel
and applies an entangling operation, ${\rm E_{C_{1}}}$, which entangles
${\rm C_{1}}$ with the ``ancillary'' qubit $|\zeta\rangle$. After
completing this operation, she forwards the qubit to Alice. Similarly,
she intercepts another qubit, ${\rm C_{2}}$, from the CB channel
and applies a corresponding entangling operation, ${\rm E_{C_{2}}}$,
to entangle ${\rm C_{2}}$ with $|\mathcal{\eta}\rangle$,
before sending the qubit onward. By analyzing the measurement outcomes
of $|\zeta\rangle$ and $|\eta\rangle$, Eve aims to reconstruct the
final key. The entangling operation ${\rm E_{C_{1}}}$ applied to
the qubit in the CA channel can be represented by a general mathematical
formulation,

\begin{equation}
{\rm \begin{array}{lcl}
|0_{C_{1}}\zeta\rangle & \stackrel{{\rm E_{C_{1}}}}{\longrightarrow} & A_{\zeta}|0_{C_{1}}\zeta_{00}\rangle+B_{\zeta}|1_{C_{1}}\zeta_{01}\rangle,\\
|1_{C_{1}}\zeta\rangle & \stackrel{{\rm E_{C_{1}}}}{\longrightarrow} & B_{\zeta}|0_{C_{1}}\zeta_{10}\rangle+A_{\zeta}|1_{C_{1}}\zeta_{11}\rangle.
\end{array}}\label{eq:Chapter4_Eq20}
\end{equation}
Here, subscripts ${\rm C_{1}}$ and ${\rm C_{2}}$ denote the respective
qubits in the CA and CB channels. Similarly, the entangling operation
${\rm E_{C_{2}}}$ on the CB channel follows a comparable expression.

\begin{equation}
{\rm \begin{array}{lcl}
|0_{C_{2}}\eta\rangle & \stackrel{{\rm E_{C_{2}}}}{\longrightarrow} & A_{\eta}|0_{C_{2}}\eta_{00}\rangle+B_{\eta}|1_{C_{2}}\eta_{01}\rangle,\\
|1_{C_{2}}\eta\rangle & \stackrel{{\rm E_{C_{2}}}}{\longrightarrow} & B_{\eta}|0_{C_{2}}\eta_{10}\rangle+A_{\eta}|1_{C_{2}}\eta_{11}\rangle.
\end{array}}\label{eq:Chapter4_Eq21}
\end{equation}
The unitary nature of operators ${\rm E_{C_{1}}}$ and ${\rm E_{C_{2}}}$
imposes the constraints ${\rm A_{\zeta}^{2}+B_{\zeta}^{2}=1}$ and
${\rm A_{\eta}^{2}+B_{\eta}^{2}=1}$, along with the conditions $\langle\zeta_{00}|\zeta_{10}\rangle+\langle\zeta_{01}|\zeta_{11}\rangle=0$
and $\langle\eta_{00}|\eta_{10}\rangle+\langle\eta_{01}|\eta_{11}\rangle=0$.
To simplify the analysis, a set of assumptions is introduced: $\langle\zeta_{00}|\zeta_{01}\rangle=\langle\zeta_{00}|\zeta_{10}\rangle=\langle\zeta_{10}|\zeta_{11}\rangle=\langle\zeta_{01}|\zeta_{11}\rangle=0$
and $\langle\eta_{00}|\eta_{01}\rangle=\langle\eta_{00}|\eta_{10}\rangle=\langle\eta_{10}|\eta_{11}\rangle=\langle\eta_{01}|\eta_{11}\rangle=0$.
Although these assumptions restrict the generality of Eve's attack,
they still represent a typical case following Eve\textquoteright s
entangling operations \cite{ZZZX_2006}. To characterize Eve's non-orthogonal
states, the relations can be expressed as $\langle\zeta_{00}|\zeta_{11}\rangle=\cos\alpha_{\zeta},\langle\zeta_{01}|\zeta_{10}\rangle=\cos\beta_{\zeta},\langle\eta_{00}|\eta_{11}\rangle=\cos\alpha_{\eta},\langle\eta_{01}|\eta_{10}\rangle=\cos\beta_{\eta}$.
By applying these expressions to the condition represented in Equation
(\ref{eq:Chapter4_Eq6}), the probability of Eve's undetected presence
can be determined. The resulting composite system after Eve's interference
is then analyzed accordingly.

\begin{equation}
\begin{array}{lcl}
|\psi_{2}^{e'}\rangle & = & {\rm E_{C_{2}}}\left[{\rm E_{C_{1}}}\left({\rm |0\rangle_{A}|\phi^{+}\rangle_{{\rm C_{1}C_{2}}}|1\rangle_{{\rm B}}}\right)|\zeta\rangle\right]|\eta\rangle\\
 & = & {\rm E_{C_{2}}}\left[{\rm E_{C_{1}}}\frac{1}{\sqrt{2}}\left(|0\rangle|00\rangle|1\rangle+|0\rangle|11\rangle|1\rangle\right)_{{\rm AC_{1}C_{2}B}}|\zeta\rangle\right]|\eta\rangle\\
 & = & \frac{1}{\sqrt{2}}\left[\left({\rm A_{\zeta}A_{\eta}}|0\rangle|00\rangle|1\rangle|\zeta_{00}\rangle|\eta_{00}\rangle+{\rm A_{\zeta}B_{\eta}}|0\rangle|01\rangle|0\rangle|\zeta_{00}\rangle|\eta_{01}\rangle\right.\right.\\
 & + & \left.{\rm B_{\zeta}A_{\eta}}|1\rangle|10\rangle|1\rangle|\zeta_{01}\rangle|\eta_{00}\rangle+{\rm B_{\zeta}B_{\eta}}|1\rangle|11\rangle|0\rangle|\zeta_{01}\rangle|\eta_{01}\rangle\right)\\
 & + & \left({\rm A_{\zeta}A_{\eta}}|1\rangle|11\rangle|0\rangle|\zeta_{11}\rangle|\eta_{11}\rangle+{\rm A_{\zeta}B_{\eta}}|1\rangle|10\rangle|1\rangle|\zeta_{11}\rangle|\eta_{10}\rangle\right.\\
 & + & {\rm \left.\left.B_{\zeta}A_{\eta}|0\rangle|01\rangle|0\rangle|\zeta_{10}\rangle|\eta_{11}\rangle+{\rm B_{\zeta}B_{\eta}}|0\rangle|00\rangle|1\rangle|\zeta_{10}\rangle|\eta_{10}\rangle\right)_{AC_{1}C_{2}B}\right]}.
\end{array}\label{eq:Chapter4_Eq22}
\end{equation}
Alice and Bob receive their respective qubits from Charlie after Eve
interacts with the channel particle and her \textit{ancillary} particle.
Following the protocol, they execute the CNOT operation and subsequently
perform Bell measurements on their two-particle system. The operations
performed and the possible measurement results obtained by Alice and
Bob can be described by the following equation.

\begin{equation}
\begin{array}{lcl}
|\psi_{2}^{e}\rangle & = & {\rm CNOT_{C_{1}\rightarrow A}CNOT_{C_{2}\rightarrow B}}|\psi_{2}^{e'}\rangle\\
 & = & \frac{1}{2\sqrt{2}}\left[\left\{ {\rm A_{\zeta}A_{\eta}\left(|\phi^{+}\rangle|\psi^{+}\rangle+|\phi^{+}\rangle|\psi^{-}\rangle+|\phi^{-}\rangle|\psi^{+}\rangle+|\phi^{-}\rangle|\psi^{-}\rangle\right)_{AC_{1}C_{2}B}|\zeta_{00}\rangle|\eta_{00}}\rangle\right.\right.\\
 & + & {\rm A_{\zeta}B_{\eta}\left(|\phi^{+}\rangle|\psi^{+}\rangle-|\phi^{+}\rangle|\psi^{-}\rangle+|\phi^{-}\rangle|\psi^{+}\rangle-|\phi^{-}\rangle|\psi^{-}\rangle\right)_{AC_{1}C_{2}B}|\zeta_{00}\rangle|\eta_{01}}\rangle\\
 & + & {\rm B_{\zeta}A_{\eta}}\left(|\phi^{+}\rangle|\psi^{+}\rangle+|\phi^{+}\rangle|\psi^{-}\rangle-|\phi^{-}\rangle|\psi^{+}\rangle-|\phi^{-}\rangle|\psi^{-}\rangle\right)_{{\rm AC_{1}C_{2}B}}|\zeta_{01}\rangle|\eta_{00}\rangle\\
 & + & \left.{\rm B_{\zeta}B_{\eta}}\left(|\phi^{+}\rangle|\psi^{+}\rangle-|\phi^{+}\rangle|\psi^{-}\rangle-|\phi^{-}\rangle|\psi^{+}\rangle+|\phi^{-}\rangle|\psi^{-}\rangle\right)_{{\rm AC_{{\rm 1}}C_{2}B}}|\zeta_{01}\rangle|\eta_{01}\rangle\right\} \\
 & + & \left\{ {\rm A_{\zeta}A_{\eta}}\left(|\phi^{+}\rangle|\psi^{+}\rangle-|\phi^{+}\rangle|\psi^{-}\rangle-|\phi^{-}\rangle|\psi^{+}\rangle+|\phi^{-}\rangle|\psi^{-}\rangle\right)_{{\rm AC_{1}C_{2}B}}|\zeta_{11}\rangle|\eta_{11}\rangle\right.\\
 & + & {\rm A_{\zeta}B_{\eta}}\left(|\phi^{+}\rangle|\psi^{+}\rangle+|\phi^{+}\rangle|\psi^{-}\rangle-|\phi^{-}\rangle|\psi^{+}\rangle-|\phi^{-}\rangle|\psi^{-}\rangle\right)_{{\rm AC_{1}C_{2}B}}|\zeta_{11}\rangle|\eta_{10}\rangle\\
 & + & {\rm B_{\zeta}A_{\eta}}\left(|\phi^{+}\rangle|\psi^{+}\rangle-|\phi^{+}\rangle|\psi^{-}\rangle+|\phi^{-}\rangle|\psi^{+}\rangle-|\phi^{-}\rangle|\psi^{-}\rangle\right)_{{\rm AC_{1}C_{2}B}}|\zeta_{10}\rangle|\eta_{11}\rangle\\
 & + & \left.\left.{\rm B_{\zeta}B_{\eta}}\left(|\phi^{+}\rangle|\psi^{+}\rangle+|\phi^{+}\rangle|\psi^{-}\rangle+|\phi^{-}\rangle|\psi^{+}\rangle+|\phi^{-}\rangle|\psi^{-}\rangle\right)_{{\rm AC_{1}C_{2}B}}|\zeta_{10}\rangle|\eta_{10}\rangle\right\} \right].
\end{array}\label{eq:Chapter4_Eq23}
\end{equation}
Eve\textquoteright s presence can be detected when Alice and Bob obtain
the measurement results $|\phi^{+}\rangle$ and $|\psi^{-}\rangle$
or $|\phi^{-}\rangle$ and $|\psi^{+}\rangle$ after conducting Bell
measurements. The probability of detecting Eve in such cases is given
by:

\begin{equation}
\begin{array}{lcl}
{\rm P_{d}^{2}} & {\rm =} & {\rm \frac{1}{2}\left[A_{\zeta}^{2}A_{\eta}^{2}\left(1+\cos\alpha_{\zeta}\cos\alpha_{\eta}\right)+A_{\zeta}^{2}B_{\eta}^{2}\left(1+\cos\alpha_{\zeta}\cos\beta_{\eta}\right)\right.}\\
 & + & {\rm \left.B_{\zeta}^{2}A_{\eta}^{2}\left(1+\cos\beta_{\zeta}\cos\alpha_{\eta}\right)+B_{\zeta}^{2}B_{\eta}^{2}\left(1+\cos\beta_{\zeta}\cos\beta_{\eta}\right)\right].}
\end{array}\label{eq:Chapter4_Eq24}
\end{equation}
Here, ${\rm P_{d}^{2}}$ represents the probability of detecting Eve\textquoteright s
interference for a specific expected outcome, as defined by Equation
(\ref{eq:Chapter4_Eq6}). The superscript $i$ (in this case, 2) denotes
the $i^{th}$ (here, second) configuration among the various possible
state choices made by Alice, Bob, and Charlie, with an associated
probability ${\rm P}_{{\rm d}}^{i}$, where $i\in\{1,2,\cdots,8\}$.
The detection probability for Eve remains the same across different
measurement outcomes, such that ${\rm P_{d}^{1}=P_{d}^{2}=\cdots=P_{d}^{8}}$.
Using these results, one can determine the average probability of
detecting Eve\textquoteright s presence as follows:

\begin{equation}
\begin{array}{lcl}
{\rm P_{d}} & = & {\rm \frac{1}{8}}\sum_{i}{\rm P}_{{\rm d}}^{i}\\
 & = & \frac{1}{2}\left[A_{\zeta}^{2}A_{\eta}^{2}\left(1+\cos\alpha_{\zeta}\cos\alpha_{\eta}\right)+A_{\zeta}^{2}B_{\eta}^{2}\left(1+\cos\alpha_{\zeta}\cos\beta_{\eta}\right)\right.\\
 & + & {\rm \left.B_{\zeta}^{2}A_{\eta}^{2}\left(1+\cos\beta_{\zeta}\cos\alpha_{\eta}\right)+B_{\zeta}^{2}B_{\eta}^{2}\left(1+\cos\beta_{\zeta}\cos\beta_{\eta}\right)\right].}
\end{array}\label{eq:Chapter4_Eq25}
\end{equation}
Now, considering that Eve aims to minimize the probability of being
detected, she must employ a strategy to lower her detection probability
as much as possible. The optimal scenario for Eve is achieved when
the condition ${\rm A_{\zeta}=A_{\eta}=1}$ holds. Additionally, an
optimal incoherent attack by Eve is assumed to be balanced, satisfying
$\alpha_{\zeta}=\alpha_{\eta}=\alpha$ \cite{LM_05}. The ``minimal
probability of detecting'' Eve is denoted as ${\rm d}$, expressed
as:

\begin{equation}
\begin{array}{lcl}
{\rm d}\equiv{\rm min(P_{d})} & = & \frac{1}{2}\left(1+\cos^{2}\alpha\right)\end{array}\label{eq:Chapter4_Eq26}
\end{equation}
The relationship described by Equation (\ref{eq:Chapter4_Eq26}) is
illustrated in Figure \ref{fig:Chapter4_Fig2} demonstrating that
the ``minimal detection probability'' of Eve reaches zero if she
utilizes an orthogonal state\footnote{To simplify the analysis, we consider the optimal conditions that
Eve must satisfy to avoid complex calculations. A more generalized
assumption can be adopted, where ${\rm A_{\zeta}^{2}+B_{\zeta}^{2}=1}$,
${\rm A_{\eta}^{2}+B_{\eta}^{2}=1}$, $\alpha_{\zeta}\neq\alpha_{\eta}$
and $\beta_{\zeta}\neq\beta_{\eta}$. This allows us to further investigate
how the minimal detection probability (d) of Eve is influenced by
the non-orthogonality of her quantum state.}.

Next, we analyze the information Eve obtains about the final one-bit
key. Eve's attack operations denoted as ${\rm E_{C_{1}}}$ and ${\rm E_{C_{2}}}$,
target the qubits transmitted through the CA and CB channels, respectively,
following an attack strategy $\mathcal{E}$. Consequently, the mutual
information between Eve's attack strategy $\mathcal{E}$ and the key
$\mathcal{K}$ is given by:

\begin{equation}
I\left(\mathcal{K}:\mathcal{E}\right)=H\left(\mathcal{E}\right)-H\left(\mathcal{E}|\mathcal{K}\right),\label{eq:Chapter4_Eq27}
\end{equation}
where $H\left(\mathcal{E}\right)$ represents the entropy associated
with Eve's attack strategy, while $H\left(\mathcal{E}|\mathcal{K}\right)$
denotes the conditional entropy given the key. Under Eve\textquoteright s
optimal attack conditions, as previously discussed, the Equation (\ref{eq:Chapter4_Eq23})
can be further simplified.

\begin{equation}
\begin{array}{lcl}
|\psi_{2}^{e}\rangle & = & \frac{1}{2\sqrt{2}}\left[\left\{ {\rm \left(|\phi^{+}\rangle|\psi^{+}\rangle+|\phi^{+}\rangle|\psi^{-}\rangle+|\phi^{-}\rangle|\psi^{+}\rangle+|\phi^{-}\rangle|\psi^{-}\rangle\right)_{1234}|\zeta_{00}\rangle|\eta_{00}}\rangle\right\} \right.\\
 & + & \left.\left\{ \left(|\phi^{+}\rangle|\psi^{+}\rangle-|\phi^{+}\rangle|\psi^{-}\rangle-|\phi^{-}\rangle|\psi^{+}\rangle+|\phi^{-}\rangle|\psi^{-}\rangle\right)_{1234}|\zeta_{11}\rangle|\eta_{11}\rangle\right\} \right].
\end{array}\label{eq:Chapter4_Eq28}
\end{equation}
From this formulation, it follows that Eve's measurement outcome will
be either $|\zeta_{00}\rangle|\eta_{00}\rangle$ or $|\zeta_{11}\rangle|\eta_{11}\rangle$,
each occurring with equal probability, i.e., $\frac{1}{2}$. Considering
various state preparations by Charlie, Alice, and Bob (such as $|\psi_{1}^{e}\rangle,|\psi_{3}^{e}\rangle,\cdots,|\psi_{8}^{e}\rangle$),
Eve's measurement outcomes under attack strategy $\mathcal{E}$ will
still have a probability of $\frac{1}{2}$. The Shannon entropy of
$\mathcal{E}$ is given by $H\left(\mathcal{E}\right)=1$, while the
conditional entropy given $\mathcal{K}$ is expressed as $H\left(\mathcal{E}|\mathcal{K}\right)=h_{2}\left(Q_{\mathcal{E\mathcal{K}}}\right)$,
where $h_{2}$ represents the binary entropy function, and $Q_{\mathcal{E}\mathcal{K}}$
denotes the QBER introduced by Eve\textquoteright s attack in an attempt
to infer the one-bit key. Thus, we need to derive an expression for
$Q_{\mathcal{E}\mathcal{K}}$. The probability of correctly distinguishing
between two quantum states with scalar product $\cos\alpha$ is given
by $\frac{1}{2}\left(1+\sin\alpha\right)$ \cite{H1969,GRT+02}. Eve's
ability to infer the final key depends on her capability to differentiate
between the quantum states $\zeta$ and $\eta$. However, errors in
distinguishing these states could occur---either misidentifying $\zeta$
or $\eta$. A single error leads to an incorrect key interpretation,
but if two errors occur in a compensatory manner, Eve may still reconstruct
the correct key. The general expression for $Q_{\mathcal{E}\mathcal{K}}$,
incorporating $\alpha_{\zeta}$ and $\alpha_{\eta}$, is derived accordingly. 

\begin{equation}
\begin{array}{lcl}
Q_{\mathcal{E}\mathcal{K}} & =1- & \left[\left(\frac{1+\sin\alpha_{\zeta}}{2}\right)\left(\frac{1+\sin\alpha_{\eta}}{2}\right)+\left(\frac{1-\sin\alpha_{\zeta}}{2}\right)\left(\frac{1-\sin\alpha_{\eta}}{2}\right)\right]\\
 & = & \frac{1-\sin\alpha_{\zeta}\sin\alpha_{\eta}}{2}.
\end{array}\label{eq:Chapter4_Eq29}
\end{equation}
Assuming the optimal case for Eve, where $\alpha_{\zeta}=\alpha_{\eta}=\alpha$,
and utilizing Equation (\ref{eq:Chapter4_Eq27}) along with Equation
(\ref{eq:Chapter4_Eq26}), we obtain:

\begin{equation}
\begin{array}{lcl}
I\left(\mathcal{K}:\mathcal{E}\right) & = & \frac{1}{2}\left[1-h_{2}\left(\frac{1-\sin^{2}\alpha}{2}\right)\right].\end{array}\label{eq:Chapter4_Eq30}
\end{equation}
Here, the factor $\frac{1}{2}$ arises from the fact that in Equation
(\ref{eq:Chapter4_Eq28}), Alice and Bob\textquoteright s outcomes
remain identical and occur with equal probability, regardless of Eve\textquoteright s
measurement results.

\begin{figure}[h]
\begin{centering}
\includegraphics[scale=0.5]{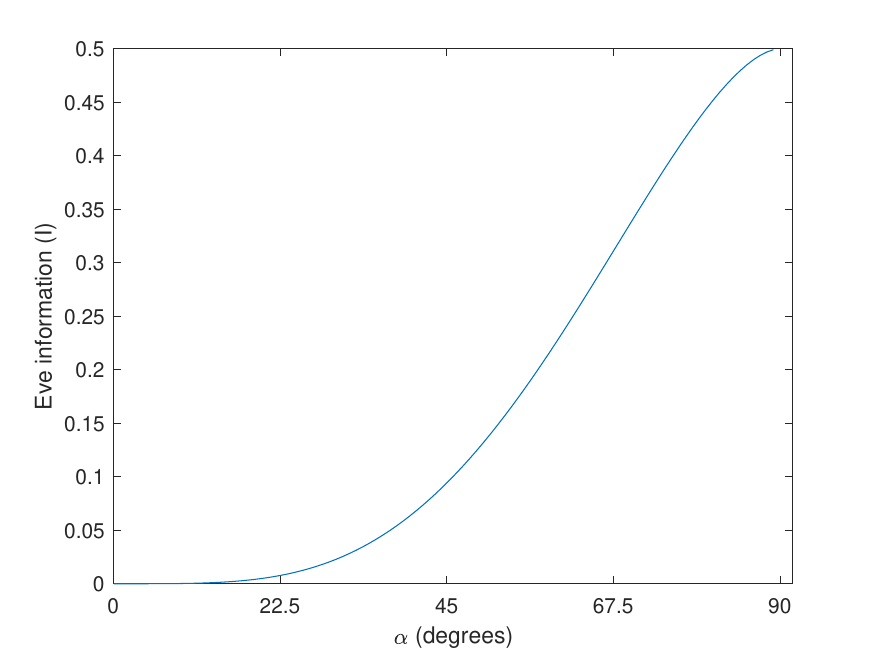} 
\par\end{centering}
\caption{The dependence of Eve\textquoteright s accessible information $({\rm I})$
on the angle ${\rm \alpha}$, representing non-orthogonality.}\label{fig:Chapter4_Fig3}
\end{figure}

Figure \ref{fig:Chapter4_Fig3} illustrates the relationship between
$I$ and $\alpha$ under optimal conditions for Eve. As observed in
the figure, when Eve employs orthogonal states ($\alpha=\frac{\pi}{2}$),
the maximum extractable information is limited to 0.5 bits. In this
scenario, Eve's detection probability is zero. However, if she utilizes
non-orthogonal states, her information gain remains suboptimal in
all cases\footnote{It is important to note that this conclusion would differ if we did
not assume$B_{\eta}=B_{\zeta}=0$ and $A_{\eta}=A_{\zeta}=1$. A more
generalized analysis could be conducted under these generalized assumptions.}. In practical quantum communication settings, noise may be introduced
when Eve applies a unitary transformation involving the ancilla state,
denoted by $A_{\eta}$ and $A_{\zeta}<1$. This factor complicates
Eve's ability to execute an attack under ideal conditions. Consequently,
Eve's presence becomes detectable, even when leveraging an orthogonal
basis. Furthermore, the use of decoy states in legitimate quantum
transmissions is a well-established method for improving channel security
(refer to the stepwise breakdown of the CQKA protocol). However, for simplicity,
our security analysis does not incorporate decoy states. If decoy
states were considered, Eve\textquoteright s detection probability
would increase compared to the scenario without them. In the next
section, we compute the tolerable error bound, demonstrating the resilience
of our protocol against collective attacks.

From Table\ref{tab:Chapter4_Tab1}, it follows that after executing
CQKA protocol, the final key agreement exhibits maximum entropy due
to classical announcements within our framework. Additionally, our
security analysis assumes a simplified model, leading to a symmetric
measurement output for Alice and Bob (refer to Equation (\ref{eq:Chapter4_Eq28}))
relative to Eve's ancillary state measurement. Despite this assumption,
we evaluate Eve\textquoteright s probability of successfully estimating
the final key in a generalized manner\footnote{Depending on the specific protocol and the nature of Eve\textquoteright s
collective attack, her success in estimating the final key can be
described in terms of detection probability rather than being treated
as a fixed value. Therefore, we adopt a more generalized analytical
approach.}.

Based on Eve's measurement results $|\zeta_{ij}\rangle|\eta_{kl}\rangle$
for$i,j,k,l\in\{0,1\}$, along with Charlie\textquoteright s announcement,
Eve attempts to infer the final one-bit key. To generalize, let us
assume Eve assigns the key as 0 with probability $e$ and as 1 with
probability\footnote{In our protocol, the final key distribution remains symmetric with
respect to Charlie\textquoteright s announcement, making the role
of $e$ particularly significant in our analysis.} $1-e$. Our objective is to determine the total probability of Eve
correctly estimating the final key.

\begin{equation}
\begin{array}{lcl}
{\rm Pr_{s}} & = & \left(\frac{1+\sin\alpha}{2}\right)^{2}\left\{ \frac{1}{2}e+\frac{1}{2}(1-e)\right\} +\left(\frac{1-\sin\alpha}{2}\right)^{2}\left\{ \frac{1}{2}e+\frac{1}{2}(1-e)\right\} \\
 & + & \frac{\left(1+\sin\alpha\right)\left(1-\sin\alpha\right)}{2}\left\{ \frac{1}{2}e+\frac{1}{2}(1-e)\right\} \\
 & = & \frac{1}{2}.
\end{array}\label{eq:Chapter4_Eq31}
\end{equation}
By employing Equation (\ref{eq:Chapter4_Eq26}) and Eq. (\ref{eq:Chapter4_Eq31}),
the probability of Eve successfully acquiring the final key denoted
as ${\rm n}$ (representing the number of final key bits), can be
determined,

\begin{equation}
\begin{array}{lcl}
{\rm Pr} & = & {\rm \left[Pr_{s}\left(1-d\right)\right]^{n}}\\
 & = & {\rm \left[\frac{1}{2}\left(1-d\right)\right]^{n}.}
\end{array}\label{eq:Chapter4_Eq32}
\end{equation}

\begin{figure}[h]
\begin{centering}
\includegraphics[scale=0.47]{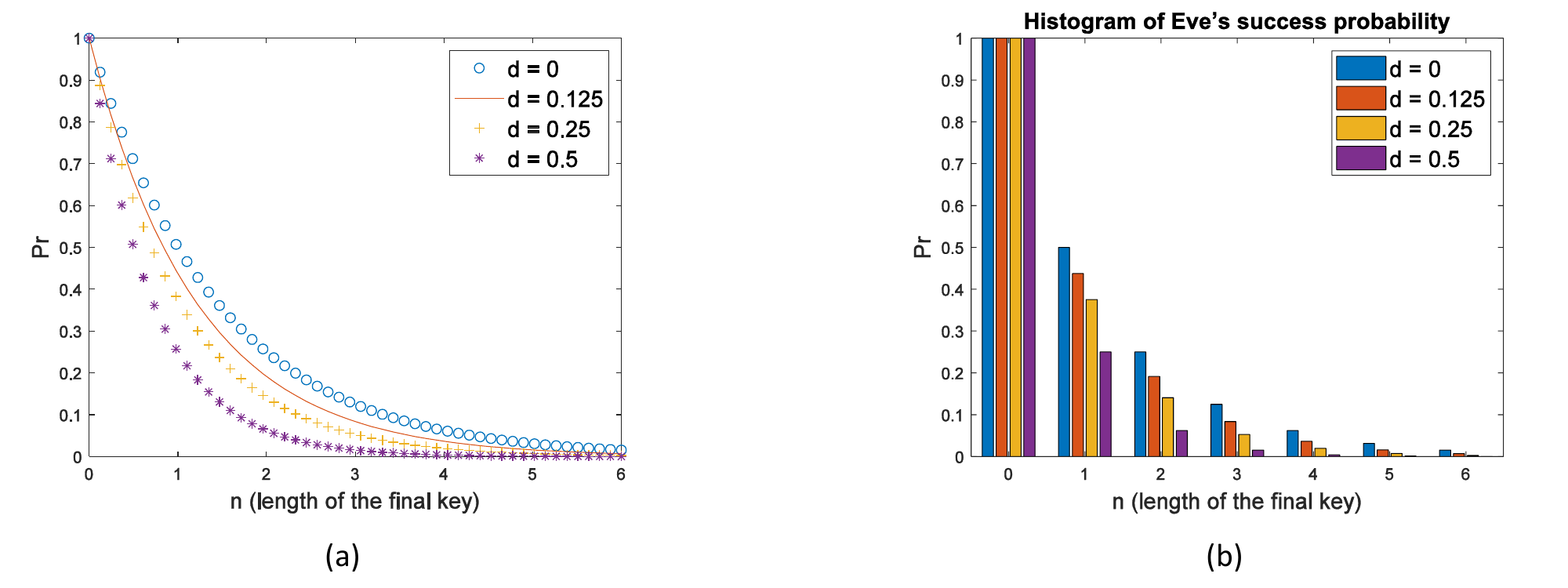} 
\par\end{centering}
\caption{The interrelation between ${\rm Pr,d}$ and ${\rm n}$ is depicted
as follows. (a) The dependence of ${\rm Pr}$ on ${\rm n}$ is examined
for varying values of ${\rm d}$. In particular, the ``circle''
line corresponds to ${\rm d=0\%}$, the ``continuous'' line represents
${\rm d=12.5\%}$, the ``plus sign'' line denotes ${\rm d=25\%}$,
and the ``asterisk'' line signifies ${\rm d=50\%}$. (b) A histogram
illustrates Eve's probability of success for ${\rm n=6}$ across different
values of ${\rm d}$.}\label{fig:Chapter4_Fig4}
\end{figure}

The correlation between ${\rm Pr}$ and ${\rm n}$ for varying values
of ${\rm d}$ is depicted in Figure\ref{fig:Chapter4_Fig4}. The plotted
results indicate that when ${\rm n=6}$, ${\rm Pr}$ is approximately
zero. For instance, if Eve\textquoteright s detection probability
(d) is set at $25\%$, the probability of her successfully obtaining
the final key is $2.629\times10^{-3}$, which is negligible. Figure
\ref{fig:Chapter4_Fig4}.a provides insight into the value of $n$
at which ${\rm Pr}$ converges to zero. Since $n$ is inherently discrete,
Figure \ref{fig:Chapter4_Fig4}.b presents a histogram illustrating
${\rm Pr}$ for integer values of ${\rm n}$. Clearly, as ${\rm n}$
increases sufficiently, the information gained by Eve through an impersonation-based
fraudulent attack becomes effectively nullified.

\textit{Bound for tolerable error under collective attack scenario}:
Due to the inherent symmetry in the composite system\textquoteright s
outcomes, we assume that the shared quantum state $|\psi_{2}^{e}\rangle$,
involving Alice, Bob, and Eve, follows the form given in Equation
(\ref{eq:Chapter4_Eq28}). In the scenario of a collective attack
\cite{PAB+09,DW05}, the asymptotic lower bound for the secret key
rate is governed by the Devetak-Winter rate equation. This equation
quantifies the difference between the mutual information shared between
Alice and Bob and the Holevo quantity describing Eve\textquoteright s
accessible information about Bob\textquoteright s data\footnote{In this framework, we evaluate the Holevo bound concerning Eve\textquoteright s
information on Bob\textquoteright s measurements. Classical one-way
post-processing is performed from Bob to Alice, directly influenced
by his measurement outcome after the Bell-state analysis.},

\begin{equation}
r\ge r_{{\rm DW}}=I\left(A:B\right)-\chi\left(B:E\right).\label{eq:Chapter4_Eq33}
\end{equation}
The secret key rate in the asymptotic limit, denoted as $r$, has
a lower bound $r_{DW}$, given by the Devetak-Winter rate equation.
In this context, the mutual information measure \cite{PAB+09} is
expressed as: $I\left(A:B\right)=H\left(B\right)-H\left(B|A\right)$
whereas the Holevo quantity is given by, $\chi\left(B:E\right)=S\left(\rho_{E}\right)-\underset{k_{B}=0,1}{\sum}p_{k_{B}}S\left(\rho_{E|k_{B}}\right)$.
Here, $S$ represents the von Neumann entropy, $\rho_{E}={\rm Tr}_{AB}|\psi_{2}^{e}\rangle\langle\psi_{2}^{e}|$
denotes Eve\textquoteright s quantum state after tracing out Alice's
and Bob\textquoteright s subsystems, and $\rho_{E|k_{B}}$ is Eve\textquoteright s
conditional quantum state given Bob\textquoteright s measurement outcome
$k_{B}$. Due to symmetry considerations, the probabilities of Bob\textquoteright s
final measurement outcomes are equal, i.e., $p_{0}=p_{1}=\frac{1}{2}$,
when incorporating errors. Assuming uniform error rates across Alice\textquoteright s
and Bob\textquoteright s measurement processes, we define this shared
error rate as $\epsilon$. Consequently, the mutual information simplifies
to $I\left(A:B\right)=1-h\left(\epsilon\right)$, where $h\left(\epsilon\right)$
denotes the binary entropy function. By utilizing Equations (\ref{eq:Chapter4_Eq6})
and (\ref{eq:Chapter4_Eq28}) can be generalized\footnote{The contribution of ancillary states ($\eta$) is omitted, as they
do not provide additional information under the optimal conditions
$B_{\eta}=B_{\zeta}=0$ and $A_{\eta}=A_{\zeta}=1$, maximizing Eve\textquoteright s
limitations in extracting key information.} while accounting for the error probability $\epsilon$, which encapsulates
the combined impact of quantum channel imperfections and potential
eavesdropping \cite{LF+11}.

\begin{equation}
\begin{array}{lcl}
|\psi_{2}^{e}\rangle & = & \sqrt{\frac{1-\epsilon}{4}}|\phi^{+}\rangle_{{\rm AC_{1}}}|\psi^{+}\rangle_{{\rm C}_{2}B}\left(|\zeta_{00}\rangle+|\zeta_{11}\rangle\right)+\sqrt{\frac{\epsilon}{4}}|\phi^{+}\rangle_{{\rm AC_{1}}}|\psi^{-}\rangle_{{\rm C_{2}B}}\left(|\zeta_{00}\rangle-|\zeta_{11}\rangle\right)\\
 & + & \sqrt{\frac{\epsilon}{4}}|\phi^{-}\rangle_{{\rm AC_{1}}}|\psi^{+}\rangle_{{\rm C_{2}B}}\left(|\zeta_{00}\rangle-|\zeta_{11}\rangle\right)+\sqrt{\frac{1-\epsilon}{4}}|\phi^{-}\rangle_{{\rm AC_{1}}}|\psi^{-}\rangle_{{\rm C_{2}B}}\left(|\zeta_{00}\rangle+|\zeta_{11}\rangle\right)
\end{array}.\label{eq:Chapter4_Eq34}
\end{equation}
By performing Bell basis measurements on the systems linked to Alice
and Bob, one can directly ascertain the state of Eve's system, which
can be represented as $|\theta^{\phi^{\pm},k_{B}}\rangle$, where
$\phi^{\pm}$ and $k_{B}$ denote Alice and Bob's respective outcomes.
The explicit form of Eve's state is given as follows:

\begin{equation}
\begin{array}{lcl}
|\theta^{\phi^{+},1}\rangle & = & \sqrt{\frac{1-\epsilon}{4}}\left(|\zeta_{00}\rangle+|\zeta_{11}\rangle\right)\\
|\theta^{\phi^{+},0}\rangle & = & \sqrt{\frac{\epsilon}{4}}\left(|\zeta_{00}\rangle-|\zeta_{11}\rangle\right)\\
|\theta^{\phi^{-},1}\rangle & = & \sqrt{\frac{\epsilon}{4}}\left(|\zeta_{00}\rangle-|\zeta_{11}\rangle\right)\\
|\theta^{\phi^{-},0}\rangle & = & \sqrt{\frac{1-\epsilon}{4}}\left(|\zeta_{00}\rangle+|\zeta_{11}\rangle\right)
\end{array},\label{eq:Chapter4_Eq35}
\end{equation}
Eve's system is formulated within the density operator framework,
with a conditional dependency on Bob's output,

\begin{equation}
\begin{array}{lcl}
\sigma_{E}^{0}=\rho_{E|0} & = & \epsilon|\theta^{\phi^{+},0}\rangle\langle\theta^{\phi^{+},0}|+\left(1-\epsilon\right)|\theta^{\phi^{-},0}\rangle\langle\theta^{\phi^{-},0}|\\
 & = & \left(\frac{\epsilon^{2}}{4}+\frac{\left(1-\epsilon\right)^{2}}{4}\right)\left[|\zeta_{00}\rangle\langle\zeta_{00}|+|\zeta_{11}\rangle\langle\zeta_{11}|\right]+\left(-\frac{\epsilon^{2}}{4}+\frac{\left(1-\epsilon\right)^{2}}{4}\right)\left[|\zeta_{00}\rangle\langle\zeta_{11}|+|\zeta_{11}\rangle\langle\zeta_{00}|\right]\\
\\\sigma_{E}^{1}=\rho_{E|1} & = & \left(1-\epsilon\right)|\theta^{\phi^{+},1}\rangle\langle\theta^{\phi^{+},1}|+\epsilon|\theta^{\phi^{-},1}\rangle\langle\theta^{\phi^{-},1}|\\
 & = & \left(\frac{\epsilon^{2}}{4}+\frac{\left(1-\epsilon\right)^{2}}{4}\right)\left[|\zeta_{00}\rangle\langle\zeta_{00}|+|\zeta_{11}\rangle\langle\zeta_{11}|\right]+\left(-\frac{\epsilon^{2}}{4}+\frac{\left(1-\epsilon\right)^{2}}{4}\right)\left[|\zeta_{00}\rangle\langle\zeta_{11}|+|\zeta_{11}\rangle\langle\zeta_{00}|\right]
\end{array}.\label{eq:Chapter4_Eq36}
\end{equation}
By tracing out Alice's and Bob's systems, the resulting density matrix
of Eve's system is obtained,

\begin{equation}
\rho_{E}=\frac{1}{2}\left[\left(|\zeta_{00}\rangle\langle\zeta_{00}|+|\zeta_{11}\rangle\langle\zeta_{11}|\right)+\left(1-2\epsilon\right)\left(|\zeta_{00}\rangle\langle\zeta_{11}|+|\zeta_{11}\rangle\langle\zeta_{00}|\right)\right].\label{eq:Chapter4_Eq37}
\end{equation}
The secret key rate is subsequently expressed using Equations (\ref{eq:Chapter4_Eq33}),
(\ref{eq:Chapter4_Eq36}), and (\ref{eq:Chapter4_Eq37}),

\begin{equation}
\begin{array}{lcl}
r\ge r_{{\rm DW}} & = & 1-h\left(\epsilon\right)-\left[S\left(\rho_{E}\right)-\left(\frac{1}{2}S\left(\sigma_{E}^{0}\right)+\frac{1}{2}S\left(\sigma_{E}^{1}\right)\right)\right]\end{array}.\label{eq:Chapter4_Eq38}
\end{equation}
To compute $r_{DW}$, we begin by defining an orthogonal basis $\left\{ |E_{00}\rangle,|E_{01}\rangle,|E_{10}\rangle,|E_{11}\rangle\right\} $
within the Hilbert space $\mathscr{H}^{E}$. This allows consideration
of the states in the orthogonal basis, specifically $|\zeta_{00}\rangle=\sum_{ij}a_{ij}E_{ij}$
and $|\zeta_{11}\rangle=\sum_{ij}f_{ij}E_{ij}$, where $i,j\in\left\{ 0,1\right\} $.
These states satisfy the constraints $|a_{00}|^{2}+|a_{01}|^{2}+|a_{10}|^{2}+|a_{11}|^{2}=1$,
$|f_{00}|^{2}+|f_{01}|^{2}+|f_{10}|^{2}+|f_{11}|^{2}=1$, and $a_{00}^{*}f_{00}+a_{01}^{*}f_{01}+a_{10}^{*}f_{10}+a_{11}^{*}f_{11}=\cos\alpha$.
Through extensive calculations, the eigenvalues of $\rho_{E}$ are
derived as $\lambda_{1,2}^{\rho_{E}}=0,0$,
\[
\lambda_{3}^{\rho_{E}}=\frac{1}{2}\left(1+{\rm cos}\alpha-2\epsilon{\rm cos}\alpha-\sqrt{\left(1-2\epsilon+{\rm cos}\alpha\right)^{2}}\right),
\]
and
\[
\lambda_{4}^{\rho_{E}}=\frac{1}{2}\left(1+{\rm cos}\alpha-2\epsilon{\rm cos}\alpha+\sqrt{\left(1-2\epsilon+{\rm cos}\alpha\right)^{2}}\right).
\]
Similarly, the eigenvalues for $\sigma_{E}^{1}$ or $\sigma_{E}^{0}$
are determined as $\lambda_{1,2}^{\sigma_{E}^{\pm}}=0,0$,
\[
\lambda_{3}^{\sigma_{E}^{\pm}}=\frac{1}{4}\left(1-2\epsilon+2\epsilon^{2}+{\rm cos}\alpha-2\epsilon{\rm cos}\alpha-\sqrt{\left(1-2\epsilon+\left(1-2\epsilon+2\epsilon^{2}\right){\rm cos}\alpha\right)^{2}}\right),
\]
and
\[
\lambda_{4}^{\sigma_{E}^{\pm}}=\frac{1}{4}\left(1-2\epsilon+2\epsilon^{2}+{\rm cos}\alpha-2\epsilon{\rm cos}\alpha+\sqrt{\left(1-2\epsilon+\left(1-2\epsilon+2\epsilon^{2}\right){\rm cos}\alpha\right)^{2}}\right).
\]
Subsequently, the entropy expressions are utilized as $S\left(\rho_{E}\right)=-\sum_{i}\lambda_{i}^{\rho_{E}}{\rm log_{2}\lambda_{i}^{\rho_{E}}}$
and $S\left(\sigma_{E}^{\pm}\right)=-\sum_{i}\lambda_{i}^{\sigma_{E}^{\pm}}{\rm log_{2}\lambda_{i}^{\sigma_{E}^{\pm}}}$
to derive the explicit form of Equation (\ref{eq:Chapter4_Eq38}).

\begin{figure}[h]
\begin{centering}
\includegraphics[scale=0.5]{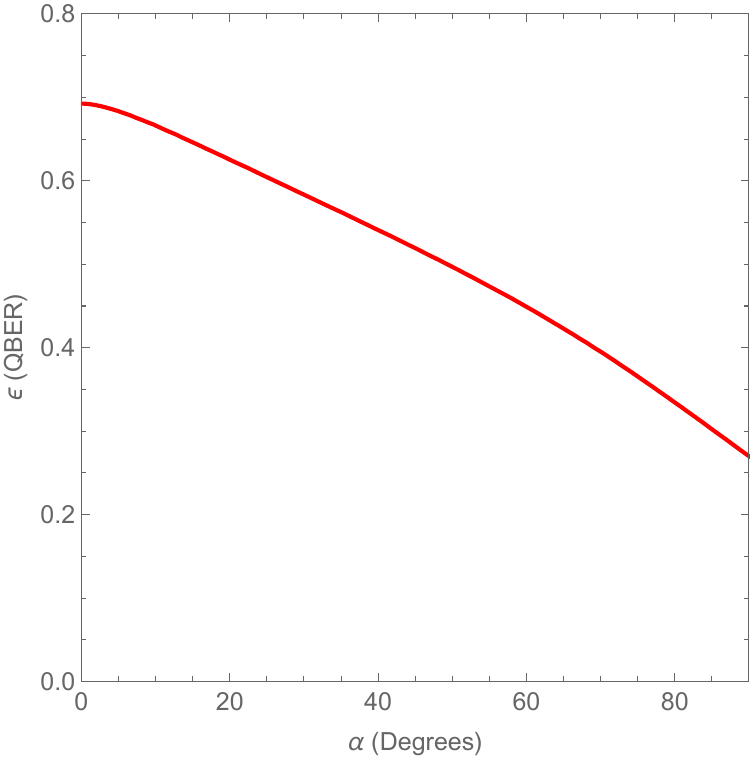} 
\par\end{centering}
\caption{The plot illustrates the relationship between tolerable QBER ($\epsilon$)
and the angle characterizing non-orthogonality $({\rm \alpha})$.}\label{fig:Chapter4_Fig5}
\end{figure}

To ascertain the tolerable error limit under secure conditions, referred
to as the tolerable QBER, we solve Equation (\ref{eq:Chapter4_Eq38})
in terms of $\alpha$. A graphical representation is produced, showcasing
the tolerable error probability $\epsilon$ as a function of the angle
$\alpha$, which signifies non-orthogonality (cf. Figure \ref{fig:Chapter4_Fig5}).
The plot demonstrates that increasing non-orthogonality in Eve's ancilla
state leads to a higher $\epsilon$. Specifically, when Eve employs
an orthogonal ancilla state ($\alpha=\frac{\pi}{2}$), the permissible
error threshold is identified as $27\%$. This threshold acts as a
crucial security benchmark in the context of collective attacks. Notably,
at the $27\%$ threshold, $r$ equals zero. However, for the same
$\alpha$ value, as illustrated in Figure \ref{fig:Chapter4_Fig3},
the error probability reaches 0.5, arising from the potential of obtaining
a nonzero $r$ value.

\section{Influence of noise on the proposed CQKA protocol}\label{sec:Chapter4_Sec4}

Decoherence induced by environmental factors \cite{walls1994atomic,BLP+16,ZG97}
is a well-documented constraint on the effectiveness of quantum-enhanced
protocols \cite{XDS+13,TPB17}. This issue presents a major hurdle
in realizing quantum communication schemes, as qubit interactions
with their surroundings can degrade accuracy, even in the absence
of an adversarial entity. Such interference may lead to erroneous
outputs, making any practical implementation inherently vulnerable
to noise from external sources \cite{GC85}. To ensure feasibility,
a protocol must retain its reliability despite the presence of a controlled
level of noise. Additionally, if an eavesdropper (Eve) attempts to
manipulate the channel, she could mask her intrusion as environmental
noise \cite{ZG97,RDR10,DRS11,SWY22}. However, distinguishing between
natural noise and intentional interference is feasible through established
detection methodologies \cite{LLM+22,LKL+25}, preserving protocol
security even under adversarial conditions \cite{HTS22}.

In this study, we explore the effects of noise on the quantum communication
protocols under investigation. Specifically, we evaluate amplitude
damping and phase damping within a Markovian framework, alongside
dephasing and depolarization in a non-Markovian reservoir context
\cite{SXL+18,NBS24}. Interestingly, while non-Markovian dephasing
can extend entanglement longevity, dissipative interactions may also
lead to entanglement revival. The quantum system\textquoteright s
behavior is represented using the Kraus formalism, providing essential
insights into the characteristics of these noise channels, which are
integral to quantum information and communication implementations.

The Kraus operators governing amplitude damping (AD) are given as
follows \cite{nielsen2010quantum,MTP+22}:

\begin{equation}
\begin{array}{ccccccccc}
\mathcal{F}_{0}^{{\rm AD}} & = & \left(\begin{array}{cc}
1 & 0\\
0 & \sqrt{1-\eta_{a}}
\end{array}\right) &  & {\rm and} &  & \mathcal{F}_{1}^{{\rm AD}} & = & \left(\begin{array}{cc}
1 & \sqrt{\eta_{a}}\\
0 & 0
\end{array}\right)\end{array}.\label{eq:Amplitude_Damping}
\end{equation}
Similarly, the Kraus operators for phase damping (PD) are:

\begin{equation}
\begin{array}{ccccccccc}
\mathcal{F}_{0}^{{\rm PD}} & = & \left(\begin{array}{cc}
1 & 0\\
0 & \sqrt{1-\eta_{p}}
\end{array}\right) &  & {\rm and} &  & \mathcal{F}_{1}^{{\rm PD}} & = & \left(\begin{array}{cc}
1 & 0\\
0 & \sqrt{\eta_{p}}
\end{array}\right)\end{array}.\label{eq:Phase_Damping}
\end{equation}
Here, $\eta_{j},j\in\left\{ a,p\right\} $ represents the damping
parameter. Likewise, the evolution of non-Markovian dephasing (NMDPH)
is determined by the Kraus operators outlined below \cite{NDB19}:

\begin{equation}
\begin{array}{lcl}
\mathcal{N}_{0}^{{\rm NMDPH}} & = & \sqrt{\left(1-\alpha p\right)\left(1-p\right)}\mathds{I},\\
\\\mathcal{N}_{0}^{{\rm NMDPH}} & = & \sqrt{p\left[1+\alpha\left(1-p\right)\right]}\sigma_{Z}.
\end{array}\label{eq:DePhasing}
\end{equation}
For the non-Markovian depolarizing (NMDPO) channel \cite{SSB18},

\begin{equation}
\begin{array}{lcl}
\mathcal{N}_{\mathds{I}}^{{\rm NMDPO}} & = & \sqrt{\left(1-3\alpha p\right)\left(1-p\right)}\mathds{I},\\
\\\mathcal{N}_{X}^{{\rm NMDPO}} & = & \sqrt{\left[1+3\alpha\left(1-p\right)\right]\frac{p}{3}}\sigma_{X},\\
\\\mathcal{N}_{Y}^{{\rm NMDPO}} & = & \sqrt{\left[1+3\alpha\left(1-p\right)\right]\frac{p}{3}}\sigma_{Y},\\
\\\mathcal{N}_{Z}^{{\rm NMDPO}} & = & \sqrt{\left[1+3\alpha\left(1-p\right)\right]\frac{p}{3}}\sigma_{Z}.
\end{array}\label{eq:DePolarizing}
\end{equation}
The parameter $\alpha$ $\left(0\leq\alpha\le1\right)$ quantifies
the extent of non-Markovianity. Specifically, when $\alpha=0$, the
channel exhibits standard dephasing, whereas larger values of $\alpha$
signify a heightened degree of non-Markovian behavior. Additionally,
$p$ serves as a time-dependent parameter constrained within $0\leq p\le\frac{1}{2}$
\cite{NDB19}, evolving monotonically over time. Here, $\mathds{I}$
denotes the identity matrix in a two-dimensional Hilbert space, while
$\sigma_{X},\sigma_{Y}\,{\rm and}\,\sigma_{Z}$ correspond to the
Pauli matrices.

\begin{figure}[h]
\begin{centering}
\includegraphics[scale=0.48]{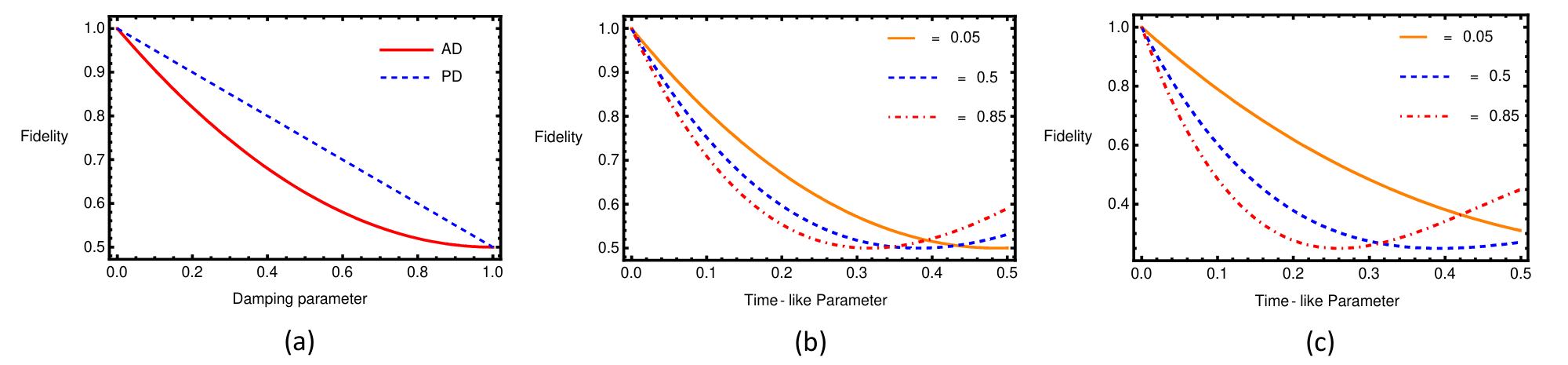}
\par\end{centering}
\caption{The variation of average fidelity under a noisy quantum channel is
analyzed. (a) Depicts the dependence of average fidelity on the channel
parameter $\eta_{j}$ for the CQKA scheme within the Markovian framework
for AD and PD channels, as described by Equations (\ref{eq:Amplitude_Damping})
and (\ref{eq:Phase_Damping}). (b) and (c) illustrate the variation
in average fidelity for NMDPH and NMDPO channels within the non-Markovian
regime, plotted against the time-like parameter $p$ for various values
of $\alpha$, respectively (refer to Equations (\ref{eq:DePhasing})
and (\ref{eq:DePolarizing})).}\label{fig:NOISY_CHANNEL}
\end{figure}

Prior to evaluating the impact of noisy channels on the proposed scheme,
it is essential to establish the initial configuration. When a $p$-qubit
composite state, $\rho_{i}=|\Phi\rangle\langle\Phi|$, serves as the
initial resource for executing the protocol---where $p$ home qubits
and $p-q$ travel qubits are represented by $h$ and $t$, respectively---the
general form of the final state before measurement can be expressed
as:

\[
\begin{array}{lcl}
\rho_{f}^{k} & = & \underset{j}{\sum}\left\{ I_{h}^{\otimes p}\otimes\left(\mathcal{F}_{1}^{k}\otimes\ldots\otimes\mathcal{F}_{j}^{k}\otimes\ldots\otimes\mathcal{F}_{p-q}^{k}\right)_{t}\right\} \rho_{i}\left\{ I_{h}^{\otimes p}\otimes\left(\mathcal{F}_{1}^{k}\otimes\ldots\otimes\mathcal{F}_{j}^{k}\otimes\ldots\otimes\mathcal{F}_{p-q}^{k}\right)_{t}\right\} ^{\dagger}.\end{array}
\]
The Kraus operators $\mathcal{F}_{j}^{k}$ characterize the various
noise channels, where $k\in\left\{ {\rm AD,PD,NMDPH,NMDPO}\right\} $.
The degradation in protocol performance due to noise can be quantitatively
assessed using a fidelity-based metric, specifically the squared fidelity
(commonly referred to as average fidelity) \cite{nielsen2010quantum},
which is defined as:

\[
\begin{array}{lcl}
F^{k} & = & \langle\Phi^{f}|\rho_{f}^{k}|\Phi^{f}\rangle.\end{array}
\]
Here, $|\Phi^{f}\rangle$ denotes the ideal final state that the initial
pure state $|\Phi\rangle$ would achieve in the absence of decoherence,
assuming perfect encoding operations performed by each participant.

The performance of the proposed scheme is analyzed by modeling the
quantum channel as a noisy medium. Particles ${\rm C_{1}}$ and ${\rm C_{2}}$
propagate through this environment, where we initially examine the
Markovian regime, considering both AD and PD channels. To assess the
average fidelity \cite{nielsen2010quantum} between the noiseless
state and the state affected by noise, we plot its variation as a
function of the damping parameter $\eta_{j}$ in Figure \ref{fig:NOISY_CHANNEL}.a.
The fidelity in the AD channel exhibits a more rapid decline with
increasing $\eta_{a}$ compared to the PD channel. The fidelity expressions
for these channels are given by $1-\frac{1}{2}\left(\eta-2\right)\eta$
for AD and $1-\frac{\eta}{2}$ for PD. It is noteworthy that the fidelity
decay in the AD channel follows a nonlinear trend, whereas a linear
decline is observed in the PD channel. At $\eta_{j}=1$, fidelity
reaches a minimum value of $0.5$ in both cases, representing the
lower bound in a highly damped environment.

For the non-Markovian regime, the temporal parameter $p$ is defined
as$p=\frac{1}{2}\left(1-e^{-\mathcal{K}t}\right)$, ranging from $0$
to $\frac{1}{2}$. The average fidelity expressions for the non-Markovian
dephasing and depolarizing channels are given by $\frac{1}{2}\left[1+\left\{ 1-2p+2\left(p-1\right)p\,\alpha\right\} \right]$
and $1+\frac{2}{3}\left[3\left(p-1\right)\alpha-1\right]\left[6\left(p-1\right)p\,\alpha-2p+3\right]$,
respectively. Figure \ref{fig:NOISY_CHANNEL}.b illustrates the impact
of the non-Markovian dephasing channel on the proposed scheme, revealing
that fidelity decreases as $p$ increases, with the rate of decline
dependent on $\alpha$. For smaller $\alpha$ values, fidelity remains
stable for an extended period, while for larger $\alpha$, a substantial
revival in fidelity is observed, with the minimum value reaching $0.5$.
A similar pattern is evident for the non-Markovian depolarizing channel
in Figure\ref{fig:NOISY_CHANNEL}.c, though the fidelity drop is more
pronounced at higher $\alpha$ values compared to the dephasing channel.
Conversely, for $\alpha=0.05$, the fidelity reduction is slower in
the depolarizing channel than in the dephasing channel, indicating
that fidelity is sustained for a longer duration in the depolarizing
channel at lower $\alpha$ values. However, at higher $\alpha$ values
$\left(\alpha=0.5,0.85\right)$, fidelity persists for a shorter duration
in the depolarizing channel than in the dephasing channel. Additionally,
the non-Markovian depolarizing channel exhibits a gradual fidelity
enhancement beyond a certain $p$ value, contingent on the specific
$\alpha$ parameter.

A crucial consideration involves examining a specific scenario: as
will be demonstrated, exploiting non-Markovianity can enhance system
performance (i.e., increase efficiency) in contrast to a Markovian
environment. Given this, an eavesdropper (denoted as Eve) may replace
a Markovian channel between Alice and Bob with a non-Markovian counterpart
to reduce the probability of detection (or alternatively, swap a noisier
channel with one exhibiting less noise). Therefore, legitimate communicating
parties must incorporate this possibility when defining the permissible
error threshold \cite{Z15,Z16}. Moreover, establishing this threshold
may necessitate characterizing the communication channel. For example,
noise in an ``amplitude-damping channel'' could be identified and mitigated
accordingly \cite{OSB15}.

\section{Comparison of the proposed CQKA protocol with existing protocols}\label{sec:Chapter4_Sec5}

This section presents a concise comparison between the proposed CQKA
and QKA protocols and other recently introduced schemes of a similar
nature. Notably, multiparty-based QKA protocols \cite{LZ+21} are
intentionally excluded, as the proposed scheme is specifically designed
for key agreement between two parties, either with or without the
involvement of a third party. Several key parameters are considered
in assessing the performance of QKA schemes, including quantum resources
(QR), quantum channel (QC) requirements, quantum efficiency (QE),
quantum memory (QM), third-party (TP) involvement, the number of parties
(NoP) required to establish the key agreement, and quantum operations
executed by legitimate parties. The comparison is conducted based
on these parameters to highlight the advantages and limitations of
the proposed scheme relative to existing alternatives.

We adopt the initial definition of QE introduced by Cabello in his
pioneering work \cite{C2000}. Specifically, QE is expressed as $\eta_{1}=\frac{b_{s}}{q_{t}+b_{t}}$,
where $b_{s}$ denotes the expected number of secret classical bits
shared among legitimate parties, $q_{t}$ represents the qubits transmitted
via the quantum channel per protocol step, and $b_{t}$ accounts for
the classical bits required to finalize the key post-protocol execution.
Since quantum resources are costlier and more challenging to preserve
in a coherent state compared to classical communication, we also employ
an alternative QE metric frequently utilized in literature: $\eta_{2}=\frac{b_{s}}{q_{t}}$
\cite{TH+11}. Notably, we exclude decoy state qubits from QE computations,
as their role is limited to channel security verification rather than
contributing to key generation. For comparative analysis, we first
consider the QKA protocol by Huang et al., which leverages quantum
correlations in EPR pairs alongside single-particle measurements to
establish identical key values between two participants, Alice and
Bob \cite{HW+14}. This method represents an entanglement-based extension
of the BB84 QKD protocol, differing primarily in the disclosure of
basis information prior to Alice and Bob\textquoteright s measurements.
Execution of this protocol necessitates quantum memory. However, since
Bob reveals the basis information to align measurement results, the
protocol becomes susceptible to a photon-number splitting attack by
an eavesdropper. A comparable approach was reported by Xu et al.,
utilizing the GHZ state, where three participants obtain identical
keys following the protocol\textquoteright s execution \cite{XW+14}.
In their scheme, a three-qubit quantum state serves as the QR, and
a subset of the initial sequence, prepared by party $A_{1}$, is used
to ensure security. To enhance QE, the legitimate users perform single-qubit
measurements exclusively in the computational basis ($B_{Z}$) once
channel security is established. The protocol's efficiency\footnote{For this protocol, the key parameter values are defined as follows:
$b_{s}=n-ns$, $q_{t}=2n$, $b_{t}=ns$, and efficiency is bounded
within $0.5\ge\eta\ge0$, with $s\in[0,1]$. Security improves as
$s$ increases.} is given by $\eta_{1}=\frac{n-ns}{2n+ns}$, where $s$ represents
the fraction of qubits from the initial sequence allocated for channel
security verification. Like the previous scheme, this protocol also
necessitates quantum memory.

In the same year, Shukla et al. introduced a QKA scheme based on EPR
pairs and a bidirectional quantum channel \cite{SA+14}. However,
this protocol is vulnerable to noise and demands quantum memory. Additionally,
Pauli operations are required at Bob\textquoteright s end. The structure
of this scheme closely resembles a modified version of the ping-pong
protocol \cite{BF02}. He et al. later proposed a more efficient protocol
($\eta=0.5$), incorporating a four-qubit cluster state that demands
additional quantum resources \cite{HM15}. Nevertheless, maintaining
coherence in a four-qubit cluster state presents significant challenges,
and the protocol also relies on Pauli operations. Furthermore, its
implementation requires a two-way quantum channel and quantum memory,
posing limitations given the current state of quantum communication
technology. A related protocol was introduced by Yang et al., employing
a four-qubit cluster state and cluster basis measurement \cite{YL+19}.
The authors integrated a classical permutation operation in a complex
manner to enhance security against adversarial attacks by Eve. However,
this approach inherently relies on quantum memory and Pauli operations\footnote{The efficiency of this scheme is expressed as$\eta_{1}=\frac{4n-nC}{4n+4n+nC}<0.5$,
where $C\in[0,1]$ represents the checking set, which does not function
as a decoy state.}.

We now examine the advantages and limitations of our two-party QKA
schemes compared to previous approaches. For the proposed QKA protocol,
we employ EPR pairs and single-photon states, which align well with
current technological capabilities. Notably, it does not require quantum
memory for implementation and utilizes a one-way quantum channel to
mitigate unnecessary channel noise, unlike earlier protocols. Additionally,
our scheme does not rely on Pauli operations and achieves quantum
efficiencies of $\eta_{1}=0.33$ and $\eta_{2}=1$. Certain scenarios
necessitate the involvement of an untrusted third party in QKA between
two legitimate participants; our CQKA protocol is designed to accommodate
this requirement. In contrast, Tang et al. proposed a CQKA protocol
that utilizes GHZ states, Hadamard gates, and Pauli operations \cite{TS+20}.
Their approach depends on a two-way quantum channel between Alice
and Bob and requires quantum memory, both of which are key drawbacks.
Our proposed CQKA scheme simplifies implementation by only requiring
EPR pairs and CNOT operations, making it feasible with current technology.
This is particularly advantageous when an untrusted third party is
involved in facilitating QKA. Furthermore, our scheme eliminates the
need for quantum memory and a two-way quantum channel, setting it
apart from existing protocols. A comparative analysis of these aspects
is presented in Table \ref{tab:Chapter4_Tab3}.

\begin{table}
\caption{Explicit comparison with earlier protocols. Y - required, N - not
required, QC - quantum channel, QR - quantum resources, TR - third
party, QM - quantum memory, NoP - number of parties, QE - quantum
efficiency.}\label{tab:Chapter4_Tab3}

\centering{}\centering{}%
\begin{tabular}{|>{\centering}p{1.75cm}|>{\centering}p{0.75cm}|>{\centering}p{2cm}|>{\centering}p{1.75cm}|>{\centering}p{0.75cm}|>{\centering}p{0.75cm}|>{\centering}p{2.5cm}|>{\centering}p{1.75cm}|}
\hline 
Protocol  & NoP  & QR  & QC  & QM  & TR  & QE $(\eta_{1})$  & QE $(\eta_{2})$\tabularnewline
\hline 
Huang et al. \cite{HW+14}  & 2  & EPR pair  & one-way  & Y  & N  & $\frac{n}{2n}=0.5$  & $\frac{n}{n}=1$\tabularnewline
\hline 
Xu et al. \cite{XW+14}  & 3  & GHZ state  & one-way  & Y  & N  & $\frac{n-ns}{2n+ns}<0.5$  & $\frac{n-s}{2n}<0.5$\tabularnewline
\hline 
Shukla et al. \cite{SA+14}  & 2  & EPR pair  & two-way  & Y  & N  & $\frac{n}{2n+n}=0.33$  & $\frac{n}{2n}=0.5$\tabularnewline
\hline 
He et al. \cite{HM15}  & 2  & four-qubit cluster state  & two-way  & Y  & N  & $\frac{4n}{4n+4n}=0.5$  & $\frac{4n}{4n}=1$\tabularnewline
\hline 
Yang et al. \cite{YL+19}  & 2  & four-qubit cluster state  & one-way  & Y  & N  & $\frac{4n-nC}{4n+4n+nC}<0.5$  & $\frac{4n-C}{4n}<1$\tabularnewline
\hline 
Tang et al. \cite{TS+20}  & 2  & GHZ state  & two-way  & Y  & Y  & $\frac{2n}{6n+n}=0.285$  & $\frac{2n}{6n}=0.33$\tabularnewline
\hline 
Our Protocol 1  & 2  & EPR pair, single qubit  & one-way  & N  & Y  & $\frac{n}{2n+3n}=0.2$  & $\frac{n}{2n}=0.5$\tabularnewline
\hline 
Our Protocol 2  & 2  & EPR pair, single qubit state  & one-way  & N  & N  & $\frac{n}{n+2n}=0.33$  & $\frac{n}{n}=1$\tabularnewline
\hline 
\end{tabular}
\end{table}

\section{Discussion}\label{sec:Chapter4_Sec6}

Key conditions that a QKA protocol must fulfill are examined in Section
\ref{sec:Chapter4_Sec2}. In that section, we have demonstrated that
the proposed schemes meet the \textit{correctness} criterion. Here,
we provide a concise discussion on the fairness condition by considering
a specific scenario within our protocol, as outlined in Tables \ref{tab:Chapter4_Tab1}
and \ref{tab:Chapter4_Tab2}. As previously stated, the \textit{fairness}
condition ensures that no individual participant can unilaterally
manipulate or exert control over the final agreement key. Specifically,
if Alice (Bob) attempts to influence the final key by selecting all
qubits in her (his) sequence $S_{{\rm A}}(S_{{\rm B}})$ to be either
$|0\rangle$ or $|1\rangle$, it does not introduce any bias in the
final key. This requirement can be validated by analyzing a particular
example from Table \ref{tab:Chapter4_Tab1} and Equations (\ref{eq:Chapter4_Eq6})-(\ref{eq:Chapter4_Eq13}).
Without loss of generality, let us assume that Alice selects all qubits
in the $|0\rangle$ state, while Bob generates his sequence randomly\footnote{Charlie\textquoteright s state is randomly chosen between $|\phi^{+}\rangle$
and $|\phi^{-}\rangle$, with bit values of 0 and 1, respectively.}. Subsequently, both Alice and Bob execute the CNOT operation on their
respective qubits using those received from Charlie. As derived from
Equations (\ref{eq:Chapter4_Eq6})-(\ref{eq:Chapter4_Eq13}), the
final agreement keys exhibit an equal probability distribution of
0s and 1s, ensuring no bias is introduced. Additionally, \textit{fairness}
mandates that all legitimate parties contribute equally to the final
agreement key. In Step 4 of the CQKA protocol, Alice and Bob independently
generate $n$-qubit sequences $S_{{\rm A}}$ and $S_{{\rm B}}$ in
the $Z$ basis, with each qubit randomly chosen as $|0\rangle$ or
$|1\rangle$. In Step 5, they perform the CNOT operation, followed
by a Bell measurement on their two-qubit systems. The final key is
determined based on their measurement results and subsequent announcements
in Steps 7 and 8, as detailed in Tables \ref{tab:Chapter4_Tab1} and
\ref{tab:Chapter4_Tab2}. This process confirms that both Alice and
Bob contribute equally to the final agreement key, satisfying the
fairness condition. Another crucial requirement is \textit{security},
which is thoroughly discussed in Section \ref{sec:Chapter4_Sec3}.

\begin{figure}
\begin{centering}
\includegraphics[scale=0.5]{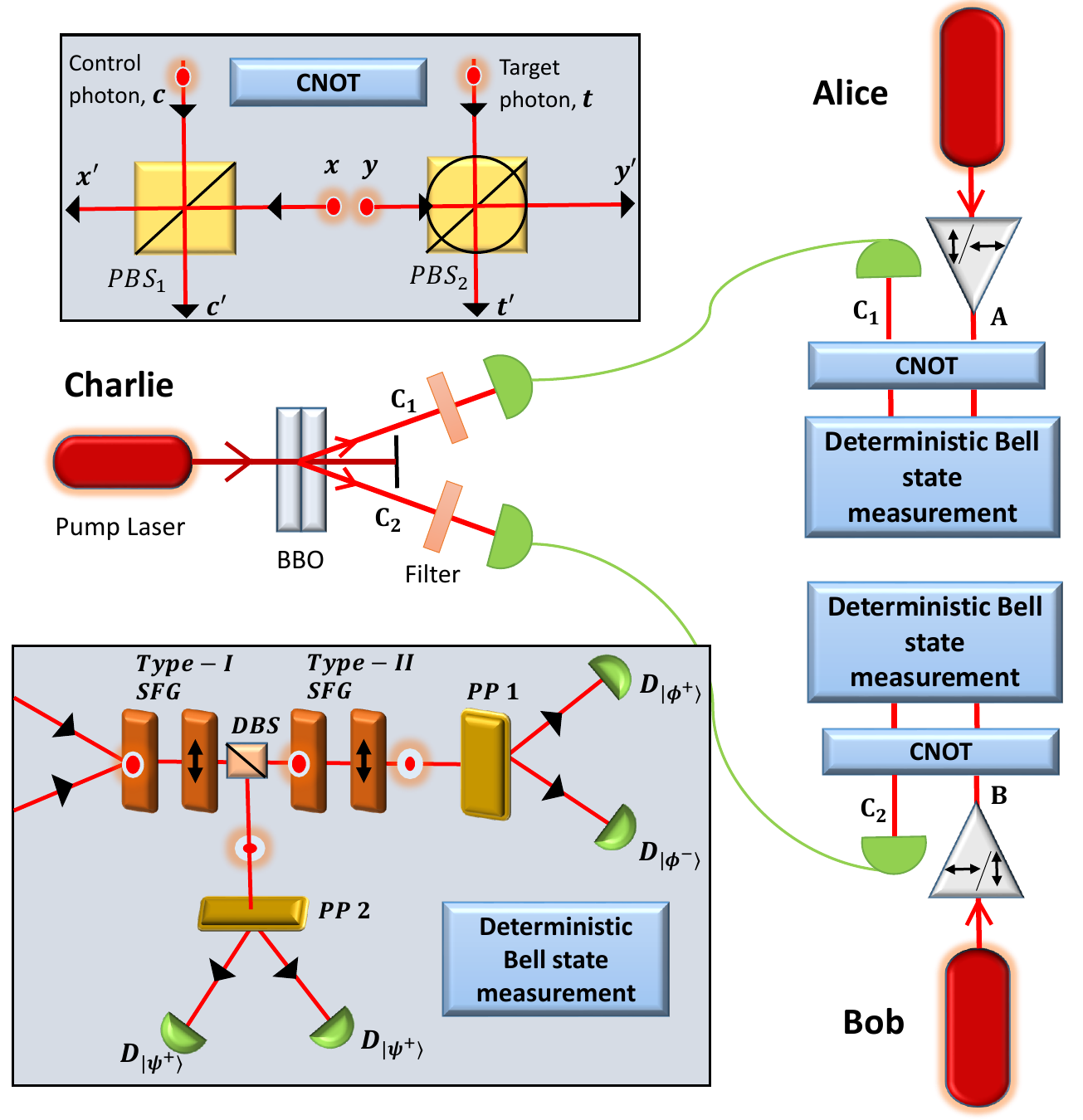}
\par\end{centering}
\caption{The optical design of the CQKA scheme integrates entangled photons,
a CNOT operation, and a full Bell state measurement. The entangled
photon pairs (Bell states) are generated using two thick $\beta$-barium
borate (BBO) nonlinear crystals arranged in a cross-configuration
\cite{PPH+22}. Bell state measurement is performed via nonlinear
interactions, specifically sum-frequency generation (SFG) of type-I
and type-II. A dichroic beam splitter (DBS) is incorporated, along
with two 45$^{\circ}$ projectors (PP1 and PP2) \cite{SKP21}. The system further
utilizes polarizing beam splitter(s) (${\rm PBS}_{1}$ and ${\rm PBS}_{2}$),
each configured at a 45$^{\circ}$ orientation with half-wave plate (HWP) integrated
at the input and output \cite{ZZC+05}.}\label{fig:Experimental_Setup}
\end{figure}

Recent experimental realizations of quantum communication have been
carried out using both free-space channels \cite{LCP+22,SJG+21} and
fiber-optic links \cite{BRM+20}. Within quantum networks, trajectories
serve as conduits for transmitting quantum systems \cite{RRE+21}.
Effective phase synchronization and robust reference signals are crucial
for these networks, prompting extensive research to develop efficient
solutions \cite{ZLX+23}{]}. Single-photon quantum communication experiments
are fundamental for the advancement of future quantum technologies
\cite{ZLX+23}. In this work, we propose an experimental implementation
of our quantum communication protocol. The setup consists of three
main components (refer to Figure \ref{fig:Experimental_Setup}): an
entangled photon source operated by Charlie, state preparation stations
managed by Alice and Bob, and a Bell-state measurement system. Charlie
generates a Bell state $\left(|\phi^{\pm}\rangle=\frac{1}{\sqrt{2}}\left(|00\rangle\pm|11\rangle\right)\right)$
using a nonlinear crystal via spontaneous parametric down-conversion
(SPDC) \cite{PPH+22}, distributing one photon to Alice and the other
to Bob through fiber or free-space links equipped with polarization
stabilization. Alice and Bob independently prepare single-photon states
in the Z basis ($|0\rangle$ or $|0\rangle$) using heralded single-photon
sources and polarization control elements, such as waveplate and polarizer.
Each station employs a linear-optics-based CNOT gate \cite{ZZC+05},
where Charlie\textquoteright s entangled photon serves as the control
qubit while their locally prepared photon acts as the target qubit.
The CNOT gate is implemented using polarizing beam splitter (${\rm PBS}_{1}$
and ${\rm PBS}_{2}$), waveplate, and postselection techniques. Bell-state
measurement is deterministically conducted through nonlinear interactions
via sum-frequency generation (SFG) in type-I and type-II processes
\cite{SKP21}. The generated sum-frequency photons are separated by
a dichroic beam splitter (DBS), and two $45^{\circ}$ polarization
projectors (PP1 and PP2) analyze the output states, ensuring reliable
Bell-state identification with high fidelity. High-efficiency single-photon
detectors and a synchronized timing system facilitate precise detection
and data acquisition, thereby validating the protocol's successful
execution and confirming entanglement correlations.

\section{Conclusion}\label{sec:Chapter4_Sec7}

In certain instances, establishing a key agreement between two parties
may necessitate the involvement of an untrusted third party. The quantum
analogue of such scenarios demands a QKA protocol incorporating a
semi-trusted third-party controller. Our proposed protocol is structured
to fulfill the requirements of a controller within a QKA framework.
Section \ref{sec:Chapter4_Sec2} provides a detailed step-by-step
exposition of our protocol, elucidating how quantum particle behavior
can be harnessed to construct such protocols for practical implementations.
However, specific conditions must be met for the successful execution
of a QKA protocol, as outlined in Section \ref{sec:Chapter4_Sec2}.
The necessary security proofs addressing potential attack scenarios
are detailed in Section \ref{sec:Chapter4_Sec3}, where we discuss
critical aspects of security analysis. To implement the CQKA protocol,
a Bell state---generated by an untrusted third party, Charlie---is
required and subsequently distributed to the legitimate participants.
Once Alice and Bob receive the first and second qubits of the Bell
state prepared by Charlie, they execute the CNOT operation as prescribed
by our protocol. Notably, if an eavesdropper (Eve) attempts an intercept-and-resend
attack, she can only infer the Bell state originally generated by
Charlie, as the channel particles are no longer entangled due to Alice
and Bob's operations at that stage. Furthermore, Charlie publicly
discloses this Bell state information $(k_{C})$ at the conclusion
of the protocol, ensuring that Eve gains no advantage from such an
attack. Additionally, we analyze the scenario where Eve attempts to
impersonate the third party, referred to as an ``impersonated
fraudulent attack''. 

Our protocol is demonstrated to be resilient against this attack,
with corresponding analytical proof supporting its security. Additionally,
a crucial security evaluation is performed against the collective
attack, establishing that our protocol remains secure even under more
generalized attack scenarios\footnote{The proposed QKA protocol represents a specific instance of the CQKA
protocol. Consequently, the security analysis conducted for the CQKA
protocol is directly applicable to the QKA protocol as well.}. We further determine that the probability of Eve successfully extracting
the final two-bit key is negligibly small (approaching elimination)
when $n=6$ (refer to Figure \ref{fig:Chapter4_Fig4}). Moreover,
our security analysis can be extended by treating the inner product
of Eve's ancillary state as a complex variable, wherein each ancilla
system is represented using a complete orthogonal basis in the Hilbert
space $\mathcal{\mathscr{H}^{\mathrm{\mathscr{E}}}}$ \cite{LF+11}.
Furthermore, we estimate the tolerable error bound of our scheme by
incorporating this approach and determine the acceptable QBER as a
function of $\alpha$. We evaluate our scheme under different noisy
channel conditions and analyze the fidelity variation in relation
to the channel parameter. Our findings indicate that the fidelity
remains largely unaffected, even at higher values of the noise parameter.
In Section \ref{sec:Chapter4_Sec5}, we further compare the advantages
and limitations of our protocol with other recently proposed protocols
of the same category. The results highlight that our protocol is not
only more practical for implementation using current technological
capabilities but also exhibits greater efficiency compared to many
existing schemes discussed earlier (see Section \ref{sec:Chapter4_Sec5}).
Notably, it introduces a novel and desirable feature involving the
participation of a third party.

\newpage



\chapter{``GAMING THE QUANTUM'' TO GET SECURE BOUND FOR A QUANTUM COMMUNICATION PROTOCOL}\label{Ch5:Chapter5_QG}
\graphicspath{{Chapter5/Chapter5Figs/}{Chapter5/Chapter5Figs/}}

\section{Introduction}

Game theory analyzes strategic decision-making, modeling how individuals
interact in competitive and cooperative settings. It plays a vital
role in optimizing choices across business, economics \cite{G92},
political science \cite{O86}, biology \cite{C13,NS99}, and military
strategy \cite{D59,H97}. Each participant selects from a set of strategies,
with preferences represented in a payoff matrix. A key concept is
the Nash equilibrium, where no player gains by unilaterally changing
their strategy \cite{N50,N51}, ensuring stability in decision-making.
Quantum mechanics is among the most impactful theories in history.
Despite early controversies, its predictions have been consistently
validated through experiments \cite{AGR82}. Quantum game theory explores
interactive decision-making involving quantum technology, serving as a quantum communication protocol and a more efficient means
of randomizing strategies compared to classical games \cite{L11}.
Introduced in 1999 by David \cite{M99}, along with Jens, Martin,
and Maciej \cite{EWL99}, this field examines games leveraging quantum
information, revealing advantages for quantum players over classical
ones. Numerous quantum games, expanding on these foundational works,
have since been extensively studied \cite{GZK08}. Experimentally,
researchers have implemented the quantum Prisoners' Dilemma using
an NMR quantum computer \cite{DLX+02}. Vaidman \cite{V99} also demonstrated
a game where players always win when sharing a GHZ state, unlike classical
counterparts who rely on probability. Quantum strategies have further
contributed to fairness in remote gambling \cite{GVW99} and the development
of secure quantum auction algorithms \cite{P07}. Flitney and Abbott
\cite{FA03} have investigated quantum variations of Parrondo's games,
highlighting their role in enhancing network security and inspiring
novel quantum algorithms. These studies also introduce a new perspective
on defining games and protocols \cite{DP+23,DP2023,DP+24}. Additionally,
eavesdropping \cite{E91,NH97} and optimal cloning \cite{W98} can
be interpreted as strategic interactions among participants, further
reinforcing the link between game theory and quantum mechanics. This
interconnection follows two main approaches: (1) leveraging quantum
resources to play classical games with enhanced advantages, referred
to as ``quantized game'', and (2) applying game-theoretic principles
to quantum scenarios, termed ``gaming the quantum''. A ``quantized
game'' is mathematically described as a unitary function mapping quantum
superpositions of players\textquoteright{} pure strategies onto the
game\textquoteright s Hilbert space while preserving its core properties
under certain conditions. Conversely, ``gaming the quantum'' \cite{KP+13}
involves utilizing game theory within quantum mechanics to derive
strategic solutions. In this chapter, we adopt the ``gaming the quantum'' approach by applying non-cooperative game theory to a quantum communication
protocol, demonstrating how Nash equilibrium can function as an effective
solution concept.

Building on these insights, Nash equilibrium points can be utilized
to define secure bounds for various quantum information parameters.
This approach involves modeling quantum schemes as game-theoretic
scenarios and analyzing their Nash equilibria, which provide stable
conditions and rational probability distributions for decision-making
in a mixed-strategy framework. These probabilities help establish
a stable gaming environment, enabling the evaluation of cryptographic
parameters such as the threshold QBER. In practical scenarios where
decisions are made autonomously, achieving stability is challenging
and may lead to increased uncertainty. Therefore, identifying the
minimum QBER from a stable game setting defines a secure threshold
for implementing quantum communication protocols. This work specifically
examines the secure threshold bound for QBER within the DL04 protocol
\cite{DL04}, a direct secure quantum communication scheme. Such protocols
generally fall into two broad categories \cite{LDW+07}. Secure quantum
communication protocols can be categorized into two types. The first
category, DSQC protocols \cite{ZXF+06,HHT11,DP22,DP23}, requires
the receiver to decode the message only after receiving at least one
bit of additional classical information per qubit. The second category,
QSDC protocols \cite{BEK+02,LL02,BF02,DLL03,DL04,DBC+04,WDL+05,ZSL20,WLY+19},
eliminates the need for classical information exchange. Beige et al.
\cite{BEK+02} proposed a QSDC scheme where message retrieval depends
on additional classical information per qubit. Bostr{\"o}m and Felbinger
introduced the ping-pong QSDC protocol \cite{BF02}, offering secure
key distribution and quasi-secure direct communication over an ideal
quantum channel. In 2004, Deng and Long \cite{DL04} developed the
DL04 QSDC protocol, which operates without entangled states. Notably,
secure transmission via the two-photon component was observed, consistent
with two-way QKD findings \cite{DL+04,LM_05,L19}. This chapter primarily
focuses on evaluating the secure threshold bound of QBER within the
DL04 protocol, chosen for its experimental viability and feasibility
in secure implementations \cite{HYJ+16,ZZS+17,QSL+19,ZSN+20,PLW+20,PSL23,NZL+18}.

The subsequent sections of this chapter are organized as follows.
Section \ref{sec:Chapter5_Sec2} introduces game theory within the
specific scope of this study, as it forms the analytical foundation
of our work. Additionally, this section conceptualizes the DL04 protocol
as a game-like scenario. Section \ref{sec:Chapter5_Sec3} delves into
the computation of information-theoretic parameters essential for
deriving payoff functions across various attack scenarios. In Section
\ref{sec:Chapter5_Sec4}, we mathematically formulate the game and
utilize a graphical approach to analyze the Nash equilibrium. Furthermore,
we examine equilibrium points to establish the secure threshold limit
of QBER for the DL04 protocol. Lastly, Section \ref{sec:Chapter5_Sec5}
summarizes our results and provides a concluding discussion.

\section{Fundamentals of Game Theory}\label{sec:Chapter5_Sec2}

Before exploring the technical aspects of this study, it is essential
to provide a concise overview of ``game theory''. Game theory consists
of mathematical frameworks used to analyze decision-making scenarios
that involve both competition and cooperation, where the choices made
by each participant (player) depend on the decisions of others. Each
player has a defined set of strategies or possible actions, which
dictate their responses to various situations within the game. The
central goal is to determine optimal strategies for players under
these conditions and to identify equilibrium states. Game outcomes
are typically expressed as payoffs, representing the utility or benefit
each player gains based on the collective strategy choices of all
participants. These payoffs can be quantified using numerical values,
rankings, or other evaluative measures. Game theory can be formulated
in multiple ways, with two primary representations: the ``normal
form'' (strategic form) and the ``extensive form'' (dynamic game).
The normal form represents payoffs in a matrix structure, displaying
outcomes for different strategy combinations, whereas the extensive
form employs a tree structure to illustrate sequential and simultaneous
decision-making processes. Games are generally classified into two
categories: ``zero-sum games'' and ``non-zero-sum games''. In
a zero-sum game, the total available payoff remains constant, meaning
one player's gain directly results in another's loss. Conversely,
non-zero-sum games allow for variable total payoffs, where players'
interests may be partially aligned rather than entirely conflicting.
A crucial concept in game theory is Nash equilibrium, which describes
a scenario in which no player has an incentive to unilaterally alter
their chosen strategy, assuming all other players maintain their strategies.
This equilibrium represents a stable outcome where no player can improve
their payoff by deviating from their current strategy. Another key
principle is ``Pareto efficient'' (or ``Pareto optimal''), where
no alternative strategy set exists that improves one player's outcome
without disadvantaging another. Additionally, a dominant strategy
is one that consistently yields the best outcome for a player, regardless
of the strategies selected by others. This concept plays a significant
role in rational decision-making. Players may follow a pure strategy,
where a single action is chosen with certainty, or a mixed strategy,
where actions are selected based on assigned probabilities. In certain
games, mixed strategies contribute to the formation of a Nash equilibrium
when no single pure strategy is superior \cite{EWL99,PS03,alonso2019quantum,BW22,KK18}.

\textit{Quantum game} In conventional games that permit mixed strategies,
players formulate their strategies by utilizing real coefficients
to construct convex linear combinations of their pure strategies \cite{O04}.
Conversely, in quantum games \cite{EWL99,EW2000}, strategic choices
involve unitary transformations and quantum states, which exist within
significantly expanded strategic spaces. This has led to discourse
suggesting that quantum games could serve as extensions of classical
game theory, potentially granting participants a quantum advantage
\cite{EP02}. As previously noted, any quantum protocol, such as those
in quantum cryptographic scenarios, can be interpreted through a game-theoretic
aspect. In such a framework, players\textquoteright{} strategies are
determined by their application of unitary operations and selection
of measurement bases, with payoffs derived from measurement outcomes.
The quantum advantage emerges from the optimal sequential application
of quantum operations on quantum states by the players. In classical
quantum game theory literature, players are typically restricted to
performing coherent permutations of standard basis states or other
similarly constrained unitary operations. In contrast, quantum players
can leverage a broader class of unitary transformations, encompassing
a more comprehensive set of operations with fewer constraints. Before
advancing further, it is essential to formally define ``pure quantum
strategies'', ``mixed quantum strategies'', and ``positive operator-valued
measure (POVM) strategies'' \cite{EW2000} within the context of
this study.

A pure quantum strategy consists of deterministic actions represented
by unitary transformations applied to a player's quantum state. This
approach begins with an initial quantum state $|\Psi\rangle$ within
a Hilbert space $\mathcal{H}$. Each participant $i$ selects a unitary
operation $U_{i}$ to apply to their respective portion of the quantum
system. Once all players have executed their chosen unitary operations,
the resultant quantum state is expressed as

\[
\begin{array}{lcl}
|\Psi_{f}\rangle & = & \left(U_{1}\otimes U_{2}\otimes\cdots\otimes U_{i}\otimes\cdots\otimes U_{n}\right)|\Psi\rangle,\end{array}
\]
where $n$ represents the total number of participants. The final
state $|\Psi_{f}\rangle$ is subsequently measured to determine the
outcome of the game. Conversely, a mixed quantum strategy incorporates
probabilistic selections of different pure strategies. Similar to
the pure strategy case, the system starts in the state $|\Psi\rangle$.
Each player $i$ has a set of unitary transformations $\left\{ U_{i}^{1},U_{i}^{2},\cdots,U_{i}^{k}\right\} $,
each associated with a probability distribution $\left\{ p_{i}^{1},p_{i}^{2},\cdots,p_{i}^{k}\right\} $,
constrained by $\sum_{j=1}^{k}p_{i}^{j}=1$. The selection
of a unitary operator $U_{i}$ follows the probability $p_{i}^{j}$.
Consequently, the ensemble of possible final states forms a probabilistic
mixture of pure states, leading to the mixed state representation:

\[
\begin{array}{lcl}
\rho_{f} & = & \underset{j}{\sum}p_{i}^{j}\left(U_{i}^{j}\otimes\cdots\otimes U_{n}^{j}\right)|\Psi\rangle\langle\Psi|\left(U_{i}^{j}\otimes\cdots\otimes U_{n}^{j}\right)^{\dagger}.\end{array}
\]
This mixed state $\rho_{f}$ undergoes measurement to determine the
game's final outcome. A POVM-based quantum strategy employs generalized
quantum measurements, defined by a set of POVM elements. Beginning
with the state $|\Psi\rangle$, each player $i$ utilizes a POVM,
consisting of positive semi-definite operators $\left\{ \mathcal{E}_{i}^{1},\mathcal{E}_{i}^{2},\cdots,\mathcal{E}_{i}^{k}\right\} $
that satisfy the completeness condition $\underset{j}{\sum}\mathcal{E}_{i}^{j}=\mathds{1}$.
The probability of outcome $j$ is given by $p_{i}^{j}=\langle\Psi|E_{i}^{j}|\Psi\rangle$.
Based on the observed measurement outcome, players may perform conditional
unitary transformations or other quantum operations. The resultant
state, dependent on both measurement results and subsequent operations,
is represented as

\[
\begin{array}{lcl}
\rho_{f} & = & \underset{j}{\sum}p_{i}^{j}\left(U_{i}^{j}\otimes\cdots\otimes U_{n}^{j}\right)|\Psi\rangle\langle\Psi|\left(U_{i}^{j}\otimes\cdots\otimes U_{n}^{j}\right)^{\dagger}.\end{array}
\]
The final mixed state $\rho_{f}$ is then measured to determine the
game's outcome. Among these strategies, POVM-based methods offer the
highest degree of generality and adaptability within quantum game
theory. In this chapter, we implement a mixed quantum strategy where
Alice, Bob, and Eve execute their quantum operations based on specific
probabilistic combinations to achieve Nash equilibrium points. The
resulting mixed state determines the respective payoff values for
each player.

The quantum game under consideration is a non-cooperative multiplayer
scenario that does not possess a pure strategy Nash equilibrium. This
absence results from the inherent structure of the multiplayer framework
and the methodology employed for computing payoffs, akin to the approach
in \cite{KP+13}. The Nash equilibrium concept is a cornerstone in
game theory, serving as a predictive tool for the behavior of independent
decision-makers. The proof of Nash\textquoteright s theorem, which
establishes the existence of an equilibrium in mixed strategies within
classical game settings are relatively straightforward and fundamentally
relies on Kakutani\textquoteright s ``fixed-point theorem'' \cite{Kakutani1941}.
In the domain of quantum games, Meyer demonstrated that Nash equilibria
in mixed strategies, represented as mixed quantum states, exist by
leveraging Glicksberg\textquoteright s extension of Kakutani\textquoteright s
``fixed-point theorem'' to topological vector spaces \cite{Glicksberg1952}.
However, Khan et al. (2019) established that the direct application
of Kakutani\textquoteright s theorem is not feasible for quantum games
utilizing pure quantum strategies \cite{KH19}. Nonetheless, under
specific conditions, Nash\textquoteright s embedding theorem, which
maps compact Riemannian manifolds into Euclidean space \cite{Nash_1956},
enables an indirect application of Kakutani\textquoteright s theorem,
thereby ensuring the existence of Nash equilibrium in pure quantum
strategies. Their work offers a rigorous mathematical exploration
of non-cooperative game theory and fixed-point theorems, as detailed
in \cite{KH19}.

\textit{Our contribution} In our game, several critical aspects must
be considered since it operates as a ``mixed quantum strategic
game''. When players select quantum strategies according to a probability
distribution---thus employing mixed quantum strategies---Meyer utilized
Glicksberg\textquoteright s fixed-point theorem \cite{Glicksberg1952}
to establish the guaranteed existence of a Nash equilibrium. His research
also provides valuable insights into equilibrium behavior within quantum
computational frameworks. The identification of quantum advantage
as a Nash equilibrium in quantum games, as proposed by Meyer, remains
an area with limited exploration \cite{KSB+18}. Furthermore, in quantum
communication protocols---where quantum processes are inherently
noisy and represented using density matrices or mixed quantum states---the
Meyer--Glicksberg theorem ensures the presence of a Nash equilibrium
\cite{M99}. Typically, in such protocols, quantum players with fewer
restrictions can leverage mixed strategies to determine their best
response functions by evaluating their respective payoffs. The Nash
equilibrium points are obtained as solutions to these best response
functions. While not all equilibrium points necessarily achieve Pareto
optimality, they provide a foundation for analyzing the probability
distributions governing mixed strategies. This analysis facilitates
the identification of ``Pareto optimal Nash equilibria'', aiding
in the determination of secure bounds for various quantum information
parameters, such as the secret key rate and QBER, by applying core
principles from quantum information theory \cite{NC10,W17,W18,P13}.
In this chapter, we examine the DL04 protocol \cite{DL04} through
a game-theoretic lens, considering different attack strategies---including
collective and IR attacks---to derive the secure threshold limit
for QBER.

In the field of quantum cryptography, QSDC utilizes quantum states
as information carriers to ensure secure transmission. Unlike conventional
cryptographic techniques, QSDC eliminates the necessity for pre-established
secret keys \cite{BEK+02,LL02,BF02,DLL03,DL04,DBC+04,WDL+05}. It
fundamentally focuses on enabling secure and reliable communication
through quantum mechanical principles. Extensive experimental research
has validated the feasibility of QSDC and highlighted its potential
applications \cite{HYJ+16,LHA+16,ZHS+17,QSL+19,MMB+19,PLW+20,ZSQ+22,LPS+22,PSL23}.
Over the past two decades, persistent advancements have contributed
to the maturation of QSDC, positioning it as a promising technology
for next-generation secure communication, including potential military
applications \cite{YWH+21,K21}. Among various QSDC protocols introduced
so far, the DL04 protocol stands out due to its prominence in recent
experimental studies \cite{HYJ+16,ZZS+17,QSL+19,ZSN+20,PLW+20,PSL23,NZL+18}.
Given its significance, our research specifically focuses on the DL04
protocol while ensuring that the developed strategy maintains general
applicability.

We begin with a brief overview of the DL04 protocol \cite{DL04},
modified slightly for applicability in a gaming context. In this protocol,
Bob generates quantum states (photons) at random, choosing between
$|0\rangle$ and $|1\rangle$ in the computational basis ($Z$ basis)
or $|+\rangle$ and $|-\rangle$ in the diagonal basis ($X$ basis),
with probabilities $p$ and $1-p$, respectively. He then transmits
this sequence to Alice. Alice operates in two distinct modes: message
mode (encoding mode) and control mode (security verification mode).
During control mode, Alice randomly selects a subset of the received
photons to detect potential eavesdropping, utilizing a beam splitter.
Each photon in this subset is measured in either the $Z$ or $X$
basis. Alice then informs Bob of the positions of these security check
photons, the chosen measurement bases, and the corresponding outcomes.
Together, they compute the initial QBER. If the QBER remains below
a predefined threshold, they proceed with the next phase; otherwise,
the transmission is discarded. In message mode, Alice applies a quantum
operation $I$ ($iY\equiv ZX$) on the remaining qubits to encode
the bit $0$ with probability $q$ (or 1 with probability $1-q$).
Additionally, she designates some photons for encoding random bits
(0 or 1) to verify the reliability of the second transmission before
forwarding them back to Bob. Most of the photons in this second transmission
contain Alice\textquoteright s encoded message, while a small fraction
is used to estimate the second QBER. Upon receipt of the photons,
Bob deciphers the classical bits encoded by Alice based on his initial
preparation bases. Alice then reveals the positions of the photons
used for random number encoding, enabling both parties to evaluate
the QBER of the second transmission. This secondary QBER estimation
ensures the integrity of the transmission. If the error rate remains
below the acceptable threshold, the transmission is considered successful.
A comparable protocol, LM05, introduced by Lucamarini et al. \cite{LM_05},
follows a similar structure. In LM05, Alice\textquoteright s execution
of the control mode mirrors that of DL04, but classical announcements
are allowed at that stage. Once Bob receives the sequence from Alice,
he performs projective measurements for decoding and subsequently
announces the classical security-check bits. In both DL04 and LM05,
the objective for legitimate users is to achieve perfect ``double
correlation'' in measurement results for both the forward and return
transmission paths.

Within the DL04 protocol, two quantum participants implement strategic
methods to uphold secure communication. Our analysis is centered on
establishing the secure threshold for the QBER while considering the
interference of an eavesdropper, Eve, who may employ either quantum
or classical attack methodologies. Specifically, we examine quantum
Eve's utilization of three attack approaches: W{\'o}jcik\textquoteright s
original attack \cite{W03}, W{\'o}jcik\textquoteright s symmetrized
attack \cite{W03}, and Pavi{\v{c}}i{\'c}\textquoteright s attack
\cite{P+13}, labeled as $E_{1}$, $E_{2}$ and $E_{3}$, respectively.
Additionally, a classical Eve executes an IR attack, denoted as $E_{4}$.
A comprehensive examination of these attack strategies is provided
in the next section, whereas a concise overview is presented here to facilitate
conceptual understanding (for further details, refer to Section
\ref{sec:Chapter5_Sec3}). In the $E_{1}$ attack, Eve applies the
unitary operator $Q_{txy}$ during the B-A attack phase and its conjugate,
$Q_{txy}^{\dagger}\left(\equiv Q_{txy}^{-1}\right)$, during the A-B
attack. This operator is represented as:

\[
\begin{array}{lcl}
Q_{txy}={\rm SWAP}_{tx}\,{\rm CPBS}_{txy}\,H_{y} & \equiv & {\rm SWAP}_{tx}\otimes I_{y}\,{\rm CPBS}_{txy}\,I_{t}\otimes I_{x}\otimes H_{y}\end{array}
\]
Here, A-B and B-A attacks correspond to Eve\textquoteright s intervention
when quantum states are transmitted from Alice to Bob and vice versa.
This attack induces the system into an extended Hilbert space through
Eve\textquoteright s unitary operations, increasing randomness due
to quantum superposition. Consequently, the joint probabilities of
Alice\textquoteright s, Bob\textquoteright s, and Eve\textquoteright s
measurement outcomes are altered as follows:

\[
\begin{array}{lcl}
p_{000}^{E_{1}}=q, &  & p_{001}^{E_{1}}=p_{010}^{E_{1}}=p_{011}^{E_{1}}=0,\\
\\p_{100}^{E_{1}}=\frac{1}{4}\left(1-q\right), &  & p_{101}^{E_{1}}=\left(1-q\right)\left(\frac{1}{4}+\frac{p}{2}\right),\\
\\p_{110}^{E_{1}}=0, &  & p_{111}^{E_{1}}=\frac{1}{2}\left(1-p\right)\left(1-q\right).
\end{array}
\]
The QBER for the $E_{1}$ attack is given by $\frac{1}{2}\left(1-q\right)\left(1+p\right)$,
and the probability of detecting Eve\textquoteright s presence is
0.1875. 

The $E_{2}$ attack resembles $E_{1}$, with the distinction that
an additional unitary operation, $S_{ty}$, is applied with a probability
of $\frac{1}{2}$ after $Q_{txy}^{-1}$ in the A-B attack phase. The
operation $S_{ty}$ is defined as $X_{t}Z_{t}{\rm CNOT}_{ty}X_{t}$.
This leads to modified measurement probabilities:

\[
\begin{array}{lcl}
p_{000}^{E_{2}}=\frac{q}{8}\left(5+p\right), &  & p_{001}^{E_{2}}=\frac{q}{8}\left(1+p\right),\\
\\p_{010}^{E_{2}}=p_{011}^{E_{2}}=\frac{q}{8}\left(1-p\right), &  & p_{100}^{E_{2}}=\frac{1}{4}\left(1-q\right),\\
\\p_{101}^{E_{2}}=\frac{1}{4}\left(1-q\right)\left(1+2p\right), &  & p_{110}^{E_{2}}=0,\\
\\p_{111}^{E_{2}}=\frac{1}{2}\left(1-p\right)\left(1-q\right).
\end{array}
\]
The QBER for the $E_{2}$ attack is $\frac{1}{4}\left(2+2p-q-3pq\right)$,
with a detection probability of 0.1875. In the $E_{3}$ attack, Eve
employs the unitary operation $Q_{txy}^{\prime}$ when the photon
moves from Bob to Alice (B-A attack), while in the A-B attack, she
applies the inverse operator $Q_{txy}^{\prime-1}$. The unitary transformation
is expressed as:

\[
\begin{array}{lcl}
Q_{txy}^{\prime} & = & {\rm CNOT}_{ty}\left({\rm CNOT}_{tx}\otimes I_{y}\right)\left(I_{t}\otimes{\rm PBS}_{xy}\right){\rm CNOT}_{ty}\left({\rm CNOT}_{tx}\otimes I_{y}\right)\left(I_{t}\otimes H_{x}\otimes H_{y}\right).\end{array}
\]
This results in the following joint probabilities:

\[
\begin{array}{lcl}
p_{000}^{E_{3}}=q, &  & p_{001}^{E_{3}}=p_{010}^{E_{3}}=p_{011}^{E_{3}}=0,\\
\\p_{100}^{E_{3}}=p_{101}^{E_{3}}=p_{110}^{E_{3}}=0, &  & p_{111}^{E_{3}}=\left(1-q\right).
\end{array}
\]
Here, the QBER remains 0, while the probability of detecting Eve\textquoteright s
presence is 0.1875. Lastly, the $E_{4}$ attack corresponds to an
IR attack, where the measurement outcome probabilities are:

\[
\begin{array}{ccccc}
p_{000}^{E_{4}}=\frac{3}{4}q, &  & p_{001}^{E_{4}}=p_{011}^{E_{4}}=0, &  & p_{010}^{E_{4}}=\frac{1}{4}q,\\
\\p_{100}^{E_{4}}=p_{110}^{E_{4}}=0, &  & p_{101}^{E_{4}}=\frac{1}{4}\left(1-q\right), &  & p_{111}^{E_{4}}=\frac{3}{4}\left(1-q\right).
\end{array}
\]
For this scenario, the QBER is 0.25, and Eve's presence can be detected
with a probability of 0.375.

\begin{figure}[h]
\begin{centering}
\includegraphics[scale=0.55]{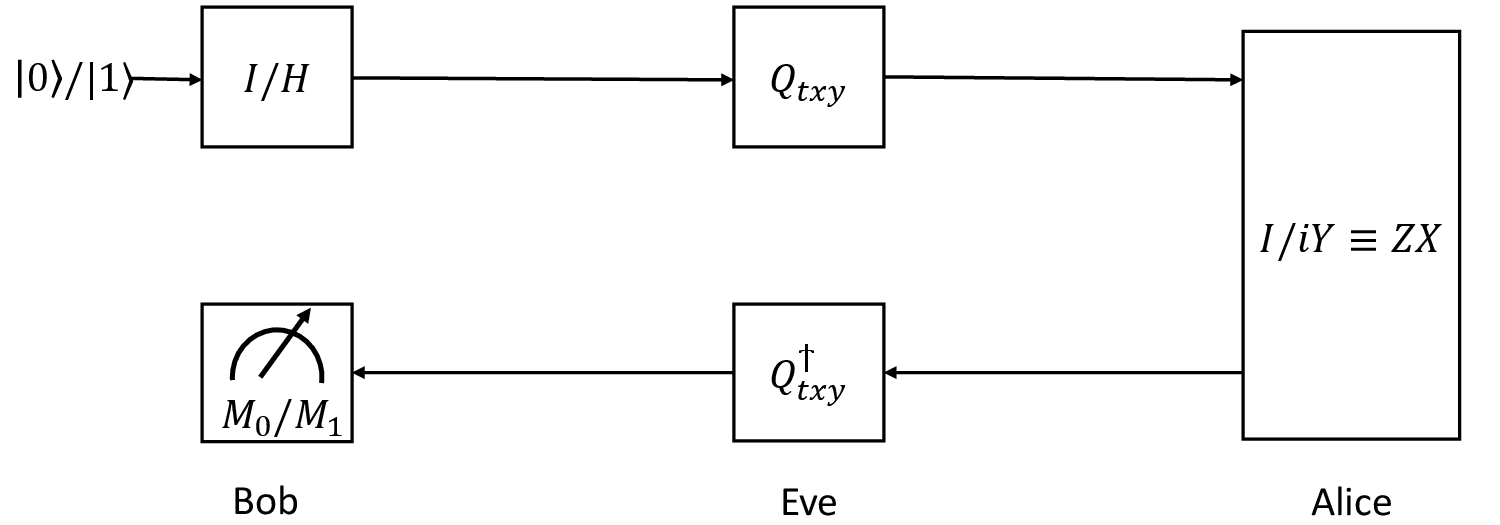} 
\par\end{centering}
\caption{The circuit representation of DL04 quantum game.}\label{fig:Chapter5_Fig1}
\end{figure}

Before introducing the conventional payoff function, it is essential
to frame the DL04 protocol within a game-theoretic context, where
principles of game theory are integrated into a quantum communication
protocol (see Figure \ref{fig:Chapter5_Fig1}). To simplify the analysis,
the DL04 quantum game is categorized into distinct game scenarios.
However, the subsequent discussion presents a generalized mapping
approach \cite{EW2000,EWL99,KP13}. Consider a game $\mathcal{G}$
modeled as a ``normal-form game'', which is mathematically defined
as a function with a specified domain and range. The players' preferences
are associated with elements of the range, influencing their rational
strategic decisions within the domain. The strategic choices available
to players in the quantum domain are now outlined. More formally,
the game $\mathcal{G}$ has an output set $O$ as its range and a
domain represented by the Cartesian product $S_{1}\times S_{2}\times\cdots\times S_{n}$,
where $S_{i}$ denotes the set of pure strategies available to the
$i^{th}$ player. Each element within this domain, $S_{1}\times S_{2}\times\cdots\times S_{n}$,
corresponds to a ``game play'' or ``strategy profile''. The
game function is expressed as $\mathcal{G}:S_{1}\times S_{2}\times\cdots\times S_{n}\longrightarrow O$,
where $n$ denotes the total number of players. In this specific case,
the game involves three participants: Alice, Bob, and Eve. Bob selects
a quantum state, Alice determines the measurement basis, and Eve executes
an attack strategy, all of which constitute elements of the domain
$S_{1}\times S_{2}\times S_{3}$, where $S_{1},S_{2}$, and $S_{3}$
correspond to the strategic choices of Alice, Bob, and Eve, respectively.
Thus, the game function can be specifically defined as $\mathcal{G}:S_{1}\times S_{2}\times S_{3}\longrightarrow O$
within the context of pure strategies. In practice, Bob encodes quantum
states in either the $Z$ or $X$ basis before transmitting them to
Alice. To ensure security, Alice measures a portion of Bob\textquoteright s
sequence and subsequently encodes a message by applying a Pauli operation
before returning it. Throughout this process, Eve attempts different
attack strategies via the quantum channel. The following sections
elaborate on how this game framework is incorporated into the quantum
communication protocol.


Alice initializes her primary sequence as a mixed quantum state, mathematically
expressed as

\[
\frac{p}{2}\left(|0\rangle\langle0|+|1\rangle\langle1|\right)+\frac{1-p}{2}\left(|+\rangle\langle+|+|-\rangle\langle-\right)
\]
Alice\textquoteright s quantum states exist within a two-dimensional
projective complex Hilbert space, denoted as $\mathcal{H}_{2}$. This
mixed state comprises four pure states, which collectively form a
superposition within $\mathcal{H}_{2}$. Consequently, when $|0\rangle$
undergoes a projective measurement in the $X$ basis\footnote{Quantum measurements in a two-dimensional Hilbert space are characterized
by the operator set: \{$M_{0}$,$M_{1}$: $M_{0}=|+\rangle\langle+|$,
$M_{1}=|-\rangle\langle-|$\}.}, it probabilistically collapses into either $|+\rangle$ or $|+\rangle$
with an equal probability of $\frac{1}{2}$, demonstrating quantum
superposition. Eve executes quantum attacks utilizing unitary transformations
and ancillary states, all operating within a projective complex Hilbert
space. Specifically, the attack strategies $E_{1}$, $E_{2}$, and
$E_{3}$ are conducted in an eight-dimensional Hilbert space $\mathcal{H}_{8}$,
whereas $E_{4}$ operates in $\mathcal{H}_{2}$. Within different
game-theoretic scenarios, Eve alternates between attack strategies
$E_{i}$ and $E_{j}$ with probabilities $r$ and $1-r$, respectively.
The interaction of Eve\textquoteright s unitary operations with Alice
and Bob\textquoteright s states enhances the quantum superposition,
effectively increasing the number of possible outcomes based on Alice's
measurement basis. This expansion in superposition influences the
measurement results for the legitimate players within the game\footnote{A comprehensive analysis of the unitary operations performed by Alice
and Bob, along with their measurement outcomes, is detailed in Section
\ref{sec:Chapter5_Sec3}.}. This also broadens the possible game outcomes $O$, where $O=\left\{ o_{1},o_{2},o_{3}\right\} $,
representing the respective outputs of Alice, Bob, and Eve. Alice
encodes information using Pauli operations, which also exist within
$\mathcal{H}_{2}$. Thus, the game framework can be described as a
mixed strategy game, denoted by the function $\mathcal{M}$. The 1-simplex
$\Delta_{1}$ represents the space of probability distributions over
each player's pure strategies, corresponding to the set of \textit{mixed
strategies}. Therefore, our mixed strategy game $\mathcal{M}$ is
formally defined over a domain that is the Cartesian product of the
probability distributions over the pure strategies of all players.

\[
\begin{array}{lcl}
\mathcal{M} & : & \Delta_{1}\times\Delta_{1}\times\Delta_{1}\longrightarrow\Delta_{7}\\
\\\mathcal{M} & : & \left(\left(p,1-p\right),\left(q,1-q\right),\left(r,1-r\right)\right)\\
 & \mapsto & \left(pqr,pq\left(1-r\right),p\left(1-q\right)r,p\left(1-q\right)\left(1-r\right),\left(1-p\right)qr,\right.\\
 &  & \left.\left(1-p\right)q\left(1-r\right),\left(1-p\right)\left(1-q\right)r,\left(1-p\right)\left(1-q\right)\left(1-r\right)\right)
\end{array},
\]
The framework encompasses probability distributions over possible
outcomes by mapping these outcomes to the vertices of the 7-simplex
$\Delta_{7}$. The necessary projective complex Hilbert space associated
with $\mathcal{M}$ is specified as follows:

\[
\begin{array}{lcl}
{\rm Bob's\,prepared\,state} & \longrightarrow & \mathcal{H}_{2}\\
{\rm Alice's\,measuremet\,basis} & \longrightarrow & \mathcal{H}_{2}\\
{\rm Alice's\,unitary\,operation} & \longrightarrow & \mathcal{H}_{2}\\
{\rm Eve's\,}E_{1},E_{2},E_{3}\,{\rm attack} & \longrightarrow & \mathcal{H}_{8}\\
{\rm Eve's\,}E_{4}\,{\rm attack} & \longrightarrow & \mathcal{H}_{2}
\end{array}.
\]
Furthermore, the superposition states are attributed to Bob, whereas
the unitary operations are performed by Alice and Eve, positioning
the overall state within an extended complex Hilbert space $\mathcal{H}_{8}$.
This mechanism introduces an elevated degree of randomness through
quantum superposition. Such enhanced randomness influences the outputs
of legitimate participants and expands the strategic scope of the
game. Consequently, these outcomes significantly impact key determinants
of the players' payoff functions, including mutual information, QBER,
and the detection probability of Eve\textquoteright s intrusion. Ultimately,
the effects of these payoff functions dictate the Nash equilibrium
conditions across different game settings. By utilizing quantum superposition
states and unitary transformations, non-cooperative game theory is
applied to the DL04 protocol, demonstrating Nash equilibrium as a
viable solution concept. Thus, the work of this chapter integrates
quantum mechanics into game-theoretic analysis, \textit{gaming the
quantum} \cite{KP13,KSB+18}.

Next, we define the payoff structure for each participant. The mutual
information shared between Alice and Bob positively contributes to
their respective payoffs while diminishing Eve\textquoteright s payoff.
Conversely, any mutual information exchanged between Alice and Eve
or Bob and Eve negatively impact the legitimate players' payoffs.
Moreover, the legitimate parties gain advantages from their ability
to detect Eve\textquoteright s presence\footnote{The impact of both parameters is of equal importance, depending on
the specific quantitative values of the QBER and the probability of
detecting Eve's presence ($P_{d}$) in both the message and control
modes of the DL04 protocol.}, which enhances their payoffs in this adversarial setting. Conversely,
Eve\textquoteright s payoff increases with the amount of information
acquired from Alice and Bob but decreases as the mutual information
between Alice and Bob grows. Additionally, Eve incurs a penalty upon
detection, making her payoff more favorable when the probability of
remaining undetected is higher. Eve may also deploy various quantum
gates to extract information from Alice and Bob, but an increased
number of such operations adds to her computational overhead, ultimately
reducing her payoff. Consequently, the payoff structure can be tailored
to capture different strategic scenarios and player incentives under
consideration. The DL04 protocol, modeled as a game, can be viewed
within the framework of a zero-sum game. The payoffs for Alice, Bob,
and Eve, under a general attack strategy $\mathcal{E}$, can thus
be formulated\footnote{It is assumed that the complexity of quantum state preparation and
measurement operations is identical for Alice, Bob, and Eve. Consequently,
this aspect is not taken into account when computing the payoffs for
these parties.} as follows \cite{KK18}.

\begin{equation}
\begin{array}{lcl}
P_{A}^{\mathcal{E}}(p,q) & = & \omega_{a}I\left({\rm A,B}\right)-\omega_{b}I\left({\rm A,E}\right)-\omega_{c}I\left({\rm B,E}\right)+\omega_{d}\left(\frac{P_{d}+{\rm QBER}}{2}\right)\\
\\P_{B}^{\mathcal{E}}(p,q) & = & \omega_{a}I\left({\rm A,B}\right)-\omega_{c}I\left({\rm A,E}\right)-\omega_{b}I\left({\rm B,E}\right)+\omega_{d}\left(\frac{P_{d}+{\rm QBER}}{2}\right)\\
\\P_{E}^{\mathcal{E}}(p,q) & = & -\omega_{e}I\left({\rm A,B}\right)+\omega_{f}I\left({\rm A,E}\right)+\omega_{g}I\left({\rm B,E}\right)+\omega_{h}\left(1-\frac{P_{d}+{\rm QBER}}{2}\right)\\
 & - & \omega_{i}n_{1}-\omega_{j}n_{2}-\omega_{k}n_{3}
\end{array}.\label{eq:Chapter5_Eq1}
\end{equation}
Here, $\omega_{a},\,\omega_{b},\,\omega_{c},\,\omega_{d},\,\omega_{e},\,\omega_{f},\,\omega_{g},\,\omega_{h},\,\omega_{i},\,\omega_{j}$,
and $\omega_{k}$ represent positive real-valued coefficients, serving
as weighting factors for different components in the payoff function.
These weights are constrained such that $\sum_{m=a}^{d}\omega_{m}=1$
and $\sum_{n=e}^{k}\omega_{n}=1$, where$m\in\{a,b,c,d\}$ and $n\in\{e,f,\ldots,k\}$.
Additionally, $I\left({\rm A,B}\right)$, $I\left({\rm A,E}\right)$,
and $I\left({\rm B},{\rm E}\right)$ denote the mutual information
shared between Alice and Bob, Alice and Eve, and Bob and Eve, respectively.
The parameter $P_{d}$ represents the probability of detecting Eve\textquoteright s
intrusion, while $n_{1}$, $n_{2}$, and $n_{3}$ correspond to the
number of single-qubit, two-qubit, and three-qubit gates, respectively.
Notably, in the defined payoff function, $P_{d}$ and $QBER$ are
treated equivalently since both provide insights into Eve\textquoteright s
presence in a similar manner.

Under the assumption that Eve possesses unlimited quantum resources
and operates with flawless quantum gates, the weights are set such
that $\omega_{i}=\omega_{j}=\omega_{k}=0$. This implies that the
number of quantum gates Eve employs for her attack does not impose
any constraints on her payoff function. To facilitate the analysis,
it is further assumed that all components contribute equally to the
payoff. Therefore, in this model, the weights are uniformly distributed
as $\omega_{a}=\omega_{b}=\omega_{c}=\omega_{d}=\omega_{e}=\omega_{f}=\omega_{g}=\omega_{h}=0.25$,
leading to a reformulation of Equation (\ref{eq:Chapter5_Eq1}),

\begin{equation}
\begin{array}{lcl}
P_{A}^{\mathcal{E}}(p,q) & = & 0.25\times\left[I\left({\rm A,B}\right)-I\left({\rm A,E}\right)-I\left({\rm B,E}\right)+\left(\frac{P_{d}+{\rm QBER}}{2}\right)\right]\\
\\P_{B}^{\mathcal{E}}(p,q) & = & 0.25\times\left[I\left({\rm A,B}\right)-I\left({\rm A,E}\right)-I\left({\rm B,E}\right)+\left(\frac{P_{d}+{\rm QBER}}{2}\right)\right]\\
\\P_{E}^{\mathcal{E}}(p,q) & = & 0.25\times\left[-I\left({\rm A,B}\right)+I\left({\rm A,E}\right)+I\left({\rm B,E}\right)+\left(1-\frac{P_{d}+{\rm QBER}}{2}\right)\right]
\end{array}.\label{eq:Chapter5_Eq2}
\end{equation}
In this scenario, it is evident that Alice and Bob receive identical
payoffs, as represented by Equation (\ref{eq:Chapter5_Eq2}). However,
for our analysis, we will consider these payoffs independently. This
distinction arises due to the differing probabilities---$q$ for
Alice and $p$ for Bob---associated with choosing encoded bit values
and employing the $Z$ basis for initializing quantum states. The
structured representation of the game $\mathcal{M}$ in Table \ref{tab:Chapter5_Tab1}
further clarifies this concept.

\begin{table}[h]
\centering{}\captionsetup{justification=justified, singlelinecheck=false}
\caption{The payoff matrix corresponding to the $\mathcal{M}$ game is presented,
where the three values within parentheses signify the payoffs of Alice,
Bob, and Eve, respectively. Here, the probabilities assigned to an
eavesdropper Eve's choice of attack strategies $E_{i}$ and $E_{j}$
are denoted by $r$ and $1-r$, respectively.}\label{tab:Chapter5_Tab1}
\begin{tabular}{|c|c|c|c|}
\hline 
 & \multicolumn{3}{c|}{Bob}\tabularnewline
\hline 
\multirow{12}{*}{Alice} &  & $Z\left(p\right)$ & $X\left(1-p\right)$\tabularnewline
\cline{2-4}
 &  & $\left(rP_{A}^{E_{i}}\left(p,q\right)+\left(1-r\right)P_{A}^{E_{j}}\left(p,q\right),\right.$ & $\left(rP_{A}^{E_{i}}\left(1-p,q\right)+\right.$\tabularnewline
 & $0\left(q\right)$ & $rP_{B}^{E_{i}}\left(p,q\right)+\left(1-r\right)P_{B}^{E_{j}}\left(p,q\right),$ & $\left(1-r\right)P_{A}^{E_{j}}\left(1-p,q\right),\,rP_{B}^{E_{i}}\left(1-p,q\right)$\tabularnewline
 &  & $\left.P_{E}^{E_{i/j}}\left(p,q\right)\right)$ & $+\left(1-r\right)P_{B}^{E_{j}}\left(1-p,q\right),$\tabularnewline
 &  &  & $\left.P_{E}^{E_{i/j}}\left(1-p,q\right)\right)$\tabularnewline
 &  &  & \tabularnewline
\cline{2-4}
 &  &  & \tabularnewline
 &  & $\left(rP_{A}^{E_{i}}\left(p,1-q\right)+\right.$ & $\left(rP_{A}^{E_{i}}\left(1-p,1-q\right)+\right.$\tabularnewline
 &  & $\left(1-r\right)P_{A}^{E_{j}}\left(p,1-q\right),$ & $\left(1-r\right)P_{A}^{E_{j}}\left(1-p,1-q\right),$\tabularnewline
 & $1\left(1-q\right)$ & $rP_{B}^{E_{i}}\left(p,1-q\right)+$ & $rP_{B}^{E_{i}}\left(1-p,1-q\right)+$\tabularnewline
 &  & $\left(1-r\right)P_{B}^{E_{j}}\left(p,1-q\right),$ & $\left(1-r\right)P_{B}^{E_{j}}\left(1-p,1-q\right),$\tabularnewline
 &  & $\left.P_{E}^{E_{i/j}}\left(p,1-q\right)\right)$ & $\left.P_{E}^{E_{i/j}}\left(1-p,1-q\right)\right)$\tabularnewline
\hline 
\end{tabular}
\end{table}

\section{Information-theoretic parameters in different attack scenarios}\label{sec:Chapter5_Sec3}

In this section, we analyze various attack scenarios to determine
the relevant information-theoretic parameters. To compute the payoffs
for all parties using Equation (\ref{eq:Chapter5_Eq2}), it is essential
to obtain the mutual information between players in the game scenario,
the detection probability of Eve\textquoteright s presence, and the
QBER for each respective attack scenario. These parameters are crucial
for deriving the best response function, which is discussed in the
following section.

\subsection{$E_{1}$ attack strategy}\label{subsec:Chapter5_Sec3.1}

We adopt W{\'o}jcik's attack methodology \cite{W03} as $E_{1}$
to exploit the DL04 protocol under both message and control configurations.
As outlined earlier, Bob initiates the process by randomly selecting
the initial quantum state from $|0\rangle$, $|1\rangle$, $|+\rangle$,
and $|-\rangle$ with respective probabilities of $\frac{p}{2}$,
$\frac{p}{2}$, $\frac{1-p}{2}$, and $\frac{1-p}{2}$. Alice then
encodes $j$ (where $j\in\{0,1\}$) onto the travel photon\footnote{Here, $t$ refers to the travel photon (qubit) sent by Bob (Alice)
in the B-A (A-B) attack scenario.} (denoted by the subscript $t$) received from Bob using the operation\footnote{Alice encodes a 0 (1) using the operation $iY_{t}^{0}\equiv I_{t}$
$\left(iY_{t}^{1}\equiv\left(ZX\right)_{t}^{1}\right)$ with respective
probabilities $q$ $\left(1-q\right)$.} $iY_{t}^{j}$. The communication channel shared between Alice and
Bob is modeled as a lossy quantum channel, characterized by a single-photon
transmission efficiency $\eta$. Meanwhile, Eve substitutes this lossy
channel with an ideal one, where $\eta=1$. To execute her attack,
Eve employs two auxiliary spatial modes, $x$ and $y$, and an ancillary
photon initially prepared in the state $|0\rangle$. She performs
eavesdropping twice: once during the photon transmission from Bob
to Alice (B-A attack) and again when it returns from Alice to Bob
(A-B attack). The attack procedure (see Figure 2 in Ref. \cite{W03})
begins with Eve initializing the auxiliary modes $x$ and $y$ in
the state $|{\rm vac}\rangle_{x}|0\rangle_{y}$, where $|{\rm vac}\rangle$
signifies an empty mode. During the B-A attack, Eve applies the unitary
transformation $Q_{txy}$, and during the A-B attack, she utilizes
its Hermitian conjugate $Q_{txy}^{\dagger}$ (or equivalently $Q_{txy}^{-1}$).
The transformation $Q_{txy}$ is defined as

\[
\begin{array}{lcl}
Q_{txy}={\rm SWAP}_{tx}\,{\rm CPBS}_{txy}\,H_{y} & \equiv & {\rm SWAP}_{tx}\otimes I_{y}\,{\rm CPBS}_{txy}\,I_{t}\otimes I_{x}\otimes H_{y}\end{array}
\]
where it consists of a Hadamard gate (a single-qubit operation), a
SWAP gate (a two-qubit operation), and a controlled polarizing beam
splitter (CPBS), which is a three-qubit gate\footnote{The polarizing beam splitter (PBS) is assumed to allow the transmission
of photons in the $|0\rangle$ state while reflecting those in the
$|1\rangle$ state.}. We will now evaluate this attack strategy under both message and
control modes to determine the relevant payoff functions for Alice,
Bob, and Eve.

\textbf{\textit{Message mode:}} To begin with a straightforward approach,
let us analyze cases where Alice encodes a 0. Specifically, we consider
the scenario in which Bob initializes the system in the state $|0\rangle_{t}$.
Subsequently, we examine Eve's attack strategy on the composite system,
initially represented as $|0\rangle_{t}|{\rm vac}\rangle_{x}|0\rangle_{y}$
and denoted by $E_{1}$,

\[
\begin{array}{lcl}
|{\rm B-A}\rangle_{|0\rangle E_{1}} & = & Q_{txy}\,|0\rangle_{t}|{\rm vac}\rangle_{x}|0\rangle_{y}\\
 & = & {\rm SWAP}_{tx}\,{\rm CPBS}_{txy}\,H_{y}\,\left(|0\rangle|{\rm vac}\rangle|0\rangle\right)_{txy}\\
 & = & {\rm SWAP}_{tx}\,{\rm CPBS}_{txy}\,\frac{1}{\sqrt{2}}\left(|0\rangle|{\rm vac}\rangle|0\rangle+|0\rangle|{\rm vac}\rangle|1\rangle\right)_{txy}\\
 & = & {\rm SWAP}_{tx}\,\frac{1}{\sqrt{2}}\left(|0\rangle|0\rangle|{\rm vac}\rangle+|0\rangle|{\rm vac}\rangle|1\rangle\right)_{txy}\\
 & = & \frac{1}{\sqrt{2}}\left[|0\rangle|0\rangle|{\rm vac}\rangle+|{\rm vac}\rangle|0\rangle|1\rangle\right]_{txy}
\end{array}.
\]
When Alice applies the operation $iY_{t}^{0}$ (corresponding to encoding
0) on the $t$-state, the composite system, denoted as $|{\rm B-A}\rangle_{|0\rangle E_{1}}$,
remains unchanged. This is symbolically expressed as $|{\rm B-A}\rangle_{|0\rangle E_{1}}^{0}$,
where the superscript indicates the bit Alice encodes. The resulting
system after Eve's A-B attack, denoted by $\left(Q_{txy}^{-1}\right)$,
is given by:

\begin{equation}
\begin{array}{lcl}
|{\rm A-B}\rangle_{|0\rangle E_{1}}^{0} & = & Q_{txy}^{-1}\,|{\rm B-A}\rangle_{|0\rangle E_{1}}^{0}\\
 & = & Q_{txy}^{-1}\,\frac{1}{\sqrt{2}}\left[|0\rangle|0\rangle|{\rm vac}\rangle+|{\rm vac}\rangle|0\rangle|1\rangle\right]_{txy}\\
 & = & H_{y}\,{\rm CPBS}_{txy}\,{\rm SWAP}_{tx}\,\frac{1}{\sqrt{2}}\left[|0\rangle|0\rangle|{\rm vac}\rangle+|{\rm vac}\rangle|0\rangle|1\rangle\right]_{txy}\\
 & = & H_{y}\,{\rm CPBS}_{txy}\,\frac{1}{\sqrt{2}}\left[|0\rangle|0\rangle|{\rm vac}\rangle+|0\rangle|{\rm vac}\rangle|1\rangle\right]_{txy}\\
 & = & H_{y}\,\frac{1}{\sqrt{2}}\left[|0\rangle|{\rm vac}\rangle|0\rangle+|0\rangle|{\rm vac}\rangle|1\rangle\right]_{txy}\\
 & = & |0\rangle_{t}|{\rm vac}\rangle_{x}|0\rangle_{y}
\end{array}.\label{eq:Chapter5_AEq1}
\end{equation}
This final state is consistent with the original composite system,
which is expected, as Alice's encoding operation acts as an identity
transformation for encoding 0. Meanwhile, Eve implements a unitary
A-B attack operation $\left(Q_{txy}\right)$ and subsequently its
inverse, the unitary B-A attack operation $\left(Q_{txy}^{-1}\right)$.
Similarly, analogous composite systems can be described for cases
where Alice encodes 0 while Bob prepares the initial states $|1\rangle_{t}$,
$|+\rangle_{t}$, and $|-\rangle_{t}$.

\begin{equation}
\begin{array}{lcl}
|{\rm A-B}\rangle_{|1\rangle E_{1}}^{0} & = & |1\rangle_{t}|{\rm vac}\rangle_{x}|0\rangle_{y}\end{array},\label{eq:Chapter5_AEq2}
\end{equation}

\begin{equation}
\begin{array}{lcl}
|{\rm A-B}\rangle_{|+\rangle E_{1}}^{0} & = & |+\rangle_{t}|{\rm vac}\rangle_{x}|0\rangle_{y}\end{array},\label{eq:Chapter5_AEq3}
\end{equation}
and

\begin{equation}
\begin{array}{lcl}
|{\rm A-B}\rangle_{|-\rangle E_{1}}^{0} & = & |-\rangle_{t}|{\rm vac}\rangle_{x}|0\rangle_{y}\end{array},\label{eq:Chapter5_AEq4}
\end{equation}
respectively.

Now, let us explore the scenario where Alice encodes a 1 by applying
the operation $iY_{t}^{1}$ on the $t$-state. First, we analyze the
system when Bob's initial state is $|0\rangle_{t}$. Previously, we
established that the composite system is represented as

\[
\begin{array}{lcl}
|{\rm B-A}\rangle_{|0\rangle} & = & \frac{1}{\sqrt{2}}\left[|0\rangle|0\rangle|{\rm vac}\rangle+|{\rm vac}\rangle|0\rangle|1\rangle\right]_{txy}.\end{array}
\]
Following Alice's encoding operation, Eve subsequently applies $Q_{txy}^{-1}$,
resulting in the composite system:

\begin{equation}
\begin{array}{lcl}
|{\rm A-B}\rangle_{|0\rangle E_{1}}^{1} & = & Q_{txy}^{-1}\,iY_{t}^{1}\,\frac{1}{\sqrt{2}}\left[|0\rangle|0\rangle|{\rm vac}\rangle+|{\rm vac}\rangle|0\rangle|1\rangle\right]_{txy}\\
 & = & Q_{txy}^{-1}\,\frac{1}{\sqrt{2}}\left[-|1\rangle|0\rangle|{\rm vac}\rangle+|{\rm vac}\rangle|0\rangle|1\rangle\right]_{txy}\\
 & = & H_{y}\,{\rm CPBS}_{txy}\,{\rm SWAP}_{tx}\,\frac{1}{\sqrt{2}}\left[-|1\rangle|0\rangle|{\rm vac}\rangle+|{\rm vac}\rangle|0\rangle|1\rangle\right]_{txy}\\
 & = & H_{y}\,{\rm CPBS}_{txy}\,\frac{1}{\sqrt{2}}\left[-|0\rangle|1\rangle|{\rm vac}\rangle+|0\rangle|{\rm vac}\rangle|1\rangle\right]_{txy}\\
 & = & H_{y}\,\frac{1}{\sqrt{2}}\left[-|0\rangle|1\rangle|{\rm vac}\rangle+|0\rangle|{\rm vac}\rangle|1\rangle\right]_{txy}\\
 & = & \left[-\frac{1}{\sqrt{2}}|0\rangle|1\rangle|{\rm vac}\rangle+\frac{1}{2}|0\rangle|{\rm vac}\rangle|0\rangle-\frac{1}{2}|0\rangle|{\rm vac}\rangle|1\rangle\right]_{txy}
\end{array}.\label{eq:Chapter5_AEq5}
\end{equation}
For the sake of brevity, we will omit explicit derivations of similar
cases and instead focus on presenting the key results. Next, we analyze
the transformation of the composite system when Bob's initial state
is $|1\rangle_{t}$ following Eve\textquoteright s B-A attack.

\[
\begin{array}{lcl}
|{\rm B-A}\rangle_{|1\rangle E_{1}} & = & Q_{txy}\,|1\rangle_{t}|{\rm vac}\rangle_{x}|0\rangle_{y}\\
 & = & \frac{1}{\sqrt{2}}\left[|{\rm vac}\rangle|1\rangle|0\rangle\rangle+|1\rangle|1\rangle|{\rm vac}\rangle\right]_{txy}
\end{array}.
\]
The composite system that emerges from Alice's operation combined
with Eve's A-B attack can be expressed as follows:

\begin{equation}
\begin{array}{lcl}
|{\rm A-B}\rangle_{|1\rangle E_{1}}^{1} & = & Q_{txy}^{-1}\,iY_{t}^{1}\,\frac{1}{\sqrt{2}}\left[|{\rm vac}\rangle|1\rangle|0\rangle\rangle+|1\rangle|1\rangle|{\rm vac}\rangle\right]_{txy}\\
 & = & \left[\frac{1}{\sqrt{2}}|1\rangle|0\rangle|{\rm vac}\rangle+\frac{1}{2}|1\rangle|{\rm vac}\rangle|0\rangle+\frac{1}{2}|1\rangle|{\rm vac}\rangle|1\rangle\right]_{txy}
\end{array}.\label{eq:Chapter5_AEq6}
\end{equation}
The outcomes corresponding to Bob's initial states, $|+\rangle$ and
$|-\rangle$, are as follows:

\[
\begin{array}{lcl}
|{\rm B-A}\rangle_{|+\rangle E_{1}} & = & \frac{1}{2}\left[|0\rangle|0\rangle|{\rm vac}\rangle+|{\rm vac}\rangle|0\rangle|1\rangle+|{\rm vac}\rangle|1\rangle|0\rangle+|1\rangle|1\rangle|{\rm vac}\rangle\right]_{txy}\end{array},
\]

\begin{equation}
\begin{array}{lcl}
|{\rm A-B}\rangle_{|+\rangle E_{1}}^{1} & = & \left[\frac{1}{2}\left\{ |+\rangle|{\rm vac}\rangle|0\rangle-|-\rangle|{\rm vac}\rangle|1\rangle\right\} +\frac{1}{2\sqrt{2}}\left\{ -|+\rangle|1\rangle|{\rm vac}\rangle-|-\rangle|1\rangle|{\rm vac}\rangle\right.\right.\\
 & + & \left.\left.|+\rangle|0\rangle|{\rm vac}\rangle-|-\rangle|0\rangle|{\rm vac}\rangle\right\} \right]_{txy}
\end{array},\label{eq:Chapter5_AEq7}
\end{equation}
and

\[
\begin{array}{lcl}
|{\rm B-A}\rangle_{|-\rangle E_{1}} & = & \frac{1}{2}\left[|0\rangle|0\rangle|{\rm vac}\rangle+|{\rm vac}\rangle|0\rangle|1\rangle-|{\rm vac}\rangle|1\rangle|0\rangle-|1\rangle|1\rangle|{\rm vac}\rangle\right]_{txy}\end{array},
\]

\begin{equation}
\begin{array}{lcl}
|{\rm A-B}\rangle_{|-\rangle E_{1}}^{1} & = & \left[\frac{1}{2}\left\{ |-\rangle|{\rm vac}\rangle|0\rangle-|+\rangle|{\rm vac}\rangle|1\rangle\right\} +\frac{1}{2\sqrt{2}}\left\{ -|+\rangle|1\rangle|{\rm vac}\rangle-|-\rangle|1\rangle|{\rm vac}\rangle\right.\right.\\
 & - & \left.\left.|+\rangle|0\rangle|{\rm vac}\rangle+|-\rangle|0\rangle|{\rm vac}\rangle\right\} \right]_{txy}
\end{array}.\label{eq:Chapter5_AEq8}
\end{equation}
In the primary discussion, we previously established that Bob can
deterministically decode Alice's message ($j$) by measuring the qubit
on the same basis on which it was initially prepared. This process
does not require a classical announcement. The information obtained
by Bob upon decoding is denoted as $m$. Given the aforementioned
results, we can directly determine Eve's optimal approach to infer
Alice's encoded information based on her measurement outcomes from
her auxiliary modes, $x$ and $y$. Eve's decoded information is represented
as $k$.

\textit{Eve's optimal strategy} Eve correctly deciphers $k=0$ if
her ancillary spatial mode $x$ is in an empty state while mode $y$
is in state $|0\rangle$, mathematically expressed as $|{\rm vac}\rangle_{x}|0\rangle_{y}$.
Conversely, Eve decodes $k=1$ when her $x$ mode is in a nonempty
state, or if it is empty while her $y$ mode is in state $|1\rangle$,
described as $|0\rangle_{x}|{\rm vac}\rangle_{y}$, $|1\rangle_{x}|{\rm vac}\rangle_{y}$,
or $|{\rm vac}\rangle_{x}|1\rangle_{y}$. Now, considering the parameter
$p_{jmk}^{E_{1}}$, which defines the joint probability distribution
where $j$, $m$, and $k$ correspond to Alice\textquoteright s encoding,
Bob\textquoteright s decoding, and Eve\textquoteright s inferred information,
respectively, we can express these probabilities using Equations (\ref{eq:Chapter5_AEq1})
$-$ (\ref{eq:Chapter5_AEq8}) as follows:

\begin{equation}
\begin{array}{lcl}
p_{000}^{E_{1}} & = & q\\
\\p_{001}^{E_{1}} & = & p_{010}^{E_{1}}=p_{011}^{E_{1}}=0\\
\\p_{100}^{E_{1}} & = & \frac{1}{4}\left(1-q\right)\\
\\p_{101}^{E_{1}} & = & \left(1-q\right)\left(\frac{1}{4}+\frac{p}{2}\right)\\
\\p_{110}^{E_{1}} & = & 0\\
\\p_{111}^{E_{1}} & = & \frac{1}{2}\left(1-p\right)\left(1-q\right)
\end{array}.\label{eq:Chapter5_AEq9}
\end{equation}
To evaluate the mutual information shared among Alice, Bob, and Eve,
we employ Equation (\ref{eq:Chapter5_AEq9}). To further simplify
the resulting expressions, we utilize Shannon entropy, defined as
$\mathbf{H}\left[x\right]=-x\log_{2}x$, where $x$ represents the
probability of a particular event occurring.

\begin{equation}
\begin{array}{lcl}
H\left({\rm B|A}\right)_{E_{1}} & = & \left(1-q\right)\left(\mathbf{H}\left[\frac{1}{2}\left(1-p\right)\right]+\mathbf{H}\left[\frac{1}{2}\left(1+p\right)\right]\right)\\
\\H\left({\rm B}\right)_{E_{1}} & = & \mathbf{H}\left[q+\frac{1}{2}\left(1-q\right)\left(1+p\right)\right]+\mathbf{H}\left[\frac{1}{2}\left(1-p\right)\left(1-q\right)\right]\\
\\I\left({\rm A,B}\right)_{E_{1}} & = & H\left({\rm B}\right)_{E_{1}}-H\left({\rm B|A}\right)_{E_{1}}
\end{array},\label{eq:Chapter5_AEq10}
\end{equation}

\begin{equation}
\begin{array}{lcl}
H\left({\rm A|E}\right)_{E_{1}} & = & \frac{1}{4}\left(1+3q\right)\left(\mathbf{H}\left[\frac{4q}{1+3q}\right]+\mathbf{H}\left[\frac{1-q}{1+3q}\right]\right)\\
\\H\left({\rm A}\right)_{E_{1}} & = & \mathbf{H}\left[q\right]+\mathbf{H}\left[1-q\right]\\
\\I\left({\rm A,E}\right)_{E_{1}} & = & H\left({\rm A}\right)_{E_{1}}-H\left({\rm A|E}\right)_{E_{1}}
\end{array},\label{eq:Chapter5_AEq11}
\end{equation}
and

\begin{equation}
\begin{array}{lcl}
H\left({\rm B|E}\right)_{E_{1}} & = & \frac{3}{4}\left(1-q\right)\left(\mathbf{H}\left[\frac{1}{3}\left(1+2p\right)\right]+\mathbf{H}\left[\frac{2}{3}\left(1-p\right)\right]\right)\\
\\H\left({\rm B}\right)_{E_{1}} & = & \mathbf{H}\left[\frac{1}{2}\left(p+q-pq+1\right)\right]+\mathbf{H}\left[\frac{1}{2}\left(1-p\right)\left(1-q\right)\right]\\
\\I\left({\rm B,E}\right)_{E_{1}} & = & H\left({\rm B}\right)_{E_{1}}-H\left({\rm B|E}\right)_{E_{1}}
\end{array}.\label{eq:Chapter5_AEq12}
\end{equation}
Additionally, it is evident that eavesdropping introduces a QBER,

\begin{equation}
\begin{array}{lcl}
{\rm QBER}_{E_{1}} & = & \underset{j\ne m}{\sum}p_{jmk}^{E_{1}}\\
 & = & \frac{1}{2}\left(1-q\right)\left(1+p\right)
\end{array}.\label{eq:Chapter5_AEq13}
\end{equation}
\textbf{\textit{Control mode:}} As described in \cite{DL04}, the
previously mentioned ``double'' control mode comprises two independent
tests conducted on the quantum channel, each analogous to the single
test performed in the one-way BB84 protocol \cite{BB84,GRT+02}. Following
Eve\textquoteright s B-A attack, when Bob initializes the state as
$|0\rangle_{t}$, the resulting composite system can be expressed
as:

\[
\begin{array}{lcl}
|{\rm B-A}\rangle_{|0\rangle E_{1}} & = & \frac{1}{\sqrt{2}}\left[|0\rangle|0\rangle|{\rm vac}\rangle+|{\rm vac}\rangle|0\rangle|1\rangle\right]_{txy}\end{array}.
\]
Within the control mode, Alice performs a random measurement on the
traveling qubit using both the computational ($Z$ basis) and diagonal
($X$ basis) bases before sending the projected qubit back to Bob.
Bob then measures the qubit using the same basis he initially selected
(in this case, the $Z$ basis). Consequently, only scenarios where
Alice measures the state $t$ in the $Z$ basis are considered. The
expression $|{\rm B-A}\rangle_{|0\rangle E_{1}}$ signifies that there
is a $\frac{1}{2}$ probability that Alice will not detect any photon.
However, if a photon is detected, its state remains identical to Bob\textquoteright s
initial state. This suggests that in the $Z$ basis, Eve\textquoteright s
B-A attack does not lead to a detectable probability of eavesdropping.
Nevertheless, eavesdropping may still be inferred by monitoring transmission
losses, provided that the channel transmittance between Alice and
Bob remains below 0.5. In this setting, Eve executes an A-B attack,
denoted as $Q_{txy}^{-1}$, on the photon detected by Alice. Bob subsequently
measures the state $t$ in the $Z$ basis and sends the projected
qubit back to Alice. The composite system following Eve's A-B attack
on Alice and Bob is:

\begin{equation}
\begin{array}{lcl}
Q_{txy}^{-1}\,|0\rangle_{t}|0\rangle_{x}|{\rm vac}\rangle_{y} & = & H_{y}\,{\rm CPBS}_{txy}\,{\rm SWAP}_{tx}\,|0\rangle_{t}|0\rangle_{x}|{\rm vac}\rangle_{y}\\
 & = & \frac{1}{\sqrt{2}}\left[|0\rangle|{\rm vac}\rangle|0\rangle+|0\rangle|{\rm vac}\rangle|1\rangle\right]_{txy}
\end{array}.\label{eq:Chapter5_AEq14}
\end{equation}
When Bob's initial state is $|1\rangle_{t}$, Alice has an equal probability
of detecting or not detecting a photon. Bob then measures the state
$t$ in the $Z$ basis and transmits the projected qubit back to Alice.
The composite system after Eve\textquoteright s A-B attack is:

\begin{equation}
\begin{array}{lcl}
Q_{txy}^{-1}\,|1\rangle_{t}|1\rangle_{x}|{\rm vac}\rangle_{y} & = & \frac{1}{\sqrt{2}}\left[|1\rangle|{\rm vac}\rangle|0\rangle-|1\rangle|{\rm vac}\rangle|1\rangle\right]_{txy}\end{array}.\label{eq:Chapter5_AEq15}
\end{equation}
The scenario undergoes a significant shift if Bob initializes the
state in the $X$ basis rather than the $Z$ basis. Specifically,
if Bob's initial state is $|+\rangle_{t}$, then following Eve\textquoteright s
B-A attack, the resulting composite system is given by:

\[
\begin{array}{lcl}
|{\rm B-A}\rangle_{|+\rangle E_{1}} & = & \frac{1}{2}\left[|0\rangle|0\rangle|{\rm vac}\rangle+|{\rm vac}\rangle|0\rangle|1\rangle+|{\rm vac}\rangle|1\rangle|0\rangle+|1\rangle|1\rangle|{\rm vac}\rangle\right]_{txy}\\
 & = & \frac{1}{2}\left[\frac{1}{\sqrt{2}}\left\{ \left(|+\rangle+|-\rangle\right)|0\rangle|{\rm vac}\rangle+\left(|+\rangle-|-\rangle\right)|1\rangle|{\rm vac}\rangle\right\} +|{\rm vac}\rangle|0\rangle|1\rangle\right.\\
 & + & \left.|{\rm vac}\rangle|1\rangle|0\rangle\right]_{txy}
\end{array}.
\]
Alice, similar to the $Z$ basis case, has a probability of $\frac{1}{2}$
of not detecting a photon. However, in instances where detection occurs,
the photon is never found in Bob\textquoteright s original state $|+\rangle_{t}$.
This indicates that in the B-A attack, $\left(Q_{txy}\right)$, Eve\textquoteright s
detection process is relevant to the $X$ basis configuration, unlike
the $Z$ basis case. Next, we rigorously evaluate Eve\textquoteright s
A-B attack $\left(Q_{txy}^{-1}\right)$ by considering all four possible
measurement outcomes from Alice. The final composite system upon which
Bob performs his measurement, after Alice\textquoteright s measurement
collapses the state, is given by $|+\rangle_{t}|0\rangle_{x}|{\rm vac}\rangle_{y}$,

\begin{equation}
\begin{array}{lcl}
Q_{txy}^{-1}\,|+\rangle_{t}|0\rangle_{x}|{\rm vac}\rangle_{y} & = & Q_{txy}^{-1}\,\frac{1}{\sqrt{2}}\left[|0\rangle|0\rangle|{\rm vac}\rangle+|1\rangle|0\rangle|{\rm vac}\rangle\right]_{txy}\\
 & = & \frac{1}{\sqrt{2}}|0\rangle_{t}\left[\frac{1}{\sqrt{2}}|{\rm vac}\rangle\left(|0\rangle+|1\rangle\right)+|1\rangle|{\rm vac}\rangle\right]_{xy}\\
 & = & \frac{1}{2}\left(|+\rangle+|-\rangle\right)_{t}\left[\frac{1}{\sqrt{2}}|{\rm vac}\rangle\left(|0\rangle+|1\rangle\right)+|1\rangle|{\rm vac}\rangle\right]_{xy}
\end{array}.\label{eq:Chapter5_AEq16}
\end{equation}
Similarly, the remaining three composite states following Alice\textquoteright s
measurement are $|+\rangle_{t}|1\rangle_{x}|{\rm vac}\rangle_{y}$,
$|-\rangle_{t}|0\rangle_{x}|{\rm vac}\rangle_{y}$, and $|-\rangle_{t}|1\rangle_{x}|{\rm vac}\rangle_{y}$.
As a result, the final composite systems take the form:

\begin{equation}
\begin{array}{lcl}
Q_{txy}^{-1}\,|+\rangle_{t}|1\rangle_{x}|{\rm vac}\rangle_{y} & = & \frac{1}{2}\left(|+\rangle-|-\rangle\right)_{t}\left[\frac{1}{\sqrt{2}}|{\rm vac}\rangle\left(|0\rangle-|1\rangle\right)+|0\rangle|{\rm vac}\rangle\right]_{xy}\end{array},\label{eq:Chapter5_AEq17}
\end{equation}

\begin{equation}
\begin{array}{lcl}
Q_{txy}^{-1}\,|-\rangle_{t}|0\rangle_{x}|{\rm vac}\rangle_{y} & = & \frac{1}{2}\left(|+\rangle+|-\rangle\right)_{t}\left[\frac{1}{\sqrt{2}}|{\rm vac}\rangle\left(|0\rangle+|1\rangle\right)-|1\rangle|{\rm vac}\rangle\right]_{xy}\end{array},\label{eq:Chapter5_AEq18}
\end{equation}
and

\begin{equation}
\begin{array}{lcl}
Q_{txy}^{-1}\,|-\rangle_{t}|1\rangle_{x}|{\rm vac}\rangle_{y} & = & \frac{1}{2}\left(|+\rangle-|-\rangle\right)_{t}\left[-\frac{1}{\sqrt{2}}|{\rm vac}\rangle\left(|0\rangle-|1\rangle\right)+|0\rangle|{\rm vac}\rangle\right]_{xy}\end{array}.\label{eq:Chapter5_AEq19}
\end{equation}
In cases where Bob's initial state is $|-\rangle_{t}$, the composite
state after Eve\textquoteright s B-A attack becomes:

\[
\begin{array}{lcl}
|{\rm B-A}\rangle_{|-\rangle E_{1}} & = & \frac{1}{2}\left[|0\rangle|0\rangle|{\rm vac}\rangle+|{\rm vac}\rangle|0\rangle|1\rangle-|{\rm vac}\rangle|1\rangle|0\rangle-|1\rangle|1\rangle|{\rm vac}\rangle\right]_{txy}\\
 & = & \frac{1}{2}\left[\frac{1}{\sqrt{2}}\left\{ \left(|+\rangle+|-\rangle\right)|0\rangle|{\rm vac}\rangle-\left(|+\rangle-|-\rangle\right)|1\rangle|{\rm vac}\rangle\right\} +|{\rm vac}\rangle|0\rangle|1\rangle\right.\\
 & - & \left.|{\rm vac}\rangle|1\rangle|0\rangle\right]_{txy}
\end{array}.
\]
From a conceptual standpoint, it is evident that the overall scenario
remains fundamentally equivalent whether Bob\textquoteright s initial
state is $|-\rangle_{t}$ or $|+\rangle_{t}$. Consequently, Eve\textquoteright s
A-B attack affects all four possible outcomes of Alice\textquoteright s
measurement in the same manner, yielding identical results as expressed
in Equations (\ref{eq:Chapter5_AEq16}) $-$ (\ref{eq:Chapter5_AEq19}).

If we disregard the possibility of photon non-detection---favoring
an optimal eavesdropping strategy for Eve---the probability of Eve
remaining undetected in the $E_{1}$ attack scenario can be determined
using Equations (\ref{eq:Chapter5_AEq14}) $-$ (\ref{eq:Chapter5_AEq19}),

\[
\begin{array}{lcl}
P_{nd}^{E_{1}} & = & \frac{1}{4}\left(1+1+\frac{1}{4}+\frac{1}{4}\right)=\frac{5}{8}\end{array}.
\]
The mean\footnote{Eve employs two attack strategies---B-A attack and A-B attack. Additionally,
there exists a ``double'' control mode for Eve\textquoteright s
detection, as described in \cite{DL04}, leading to the emergence
of the $\frac{1}{2}$ factor.} probability of detecting Eve's presence in the $E_{1}$ attack scenario
is as follows:

\begin{equation}
\begin{array}{lcl}
P_{d}^{E_{1}} & = & \frac{1}{2}\left(1-\frac{5}{8}\right)=\frac{3}{16}=0.1875\end{array}.\label{eq:Chapter5_AEq20}
\end{equation}
The upper limit of Alice and Bob's tolerance for the probability of
detecting Eve in the control mode within the $E_{1}$ attack scenario
is 0.1875. This implies that the threshold probability for Eve's detection
is $18.75\%$.

\subsection{$E_{2}$ attack strategy}\label{subsec:Chapter5_Sec3.2}

\textbf{\textit{Message mode:}} W{\'o}jcik proposed an alternative
approach for symmetry-based attacks, designated as $E_{2}$, where
an additional unitary operation, $S_{ty}$, is applied with a probability
of $\frac{1}{2}$ immediately after executing $Q_{txy}^{-1}$ during
the A-B attack \cite{W03}. The $S_{ty}$ operation is characterized
as $X_{t}Z_{t}{\rm CNOT}_{ty}X_{t}$, integrating the Pauli-$Z$ operation,
Pauli-$X$, and the controlled-NOT (CNOT) gate. In this scenario,
the B-A attack remains identical to that in $E_{1}$, with modifications
exclusively affecting the A-B attack. The notation conventions remain
unchanged, apart from substituting $E_{1}$ with $E_{2}$. The final
state of the composite system, following the application of the $S_{ty}$
operation in an A-B attack---where Alice encodes a 0 and Bob's initial
state is $|0\rangle_{t}$---is determined as follows:

\begin{equation}
\begin{array}{lcl}
|{\rm A-B}\rangle_{|0\rangle E_{2}}^{0} & = & S_{ty}\,|{\rm A-B}\rangle_{|0\rangle E_{1}}^{0}\\
 & = & X_{t}\,Z_{t}\,{\rm CNOT}_{ty}\,X_{t}\,|0\rangle_{t}|{\rm vac}\rangle_{x}|0\rangle_{y}\\
 & = & X_{t}\,Z_{t}\,{\rm CNOT}_{ty}\,|1\rangle_{t}|{\rm vac}\rangle_{x}|0\rangle_{y}\\
 & = & X_{t}\,Z_{t}\,|1\rangle_{t}|{\rm vac}\rangle_{x}|0\rangle_{y}\\
 & = & X_{t}\,\left(-|1\rangle_{t}|{\rm vac}\rangle_{x}|0\rangle_{y}\right)\\
 & =- & |0\rangle_{t}|{\rm vac}\rangle_{x}|0\rangle_{y}
\end{array}.\label{eq:Chapter5_BEq1}
\end{equation}
Similarly, the resulting composite system states can be obtained for
cases where Alice encodes 0 while Bob's initial states are $|1\rangle$,
$|+\rangle$, and $|-\rangle$.

\begin{equation}
\begin{array}{lcl}
|{\rm A-B}\rangle_{|1\rangle E_{2}}^{0} & = & |1\rangle_{t}|{\rm vac}\rangle_{x}|0\rangle_{y}\end{array},\label{eq:Chapter5_BEq2}
\end{equation}

\begin{equation}
\begin{array}{lcl}
|{\rm A-B}\rangle_{|+\rangle E_{2}}^{0} & = & \frac{1}{2}\left[-|+\rangle|{\rm vac}\rangle|1\rangle-|-\rangle|{\rm vac}\rangle|1\rangle+|+\rangle|{\rm vac}\rangle|0\rangle-|-\rangle|{\rm vac}\rangle|0\rangle\right]_{txy}\end{array},\label{eq:Chapter5_BEq3}
\end{equation}
and

\begin{equation}
\begin{array}{lcl}
|{\rm A-B}\rangle_{|-\rangle E_{2}}^{0} & = & \frac{1}{2}\left[-|+\rangle|{\rm vac}\rangle|1\rangle-|-\rangle|{\rm vac}\rangle|1\rangle-|+\rangle|{\rm vac}\rangle|0\rangle+|-\rangle|{\rm vac}\rangle|0\rangle\right]_{txy}\end{array},\label{eq:Chapter5_BEq4}
\end{equation}
respectively.

The final state of the composite system, after executing the $S_{ty}$
operation in an A-B attack where Alice encodes 1 and Bob\textquoteright s
initial state is $|0\rangle_{t}$, is given as follows:

\begin{equation}
\begin{array}{lcl}
|{\rm A-B}\rangle_{|0\rangle E_{2}}^{1} & = & S_{ty}\,|{\rm A-B}\rangle_{|0\rangle E_{1}}^{1}\\
 & = & X_{t}\,Z_{t}\,{\rm CNOT}_{ty}\,X_{t}\,\left[-\frac{1}{\sqrt{2}}|0\rangle|1\rangle|{\rm vac}\rangle+\frac{1}{2}|0\rangle|{\rm vac}\rangle|0\rangle-\frac{1}{2}|0\rangle|{\rm vac}\rangle|1\rangle\right]_{txy}\\
 & = & X_{t}\,Z_{t}\,{\rm CNOT}_{ty}\,\left[-\frac{1}{\sqrt{2}}|1\rangle|1\rangle|{\rm vac}\rangle+\frac{1}{2}|1\rangle|{\rm vac}\rangle|0\rangle-\frac{1}{2}|1\rangle|{\rm vac}\rangle|1\rangle\right]_{txy}\\
 & = & X_{t}\,Z_{t}\,\left[-\frac{1}{\sqrt{2}}|1\rangle|1\rangle|{\rm vac}\rangle+\frac{1}{2}|1\rangle|{\rm vac}\rangle|1\rangle-\frac{1}{2}|1\rangle|{\rm vac}\rangle|0\rangle\right]_{txy}\\
 & = & X_{t}\,\left[\frac{1}{\sqrt{2}}|1\rangle|1\rangle|{\rm vac}\rangle-\frac{1}{2}|1\rangle|{\rm vac}\rangle|1\rangle+\frac{1}{2}|1\rangle|{\rm vac}\rangle|0\rangle\right]_{txy}\\
 & = & \left[\frac{1}{\sqrt{2}}|0\rangle|1\rangle|{\rm vac}\rangle-\frac{1}{2}|0\rangle|{\rm vac}\rangle|1\rangle+\frac{1}{2}|0\rangle|{\rm vac}\rangle|0\rangle\right]_{txy}
\end{array}.\label{eq:Chapter5_BEq5}
\end{equation}
Similarly, we can characterize the remaining composite systems in
which Alice encodes the bit 1 when Bob's initial states are $|1\rangle$,
$|+\rangle$, and $|-\rangle$.

\begin{equation}
\begin{array}{lcl}
|{\rm A-B}\rangle_{|1\rangle E_{2}}^{1} & = & \left[\frac{1}{\sqrt{2}}|1\rangle|0\rangle|{\rm vac}\rangle+\frac{1}{2}|1\rangle|{\rm vac}\rangle|0\rangle+\frac{1}{2}|1\rangle|{\rm vac}\rangle|1\rangle\right]\end{array},\label{eq:Chapter5_BEq6}
\end{equation}

\begin{equation}
\begin{array}{lcl}
|{\rm A-B}\rangle_{|+\rangle E_{2}}^{1} & = & \frac{1}{2}\left[-|-\rangle|{\rm vac}\rangle|1\rangle+|+\rangle|{\rm vac}\rangle|0\rangle\right]_{txy}+\frac{1}{2\sqrt{2}}\left[|-\rangle|1\rangle|{\rm vac}\rangle+|+\rangle|1\rangle|{\rm vac}\rangle\right.\\
 & - & \left.|-\rangle|0\rangle|{\rm vac}\rangle+|+\rangle|0\rangle|{\rm vac}\rangle\right]_{txy}
\end{array},\label{eq:Chapter5_BEq7}
\end{equation}
and

\begin{equation}
\begin{array}{lcl}
|{\rm A-B}\rangle_{|-\rangle E_{2}}^{1} & = & \frac{1}{2}\left[-|+\rangle|{\rm vac}\rangle|1\rangle+|-\rangle|{\rm vac}\rangle|0\rangle\right]_{txy}+\frac{1}{2\sqrt{2}}\left[|-\rangle|1\rangle|{\rm vac}\rangle+|+\rangle|1\rangle|{\rm vac}\rangle\right.\\
 & + & \left.|-\rangle|0\rangle|{\rm vac}\rangle-|+\rangle|0\rangle|{\rm vac}\rangle\right]_{txy}
\end{array},\label{eq:Chapter5_BEq8}
\end{equation}
respectively.

\textit{Eve's optimal strategy} Eve's optimal strategy for extracting
Alice's encoded information, after measuring the ancillary spatial
modes $x$ and $y$, corresponds to the $E_{1}$ attack scenario.
As previously mentioned, there exists a $\frac{1}{2}$ probability
that an additional unitary transformation $S_{ty}$ is applied. Consequently,
Equation (\ref{eq:Chapter5_AEq9}) remains valid for calculating joint
probabilities in half of all possible cases. These joint probabilities,
expressed as $p_{jmk}^{E_{2}}$, where $j$, $m$, and $k$ respectively
denote Alice\textquoteright s encoding, Bob\textquoteright s decoding,
and Eve\textquoteright s inference can be determined using Equations
(\ref{eq:Chapter5_AEq9}) and (\ref{eq:Chapter5_BEq1}) $-$ (\ref{eq:Chapter5_BEq8}).

\begin{equation}
\begin{array}{lcl}
p_{000}^{E_{2}} & = & \frac{1}{2}\left[\frac{q}{4}\left(1+p\right)+q\right]=\frac{q}{8}\left(5+p\right)\\
\\p_{001}^{E_{2}} & = & \frac{q}{8}\left(1+p\right)\\
\\p_{010}^{E_{2}} & = & p_{011}^{E_{2}}=\frac{q}{8}\left(1-p\right)\\
\\p_{100}^{E_{2}} & = & \frac{1}{4}\left(1-q\right)\\
\\p_{101}^{E_{2}} & =\frac{1}{4} & \left(1-q\right)\left(1+2p\right)\\
\\p_{110}^{E_{2}} & = & 0\\
\\p_{111}^{E_{2}} & = & \frac{1}{2}\left(1-p\right)\left(1-q\right)
\end{array}.\label{eq:Chapter5_BEq9}
\end{equation}
By employing Equation (\ref{eq:Chapter5_BEq9}), the mutual information
shared among Alice, Bob, and Eve can be formulated as follows:

\begin{equation}
\begin{array}{lcl}
H\left({\rm B|A}\right)_{E_{2}} & = & q\left(\mathbf{H}\left[\frac{1}{4}\left(3+p\right)\right]+\mathbf{H}\left[\frac{1}{4}\left(1-p\right)\right]\right)+\left(1-q\right)\left(\mathbf{H}\left[\frac{1}{2}\left(1+p\right)\right]+\mathbf{H}\left[\frac{1}{2}\left(1-p\right)\right]\right)\\
\\H\left({\rm B}\right)_{E_{2}} & = & \mathbf{H}\left[\frac{1}{4}\left(2+2p+q-pq+2\right)\right]+\mathbf{H}\left[\frac{1}{4}\left(1-p\right)\left(2-q\right)\right]\\
\\I\left({\rm A,B}\right)_{E_{2}} & = & H\left({\rm B}\right)_{E_{2}}-H\left({\rm B|A}\right)_{E_{2}}
\end{array},\label{eq:Chapter5_BEq10}
\end{equation}

\begin{equation}
\begin{array}{lcl}
H\left({\rm E|A}\right)_{E_{2}} & = & 0.811278\\
\\H\left({\rm E}\right)_{E_{2}} & = & \mathbf{H}\left[\frac{1}{4}\left(1+2q\right)\right]+\mathbf{H}\left[\frac{1}{4}\left(3-2q\right)\right]\\
\\I\left({\rm A,E}\right)_{E_{2}} & = & H\left({\rm E}\right)_{E_{2}}-H\left({\rm E|A}\right)_{E_{2}}
\end{array},\label{eq:Chapter5_BEq11}
\end{equation}
and

\begin{equation}
\begin{array}{lcl}
H\left({\rm E|B}\right)_{E_{2}} & = & \frac{1}{4}\left(2+2p+q-pq\right)\left(\mathbf{H}\left[\frac{1}{2}\left(\frac{2+3q+pq}{2+2p+q-pq}\right)\right]+\mathbf{H}\left[\frac{1}{2}\left(\frac{2+4p-q-3pq}{2+2p+q-pq}\right)\right]\right)\\
 & + & \frac{1}{4}\left(1-p\right)\left(2-q\right)\left(\mathbf{H}\left[\frac{1}{2}\left(\frac{q}{2-q}\right)\right]+\mathbf{H}\left[\frac{1}{2}\left(\frac{4-3q}{2-q}\right)\right]\right)\\
\\H\left({\rm E}\right)_{E_{2}} & = & \mathbf{H}\left[\frac{1}{4}\left(1+2q\right)\right]+\mathbf{H}\left[\frac{1}{4}\left(3-2q\right)\right]\\
\\I\left({\rm B,E}\right)_{E_{2}} & = & H\left({\rm E}\right)_{E_{2}}-H\left({\rm E|B}\right)_{E_{2}}
\end{array}.\label{eq:Chapter5_BEq12}
\end{equation}
Additionally, it is evident that eavesdropping contributes to the
QBER.

\begin{equation}
\begin{array}{lcl}
{\rm QBER}_{E_{2}} & = & \underset{j\ne m}{\sum}p_{jmk}^{E_{2}}\\
 & = & \frac{1}{4}\left(2+2p-q-3pq\right)
\end{array}.\label{eq:Chapter5_BEq13}
\end{equation}
\textbf{\textit{Control mode:}} The fundamental idea for the control
mode has been previously illustrated in the context of the $E_{1}$
attack. The primary distinction arises from Eve's action of applying
the $S_{ty}$ operator with a $\frac{1}{2}$ probability after executing
$Q_{txy}^{-1}$ in the A-B attack scenario. Consider the case where
Bob prepares the state $|0\rangle_{t}$ and Alice performs a measurement
in the $Z$ basis. In this situation, the composite system $|B-A\rangle_{|0\rangle E_{1}}$
collapses to the state $|0\rangle_{t}|0\rangle_{x}|{\rm vac}\rangle_{y}$
with a $\frac{1}{2}$ probability when Alice detects a photon. Following
Eve\textquoteright s $E_{2}$ attack, the resulting composite system
transforms into:

\begin{equation}
\begin{array}{lcl}
S_{ty}Q_{txy}^{-1}\,|0\rangle_{t}|0\rangle_{x}|{\rm vac}\rangle_{y} & = & S_{ty}\,H_{y}\,{\rm CPBS}_{txy}\,{\rm SWAP}_{tx}\,|0\rangle_{t}|0\rangle_{x}|{\rm vac}\rangle_{y}\\
 & = & S_{ty}\,\frac{1}{\sqrt{2}}\left[|0\rangle|{\rm vac}\rangle|0\rangle+|0\rangle|{\rm vac}\rangle|1\rangle\right]_{txy}\\
 & = & X_{t}\,Z_{t}\,{\rm CNOT}_{ty}\,X_{t}\,\frac{1}{\sqrt{2}}\left[|0\rangle|{\rm vac}\rangle|0\rangle+|0\rangle|{\rm vac}\rangle|1\rangle\right]_{txy}\\
 & = & X_{t}\,Z_{t}\,{\rm CNOT}_{ty}\,\frac{1}{\sqrt{2}}\left[|1\rangle|{\rm vac}\rangle|0\rangle+|1\rangle|{\rm vac}\rangle|1\rangle\right]_{txy}\\
 & = & X_{t}\,Z_{t}\,\frac{1}{\sqrt{2}}\left[|1\rangle|{\rm vac}\rangle|1\rangle+|1\rangle|{\rm vac}\rangle|0\rangle\right]_{txy}\\
 & = & X_{t}\,\frac{1}{\sqrt{2}}\left[-|1\rangle|{\rm vac}\rangle|1\rangle-|1\rangle|{\rm vac}\rangle|0\rangle\right]_{txy}\\
 & = & \frac{1}{\sqrt{2}}\left[-|0\rangle|{\rm vac}\rangle|1\rangle-|0\rangle|{\rm vac}\rangle|0\rangle\right]_{txy}
\end{array}.\label{eq:Chapter5_BEq14}
\end{equation}
If Bob's initial state is $|1\rangle_{t}$, Alice experiences an equal
probability of detecting or not detecting a photon. She then measures
the state $t$ in the $Z$ basis and subsequently transmits the projected
qubit back to Bob. After Eve's A-B attack, the composite system is
given by:

\begin{equation}
\begin{array}{lcl}
S_{ty}Q_{txy}^{-1}\,|1\rangle_{t}|1\rangle_{x}|{\rm vac}\rangle_{y} & = & \frac{1}{\sqrt{2}}\left[|1\rangle|{\rm vac}\rangle|0\rangle-|1\rangle|{\rm vac}\rangle|1\rangle\right]_{txy}\end{array}.\label{eq:Chapter5_BEq15}
\end{equation}
Consider Bob's initial state, represented as $|+\rangle_{t}$. Following
Eve\textquoteright s B-A attack, the composite system remains in the
state $|{\rm B-A}\rangle_{|+\rangle E_{1}}$. When
Alice performs a collapse on measurement in the $X$ basis, four possible
composite states emerge: $|+\rangle_{t}|0\rangle_{x}|{\rm vac}\rangle_{y}$,
$|-\rangle_{t}|0\rangle_{x}|{\rm vac}\rangle_{y}$, $|+\rangle_{t}|1\rangle_{x}|{\rm vac}\rangle_{y}$,
and $|-\rangle_{t}|1\rangle_{x}|{\rm vac}\rangle_{y}$, as expressed
in Equations (\ref{eq:Chapter5_AEq16}) $-$ (\ref{eq:Chapter5_AEq19}).

\begin{equation}
\begin{array}{lcl}
S_{ty}\,Q_{txy}^{-1}\,|+\rangle_{t}|0\rangle_{x}|{\rm vac}\rangle_{y} & = & S_{ty}\,\frac{1}{2}\left(|+\rangle+|-\rangle\right)_{t}\left[\frac{1}{\sqrt{2}}|{\rm vac}\rangle\left(|0\rangle+|1\rangle\right)+|1\rangle|{\rm vac}\rangle\right]_{xy}\\
 & = & X_{t}\,Z_{t}\,{\rm CNOT}_{ty}\,X_{t}\,\left[\frac{1}{2}\left(|0\rangle|{\rm vac}\rangle|0\rangle+|0\rangle|{\rm vac}\rangle|1\rangle\right)+\frac{1}{\sqrt{2}}|0\rangle|1\rangle|{\rm vac}\rangle\right]_{txy}\\
 & = & X_{t}\,Z_{t}\,\left[\frac{1}{2}\left(|1\rangle|{\rm vac}\rangle|1\rangle+|1\rangle|{\rm vac}\rangle|0\rangle\right)+\frac{1}{\sqrt{2}}|1\rangle|1\rangle|{\rm vac}\rangle\right]_{txy}\\
 & = & \left[\frac{1}{2}\left(-|0\rangle|{\rm vac}\rangle|1\rangle-|0\rangle|{\rm vac}\rangle|0\rangle\right)-\frac{1}{\sqrt{2}}|0\rangle|1\rangle|{\rm vac}\rangle\right]_{txy}\\
 & = & \left[\frac{1}{2\sqrt{2}}\left(-|+\rangle|{\rm vac}\rangle|1\rangle-|-\rangle|{\rm vac}\rangle|1\rangle-|+\rangle|{\rm vac}\rangle|0\rangle-|-\rangle|{\rm vac}\rangle|0\rangle\right)\right.\\
 & - & \left.\frac{1}{2}\left(|+\rangle|1\rangle|{\rm vac}\rangle+|-\rangle|1\rangle|{\rm vac}\rangle\right)\right]_{txy}
\end{array},\label{eq:Chapter5_BEq16}
\end{equation}

\begin{equation}
\begin{array}{lcl}
S_{ty}\,Q_{txy}^{-1}\,|+\rangle_{t}|1\rangle_{x}|{\rm vac}\rangle_{y} & = & S_{ty}\,\frac{1}{2}\left(|+\rangle-|-\rangle\right)_{t}\left[\frac{1}{\sqrt{2}}|{\rm vac}\rangle\left(|0\rangle-|1\rangle\right)+|0\rangle|{\rm vac}\rangle\right]_{xy}\\
 & = & X_{t}\,Z_{t}\,{\rm CNOT}_{ty}\,X_{t}\,\left[\frac{1}{2}\left(|1\rangle|{\rm vac}\rangle|0\rangle+|1\rangle|{\rm vac}\rangle|1\rangle\right)+\frac{1}{\sqrt{2}}|1\rangle|0\rangle|{\rm vac}\rangle\right]_{txy}\\
 & = & X_{t}\,Z_{t}\,\left[\frac{1}{2}\left(|0\rangle|{\rm vac}\rangle|0\rangle+|0\rangle|{\rm vac}\rangle|1\rangle\right)+\frac{1}{\sqrt{2}}|0\rangle|0\rangle|{\rm vac}\rangle\right]_{txy}\\
 & = & \left[\frac{1}{2}\left(|1\rangle|{\rm vac}\rangle|0\rangle+|1\rangle|{\rm vac}\rangle|1\rangle\right)+\frac{1}{\sqrt{2}}|1\rangle|0\rangle|{\rm vac}\rangle\right]_{txy}\\
 & = & \left[\frac{1}{2\sqrt{2}}\left(|+\rangle|{\rm vac}\rangle|0\rangle-|-\rangle|{\rm vac}\rangle|0\rangle+|+\rangle|{\rm vac}\rangle|1\rangle-|-\rangle|{\rm vac}\rangle|1\rangle\right)\right.\\
 & + & \left.\frac{1}{2}\left(|+\rangle|0\rangle|{\rm vac}\rangle-|-\rangle|0\rangle|{\rm vac}\rangle\right)\right]_{txy}
\end{array},\label{eq:Chapter5_BEq17}
\end{equation}

\begin{equation}
\begin{array}{lcl}
S_{ty}\,Q_{txy}^{-1}\,|-\rangle_{t}|0\rangle_{x}|{\rm vac}\rangle_{y} & = & S_{ty}\,\frac{1}{2}\left(|+\rangle+|-\rangle\right)_{t}\left[\frac{1}{\sqrt{2}}|{\rm vac}\rangle\left(|0\rangle+|1\rangle\right)-|1\rangle|{\rm vac}\rangle\right]_{xy}\\
 & = & \left[\frac{1}{2\sqrt{2}}\left(-|+\rangle|{\rm vac}\rangle|1\rangle-|-\rangle|{\rm vac}\rangle|1\rangle-|+\rangle|{\rm vac}\rangle|0\rangle-|-\rangle|{\rm vac}\rangle|0\rangle\right)\right.\\
 & + & \left.\frac{1}{2}\left(|+\rangle|1\rangle|{\rm vac}\rangle+|-\rangle|1\rangle|{\rm vac}\rangle\right)\right]_{txy}
\end{array},\label{eq:Chapter5_BEq18}
\end{equation}
and

\begin{equation}
\begin{array}{lcl}
S_{ty}\,Q_{txy}^{-1}\,|-\rangle_{t}|1\rangle_{x}|{\rm vac}\rangle_{y} & = & S_{ty}\,\frac{1}{2}\left(|+\rangle-|-\rangle\right)_{t}\left[-\frac{1}{\sqrt{2}}|{\rm vac}\rangle\left(|0\rangle-|1\rangle\right)+|0\rangle|{\rm vac}\rangle\right]_{xy}\\
 & = & \left[-\frac{1}{2\sqrt{2}}\left(|+\rangle|{\rm vac}\rangle|0\rangle-|-\rangle|{\rm vac}\rangle|0\rangle-|+\rangle|{\rm vac}\rangle|1\rangle+|-\rangle|{\rm vac}\rangle|1\rangle\right)\right.\\
 & + & \left.\frac{1}{2}\left(|+\rangle|0\rangle|{\rm vac}\rangle-|-\rangle|0\rangle|{\rm vac}\rangle\right)\right]_{txy}
\end{array},\label{eq:Chapter5_BEq19}
\end{equation}
respectively.

Now, let us examine the case where Bob's initial state is $|-\rangle_{t}$.
The composite system remains in the state $|{\rm B-A}\rangle_{|-\rangle E_{1}}$
after Eve\textquoteright s B-A attack. Intuitively, the scenario remains
symmetric whether Bob\textquoteright s initial state is $|+\rangle_{t}$
or $|-\rangle_{t}$, implying that Eve\textquoteright s A-B attack
affects all four possible measurement outcomes of Alice in the same
manner. Consequently, the results are identical to those derived in
Equations (\ref{eq:Chapter5_BEq16}) $-$ (\ref{eq:Chapter5_BEq19}).

If we disregard cases where no photon detection occurs under the optimal
eavesdropping strategy (favoring Eve), the probability of Eve remaining
undetected in the $E_{2}$ attack scenario can be determined using
Equations (\ref{eq:Chapter5_BEq14}) $-$ (\ref{eq:Chapter5_BEq19}),

\[
\begin{array}{lcl}
P_{nd}^{E_{2}} & = & \frac{1}{4}\left(1+1+\frac{1}{4}+\frac{1}{4}\right)=\frac{5}{8}\end{array}.
\]
The average probability of detecting Eve in the $E_{2}$ attack scenario
is calculated,

\begin{equation}
\begin{array}{lcl}
P_{d}^{E_{2}} & = & \frac{1}{2}\left(1-\frac{5}{8}\right)=\frac{3}{16}=0.1875\end{array}.\label{eq:Chapter5_BEq20}
\end{equation}
The maximum tolerable detection probability in the control mode for this
scenario is 0.1875. In other words, the upper threshold for detecting
Eve is $18.75\%$.

\subsection{$E_{3}$ attack strategy}\label{subsec:Chapter5_Sec3.3}

In this scenario, we employ an alternative attack method, denoted
as $E_{3}$, originally introduced by Pavi{\v{c}}i{\'c} \cite{P+13}.
Eve initializes two auxiliary modes, labeled $x$ and $y$, each beginning
with an ancilla photon in the state $|{\rm vac}\rangle_{x}|0\rangle_{y}$,
where $|{\rm vac}\rangle$ signifies an empty mode. She then facilitates
an interaction between the traveling photon and these auxiliary modes\footnote{The original work \cite{P+13} explores the qutrit case, our focus
is restricted to the qubit scenario. To simplify notation, we represent
the basis states as $|0\rangle$ and $|1\rangle$ rather than $|H\rangle$
and $|V\rangle$.}. Similar to the preceding two attacks, Eve applies the unitary operation
$Q_{txy}^{\prime}$ when the photon is in transit from Bob to Alice
during a B-A attack. Conversely, in the A-B attack setting, Eve employs
the inverse operation $Q_{txy}^{\prime-1}$ on the traveling photon
$t$.

\[
\begin{array}{lcl}
Q_{txy}^{\prime} & = & {\rm CNOT}_{ty}\left({\rm CNOT}_{tx}\otimes I_{y}\right)\left(I_{t}\otimes{\rm PBS}_{xy}\right){\rm CNOT}_{ty}\left({\rm CNOT}_{tx}\otimes I_{y}\right)\left(I_{t}\otimes H_{x}\otimes H_{y}\right)\end{array}.
\]
Eve's strategy consists of sequentially applying Hadamard gates to
the $x$ and $y$ modes, followed by two CNOT gates, a PBS operation,
and two additional CNOT gates. Further details on this approach can
be found in \cite{P+13}.

\textbf{\textit{Message mode:}} Now, consider the scenario where Alice
encodes a bit value of 1. Specifically, assume Bob prepares the initial
state $|0\rangle_{t}$. Eve then executes the B-A attack on the composite
system $|0\rangle_{t}|{\rm vac}\rangle_{x}|0\rangle_{y}$, following
the $E_{3}$ attack method,

\[
\begin{array}{lcl}
|{\rm B-A}\rangle_{|0\rangle E_{3}} & = & Q_{txy}^{\prime}\,|0\rangle_{t}|{\rm vac}\rangle_{x}|0\rangle_{y}\\
 & = & {\rm CNOT}_{ty}\left({\rm CNOT}_{tx}\otimes I_{y}\right)\left(I_{t}\otimes{\rm PBS}_{xy}\right){\rm CNOT}_{ty}\left({\rm CNOT}_{tx}\otimes I_{y}\right)\left(I_{t}\otimes H_{x}\otimes H_{y}\right)\\
 & \times & \left(|0\rangle|{\rm vac}\rangle|0\rangle\right)_{txy}\\
 & = & {\rm CNOT}_{ty}\left({\rm CNOT}_{tx}\otimes I_{y}\right)\left(I_{t}\otimes{\rm PBS}_{xy}\right){\rm CNOT}_{ty}\left({\rm CNOT}_{tx}\otimes I_{y}\right)\\
 & \times & \frac{1}{\sqrt{2}}\left[|0\rangle|{\rm vac}\rangle\left(|0\rangle+|1\rangle\right)\right]_{txy}\\
 & = & {\rm CNOT}_{ty}\left({\rm CNOT}_{tx}\otimes I_{y}\right)\left(I_{t}\otimes{\rm PBS}_{xy}\right){\rm CNOT}_{ty}\,\frac{1}{\sqrt{2}}\left[|0\rangle|{\rm vac}\rangle\left(|0\rangle+|1\rangle\right)\right]_{txy}\\
 & = & {\rm CNOT}_{ty}\left({\rm CNOT}_{tx}\otimes I_{y}\right)\left(I_{t}\otimes{\rm PBS}_{xy}\right)\,\frac{1}{\sqrt{2}}\left[|0\rangle|{\rm vac}\rangle\left(|0\rangle+|1\rangle\right)\right]_{txy}\\
 & = & {\rm CNOT}_{ty}\left({\rm CNOT}_{tx}\otimes I_{y}\right)\,\frac{1}{\sqrt{2}}\left[|0\rangle|0\rangle|{\rm vac}\rangle+|0\rangle|{\rm vac}\rangle|1\rangle\right]_{txy}\\
 & = & {\rm CNOT}_{ty}\,\frac{1}{\sqrt{2}}\left[|0\rangle|0\rangle|{\rm vac}\rangle+|0\rangle|{\rm vac}\rangle|1\rangle\right]_{txy}\\
 & = & \frac{1}{\sqrt{2}}\left[|0\rangle|0\rangle|{\rm vac}\rangle+|0\rangle|{\rm vac}\rangle|1\rangle\right]_{txy}
\end{array},
\]
As Alice applies the operation $iY_{t}^{1}$ to encode the bit value
1 in state $t$, the composite system undergoes a transformation,
leading to a new state $|{\rm B-A}\rangle_{|0\rangle E_{3}}$. The
superscript 1 indicates that Alice's encoding has been incorporated.
Subsequently, after Eve\textquoteright s interference using $Q_{txy}^{\prime-1}$,
the system is modified accordingly,

\begin{equation}
\begin{array}{lcl}
|{\rm A-B}\rangle_{|0\rangle E_{3}}^{1} & = & Q_{txy}^{\prime-1}\,\frac{-1}{\sqrt{2}}\left[|1\rangle|0\rangle|{\rm vac}\rangle+|1\rangle|{\rm vac}\rangle|1\rangle\right]_{txy}\\
 & = & \left(I_{t}\otimes H_{x}\otimes H_{y}\right)\left({\rm CNOT}_{tx}\otimes I_{y}\right){\rm CNOT}_{ty}\left(I_{t}\otimes{\rm PBS}_{xy}\right)\\
 & \times & \left({\rm CNOT}_{tx}\otimes I_{y}\right){\rm CNOT}_{ty}\,\frac{-1}{\sqrt{2}}\left[|1\rangle|0\rangle|{\rm vac}\rangle+|1\rangle|{\rm vac}\rangle|1\rangle\right]_{txy}\\
 & = & \left(I_{t}\otimes H_{x}\otimes H_{y}\right)\left({\rm CNOT}_{tx}\otimes I_{y}\right){\rm CNOT}_{ty}\left(I_{t}\otimes{\rm PBS}_{xy}\right)\\
 & \times & \left({\rm CNOT}_{tx}\otimes I_{y}\right)\,\frac{-1}{\sqrt{2}}\left[|1\rangle|0\rangle|{\rm vac}\rangle+|1\rangle|{\rm vac}\rangle|0\rangle\right]_{txy}\\
 & = & \left(I_{t}\otimes H_{x}\otimes H_{y}\right)\left({\rm CNOT}_{tx}\otimes I_{y}\right){\rm CNOT}_{ty}\left(I_{t}\otimes{\rm PBS}_{xy}\right)\\
 & \times & \frac{-1}{\sqrt{2}}\left[|1\rangle|1\rangle|{\rm vac}\rangle+|1\rangle|{\rm vac}\rangle|0\rangle\right]_{txy}\\
 & = & \left(I_{t}\otimes H_{x}\otimes H_{y}\right)\left({\rm CNOT}_{tx}\otimes I_{y}\right){\rm CNOT}_{ty}\,\frac{-1}{\sqrt{2}}\left[|1\rangle|1\rangle|{\rm vac}\rangle+|1\rangle|0\rangle|{\rm vac}\rangle\right]_{txy}\\
 & = & \left(I_{t}\otimes H_{x}\otimes H_{y}\right)\left({\rm CNOT}_{tx}\otimes I_{y}\right)\,\frac{-1}{\sqrt{2}}\left[|1\rangle|1\rangle|{\rm vac}\rangle+|1\rangle|0\rangle|{\rm vac}\rangle\right]_{txy}\\
 & = & \left(I_{t}\otimes H_{x}\otimes H_{y}\right)\,\frac{-1}{\sqrt{2}}\left[|1\rangle|0\rangle|{\rm vac}\rangle+|1\rangle|1\rangle|{\rm vac}\rangle\right]_{txy}\\
 & = & \frac{-1}{2}\left[|1\rangle\left(|0\rangle+|1\rangle\right)|{\rm vac}\rangle+|1\rangle\left(|0\rangle-|1\rangle\right)|{\rm vac}\rangle\right]_{txy}\\
 & = & -|1\rangle_{t}|0\rangle_{x}|{\rm vac}\rangle_{y}
\end{array}.\label{eq:Chapter5_CEq1}
\end{equation}
For brevity, we present only the key results. Consider the cases where
Bob initializes the states as $|1\rangle_{t}$, $|+\rangle_{t}$,
and $|-\rangle_{t}$. After undergoing the B-A attack, the composite
system evolves into a transformed state as a consequence of Eve's
intervention.

\[
\begin{array}{lcl}
|{\rm B-A}\rangle_{|1\rangle E_{3}} & = & \frac{1}{\sqrt{2}}\left[|1\rangle|{\rm vac}\rangle|0\rangle+|1\rangle|1\rangle|{\rm vac}\rangle\right]_{txy}\end{array},
\]

\[
\begin{array}{lcl}
|{\rm B-A}\rangle_{|+\rangle E_{3}} & = & \frac{1}{2}\left[|0\rangle|0\rangle|{\rm vac}\rangle+|0\rangle|{\rm vac}\rangle|1\rangle+|1\rangle|{\rm vac}\rangle|0\rangle+|1\rangle|1\rangle|{\rm vac}\rangle\right]_{txy}\end{array},
\]
and

\[
\begin{array}{lcl}
|{\rm B-A}\rangle_{|-\rangle E_{3}} & = & \frac{1}{2}\left[|0\rangle|0\rangle|{\rm vac}\rangle+|0\rangle|{\rm vac}\rangle|1\rangle-|1\rangle|{\rm vac}\rangle|0\rangle-|1\rangle|1\rangle|{\rm vac}\rangle\right]_{txy}\end{array},
\]
respectively.

Following Eve's attack operation, $Q_{txy}^{\prime-1}$ (A-B attack),
the composite system evolves as follows: When Alice encodes a bit
value of 1, she applies the operation $iY_{t}^{1}$ to Bob\textquoteright s
initial states $|1\rangle_{t}$, $|+\rangle_{t}$, and $|-\rangle_{t}$,

\begin{equation}
\begin{array}{lcl}
|{\rm A-B}\rangle_{|1\rangle E_{3}}^{1} & = & |0\rangle_{t}|0\rangle_{x}|{\rm vac}\rangle_{y}\end{array},\label{eq:Chapter5_CEq2}
\end{equation}

\begin{equation}
\begin{array}{lcl}
|{\rm A-B}\rangle_{|+\rangle E_{3}}^{1} & = & |-\rangle_{t}|0\rangle_{x}|{\rm vac}\rangle_{y}\end{array},\label{eq:Chapter5_CEq3}
\end{equation}
and

\begin{equation}
\begin{array}{lcl}
|{\rm A-B}\rangle_{|-\rangle E_{3}}^{1} & = & -|+\rangle_{t}|0\rangle_{x}|{\rm vac}\rangle_{y}\end{array},\label{eq:Chapter5_CEq4}
\end{equation}
respectively.

The final composite states resulting from Eve\textquoteright s bidirectional
attacks (B-A and A-B attacks) are obtained when Alice encodes a 0
using the operation $I_{t}$, with Bob\textquoteright s initial states
being $|0\rangle_{t}$, $|1\rangle_{t}$, $|+\rangle_{t}$, and $|-\rangle_{t}$.

\begin{equation}
\begin{array}{lcl}
|{\rm A-B}\rangle_{|0\rangle E_{3}}^{0} & = & |0\rangle_{t}|{\rm vac}\rangle_{x}|0\rangle_{y}\end{array},\label{eq:Chapter5_CEq5}
\end{equation}

\begin{equation}
\begin{array}{lcl}
|{\rm A-B}\rangle_{|1\rangle E_{3}}^{0} & = & |1\rangle_{t}|{\rm vac}\rangle_{x}|0\rangle_{y}\end{array},\label{eq:Chapter5_CEq6}
\end{equation}

\begin{equation}
\begin{array}{lcl}
|{\rm A-B}\rangle_{|+\rangle E_{3}}^{0} & = & |+\rangle_{t}|{\rm vac}\rangle_{x}|0\rangle_{y}\end{array},\label{eq:Chapter5_CEq7}
\end{equation}
and

\begin{equation}
\begin{array}{lcl}
|{\rm A-B}\rangle_{|-\rangle E_{3}}^{0} & = & |-\rangle_{t}|{\rm vac}\rangle_{x}|0\rangle_{y}\end{array},\label{eq:Chapter5_CEq8}
\end{equation}
respectively.

\textit{Eve's optimal strategy }Eve\textquoteright s optimal strategy
for decoding is characterized by two distinct cases: (1) she deciphers
$k=0$ when her ancillary spatial modes $x$ and $y$ are in the vacuum
state and $|0\rangle$ state, respectively, i.e., $|{\rm vac}\rangle_{x}|0\rangle_{y}$,
and (2) she deciphers $k=1$ when the $x$ mode is in the $|0\rangle$
state and the $y$ mode is in the vacuum state, i.e., $|0\rangle_{x}|{\rm vac}\rangle_{y}$.
These scenarios are quantified by the parameter $p_{jmk}^{E_{3}}$,
which represents the joint probability, where $j$, $m$, and $k$
correspond to Alice's encoding, Bob's decoding, and Eve's decoding
outcomes, respectively. The joint probabilities are derived from Equations
(\ref{eq:Chapter5_CEq1}) $-$ (\ref{eq:Chapter5_CEq8}).

\begin{equation}
\begin{array}{lcl}
p_{000}^{E_{3}} & = & q\\
\\p_{001}^{E_{3}} & = & p_{010}^{E_{3}}=p_{011}^{E_{3}}=0\\
\\p_{100}^{E_{3}} & = & p_{101}^{E_{3}}=p_{110}^{E_{3}}=0\\
\\p_{111}^{E_{3}} & = & \left(1-q\right)
\end{array}.\label{eq:Chapter5_CEq9}
\end{equation}
Using Equation (\ref{eq:Chapter5_CEq9}), the mutual information shared
among Alice, Bob, and Eve can be formulated,

\begin{equation}
\begin{array}{lcl}
H\left({\rm B|A}\right)_{E_{3}} & = & 0\\
\\H\left({\rm B}\right)_{E_{3}} & = & \mathbf{H}\left[q\right]+\mathbf{H}\left[1-q\right]\\
\\I\left({\rm A,B}\right)_{E_{3}} & = & \mathbf{H}\left[q\right]+\mathbf{H}\left[1-q\right]
\end{array},\label{eq:Chapter5_CEq10}
\end{equation}

\begin{equation}
\begin{array}{lcl}
H\left({\rm E|A}\right)_{E_{3}} & = & 0\\
\\H\left({\rm E}\right)_{E_{3}} & = & \mathbf{H}\left[q\right]+\mathbf{H}\left[1-q\right]\\
\\I\left({\rm A,E}\right)_{E_{3}} & = & \mathbf{H}\left[q\right]+\mathbf{H}\left[1-q\right]
\end{array},\label{eq:Chapter5_CEq11}
\end{equation}
and

\begin{equation}
\begin{array}{lcl}
H\left({\rm E|B}\right)_{E_{3}} & = & 0\\
\\H\left({\rm E}\right)_{E_{3}} & = & \mathbf{H}\left[q\right]+\mathbf{H}\left[1-q\right]\\
\\I\left({\rm B,E}\right)_{E_{3}} & = & \mathbf{H}\left[q\right]+\mathbf{H}\left[1-q\right]
\end{array}.\label{eq:Chapter5_CEq12}
\end{equation}
Furthermore, it is evident that Eve\textquoteright s eavesdropping
introduces QBER, thereby affecting the overall security of the system.
\begin{equation}
\begin{array}{lcl}
{\rm QBER}_{E_{3}} & = & \underset{j\ne m}{\sum}p_{jmk}^{E_{3}}\\
 & = & 0
\end{array}.\label{eq:Chapter5_CEq13}
\end{equation}
\textbf{\textit{Control mode:}} In the given scenario, if Bob initializes
the system in state $|0\rangle_{t}$ and Alice subsequently measures
it using the $Z$ basis, the composite system $|B-A\rangle_{|0\rangle E_{3}}$
probabilistically collapses into one of two possible states with equal
probability: $|0\rangle_{t}|0\rangle_{x}|{\rm vac}\rangle_{y}$ or
$|0\rangle_{t}|{\rm vac}\rangle_{x}|1\rangle_{y}$. Following Eve's
$Q_{txy}^{\prime-1}$ (A-B) attack, the composite system undergoes
further evolution,

\begin{equation}
\begin{array}{lcl}
Q_{txy}^{\prime-1}\,|0\rangle_{t}|0\rangle_{x}|{\rm vac}\rangle_{y} & = & \frac{1}{\sqrt{2}}\left[|0\rangle|{\rm vac}\rangle|0\rangle+|0\rangle|{\rm vac}\rangle|1\rangle\right]_{txy}\\
\\Q_{txy}^{\prime-1}\,|0\rangle_{t}|{\rm vac}\rangle_{x}|1\rangle_{y} & = & \frac{1}{\sqrt{2}}\left[|0\rangle|{\rm vac}\rangle|0\rangle-|0\rangle|{\rm vac}\rangle|1\rangle\right]_{txy}
\end{array}.\label{eq:Chapter5_CEq14}
\end{equation}
When Bob initializes the system in state $|1\rangle_{t}$ and Alice
performs a $Z$-basis measurement, the composite state $|B-A\rangle_{|1\rangle E_{3}}$
collapses into either $|1\rangle_{t}|{\rm vac}\rangle_{x}|0\rangle_{y}$
and $|1\rangle_{t}|1\rangle_{x}|{\rm vac}\rangle_{y}$, each occurring
with equal probability. The application of Eve's $Q_{txy}^{\prime-1}$
(A-B) attack subsequently modifies the composite system,

\begin{equation}
\begin{array}{lcl}
Q_{txy}^{\prime-1}\,|1\rangle_{t}|{\rm vac}\rangle_{x}|0\rangle_{y} & = & \frac{1}{\sqrt{2}}\left[|1\rangle|{\rm vac}\rangle|0\rangle+|1\rangle|{\rm vac}\rangle|1\rangle\right]_{txy}\\
\\Q_{txy}^{\prime-1}\,|1\rangle_{t}|1\rangle_{x}|{\rm vac}\rangle_{y} & = & \frac{1}{\sqrt{2}}\left[|1\rangle|{\rm vac}\rangle|0\rangle-|1\rangle|{\rm vac}\rangle|1\rangle\right]_{txy}
\end{array}.\label{eq:Chapter5_CEq15}
\end{equation}
Similarly, if Bob prepares the system in the state $|+\rangle_{t}$
and Alice measures using the $X$ basis, the composite system $|B-A\rangle_{|+\rangle E_{3}}$
collapses into one of the following states with equal probability:
$|+\rangle_{t}|0\rangle_{x}|{\rm vac}\rangle_{y}$, $|-\rangle_{t}|0\rangle_{x}|{\rm vac}\rangle_{y}$,
$|+\rangle_{t}|{\rm vac}\rangle_{x}|1\rangle_{y}$, $|-\rangle_{t}|{\rm vac}\rangle_{x}|1\rangle_{y}$,
$|+\rangle_{t}|{\rm vac}\rangle_{x}|0\rangle_{y}$, $|-\rangle_{t}|{\rm vac}\rangle_{x}|0\rangle_{y}$,
$|+\rangle_{t}|1\rangle_{x}|{\rm vac}\rangle_{y}$, or $|-\rangle_{t}|1\rangle_{x}|{\rm vac}\rangle_{y}$.
Following Eve's $Q_{txy}^{\prime-1}$ (A-B) attack, the composite
system undergoes further transformations,

\begin{equation}
\begin{array}{lcl}
Q_{txy}^{\prime-1}\,|+\rangle_{t}|0\rangle_{x}|{\rm vac}\rangle_{y} & = & \frac{1}{2\sqrt{2}}\left[\left(|+\rangle+|-\rangle\right)_{t}|{\rm vac}\rangle_{x}\left(|0\rangle+|1\rangle\right)_{y}+\left(|+\rangle-|-\rangle\right)_{t}\left(|0\rangle+|1\rangle\right)_{x}|{\rm vac}\rangle_{y}\right]\\
\\Q_{txy}^{\prime-1}\,|+\rangle_{t}|{\rm vac}\rangle_{x}|1\rangle_{y} & = & \frac{1}{2\sqrt{2}}\left[\left(|+\rangle+|-\rangle\right)_{t}|{\rm vac}\rangle_{x}\left(|0\rangle-|1\rangle\right)_{y}+\left(|+\rangle-|-\rangle\right)_{t}\left(|0\rangle-|1\rangle\right)_{x}|{\rm vac}\rangle_{y}\right]\\
\\Q_{txy}^{\prime-1}\,|+\rangle_{t}|{\rm vac}\rangle_{x}|0\rangle_{y} & = & \frac{1}{2\sqrt{2}}\left[\left(|+\rangle+|-\rangle\right)_{t}\left(|0\rangle+|1\rangle\right)_{x}|{\rm vac}\rangle_{y}+\left(|+\rangle-|-\rangle\right)_{t}|{\rm vac}\rangle_{x}\left(|0\rangle+|1\rangle\right)_{y}\right]\\
\\Q_{txy}^{\prime-1}\,|+\rangle_{t}|1\rangle_{x}|{\rm vac}\rangle_{y} & = & \frac{1}{2\sqrt{2}}\left[\left(|+\rangle+|-\rangle\right)_{t}\left(|0\rangle-|1\rangle\right)_{x}|{\rm vac}\rangle_{y}+\left(|+\rangle-|-\rangle\right)_{t}|{\rm vac}\rangle_{x}\left(|0\rangle-|1\rangle\right)_{y}\right]\\
\\Q_{txy}^{\prime-1}\,|-\rangle_{t}|0\rangle_{x}|{\rm vac}\rangle_{y} & = & \frac{1}{2\sqrt{2}}\left[\left(|+\rangle+|-\rangle\right)_{t}|{\rm vac}\rangle_{x}\left(|0\rangle+|1\rangle\right)_{y}-\left(|+\rangle-|-\rangle\right)_{t}\left(|0\rangle+|1\rangle\right)_{x}|{\rm vac}\rangle_{y}\right]\\
\\Q_{txy}^{\prime-1}\,|-\rangle_{t}|{\rm vac}\rangle_{x}|1\rangle_{y} & = & \frac{1}{2\sqrt{2}}\left[\left(|+\rangle+|-\rangle\right)_{t}|{\rm vac}\rangle_{x}\left(|0\rangle-|1\rangle\right)_{y}-\left(|+\rangle-|-\rangle\right)_{t}\left(|0\rangle-|1\rangle\right)_{x}|{\rm vac}\rangle_{y}\right]\\
\\Q_{txy}^{\prime-1}\,|-\rangle_{t}|{\rm vac}\rangle_{x}|0\rangle_{y} & = & \frac{1}{2\sqrt{2}}\left[\left(|+\rangle+|-\rangle\right)_{t}\left(|0\rangle+|1\rangle\right)_{x}|{\rm vac}\rangle_{y}-\left(|+\rangle-|-\rangle\right)_{t}|{\rm vac}\rangle_{x}\left(|0\rangle+|1\rangle\right)_{y}\right]\\
\\Q_{txy}^{\prime-1}\,|-\rangle_{t}|1\rangle_{x}|{\rm vac}\rangle_{y} & = & \frac{1}{2\sqrt{2}}\left[\left(|+\rangle+|-\rangle\right)_{t}\left(|0\rangle-|1\rangle\right)_{x}|{\rm vac}\rangle_{y}-\left(|+\rangle-|-\rangle\right)_{t}|{\rm vac}\rangle_{x}\left(|0\rangle-|1\rangle\right)_{y}\right]
\end{array}.\label{eq:Chapter5_CEq16}
\end{equation}
For the case where Bob prepares the state $|-\rangle_{t}$, the corresponding
composite state post Eve's B-A attack is determined accordingly.

\[
\begin{array}{lcl}
|{\rm B-A}\rangle_{|-\rangle E_{3}} & = & \frac{1}{2}\left[|0\rangle|0\rangle|{\rm vac}\rangle+|0\rangle|{\rm vac}\rangle|1\rangle-|1\rangle|{\rm vac}\rangle|0\rangle-|1\rangle|1\rangle|{\rm vac}\rangle\right]_{txy}\\
 & = & \frac{1}{2\sqrt{2}}\left[\left(|+\rangle+|-\rangle\right)\left(|0\rangle|{\rm vac}\rangle+|{\rm vac}\rangle|1\rangle\right)-\left(|+\rangle-|-\rangle\right)\left(|{\rm vac}\rangle|0\rangle-|1\rangle|{\rm vac}\rangle\right)\right]_{txy}
\end{array}.
\]
From an intuitive standpoint, it is apparent that the overall scenario
remains equivalent regardless of whether Bob's initial state is $|-\rangle_{t}$
or $|+\rangle_{t}$. Consequently, Eve's A-B attack influences all
four possible outcomes of Alice's measurement in an identical manner,
yielding the same results as given in Equation (\ref{eq:Chapter5_CEq16}).
The probability of Eve remaining undetected in the $E_{3}$ attack
scenario can be determined using Equations (\ref{eq:Chapter5_CEq14})
$-$ (\ref{eq:Chapter5_CEq16}),

\[
\begin{array}{lcl}
P_{nd}^{E_{3}} & = & \frac{1}{4}\left(1+1+\frac{1}{4}+\frac{1}{4}\right)=\frac{5}{8}\end{array}.
\]
The average probability of detecting Eve in the $E_{3}$ attack scenario
is calculated as follows:

\begin{equation}
\begin{array}{lcl}
P_{d}^{E_{3}} & = & \frac{1}{2}\left(1-\frac{5}{8}\right)=\frac{3}{16}=0.1875\end{array}.\label{eq:Chapter5_CEq17}
\end{equation}
Alice and Bob can tolerate a maximum detection probability of 0.1875
for Eve's presence in control mode under the $E_{3}$ attack scenario.
This establishes an upper bound of 18.75\% for the probability of
detecting Eve's intervention.

\subsection{$E_{4}$ attack strategy}\label{subsec:Chapter5_Sec3.4}

Next, we define Eve's IR ($E_{4}$) attack and subsequently compute
the mutual information among Alice, Bob, and Eve. Similar to the B-A
attack, Eve performs a uniform (random) measurement in both the $Z$
and $X$ bases on Bob\textquoteright s qubit before forwarding the
projected qubit to Alice. Depending on the mode of operation---either
encoding (message mode) or measurement (control mode)---Eve then
acts on the received qubit and sends it back to Bob. In the A-B attack,
she applies the same measurement operation that was previously executed
in the B-A attack.

\textbf{\textit{Message mode:}} Through an extensive calculation,
the joint probability $p_{jmk}$ is obtained, where $j$, $m$, and
$k$ correspond to Alice\textquoteright s encoding, Bob\textquoteright s
decoding, and Eve\textquoteright s decoding information, respectively.

\begin{equation}
\begin{array}{lcl}
p_{000}^{E_{4}} & = & \frac{3}{4}q\\
\\p_{001}^{E_{4}} & = & p_{011}^{E_{4}}=0\\
\\p_{010}^{E_{4}} & = & \frac{1}{4}q\\
\\p_{100}^{E_{4}} & = & p_{110}^{E_{4}}=0\\
\\p_{101}^{E_{4}} & = & \frac{1}{4}\left(1-q\right)\\
\\p_{111}^{E_{4}} & = & \frac{3}{4}\left(1-q\right)
\end{array}.\label{eq:Chapter5_DEq1}
\end{equation}
By employing Equation (\ref{eq:Chapter5_DEq1}), the expression for
mutual information among Alice, Bob, and Eve can be derived as follows:

\begin{equation}
\begin{array}{lcl}
H\left({\rm B|A}\right)_{E_{4}} & = & 0.811278\\
\\H\left({\rm B}\right)_{E_{4}} & = & \mathbf{H}\left[\frac{1}{4}\left(1+2q\right)\right]+\mathbf{H}\left[\frac{1}{4}\left(3-2q\right)\right]\\
\\I\left({\rm A,B}\right)_{E_{4}} & = & H\left({\rm B}\right)_{E_{4}}-H\left({\rm B|A}\right)_{E_{4}}
\end{array},\label{eq:Chapter5_DEq2}
\end{equation}

\begin{equation}
\begin{array}{lcl}
H\left({\rm E|A}\right)_{E_{4}} & = & 0\\
\\H\left({\rm E}\right)_{E_{4}} & = & \mathbf{H}\left[q\right]+\mathbf{H}\left[\left(1-q\right)\right]\\
\\I\left({\rm A,E}\right)_{E_{4}} & = & \mathbf{H}\left[q\right]+\mathbf{H}\left[\left(1-q\right)\right]
\end{array},\label{eq:Chapter5_DEq3}
\end{equation}
and

\begin{equation}
\begin{array}{lcl}
H\left({\rm B|E}\right)_{E_{4}} & = & 0.811278\\
\\H\left({\rm B}\right)_{E_{4}} & = & \mathbf{H}\left[\frac{1}{4}\left(1+2q\right)\right]+\mathbf{H}\left[\frac{1}{4}\left(3-2q\right)\right]\\
\\I\left({\rm B,E}\right)_{E_{4}} & = & H\left({\rm B}\right)_{E_{4}}-H\left({\rm B|E}\right)_{E_{4}}
\end{array}.\label{eq:Mutual_Information_Bob_Eve_E4_Attack}
\end{equation}
Eavesdropping is a contributing factor to QBER,

\begin{equation}
\begin{array}{lcl}
{\rm QBER}_{E_{4}} & = & \underset{j\ne m}{\sum}p_{jmk}^{E_{4}}\\
 & = & \frac{1}{4}q+\frac{1}{4}\left(1-q\right)\\
 & = & 0.25
\end{array}.\label{eq:Chapter5_DEq4}
\end{equation}
\textbf{\textit{Control mode:}} In a bidirectional BB84-type security
framework, the probability of Eve remaining undetected throughout
both transmission directions is given by:

\[
\begin{array}{lcl}
P_{nd}^{E_{4}} & = & \frac{1}{2}\times\frac{1}{2}=\frac{1}{4}\end{array}.
\]
Consequently, the mean probability of detecting Eve's presence in
the $E_{4}$ attack scenario is:

\begin{equation}
\begin{array}{lcl}
P_{d}^{E_{4}} & = & \frac{1}{2}\left(1-\frac{1}{4}\right)=\frac{3}{8}=0.375\end{array}.\label{eq:Chapter5_DEq5}
\end{equation}
Alice and Bob can tolerate a maximum detection probability of 0.375
in control mode under the $E_{4}$ attack scenario. This implies that
the threshold for detecting Eve's presence is $37.5\%$.

\section{Analyzing the QBER threshold bound of using Nash equilibrium}\label{sec:Chapter5_Sec4}

A Nash equilibrium represents a strategic scenario in which no player
can enhance their expected payoff by unilaterally changing their strategy,
provided that other participants maintain their respective choices.
In its generalized form, Nash equilibrium characterizes a stochastic
steady-state condition in strategic interactions, wherein players
are permitted to choose a probability distribution over their available
strategies rather than being restricted to a single deterministic
choice. This probabilistic approach is known as a mixed strategy.
In a well-structured game, it is typically assumed that all players
act rationally, striving to optimize (maximize, in this context) their
expected payoffs. To simplify the analysis, we evaluate the ``mixed-strategy
Nash equilibrium''\footnote{A ``mixed strategy Nash equilibrium'' in a ``normal-form game''
consists of a set of mixed strategies where no player has an incentive
to unilaterally deviate, given the strategies adopted by others, ensuring
mutual optimality.} by examining three different game scenarios: $E_{1}$-$E_{2}$, $E_{1}$-$E_{3}$,
and $E_{2}$-$E_{3}$. Each player aims to adopt a mixed quantum strategy\footnote{All payoff elements for each player under the attack scenarios $E_{1}$,
$E_{2}$, $E_{3}$, and $E_{4}$ are computed in Appendices A, B,
C, and D, respectively.} that renders their opponent indifferent between pure quantum strategies.
Utilizing this principle, we compare the equilibrium points to establish
a secure bound for the QBER (refer to Section \ref{sec:Chapter5_Sec3}).

Firstly, we define the best response function and then apply it to
different game scenarios. For each scenario, we derive the best response
function for each participant. An individual's best response corresponds
to the probability of selecting their rational choice in a mixed strategy
framework, optimizing their utility under the assumption that other
players make their choices independently and unpredictably. Consider
Alice's best response function in the game $E_{i}$ - $E_{j}$. Assume
Alice selects her classical bit information with probability $q$
while Bob and Eve make their independent decisions with probabilities
$p$ and\footnote{Here, $r$ and $1-r$ represent the probabilities that an eavesdropper,
Eve, assigns to selecting attacks $E_{i}$ and $E_{j}$, respectively.} $r$, respectively. In the case of a pure strategy, Alice\textquoteright s
best response function---where $q=1$ or $q=0$---can be expressed
as follows \cite{HHM10}:

\[
B_{A}\left(p,r\right)\coloneqq rP_{A}^{E_{i}}\left(p,1\right)+\left(1-r\right)P_{A}^{E_{j}}\left(p,1\right)>rP_{A}^{E_{i}}\left(p,0\right)+\left(1-r\right)P_{A}^{E_{j}}\left(p,0\right).
\]
and

\[
B_{A}\left(p,r\right)\coloneqq rP_{A}^{E_{i}}\left(p,1\right)+\left(1-r\right)P_{A}^{E_{j}}\left(p,1\right)<rP_{A}^{E_{i}}\left(p,0\right)+\left(1-r\right)P_{A}^{E_{j}}\left(p,0\right),
\]
respectively. For Alice\textquoteright s mixed strategy best response
function, we define:

\[
B_{A}\left(p,r\right)\coloneqq rP_{A}^{E_{i}}\left(p,1\right)+\left(1-r\right)P_{A}^{E_{j}}\left(p,1\right)=rP_{A}^{E_{i}}\left(p,0\right)+\left(1-r\right)P_{A}^{E_{j}}\left(p,0\right).
\]
By extending the definition of the best response function, we similarly
define Bob's and Eve's best response functions as $B_{B}\left(q,r\right)$
and $B_{E}\left(p,q\right)$, where Bob and Eve assign their probabilities
$p$ and $r$ while optimizing their responses based on $q,r$ and
$p,q$, respectively.

\textit{$E_{1}$-$E_{2}$ game scenario}: The previously defined probabilities
$p$ and $q$ apply here as well. Suppose Eve selects the $E_{1}$($E_{2}$)
attack with probability $r$ ($1-r$). Alice\textquoteright s ``best
response function'', denoted as $B_{A}\left(p,r\right)$, represents
the set of probabilities she assigns to choosing the $Z$ basis as
her optimal response to $p$ and $r$ \cite{HHM10}. Similarly, the
best response functions for Bob and Eve are given by $B_{B}\left(q,r\right)$
and $B_{E}\left(p,q\right)$, respectively. We get as following:

\begin{equation}
\begin{array}{lcl}
B_{A}\left(p,r\right) & = & \begin{cases}
\left\{ q=1\right\}  & {\rm if}\,\,\,rP_{A}^{E_{1}}\left(p,1\right)+\left(1-r\right)P_{A}^{E_{2}}\left(p,1\right)\\
 & >rP_{A}^{E_{1}}\left(p,0\right)+\left(1-r\right)P_{A}^{E_{2}}\left(p,0\right)\\
\left\{ q:0\le q\le1\right\}  & {\rm if}\,\,\,rP_{A}^{E_{1}}\left(p,1\right)+\left(1-r\right)P_{A}^{E_{2}}\left(p,1\right)\\
 & =rP_{A}^{E_{1}}\left(p,0\right)+\left(1-r\right)P_{A}^{E_{2}}\left(p,0\right)\\
\left\{ q=0\right\}  & {\rm if}\,\,\,rP_{A}^{E_{1}}\left(p,1\right)+\left(1-r\right)P_{A}^{E_{2}}\left(p,1\right)\\
 & <rP_{A}^{E_{1}}\left(p,0\right)+\left(1-r\right)P_{A}^{E_{2}}\left(p,0\right)
\end{cases}\\
\\B_{B}\left(q,r\right) & = & \begin{cases}
\left\{ p=1\right\}  & {\rm if}\,\,\,rP_{B}^{E_{1}}\left(1,q\right)+\left(1-r\right)P_{B}^{E_{2}}\left(1,q\right)\\
 & >rP_{B}^{E_{1}}\left(0,q\right)+\left(1-r\right)P_{B}^{E_{2}}\left(0,q\right)\\
\left\{ p:0\le p\le1\right\}  & {\rm if}\,\,\,rP_{B}^{E_{1}}\left(1,q\right)+\left(1-r\right)P_{B}^{E_{2}}\left(1,q\right)\\
 & =rP_{B}^{E_{1}}\left(0,q\right)+\left(1-r\right)P_{B}^{E_{2}}\left(0,q\right)\\
\left\{ p=0\right\}  & {\rm if}\,\,\,rP_{B}^{E_{1}}\left(1,q\right)+\left(1-r\right)P_{B}^{E_{2}}\left(1,q\right)\\
 & <rP_{B}^{E_{1}}\left(0,q\right)+\left(1-r\right)P_{B}^{E_{2}}\left(0,q\right)
\end{cases}\\
\\B_{E}\left(p,q\right) & = & \begin{cases}
\left\{ r=1\right\}  & {\rm if}\,\,P_{E}^{E_{1}}\left(p,q\right)>P_{E}^{E_{2}}\left(p,q\right)\\
\left\{ r:0\le r\le1\right\}  & {\rm if}\,\,P_{E}^{E_{1}}\left(p,q\right)=P_{E}^{E_{2}}\left(p,q\right)\\
\left\{ r=0\right\}  & {\rm if}\,\,P_{E}^{E_{1}}\left(p,q\right)<P_{E}^{E_{2}}\left(p,q\right)
\end{cases}
\end{array}.\label{eq:Chapter5_Eq3}
\end{equation}
The best response functions corresponding to Alice, Bob, and Eve in
the game scenarios \textit{$E_{1}$-$E_{3}$} and \textit{$E_{2}$-$E_{3}$}
are defined as follows:

\begin{equation}
\begin{array}{lcl}
B_{A}\left(p,r\right) & = & \begin{cases}
\left\{ q=1\right\}  & {\rm if}\,\,\,rP_{A}^{E_{1}}\left(p,1\right)+\left(1-r\right)P_{A}^{E_{3}}\left(p,1\right)\\
 & >rP_{A}^{E_{1}}\left(p,0\right)+\left(1-r\right)P_{A}^{E_{3}}\left(p,0\right)\\
\left\{ q:0\le q\le1\right\}  & {\rm if}\,\,\,rP_{A}^{E_{1}}\left(p,1\right)+\left(1-r\right)P_{A}^{E_{3}}\left(p,1\right)\\
 & =rP_{A}^{E_{1}}\left(p,0\right)+\left(1-r\right)P_{A}^{E_{3}}\left(p,0\right)\\
\left\{ q=0\right\}  & {\rm if}\,\,\,rP_{A}^{E_{1}}\left(p,1\right)+\left(1-r\right)P_{A}^{E_{3}}\left(p,1\right)\\
 & <rP_{A}^{E_{1}}\left(p,0\right)+\left(1-r\right)P_{A}^{E_{3}}\left(p,0\right)
\end{cases}\\
\\B_{B}\left(q,r\right) & = & \begin{cases}
\left\{ p=1\right\}  & {\rm if}\,\,\,rP_{B}^{E_{1}}\left(1,q\right)+\left(1-r\right)P_{B}^{E_{3}}\left(1,q\right)\\
 & >rP_{B}^{E_{1}}\left(0,q\right)+\left(1-r\right)P_{B}^{E_{3}}\left(0,q\right)\\
\left\{ p:0\le p\le1\right\}  & {\rm if}\,\,\,rP_{B}^{E_{1}}\left(1,q\right)+\left(1-r\right)P_{B}^{E_{3}}\left(1,q\right)\\
 & =rP_{B}^{E_{1}}\left(0,q\right)+\left(1-r\right)P_{B}^{E_{3}}\left(0,q\right)\\
\left\{ p=0\right\}  & {\rm if}\,\,\,rP_{B}^{E_{1}}\left(1,q\right)+\left(1-r\right)P_{B}^{E_{3}}\left(1,q\right)\\
 & <rP_{B}^{E_{1}}\left(0,q\right)+\left(1-r\right)P_{B}^{E_{3}}\left(0,q\right)
\end{cases}\\
\\B_{E}\left(p,q\right) & = & \begin{cases}
\left\{ r=1\right\}  & {\rm if}\,\,P_{E}^{E_{1}}\left(p,q\right)>P_{E}^{E_{3}}\left(p,q\right)\\
\left\{ r:0\le r\le1\right\}  & {\rm if}\,\,P_{E}^{E_{1}}\left(p,q\right)=P_{E}^{E_{3}}\left(p,q\right)\\
\left\{ r=0\right\}  & {\rm if}\,\,P_{E}^{E_{1}}\left(p,q\right)<P_{E}^{E_{3}}\left(p,q\right)
\end{cases}
\end{array}.\label{eq:Chapter5_Eq4}
\end{equation}
and

\begin{equation}
\begin{array}{lcl}
B_{A}\left(p,r\right) & = & \begin{cases}
\left\{ q=1\right\}  & {\rm if}\,\,\,rP_{A}^{E_{2}}\left(p,1\right)+\left(1-r\right)P_{A}^{E_{3}}\left(p,1\right)\\
 & >rP_{A}^{E_{2}}\left(p,0\right)+\left(1-r\right)P_{A}^{E_{3}}\left(p,0\right)\\
\left\{ q:0\le q\le1\right\}  & {\rm if}\,\,\,rP_{A}^{E_{2}}\left(p,1\right)+\left(1-r\right)P_{A}^{E_{3}}\left(p,1\right)\\
 & =rP_{A}^{E_{2}}\left(p,0\right)+\left(1-r\right)P_{A}^{E_{3}}\left(p,0\right)\\
\left\{ q=0\right\}  & {\rm if}\,\,\,rP_{A}^{E_{2}}\left(p,1\right)+\left(1-r\right)P_{A}^{E_{3}}\left(p,1\right)\\
 & <rP_{A}^{E_{2}}\left(p,0\right)+\left(1-r\right)P_{A}^{E_{3}}\left(p,0\right)
\end{cases}\\
\\B_{B}\left(q,r\right) & = & \begin{cases}
\left\{ p=1\right\}  & {\rm if}\,\,\,rP_{B}^{E_{2}}\left(1,q\right)+\left(1-r\right)P_{B}^{E_{3}}\left(1,q\right)\\
 & >rP_{B}^{E_{2}}\left(0,q\right)+\left(1-r\right)P_{B}^{E_{3}}\left(0,q\right)\\
\left\{ p:0\le p\le1\right\}  & {\rm if}\,\,\,rP_{B}^{E_{2}}\left(1,q\right)+\left(1-r\right)P_{B}^{E_{3}}\left(1,q\right)\\
 & =rP_{B}^{E_{2}}\left(0,q\right)+\left(1-r\right)P_{B}^{E_{3}}\left(0,q\right)\\
\left\{ p=0\right\}  & {\rm if}\,\,\,rP_{B}^{E_{2}}\left(1,q\right)+\left(1-r\right)P_{B}^{E_{3}}\left(1,q\right)\\
 & <rP_{B}^{E_{2}}\left(0,q\right)+\left(1-r\right)P_{B}^{E_{3}}\left(0,q\right)
\end{cases}\\
\\B_{E}\left(p,q\right) & = & \begin{cases}
\left\{ r=1\right\}  & {\rm if}\,\,P_{E}^{E_{2}}\left(p,q\right)>P_{E}^{E_{3}}\left(p,q\right)\\
\left\{ r:0\le r\le1\right\}  & {\rm if}\,\,P_{E}^{E_{2}}\left(p,q\right)=P_{E}^{E_{3}}\left(p,q\right)\\
\left\{ r=0\right\}  & {\rm if}\,\,P_{E}^{E_{2}}\left(p,q\right)<P_{E}^{E_{3}}\left(p,q\right)
\end{cases}
\end{array}.\label{eq:Chapter5_Eq5}
\end{equation}

\begin{figure}[h]
\centering{}\includegraphics[scale=0.4]{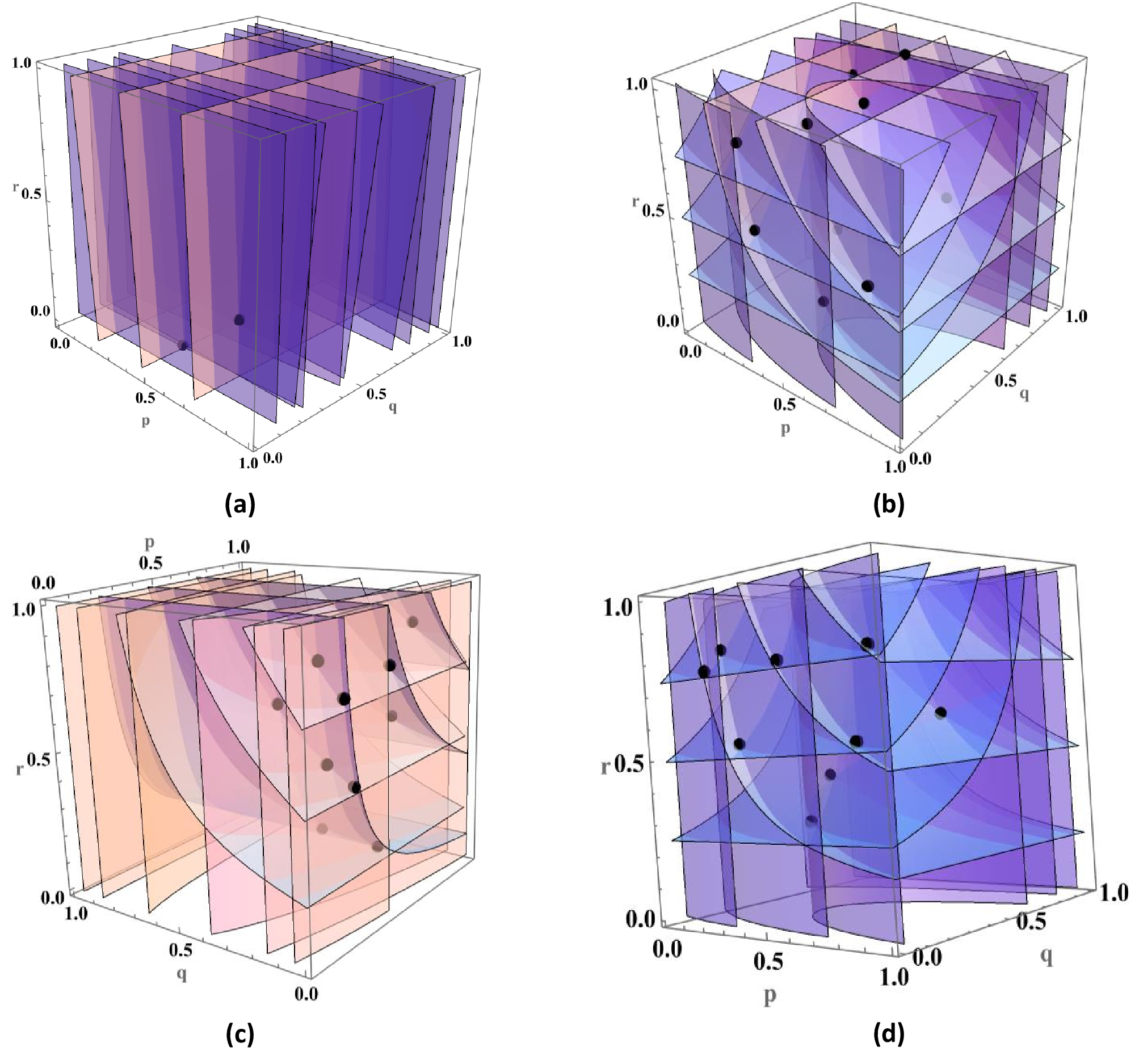}
\caption{ The figure illustrates the Nash equilibrium points where the optimal
response functions of the three participants intersect in our mixed-strategy
game framework. The response functions are represented as follows:
the low-density layer for Alice, the medium-density layer for Bob,
and the high-density layer for Eve. The subfigures depict different
game scenarios: (a) $E_{1}$-$E_{2}$, (b) $E_{1}$-$E_{3}$, (c)
$E_{2}$-$E_{3}$, and (d) $E_{1}$-$E_{4}$.}\label{fig:Chapter5_Fig2}
\end{figure}

Within the framework of a quantum game, Figure \ref{fig:Chapter5_Fig2}
illustrates the best response functions corresponding to various game
scenarios. These functions characterize the optimal strategic choices
for each player, contingent on the strategies adopted by the other
participants. The ``mixed-strategy Nash equilibrium'' is identified
at the intersection points of these best response functions, signifying
that each player's strategy is optimal given the strategies of the
others. Notably, multiple intersection points may arise, each corresponding
to distinct game scenarios. These intersections denote the Nash equilibrium
states for the players within the specified game scenarios. Subfigures
(a), (b), (c), and (d) of Figure \ref{fig:Chapter5_Fig2} explicitly
highlight the Nash equilibrium points for the $E_{1}$-$E_{2}$, $E_{1}$-$E_{3}$,
$E_{2}$-$E_{3}$, and $E_{1}$-$E_{4}$ game scenarios, respectively.
Each subfigure comprises three distinct curves, each representing
the best response function of an individual player within the mixed
quantum strategy framework. These curves delineate the optimal strategies
(by considering the best response) of each player in reaction to the strategic
choices of their counterparts. The intersections of these curves indicate
the most stable strategic configurations within each game scenario.
From these intersection points, the most stable probability distributions
in the mixed quantum strategy game can be inferred, ensuring optimality
for all participating players.

A comprehensive summary of the data, including each player's payoffs,
is presented in Table \ref{tab:NE_Points_Payoff}. This summary evaluates
the mixed-strategy Nash equilibrium points across different game scenarios
and examines the secure QBER bound. Referring to Equation (\ref{eq:Chapter5_Eq2}),
it becomes apparent that the payoff functions for the three entities
are influenced by the variables $p$ and $q$, which dictate Eve\textquoteright s
attack strategies $E_{1}$, $E_{2}$ (or $E_{3}$, $E_{4}$). Additionally,
the game can be categorized as a normal-form (strategic-form) game,
as players lack prior knowledge of their opponents' choices at the
moment of decision-making. Under this framework, all participants
independently determine their strategies: Bob opts for the $Z$ basis
with probability $p$, Eve selects an attack strategy with probability
$r$, and Alice, when operating in message mode, encodes bit 0 with
probability $q$. Both Alice and Bob share identical payoff functions
as defined in Equation (\ref{eq:Chapter5_Eq2}). In the $E_{1}$-$E_{2}$
scenario, the expected payoffs for Alice/Bob and Eve are represented
as $rP_{A/B}^{E_{1}}(p,q)+(1-r)P_{A/B}^{E_{2}}(p,q)$ and $rP_{E}^{E_{1}}(p,q)+(1-r)P_{E}^{E_{2}}(p,q)$,
respectively. A comparative analysis of the payoff differences between
Eve and Alice (refer to Table \ref{tab:NE_Points_Payoff}) enables
the identification of Nash equilibrium points where either Eve or
Alice/Bob derive the greatest advantage. Specifically, in the $E_{1}$-$E_{2}$
game scenario, Eve attains the highest individual benefit at the equilibrium
point $(0.45,0.195,0.005)$, whereas Alice/Bob gains the most at $(0.72,0.208,0.225)$.
Similarly, in the $E_{1}$-$E_{3}$ scenario, the equilibrium points
maximizing individual benefits for Eve and Alice/Bob are $\left(0.41,0.39,0.412\right)$
and $\left(0.84,0.047,0.525\right)$, respectively. In the $E_{2}$-$E_{3}$
case, Eve benefits the most at $\left(0.385,0.215,0.262\right)$,
while Alice/Bob attain the highest payoff at $\left(0.80,0.115,0.885\right)$.
Ultimately, no Nash equilibrium point in these game scenarios achieves
Pareto optimality, indicating that there is no scenario where all
players simultaneously receive the highest possible payoffs.

We compute the QBER for all Nash equilibrium points across various
game scenarios. Notably, the expected QBER for the $E_{i}$-$E_{j}$
game scenario is given by $\epsilon_{E_{i}-E_{2}}=r\,{\rm QBER}_{E_{i}}+\left(1-r\right){\rm QBER}_{E_{j}}$,
where $r$ represents the probability of Eve opting for the $E_{i}$
attack. In any specific game scenario, Eve's optimal strategy is dictated
either by the maximum payoff disparity---measuring the difference
between Eve's and Alice's payoffs---or by the minimum QBER. Our primary
goal is to establish the QBER threshold. Therefore, we determine the
minimum QBER values across all game scenarios presented in Table \ref{tab:NE_Points_Payoff},
which correspond to Nash equilibrium points where Eve is best positioned
to remain undetected. The minimum QBER values at these equilibrium
points are 0.610303 is for $E_{1}$-$E_{2}$, 0.152451 is for $E_{1}$-$E_{3}$,
and 0.143882 is for $E_{2}$-$E_{3}$ game scenario. A lower minimum
QBER\footnote{Since a lower QBER value maximally benefits Eve, establishing this
bound ensures the highest level of security for a quantum protocol.} (in message mode) signifies that Eve is employing a more effective
attack strategy in the given scenario while maintaining the same or
minimal detection probability $P_{d}$ in control mode. These reduced
QBER values serve as indicators of more advanced quantum attack strategies
by Eve within the game-theoretic framework. Through an evaluation
of the QBER threshold values, we infer that $E_{3}$ constitutes the
most formidable quantum attack, followed by $E_{2}$ and then $E_{1}$.
This suggests that as Eve deploys increasingly potent attacks, she
secures higher payoffs, thereby posing a greater risk to Alice and
Bob. To counteract this threat within the DL04 protocol game scenario,
we establish a lower minimum QBER threshold corresponding to Eve\textquoteright s
most aggressive attack. This threshold is derived by analyzing all
Nash equilibrium points encompassing the specified attacks.

As previously stated, a player with constrained access to quantum
resources effectively behaves as a classical participant. In our evaluation,
the $E_{4}$ attack is categorized as a classical strategy executed
by Eve, indicating that she functions as a classical eavesdropper
when implementing the $E_{4}$ attack. Additionally, we compare this
classical $E_{4}$ attack to the simplest quantum-based attack, $E_{1}$.
Moreover, we examine the $E_{4}$ attack, also known as the IR attack,
in relation to the $E_{1}$ attack within the framework of an $E_{1}$-$E_{4}$
game scenario. Within this strategic model, the minimum QBER is determined
to be 0.323478, which is lower than the minimum QBER observed in the
$E_{1}$-$E_{2}$ game scenario. Despite this, the $E_{4}$ attack
is considered weaker than $E_{4}$ due to its higher detection probability
($P_{d}$), recorded at 37.5\%, compared to 18.75\% for $E_{1}$ in
control mode.

To establish the secure upper bound of QBER, represented as $\epsilon$,
across all attack scenarios within the game framework, we focus on
the game scenario involving the most powerful attacks, namely the
$E_{2}$-$E_{3}$ game. In this scenario, the upper bound of QBER
is determined to be $\epsilon=0.143882$, as this corresponds to the
lowest QBER value associated with the scenario where the strongest
attack, $E_{3}$, is active. The lower bound of QBER is 0, as an isolated
$E_{3}$ attack would not introduce detectable errors in message mode
(refer to Section \ref{subsec:Chapter5_Sec3.3}). Furthermore, the
bounds for Eve\textquoteright s detection probability ($P_{d}$) in
control mode are established between 0.1875 and 0.375. Consequently,
it can be concluded that the DL04 protocol maintains $\epsilon$-security
under the set of individual and collective IR attacks, specifically
$E_{1}$, $E_{2}$, $E_{3}$ and $E_{4}$.

\begin{table}[h]
\caption{The table presents Nash equilibrium points for different strategic
interactions and provides qualitative payoff assessments for all involved
entities.}\label{tab:NE_Points_Payoff}

\begin{centering}
\centering{}%
\begin{tabular}{|>{\centering}p{2.0cm}|>{\centering}p{3.5cm}|>{\centering}p{2.5cm}|>{\centering}p{1.8cm}|>{\centering}p{1.8cm}|>{\centering}p{1.8cm}|}
\hline 
\multirow{4}{*}{\shortstack{$E_{1}$-$E_{2}$ \\ game\\ scenario}} & Nash equilibrium point ($p,q,r$)  & Alice's/Bob's payoff  & Eve's payoff  & Payoff difference & $\epsilon_{E_{1}-E_{2}}$\tabularnewline
\cline{2-6} \cline{3-6} \cline{4-6} \cline{5-6} \cline{6-6} 
 & $\left(0.72,0.208,0.225\right)$ & 0.055457 & 0.194543 & 0.13908 & 0.692404\tabularnewline
 & $\left(0.45,0.195,0.005\right)$ & 0.0446318 & 0.205368 & 0.16073 & 0.610303\tabularnewline
 &  &  &  &  & \tabularnewline
\hline 
\multirow{12}{*}{\shortstack{$E_{1}$-$E_{3}$ \\ game\\ scenario}} & Nash equilibrium point ($p,q,r$)  & Alice's/Bob's payoff  & Eve's payoff  & Payoff difference & $\epsilon_{E_{1}-E_{3}}$\tabularnewline
\cline{2-6} \cline{3-6} \cline{4-6} \cline{5-6} \cline{6-6} 
 & $\left(0.22,0.716,0.88\right)$ & -0.110497 & 0.360497 & 0.47099 & 0.152451\tabularnewline
 & $\left(0.442,0.75,0.999\right)$ & -0.0862188 & 0.336219 & 0.42243 & 0.18007\tabularnewline
 & $\left(0.41,0.39,0.412\right)$ & -0.157149 & 0.407149 & 0.56429 & 0.177181\tabularnewline
 & $\left(0.76,0.577,0.585\right)$ & -0.136264 & 0.386264 & 0.52252 & 0.21776\tabularnewline
 & $\left(0.56,0.14,0.292\right)$ & -0.0796824 & 0.329682 & 0.40936 & 0.195874\tabularnewline
 & $\left(0.325,0.064,0.532\right)$ & -0.0134987 & 0.263499 & 0.27699 & 0.329893\tabularnewline
 & $\left(0.84,0.047,0.525\right)$ & 0.0324084 & 0.217592 & 0.18518 & 0.460299\tabularnewline
 & $\left(0.485,0.465,0.915\right)$ & -0.090828 & 0.340828 & 0.43165 & 0.363472\tabularnewline
 & $\left(0.235,0.096,0.83\right)$ & -0.013356 & 0.263356 & 0.27671 & 0.463323\tabularnewline
 & $\left(0.47,0.195,0.93\right)$ & -0.0182231 & 0.268223 & 0.28644 & 0.550258\tabularnewline
 &  &  &  &  & \tabularnewline
\hline 
\multirow{11}{*}{\shortstack{$E_{2}$-$E_{3}$ \\ game\\ scenario}} & Nash equilibrium point ($p,q,r$)  & Alice's/Bob's payoff  & Eve's payoff  & Payoff difference  & $\epsilon_{E_{2}-E_{3}}$\tabularnewline
\cline{2-6} \cline{3-6} \cline{4-6} \cline{5-6} \cline{6-6} 
 & $\left(0.385,0.215,0.262\right)$ & -0.111965 & 0.361965 & 0.47393 & 0.151087\tabularnewline
 & $\left(0.47,0.055,0.205\right)$ & -0.0276507 & 0.277651 & 0.3053 & 0.143882\tabularnewline
 & $\left(0.25,0.096,0.54\right)$ & -0.0216673 & 0.271667 & 0.29633 & 0.31482\tabularnewline
 & $\left(0.24,0.268,0.71\right)$ & -0.0436386 & 0.293639 & 0.33727 & 0.35838\tabularnewline
 & $\left(0.70,0.138,0.58\right)$ & -0.00442078 & 0.254421 & 0.25884 & 0.430969\tabularnewline
 & $\left(0.284,0.02,0.472\right)$ & 0.0188573 & 0.231143 & 0.21228 & 0.298653\tabularnewline
 & $\left(0.235,0.02,0.758\right)$ & 0.0320242 & 0.217976 & 0.18595 & 0.461603\tabularnewline
 & $\left(0.222,0.10,0.865\right)$ & 0.015688 & 0.234312 & 0.21862 & 0.492488\tabularnewline
 & $\left(0.54,0.048,0.795\right)$ & 0.0558727 & 0.194127 & 0.13825 & 0.587155\tabularnewline
 & $\left(0.80,0.115,0.885\right)$ & 0.0722149 & 0.177785 & 0.10557 & 0.709991\tabularnewline
\hline 
\multirow{10}{*}{\shortstack{$E_{1}$-$E_{4}$ \\ game\\ scenario}} & Nash equilibrium point ($p,q,r$)  & Alice's/Bob's payoff  & Eve's payoff  & Payoff difference  & $\epsilon_{E_{1}-E_{4}}$\tabularnewline
\cline{2-6} \cline{3-6} \cline{4-6} \cline{5-6} \cline{6-6} 
 & $(0.23,0.095,0.825)$ & -0.00433851 & 0.254339 & 0.25867 & 0.502924\tabularnewline
 & $(0.245,0.008,0.76)$ & 0.0492999 & 0.2007 & 0.1514 & 0.529315\tabularnewline
 & $(0.572,0.02,0.765)$ & 0.0750153 & 0.174985 & 0.0999 & 0.648014\tabularnewline
 & $(0.928,0.032,0.774)$ & 0.0997124 & 0.150288 & 0.0505 & 0.77876\tabularnewline
 & $(0.324,0.065,0.535)$ & 0.0114311 & 0.238569 & 0.22713 & 0.447399\tabularnewline
 & $(0.85,0.045,0.522)$ & 0.0603314 & 0.189669 & 0.12933 & 0.580622\tabularnewline
 & $(0.405,0.387,0.415)$ & -0.124349 & 0.374349 & 0.49869 & 0.324962\tabularnewline
 & $(0.54,0.15,0.295)$ & -0.0471361 & 0.297136 & 0.34427 & 0.369328\tabularnewline
 & $(0.75,0.57,0.582)$ & -0.114078 & 0.364078 & 0.47815 & 0.323478\tabularnewline
\hline 
\end{tabular}
\par\end{centering}
\end{table}

\section{Discussion}\label{sec:Chapter5_Sec5}

This chapter introduces a novel security framework for quantum communication
protocols, specifically addressing collective attacks and IR attacks
through Nash equilibrium. This approach provides an alternative perspective,
framing security as a game-theoretic bound against collective attacks
using Nash equilibrium principles. The Nash equilibrium points represent
stable configurations that yield rational probability distributions
for decision-making within a mixed-strategy quantum game. These probabilities
define an equilibrium state for all involved entities, facilitating
the assessment of cryptographic parameters such as the QBER threshold.
In practical implementations, where participants make decisions independently,
achieving such equilibrium is improbable, potentially resulting in
a more unstable system. Consequently, identifying the lowest QBER
value within the stable game configuration establishes a secure boundary
for practical quantum protocol deployment. Our analysis demonstrates
that the security level ($\epsilon$-secure) of any quantum communication
protocol is contingent on the attack strategies employed by the eavesdropper,
Eve. A smaller $\epsilon$ value signifies a stronger attack strategy
by Eve. Moreover, findings suggest that a quantum-capable Eve possesses
significantly greater adversarial power compared to a classical Eve\footnote{Additionally, it is crucial to ensure that Eve\textquoteright s choice
of unitary operations and execution of measurement procedures remain
synchronized within the quantum framework. }. In our model, we assume Eve has access to unlimited quantum resources,
allowing us to disregard the last three terms in Eve\textquoteright s
payoff equation, as presented in Equation (\ref{eq:Chapter5_Eq1}).
To streamline our evaluation, we also assume uniform contributions
from all payoff components for each party, assigning them equal weight
($\frac{1}{4}$). The analysis is structured into four distinct game
scenarios, for which Nash equilibrium points are determined individually.
We then identify the minimum QBER value corresponding to each scenario,
which serves as the secure QBER bound for that specific case. Ultimately,
the overall threshold QBER ($\epsilon$) is established by identifying
the lowest QBER within the game scenario that corresponds to the most
adversarial attack. Furthermore, our study reveals that the DL04 protocol
exhibits vulnerability to Pavi{\v{c}}i{\'c} attack in message mode,
as the QBER in this case is zero. However, security is maintained
through the control mode, which is collaboratively executed by both
legitimate parties. Thus, while this attack poses a security risk
in message mode, it is effectively mitigated by the control mode mechanism.

Quantum communication protocols are designed for various objectives,
often balancing security against efficiency. To address these trade-offs,
our analysis can be extended by introducing distinct weighting factors
$\omega_{i},\,\omega_{j},\,\omega_{k}\neq0$, ensuring$\omega_{m}\neq\omega_{n}$
for $m\neq n$. These weights correspond to specific payoffs, with
their sum normalized to 1, thereby maintaining normalized player weight
assignments. Given $\omega_{i},\,\omega_{j},\,\omega_{k}\neq0$, Equation
(\ref{eq:Chapter5_Eq1}) results in $n_{1}=1$, $n_{2}=1$, $n_{3}=1$
for $E_{1}$, $n_{1}=2$, $n_{2}=5$, $n_{3}=0$ for $E_{3}$, $n_{1}=4$,
$n_{2}=2$, $n_{3}=1$ for $E_{2}$, and $n_{1}=2$, $n_{2}=n_{3}=0$
for $E_{4}$ attack strategies. Notably, the computational cost of
multi-qubit gates escalates with an increase in the number of qubits
involved. Furthermore, our framework can be extended to encapsulate
a comprehensive game-theoretic scenario where all attack strategies
coexist rather than treating them as separate cases. Under this paradigm,
eavesdroppers may employ different attack strategies with varying
probabilities, enabling the determination of ``Pareto-optimal
Nash equilibrium'' under specific conditions. Consequently, the formulation
of the QBER bound ($\epsilon$) inherently depends on these generalized
parameters.

Our methodology for determining a secure QBER threshold is applied
within the DL04 protocol. This adaptable approach is compatible with
both current and forthcoming quantum communication protocols, facilitating
the derivation of diverse information-theoretic bounds. This flexibility
accommodates multiple strategic implementations by participants, ensuring
alignment with the intended applications of quantum protocols. Future
research will integrate additional game-theoretic aspects tailored
to different quantum communication scenarios, considering various
quantum mechanical properties. This collaborative effort aims to refine
the analysis of quantum information bounds for more comprehensive
insights.

\newpage

\chapter{CONCLUSION AND SCOPE OF FUTURE WORKS}\label{Ch6:asd}
\graphicspath{{Chapter6/Chapter6Figs/}{Chapter6/Chapter6Figs/}}


The field of secure quantum communication is advancing rapidly. This
thesis investigates various aspects of secure quantum communication
by introducing novel protocols, extending existing frameworks, and
improving the efficiency of specific quantum communication processes.
The security of communication fundamentally relies on the security
of the cryptographic key used. Quantum technologies enable key distribution
with unconditional security by leveraging quantum resources. However,
the initialization of a QKD protocol necessitates the authentication of
the communicating parties. In many implementations, classical authentication
methods are employed, which, due to their classical nature, lack unconditional
security. This vulnerability allows an adversary (Eve) to impersonate
a legitimate party, creating a significant security loophole in otherwise
quantum-secure communication protocols. 

To address this challenge, we analyze existing QIA schemes and propose
new QIA protocols to enhance security. These protocols utilize various
quantum resources, making them feasible for implementation with current
technology. Once legitimate user identities are authenticated, protocols
such as QKD can be executed to distribute symmetric cryptographic
keys. Our analysis of prior QKD schemes reveals that information exchange
typically involves a combination of quantum and classical components
(e.g., BB84 and B92 protocols) or multiple quantum components (e.g.,
Goldenberg-Vaidman protocol). This partitioning of information fundamentally
contributes to security and necessitates the exchange of both classical
and quantum information for complete protocol implementation. We hypothesize
that increasing the proportion of quantum information exchange while
reducing reliance on classical communication may enhance both efficiency
and security bounds. To evaluate this hypothesis, we propose new QKD
protocols and analyze their security under various attack scenarios,
determining their tolerable security thresholds. Additionally, we
examine potential biases introduced by one party in the distributed
symmetric key within QKD schemes. Extending QKD to multiparty settings
also presents significant challenges. To address this, we propose
a novel QKA protocol that operates without the need for quantum memory,
improving feasibility and scalability. Furthermore, we conceptualize
quantum communication protocols as strategic games, where legitimate
parties seek to secure information from Eve while increasing her probability
of detection, whereas Eve aims to maximize her information gain while
minimizing her detectability. Applying game-theoretic principles,
we develop a method to determine the tolerable QBER based on Nash
equilibrium. This approach, termed ``gaming the quantum,'' provides
a novel framework for analyzing quantum protocols. The remainder of
this chapter presents a concise summary of the key findings of this
thesis and outlines potential directions for future research.

\section{Summary of the thesis}\label{sec:Chapter6_Sec1}

In this section, we provide a concise summary of the chapters in this thesis and discuss the key findings of our research. Initially, we outline the major results, followed by a chapter-wise discussion of the findings presented throughout the thesis.
\begin{enumerate}
\item The QIA schemes can be classified based on the quantum resources employed and the specific communication or computation tasks they address. These tasks can be adapted to develop QIA protocols utilizing various quantum resources. In the context of quantum communication, authentication can be incorporated into the initial phase of the primary protocol through a pre-shared authentication key. This thesis explicitly demonstrates how QSDC schemes can be utilized to design novel QIA protocols. Furthermore, we establish that existing schemes for other quantum communication and computation tasks can be systematically transformed into QIA schemes.
\item The proposed QIA protocols utilize single-qubit states, making them
feasible for implementation with current technology. Additionally,
decoy states are employed to verify the security of the quantum channel.
The positioning and polarization of these decoy particles depend on
the pre-shared authentication key. These assumptions contribute to
the security of the proposed protocols, which are analyzed in Chapter
2.
\item Another set of controlled QIA protocols is designed using Bell states,
permutation operations, and single-qubit state preparation by a third
party acting as the controller. These schemes require a quantum register
for implementation, with the controller being an untrusted quantum
third party. The security of these protocols is analyzed against specific
attack scenarios.
\item Once authentication is completed, the legitimate parties aim to generate
a symmetric key. To explore the benefits of increasing the quantum
component and reducing classical elements in QKD protocols, we propose
two new schemes incorporating classical reconciliation with BB84 and
SARG04 protocols. These schemes offer improved efficiency and critical
distance against PNS attack. This approach emphasizes using a higher
quantum component and fewer classical elements to develop more secure
quantum protocols, enhancing security against attacks performed by
an eavesdropper (Eve) after classical reconciliation.
\item In QKD schemes, there is a possibility of one party biasing the final
key, which should ideally remain symmetric. To address this issue
while introducing additional security features, we propose a QKA scheme.
Without requiring quantum memory or involving a trusted third party,
we introduce a novel QIA scheme.
\item The quantum advantage is utilized to analyze the security of the QKA
scheme. In our analysis, we observe that an adversary executes a collective
attack using her ancilla states and unitary operations. When the information
gained by Eve reaches its maximum (0.5 bits), the probability of detecting
Eve's presence becomes 1. This result demonstrates that quantum technology
inherently detects Eve\textquoteright s presence whenever she obtains
useful information from the quantum channel. The trace left by Eve
helps legitimate parties decide whether to continue or abort the protocol.
The security analysis highlights how the fundamental properties of
quantum particles enable the detection of Eve when she acquires useful
information.
\item This QKA scheme is secure against the collective attack considered
in our analysis. Additionally, we establish a specific secure error
bound against this attack. However, if a different set of collective
attacks is considered, this bound will change accordingly.
\item In this thesis, we propose a novel method to establish a secure bound
for any quantum protocol. To develop this approach, we model the legitimate
parties and the adversary as players in a strategic game. They formulate
strategies and engage in a mixed-strategy game without direct communication
(a non-cooperative game). Using Nash equilibrium and best response
functions, we demonstrate how game theory can be applied to determine
a secure bound for information-theoretic parameters such as the QBER.
\end{enumerate}
Thus, this thesis discusses the significance of QIA in securing quantum
communication. We comprehensively review various QIA protocols proposed
in previous research, analyzing their advantages and limitations.
Additionally, we explore how different quantum communication schemes
can be adapted into QIA protocols and propose new schemes utilizing
single and entangled quantum sources. After performing an identity
test for legitimate parties, they proceed with a quantum communication
protocol. We propose novel QKD and QKA schemes, highlighting the benefits
of incorporating additional quantum resources for QKD. In the proposed
QKA scheme, we introduce an untrusted third party and implement the
protocol without requiring quantum memory. Furthermore, our thesis
introduces a new method that integrates game theory into quantum communication,
referred to as \textquotedblleft gaming the quantum\textquotedblright ,
to establish a secure QBER for quantum protocols. The following section
presents a summary of the chapters in this thesis.
\begin{description}
\item [{Chapter~1:}] This chapter has examined existing protocols for
QIA to explore their inherent symmetry. The analysis has demonstrated
the presence of symmetry among various protocols, enabling their classification
and facilitating the identification of straightforward strategies
for adapting existing protocols to new QIA schemes applicable to different
quantum computation and communication tasks. To substantiate this
approach, two novel QIA protocols are introduced and shown to be secure
against a range of potential attacks. Notably, many of the protocols
discussed, including several existing ones, are not currently feasible
for implementation due to technological limitations. For instance,
certain protocols require a user to retain a photon until the travel
photon(s) return(s), such as in the ping-pong QIA scheme described
in \cite{ZZZX_2006} or the authentication protocols outlined in \cite{T.Mihara_2002,LB_2004,ZLG_2000,ZZZZ_2005,WZT_2006,TJ_2014,KHHYHM_2018,WZGZ_2019}.
Implementing such QIA protocols necessitates quantum memory, which
remains unavailable with current technology. Additionally, most entangled-state-based
QSDC protocols for QIA, as well as entangled-state-based quantum private
comparison schemes and QIA protocols involving a trusted third-party
(Trent) who retains a photon, encounter the same technological constraint.
At present, QIA protocols that do not necessitate quantum memory should
be prioritized. Notably, the protocols introduced here operate without
quantum memory, making them feasible with existing technology. Although
numerous approaches have been proposed for developing quantum memory,
which may prove beneficial in the future, direct teleportation-based
QIA schemes (such as those discussed in \cite{ZZZZ_2005,TJ_2014})
are unlikely to have widespread applications, even in the long run.
This limitation arises because teleportation is only viable for secure
quantum communication under ideal, noise-free quantum channels. Moreover,
any protocol reliant on shared entanglement (e.g., those outlined
in \cite{CS_01,CSPF_02,ZLG_2000,T.Mihara_2002,LB_2004,ZZZZ_2005})
may require entanglement purification or concentration due to decoherence-induced
degradation of shared entanglement. This, in turn, would necessitate
additional and potentially insecure interactions between Alice and
Bob. Despite these technical constraints, research in QIA continues
to expand rapidly, driven by the fact that the purported unconditional
security of quantum cryptographic schemes fundamentally depends on
the security of the identity authentication mechanisms employed.\\
This chapter would be incomplete without acknowledging the numerous
post-quantum authentication schemes proposed in recent years \cite{WZW+21,MGB+20}.
These schemes have not been examined here, as they are fundamentally
classical and only conditionally secure. A key assumption underlying
these approaches is that quantum computers will be incapable of efficiently
solving problems beyond the bounded-error quantum polynomial time
(BQP) complexity class. However, no formal proof supports this claim,
making it equivalent to assuming that if no efficient algorithm exists
for a given computational problem today, none will be developed in
the future. The primary content of this chapter is published on Ref. \cite{DP22}.
\item [{Chapter~2:}] In Chapter \ref{Ch6:asd}, we analyzed existing QIA protocols
to identify their inherent symmetries, enabling their classification
and adaptation for various quantum computing and communication applications,
leading to novel QIA schemes. To validate this approach, four new
QIA protocols were introduced and shown to be secure against specific
attacks. However, many existing and proposed protocols remain impractical
due to current technological limitations. For example, some require
users to store photons until travel photons return, as seen in ping-pong-based
QIA schemes \cite{ZZZX_2006} and other authentication methods \cite{T.Mihara_2002,LB_2004,ZLG_2000,ZZZZ_2005,WZT_2006,TJ_2014,KHHYHM_2018,WZGZ_2019},
which depend on quantum memory---currently unfeasible commercially.
Similar challenges affect entangled-state-based QSDC protocols, quantum
private comparison schemes, and third-party-assisted authentication
protocols. Prioritizing QIA protocols that do not require quantum
memory is preferable, as they are compatible with current technology.
The protocols proposed in this work fall into this category. While
quantum memory-dependent schemes may become viable in the future,
teleportation-based QIA approaches \cite{ZZZZ_2005,TJ_2014} are unlikely
to see widespread adoption due to their reliance on noise-free quantum
channels. Additionally, protocols leveraging shared entanglement \cite{CS_01,CSPF_02,ZLG_2000,T.Mihara_2002,LB_2004,ZZZZ_2005}
may require purification or concentration to mitigate decoherence,
potentially introducing insecure interactions between Alice and Bob.
Despite these limitations, QIA research is progressing rapidly, driven
by the critical role of identity authentication in ensuring the purported
unconditional security of quantum cryptographic protocols.\\
This chapter, alongside references \cite{DP23,DP+24}, explores the
potential for developing diverse QIA schemes. A thorough investigation
could help identify the most practical QIA approach using existing
technology. The proposed entangled-state-based QIA protocol leverages
Bell states, allowing Alice to securely transmit information to Bob
with authorization from controller Charlie, without pre-shared keys.
While it relies solely on Bell states, its implementation requires
quantum memory, which is not yet commercially available---a common
challenge across QIA, QSDC, DSQC, and quantum dialogue protocols.
However, ongoing research and recent proposals \cite{LLY+24} suggest
imminent advancements. Until then, a delay mechanism can serve as
an interim solution. Additionally, two other proposed protocols use
single-photon sources as qubits, making them feasible with current
technology for bi-directional identity authentication. The works presented in this chapter are published in Refs. \cite{DP22,DP23,DP+24}.
\item [{Chapter~3:}] In this chapter, we have proposed a new protocol
for QKD and a variant of it. The protocols consume more quantum resources
compared to SARG04 or similar protocols but transmit less classical
information over the public channel, thereby reducing the probability
of some side channel attacks. Additionally, we conduct a rigorous
security analysis of the proposed protocols and calculate the tolerable
error limit for the upper and lower bounds of the secret key rate
under a set of collective attacks. It is shown that by applying a
certain type of classical pre-processing, the tolerable error limit
can be increased. The same is illustrated through the graphs. Before
concluding, we may emphasize some important observations of our
analysis. In the seminal paper \cite{RGK_05}, the authors computed
density operators of Eve's final state in the six-state QKD protocol.
Interestingly, for our protocols, we obtained the same expressions
for the density operators describing Eve's final system, despite the
fact that in our Protocol 2 (1), neither Alice nor Bob (Alice never)
discloses the results of the measurements performed by them (her)
in cases where they (she) used different bases for preparation and
measurement. The reason for obtaining the same density operators for
Eve's system is that the terms that appear in the density matrix in
cases of basis mismatch happen, cancel each other. Additionally,
we establish that for the proposed protocols, the tolerable error
limit of QBER $\mathcal{E}\le0.124$ for the lower bound of the key
rate and $\mathcal{E}\ge0.114$ for the upper bound of the key rate
if classical pre-processing is used. In our case, the tolerable error
limits are expected to decrease in the absence of classical pre-processing.
In the practical implementation of cryptography, various types of
errors may occur during the transmission of qubits. Considering $QBER>0$,
Eve can attempt an attack using partial cloning machines \cite{CI00,NG98,C00}.
Acin et al., have shown that legitimate users of SARG04 can tolerate
errors up to 15\% when Eve uses a best-known partial cloning machine.
They also found that this tolerable error limit is higher than that
for the BB84 protocol. In our case, for $QBER>0$, the tolerable error
limit is also computed to be 15\% , which is better than the BB84
protocol and its variants. We have also performed a security-efficiency
trade-off analysis for the proposed schemes and compared their efficiency
with the SARG04 protocol, as detailed in Chapter \ref{Ch3:Chapter3_QKD}. The result in
this chapter is presented in \cite{DP+23}.
\item [{Chapter~4:}] In certain cases, key agreement between two parties
may require an untrusted third party. The quantum counterpart of such
scenarios necessitates a QKA protocol with a semi-trusted third-party
controller. Our proposed protocol is designed to meet this requirement,
leveraging quantum particle behavior for practical implementation,
as detailed in Chapter \ref{Ch4:Chapter4_QKA}. Security proofs addressing attack scenarios
are provided. For CQKA implementation, a Bell state---generated by
an untrusted third party, Charlie---is distributed to Alice and Bob.
Upon receiving their respective qubits, they execute the CNOT operation
per protocol. If an eavesdropper (Eve) attempts an intercept-and-resend
attack, she can only infer the initial Bell state, as entanglement
is lost after Alice and Bob's operations. Additionally, Charlie publicly
discloses the Bell state information $k_{C}$ at the protocol\textquoteright s
conclusion, preventing Eve from gaining any advantage. We also examine
the case where Eve attempts to impersonate the third party, known
as an ``impersonated fraudulent attack''. Our proposed protocol
is proven resilient against this attack, with analytical proof
validating its security. Additionally, a critical evaluation of the collective attack confirms its robustness under broader attack
scenarios. We show that Eve\textquoteright s probability of successfully
extracting the final two-bit key becomes negligibly small (approaching
zero) when $n=6$. Our security analysis extends by modeling Eve\textquoteright s
ancillary state as a complex variable, where each ancilla system is
represented in a complete orthogonal basis within the Hilbert space
$\mathcal{\mathscr{H}^{\mathrm{\mathscr{E}}}}$ \cite{LF+11}. Using
this framework, we estimate the tolerable error bound and determine
the acceptable QBER as a function of $\alpha$. We further evaluate
our scheme under varying noisy channel conditions, analyzing fidelity
variations concerning the channel parameter. Results indicate that
fidelity remains largely stable, even at higher noise levels. We also
compare our protocol\textquoteright s strengths and limitations with
other recent protocols in this category. Findings reveal that our
scheme is not only more practical with current technology but also
more efficient than many existing approaches. Notably, it introduces
a novel and beneficial feature involving third-party participation.
The work is published in Ref. \cite{DP2023}.
\item [{Chapter~5:}] The relationship between game theory and quantum
mechanics has been previously explored, revealing two primary approaches
that illustrate their connection. The first approach involves using
quantum resources to play a conventional game, which could also be
played without any quantum resources; however, incorporating quantum
elements offers certain advantages. The second approach applies game
theory concepts to describe a quantum mechanical scenario. For clarity,
we refer to the first approach as ``quantized game'' and the second
as ``gaming the quantum''. A quantized game is defined as a unitary
function that maps the Cartesian product of quantum superpositions
of players' pure strategies to the entire Hilbert space of the game.
This approach preserves the characteristics of the original classical
game under certain conditions. In contrast, ``gaming the quantum''
involves applying game theory principles to quantum mechanical problems
to derive solutions based on game-theoretic analysis. In Chapter \ref{Ch5:Chapter5_QG},
we utilize the gaming the quantum approach by applying non-cooperative
game theory to a quantum communication protocol, demonstrating how
Nash equilibrium can be employed as an effective solution concept.\\
The Nash equilibrium framework is utilized to determine a robust,
game-theoretic security bound on the QBER for the DL04 protocol, a
recently implemented scheme for quantum secure direct communication.
The analysis in Chapter \ref{Ch5:Chapter5_QG} considers the sender, receiver, and eavesdropper
(Eve) as quantum players, capable of executing quantum operations.
Notably, Eve is assumed to possess the ability to perform various
quantum attacks, such as W{\'o}jcik's original attack, its symmetrized
variant, and Pavi{\v{c}}i{\'c}'s attack, as well as a classical intercept-and-resend
strategy. A game-theoretic evaluation of the DL04 protocol's security
under these conditions is conducted by examining multiple game scenarios.
The findings indicate the absence of a Pareto optimal Nash equilibrium
in these cases. Therefore, mixed strategy Nash equilibrium points
are identified and employed to derive upper and lower bounds for QBER.
Additionally, the analysis highlights the susceptibility of the DL04
protocol to Pavi{\v{c}}i{\'c}'s attack in the message transmission
mode. It is also observed that the quantum attacks by Eve are more
effective than the classical attacks, as they result in lower QBER
values and a decreased probability of detecting Eve's intrusion. The
results reported in this chapter are published in Ref. \cite{DP24}.
\end{description}

\section{Limitations and Future Scope}

The limitations of any research work serve as a motivation for future
advancements toward more feasible and effective results. In this study,
we identified the inherent symmetry in previously proposed QIA schemes
and introduced new QIA protocols utilizing QSDC/DSQC. Furthermore, we have demonstrated that various communication and computational tasks can be utilized to design novel QIA schemes. While we have not explicitly outlined all possible protocols, future research could focus on systematically constructing and analyzing these protocols in terms of security and feasibility. A comparative study may be conducted to identify the most optimal QIA scheme for specific secure quantum communication tasks. Moreover, integrating blind quantum computation (BQC) presents a promising avenue for advancing QIA protocols. Additionally, the incorporation of QIA schemes into fundamental quantum protocols may enhance resource efficiency and overall performance. In this thesis, we propose a new QKD scheme that increases quantum resource usage while reducing classical
information exchange. Our protocol can be further refined, maintaining
its core motivation, to enhance efficiency and extend tolerable communication
distances. A similar approach can be applied to QKA protocols. Furthermore,
while our proposed schemes are based on DV protocols, they can be
adapted into CV frameworks, with future work focusing on performance
analysis. Noise analysis is only briefly addressed in this work, with
a focus on specific regimes. A more comprehensive analysis across
diverse environmental conditions would provide deeper insights into
the performance of these schemes. As our proposed schemes align with
current technological capabilities, they can be experimentally implemented
in the near future. Finally, we explore the concept of ``gaming the
quantum'', which provides a means to determine the QBER bound of
a quantum protocol. This approach can be further generalized to derive other information-theoretic parameters in future research. In conclusion, it is important to acknowledge that the work presented in this thesis is primarily constrained by its entirely theoretical nature. However, we anticipate that some of the proposed schemes, either in their original form or with suitable modifications, will be experimentally realized in the near future. Additionally, during the course of this study, we have observed that certain QKD protocols are frequently implemented due to their practical feasibility, despite the lack of rigorous security proofs against coherent and collective attacks. In the future, the security analysis framework developed in this work, including the game-theoretic approach, may be utilized to assess the security of such protocols. Furthermore, all the schemes proposed and analyzed in this thesis fall within the domain of discrete-variable protocols. An interesting direction for future research would be to explore the feasibility of developing continuous-variable counterparts for these schemes.

\newpage


\addcontentsline{toc}{chapter}{\fontsize{12}{10}\selectfont REFERENCES}
\normalem
\bibliographystyle{JIITREFStyle}
\renewcommand{\bibname}{\fontsize{16}{16} \selectfont REFERENCES}
\bibliography{References/references} 
\newpage


\addcontentsline{toc}{chapter}{\fontsize{12}{10}\selectfont LIST OF AUTHOR'S PUBLICATIONS}
\addtocontents{toc}{\protect\contentsline {chapter}{\fontsize{12}{10}\selectfont SYNOPSIS }{Synopsis-1}}

\chapter*{\fontsize{16}{16} \selectfont List of Author's Publications}
\vspace{-0.25in}
\begin{itemize}
\item \textbf{Published in International Journals}
\begin{enumerate}
\item Dutta A., and Pathak A., \emph{``A short review on quantum identity authentication protocols: How would Bob know that he is talking with Alice"}, Quantum Information Processing, vol. 21, pp.  369, 2022.

\item Dutta A., and Pathak A., \emph{``Controlled secure direct quantum communication inspired scheme for quantum identity authentication''}, Quantum Information Processing, vol. 22, pp. 13, 2023.

\item Dutta A., and Pathak A., \emph{``Use of Nash equilibrium in finding game theoretic robust security bound on quantum bit error rate''}, Physica Scripta, vol. 99, pp. 095106, 2024.

\item Dutta A., and  Pathak A., \emph{``Simultaneous quantum identity authentication scheme utilizing entanglement swapping with secret key preservation''}, Modern Physics Letters A, pp. 2450196, 2025.

\item Dutta A., and  Pathak A., \emph{``Collective attack free controlled quantum key agreement without quantum memory''}, Physica Scripta, vol. 100, pp. 035101, 2025.

\item Dutta A. and Pathak A.,  \emph{``New protocols for quantum key distribution with explicit upper and lower bound on secret key rate''}, The European Physical Journal D, vol. 79, no. 5,
p. 48, 2025.

\end{enumerate}

\end{itemize}

\thispagestyle{empty}
\mbox{}
\newpage

\end{document}